\documentclass[11pt,a4paper]{book}
\usepackage{valerio}
\addtocounter{secnumdepth}{1}
\addtocounter{tocdepth}{1}
\begin{document} 
\clearpage
\thispagestyle{empty}
\begin{center}
\begin{picture}(150,36)(0,0) 
\put(0,20){\includegraphics[width=3.6cm]{logo_units.epsi}}
\put(36,44){\makebox(114,9)[cc]%
{\Large \bf UNIVERSIT\`{A} DEGLI STUDI DI TRIESTE}}
\put(36,33){\makebox(114,9)[cc]%
{\large \bf Facolt\`{a} di Scienze matematiche, fisiche e naturali}}
\put(36,27){\makebox(114,6)[cc]%
{\large \bf Dipartimento di Fisica Teorica}}
\end{picture}\\[-1.5cm]
{\Large XVI CICLO DEL\\[1.5ex]
DOTTORATO DI RICERCA IN FISICA}\\
\vspace{\stretch{50}}
{\huge \bf Quantum Dynamical Entropies and\\[1.5ex]
Complexity in Dynamical Systems}\\
\vspace{\stretch{50}}
{\LARGE \bf Valerio Cappellini}{\huge \\[1.5ex]}
{\large Ph.D. Thesis}\\
\vspace{\stretch{110}}
\end{center}
\begin{minipage}[t]{4cm}
\large
\begin{flushright}
TUTORE:\\[.3cm]
COORDINATORE:
\end{flushright}
\end{minipage}
\hspace{4mm}
\begin{minipage}[t]{10cm}
\large
\begin{flushleft}
Dott. Fabio Benatti (Universit\`{a} di Trieste)\\[.3cm]
Prof. Gaetano Senatore (Universit\`{a} di Trieste)
\end{flushleft}
\end{minipage}
\newpage
\thispagestyle{empty}
\clearpage \ \clearpage
\clearpage
\thispagestyle{empty}

\begin{flushright}
\rule{0pt}{1ex}\\
\vspace{2cm}
\rule{0pt}{1ex}\\
\begin{minipage}[t]{6.2cm}
\small
\textit{All'inizio era il Caos e Dio fece l'Ordine.\\[1ex]
--- ``O che c'era prima?!\ldots''\\[1ex]
\begin{flushright}
(Roberto Benigni)
\end{flushright}} 
\end{minipage}
\end{flushright}
\newpage
\thispagestyle{empty}
\clearpage \ \clearpage
\lhead[\fancyplain{}{\bfseries\boldmath\thepage}]%
      {\fancyplain{}{\bfseries\boldmath\rightmark}}
\rhead[\fancyplain{}{\bfseries\boldmath\leftmark}]%
      {\fancyplain{}{\bfseries\boldmath\thepage}}
\chapter*{Acknowledgements}
I take this opportunity to express my gratitude for my supervisor,
Fabio Benatti.
He has inspired and motivated me tremendously and I am also very thankful
to him for his advice, encouragement and assistance during my Ph.D.
career.

\thispagestyle{empty}
\vspace{1mm}
\newpage
\thispagestyle{empty}
\clearpage \ \clearpage
\newpage

\pagenumbering{roman}
\setcounter{page}{1}
\addcontentsline{toc}{chapter}{Table of Contents}
\tableofcontents
\vspace{1mm}
\newpage
\vspace{1mm} \
\newpage
\listoffigures
\addcontentsline{toc}{chapter}{List of Figures}
\newpage
\vspace{1mm} \ 
\newpage 
\onehalfspacing
\chapter*{Introduction}
\lhead[\fancyplain{}{\bfseries\thepage}]%
      {\fancyplain{}{}}
\rhead[\fancyplain{}{\bfseries Introduction}]%
      {\fancyplain{}{\bfseries\thepage}}
\addcontentsline{toc}{chapter}{Introduction}
Classical chaos is associated with motion on a compact phase--space
with high sensitivity to initial conditions: trajectories diverge
exponentially fast and nevertheless remain confined to bounded 
regions~\cite{Dev89:1,Wig90:1,Kat99:1,Sch95:1,Gia89:1,Cas95:1,Zas85:1}.

In discrete time, such a behaviour is characterized by
a positive Lyapounov exponent $\log\lambda$, $\lambda>1$, and by
a consequent spreading of initial errors $\delta$ such that, after
$n$ time--steps,
$\delta\mapsto\delta_n\simeq\delta\,\lambda^n$.
Exponential amplification on a compact phase--space cannot grow
indefinitely, therefore the Lyapounov exponent can only be obtained as:
\begin{equation}
\log\lambda \coleq \lim_{t\to\infty}\frac{1}{n}\lim_{\delta\to0}\log\left(\frac
{\delta_n}{\delta}\right)\label{Lyap1}\ ,
\end{equation}
that is by first letting $\delta\to0$ and only
afterwards $n\to\infty$.

In quantum mechanics non--commutativity 
entails absence of continuous trajectories or, semi--classically,
an intrinsic coarse--graining of phase--space determined by Planck's constant
$\hbar$: this forbids $\delta$ (the minimal error possible) to go to
zero. Indeed, nature is fundamentally quantal and, according to the
correspondence principle, classical behaviour emerges in the limit
$\hbar\to0$.

Thus, if chaotic behaviour is identified with
$\log\lambda>0$, then it is quantally suppressed, unless,
performing the classical limit first, we let room
for $\delta\to0$~\cite{Cas95:1}.

Another way to appreciate the regularity emerging from quantization,
is to observe that quantization on compacts yields discrete energy
spectra which in term entail quasi--periodic
time--evolution~\cite{For92:1}. 

In discrete classical systems, one deals with discretized versions 
of continuous classical systems, or with cellular
automata~\cite{Cri93:1,Cri94:1,Bof02:1} and neural
networks~\cite{Her91:1} with finite
number of states. In this case, roughly speaking, the minimal distance
between two 
states or configurations is strictly larger than zero; therefore, the
reason why $\log\lambda$ is trivially zero 
is very much similar to the one encountered  
in the field of quantum chaos, its origin being now not in
non--commutativity but in the lack of a continuous structure. 
Alternative methods have thus to be
developed in order to deal with the granularity of
phase--space~\cite{Zer94:1,Zer99:1,Cri93:1,Cri94:1,Bof02:1}. 

A signature of chaotic properties of quantized/discretized dynamical
systems is  
the presence of a so called \enfasi{breaking--time} $\tau_B$, that is a
time (depending on the quantization parameter $\hbar$) 
fixing 
the time--scale where quantum and classical mechanics are expected to
almost coincide. Usually $\tau_B$ scales as
$\hbar^{-\alpha}$ for some $\alpha>0$~\cite{Zas85:1} for regular
classical limits, that is for systems that are (classically) regular;
conversely, for chaotic systems, the 
semi-classical regime typically scales as 
$-\!\log\hbar$~\cite{Gia89:1,Cas95:1,Zas85:1}. Both time scales diverge  when
$\hbar\to0$, but the shortness of the latter means that classical
mechanics has to be replaced by quantum mechanics much sooner for
quantum systems with chaotic classical behaviour. The \enfasi{logarithmic
breaking time} $-\!\log{\hbar}$ has been considered  by some as a
violation of the correspondence principle~\cite{For91:1,For92:2}, by others,
see~\cite{Cas95:1} and Chirikov in~\cite{Gia89:1}, as the evidence that time
and classical limits do not commute.

This phenomenon has also been studied for quantized/discretized
dynamical systems with  
finite number of states, possessing a well--defined
classical/continuous limit. For
instance, consider a discretized classical dynamical system: the
breaking--time 
can be heuristically estimated as the time when
the minimal error permitted, ($\delta$: in the present case coincide
with the $\hbar$--like parameter, that is the lattice spacing of the
grid on which we discretize the system) , becomes of the order of the 
phase--space bound $\Delta$. Therefore, when, in the continuum, a 
Lyapounov exponent $\log\lambda>0$ is present, the breaking--time scales as  
$\displaystyle  
\tau_B=\frac{1}{\log\lambda}\log\frac{\Delta}{\delta}$.

In order to inquire how long the classical and quantum behaviour
mimic 
each other, we need a witness of such ``classicality'', that be
related 
(as we have seen) to the presence of positive Lyapounov exponent.
By the theorems of Ruelle and Pesin~\cite{Man87:1}, the positive Lyapounov
exponents of smooth, classical dynamical systems  are related to the
dynamical entropy of Kolmogorov~\cite{Kat99:1} (\co{KS}--entropy or
\enfasi{metric entropy}) which measures the 
information per time step provided by the dynamics.
The phase--space is partitioned
into cells by means of which any trajectory is encoded into a sequence
of symbols. 
As times goes on, the richness in different symbolic trajectories  
reflects the irregularity of the motion and is
associated with strictly positive dynamical entropy~\cite{Ale81:1}.

So, since the metric entropy is related to the positive Lyapounov
exponent, and the  
latter are indicators of chaos in the semi--classical regime (for
classically 
chaotic systems), it is evident how the \co{KS}--entropy could be
profitably used to our purpose. However, the metric entropy can be
defined only on measurable classical systems, and we need then to
replace it with some tool more appropriate to finite, discrete
context. 
In view of the similarities between quantization and discretization,
our proposal is to use quantum extension of the metric entropy.

There are several
candidates for non--commutative extensions of the
latter~\cite{Con87:1,Ali94:1,Voi92:1,Acc97:1,Slo94:1}: in the
following we shall use two of 
them~\cite{Con87:1,Ali94:1} and study their classical/continuous limits. 

The most powerful tools in studying the semi--classical regime consist
essentially in focusing,
via coherent state (C.S.) techniques,  on the phase space localization  of
specific time evolving quantum observables. For this reason we will make
use of an Anti--Wick procedure of quantization, based on
C.S. states, which can be applied also to algebraically
discretize of classical continuous systems. Developing
discretization--methods, mimicking quantization procedures, allow us
to compute quantum dynamical entropies in both quantum and
classical--discrete systems.

The entropies we will use are the \co{CNT}--entropy (Connes,
Narnhofer and Thirring) and the \co{ALF}--entropy (Alicky, Lindblad and
Fannes) generically which differ on quantum systems but coincide with
the Kolmogorov metric entropy on classical ones. All these dynamical
entropies are long--time entropy rates and therefore all vanish in
systems with finite number of states. However, this does not mean that
on finite--time scales, there might not be 
an 
entropy production, but only that sometimes it has to stop.

It is exactly the analytical/numerical study of this phenomenon of
finite--time chaos that we will be concerned within this
work~\cite{Ben03:1,Ben04:1}.

As particular examples of quantum dynamical systems with chaotic
classical limit, we shall
consider finite 
dimensional quantizations of hyperbolic automorphisms of the 2-torus,
which are prototypes of chaotic behaviour; indeed, their trajectories
separate exponentially fast with a Lyapounov exponent
$\log\lambda>0$~\cite{Arn68:1,Wal82:1}. Standard quantization, \`a la Berry,
of hyperbolic automorphisms~\cite{Ber79:1,Deg93:1} yields Hilbert spaces of a
finite dimension~$N$. This dimension plays the role of semi--classical
parameter and sets the minimal size $1/N$ of quantum phase space
cells. 

On this family of quantum dynamical systems
we will compute the two entropies mentioned above, showing
that, from both of them, one recovers the Kolmogorov entropy
by  computing the average quantum entropy produced over a logarithmic
time scale and then taking the classical limit~\cite{Ben03:1}. This
confirms the 
numerical results in~\cite{Ali96:2}, where the dynamical
entropy~\cite{Ali94:1} is applied to the study of  the quantum kicked
top. In this approach, the presence of logarithmic time scales
indicates the  typical scaling for a joint time--classical limit
suited to preserve positive  entropy production in quantized
classically chaotic quantum systems.

For what concerns discrete systems, we will enlarge the set of classical
systems from the hyperbolic automorphisms of the 2-torus to
the larger class of Sawtooth Maps. For such systems, in general
singular contrary to smooth hyperbolic ones, we will provide a
rigorous discretization scheme with corresponding continuous limit in
which we will study the behavior of the \co{ALF}--entropy.

The \co{ALF}--entropy is based on the algebraic properties of dynamical
systems, that is on the fact that, independently on whether they are
commutative or not, they are describable by suitable algebras of
observables, their time evolution by linear maps on these algebras and
their states by expectations over them.

Profiting from the powerful algebraic methods to inquire finite--time
chaos, we will numerically compute the \co{ALF}--entropy in discrete
systems, and the performed analysis~\cite{Ben04:1} 
clearly show the consistency between the achieved results and our
expectations of finding a logarithmic breaking--time.\newpage
\pagenumbering{arabic}
\lhead[\fancyplain{}{\bfseries\boldmath\thepage}]%
      {\fancyplain{}{\bfseries\boldmath\rightmark}}
\rhead[\fancyplain{}{\bfseries\boldmath\leftmark}]%
      {\fancyplain{}{\bfseries\boldmath\thepage}}
\setcounter{page}{1}
\pagestyle{fancyplain}
\chapter{Quantization and Discretization on the Torus}\vspace{9mm}
\section{Algebraic settings}\label{Aset}\vspace{6mm}
\subsection{Dynamical Systems}\label{CDS}\vspace{3mm}
Usually, continuous classical motion is described by means of a
measure space ${\cal X}$, the phase--space, endowed with the Borel
$\sigma$--algebra and a normalized 
measure $\mu$, $\mu({\cal X})=1$.
The ``volumes'' 
\begin{equation*}
\mu(E)=\int_E{\rm d}\mu(\bs{x})
\end{equation*}
of measurable subsets 
$E\subseteq{\cal X}$ represent the probabilities that a
phase--point $\bs{x}\in{\cal X}$ belong to them.
By specifying the statistical properties of the system, the measure $\mu$ 
defines a ``state'' of it.

In such a scheme, a reversible discrete time dynamics amounts to an
invertible measurable 
map $T:{\cal X}\mapsto{\cal X}$ such that $\mu\circ T=\mu$ and to its
iterates $\{T^k \mid k\in\Ir\}$.
Phase--trajectories passing through $\bs{x}\in{\cal X}$ at time $0$
are then sequences ${\pg{T^k\,\bs{x}}}_{k\in\IZ}$~\cite{Kat99:1}.

Classical dynamical systems are thus conveniently described by
triplets $({\cal X},\mu,T)$.

In the present work we shall focus upon the following:
\begin{itemize}
\item
 $\c X$ -- a compact metric space:\\ the $2$--dimensional
 torus $\IT={\IR}^2/{\IZ}^2 \-=\left\{\bs{x}=(x_1,x_2)\in\IR^2\
 \pmod{1} \right\}$. We will use 
the symbol $\c X$ to refer to generic compact measure spaces,
otherwise the specific symbol $\IT$;
\item
 $\mu$ -- the Lebesgue measure
 $\mu(\ud\bs{x})=\ud x_1\; \ud x_2$ on $\IT$;
\item
 $T$ -- invertible measurable transformations from $\c X$ to itself
 such  that $T^{-1}$ are also measurable.
\end{itemize}
For this kind of systems we also provide an algebraic description, 
consisting in associating to them algebraic triplets $({\cal
M},\omega,\Theta)$, where: 
\begin{itemize}
\item
 ${\cal M}$ -- is a C* or a Von~Neumann -algebra. Non--commutative
 algebras characterize quantum dynamical systems and the
 elements of ${\cal M}$ are nothing but the observables, usually
 acting as bounded operators on a
 suitably defined Hilbert space $\cal H$.

 Commutativity will be characteristic of algebras describing classical
 systems, 
 as the ones that we are going to introduce in the next Section~\ref{TuAt}.
\item
 $\omega$ -- denotes a reference state on ${\cal M}$, that is a
 positive linear and normalized functional on it. 
\item
 $\{\Theta^k \mid k\in\Ir\}$ -- is the discrete group of *-automorphisms%
\footnote{A *-automorphism $\Theta$ of a C* algebra ${\cal M}$ is
defined to be a *-isomorphism of ${\cal M}$ into itself, i.e., $\Theta$
is a *-morphism of ${\cal M}$ with range equal to ${\cal M}$ and kernel
equal to zero. In order to be defined as a *-automorphism, a map $\Theta$
has to preserve the algebraic structure of ${\cal M}$, namely for all
$m_1,m_2\in{\cal M}$ it must hold:
$\Theta\pt{m_1+m_2}=\Theta\pt{m_1}+\Theta\pt{m_2}$, $\Theta\pt{m_1
m_2}=\Theta\pt{m_1} \Theta\pt{m_2}$ and $\Theta\pt{m_1^*}=\Theta^*\pt{m_1}$
}
 of ${\cal M}$ implementing the dynamics that leave the state
 $\omega$ invariant, i.e. $\omega\circ\Theta=\omega$. 
\end{itemize}
\subsection{Two useful algebras on the torus}\label{TuAt}\vspace{3mm}
We introduce now two functional spaces, that will be profitably used
for later purpose.

The first one is the Abelian C*--algebra $\Cspace{0}{\cal X}$ 
of complex valued continuous functions with
respect to the topology given by the \enfasi{uniform norm} 
\begin{equation}
\displaystyle \norm{f}{0}=\sup_{\bs{x}\in{\cal
X}}\Big|f\pt{\bs{x}}\Big|\ \cdot\label{uninorm}
\end{equation}

The second functional space we are going to introduce is the
Abelian (Von~Neumann) algebra $\Lspace{\infty}{\cal X}$ of essentially bounded
functions on $\cal X$. The meaning of ``essentially'' is that these
function have to be bounded with respect to the so called ``essential
norm'' $\norm{\cdot}{\infty}$, namely the essential
supremum defined by~\cite{Rud87:1}: 
\begin{equation}
\norm{f}{\infty}\coleq\esssup_{\bs{x}\in{\cal X}} \abs{f}
= \inf
\bigg\{\ \ a\in\IR
\ \ \bigg| \ \ \mu
\Big(
\big\{ \ \bs{x}
\ :\  
| f\pt{\bs{x}}
| > a 
\ \big\}
\Big) = 0
\ \ \bigg\}\ \cdot\label{AdoUMG_1}
\end{equation}
This norm is slightly different from the one defined
in~\eqref{uninorm}, when taken on functions belonging to
$\Lspace{\infty}{\cal X}$;
for instance, two functions that differ only on a set of null measure
(for instance on a single point),
will have the same norm given by~\eqref{AdoUMG_1}. Also, if
$f\in\Cspace{0}{\cal X}$, then $\norm{f}{\infty}=\norm{f}{0}$. 

From now on we adopt the symbol 
${\cal A}_{{\cal X}}$ to denote both algebras distinguishing them when
necessary. 

\noindent The Lebesgue measure $\mu$ defines a state $\omega_\mu$ on ${\cal
A}_{\cal X}$ via integration
\begin{equation}
\om:{\cal A}_{\cal X}\ni f\longmapsto\omega_\mu(f)\coleq\int_{{\cal X}} \ud\mu(\bs{x})\ f(x)\in \IR^{+}\ ;
\label{omegamu}
\end{equation}
this will be our reference state for the algebras of ${\cal A}_{\cal
X}$--types.
\section{Quantization procedures}\label{QUAPRO}\vspace{6mm}
Once the
algebraic triplet $\tripAAoT$ has been fixed, the approach of
Section~\ref{CDS} provide a general formalism  
that allows us to deal with generic dynamical systems.\\[-3ex]
\begin{quote}
\begin{NNS}{}\ \\[-7.5ex]
\begin{Ventry}{ii)}
\item[i)] Of course we could provide different triplets describing
systems that, in a suitable classical limit (argument of next
Chapter~\ref{Chap2}), ``correspond'' to the same classical dynamical
system $\tripCT$. In particular, different quantum systems (mimicking
each other in the semi classical limit) can be constructed by using
different algebras ${\cal 
M}$, with the latter chosen among commutative or not, finite or
infinite dimensional, and so on.
\item[ii)]In the future we will restrict ourselves to consider finite
dimensional 
algebras ${\cal 
M}_N$, but even with this restriction, the set of possible
choice is quite large. Intuitively we can think
that a classical dynamical system is supposed to be described by using
an abelian algebra (and this is the case), nevertheless it is not
enough to say that a non--abelian algebra provide a ``good''
description of quantum systems. 
\end{Ventry}
\end{NNS}
\end{quote}
\noindent
Assigned a classical dynamical system $\pt{{\cal A}_{\cal
X},\omega_\mu,\Theta}$, the aim of a 
quantization--dequantization procedure (specifically an
$N$--dimensional quantization) is twofold:
\begin{itemize}
\item to find  
a couple of *-morphism, ${\cal J}_{N,\infty}$ mapping ${\cal A}_{\cal
X}$ into a non  
abelian finite dimensional algebra ${\cal M}_N$ and ${\cal
J}_{\infty,N}$ mapping backward ${\cal M}_N$ into ${\cal A}_{\cal X}$;
\item to provide an automorphism $\Theta_N$ acting on ${\cal M}_N$ 
representing the quantized classical evolution $\Theta$
such that the two dynamics, the classical one on ${\cal A}_{\cal
X}$ and the quantum one on ${\cal M}_N$, commute with the action of
the two *-morphisms connecting the two algebras, that is
\begin{equation}
{\cal J}_{N,\infty}^{\phantom{j}}\circ\Theta_{\phantom{N}}^j\simeq
\Theta_N^j\circ {\cal J}_{\infty,N}^{\phantom{j}}
\label{commJT}
\end{equation}
The latter requirement can be seen as a modification of the so called Egorov's property (see~\cite{Mar99:1}). 
\end{itemize}
The difficulties in finding a convenient quantization procedure are
due to two (equivalent) facts:
\begin{itemize}
\item as far as we know from quantum mechanics, once we assign in
the algebra ${\cal M}_N$ the operators corresponding to classical
observables, some relations have to be respected. These relation
connected with the physics underlying our 
system. For instance, in our work, we will impose Canonical
Commutation Relations (CCR for short); 
\item once a quantization parameter
(something playing the role of $\hbar$, on which the two *-morphism
${\cal J}_{N,\infty}$ and ${\cal 
J}_{\infty,N}$ have to be dependent) is let to go to zero, 
the correspondence between classical and
quantum observables has to be fixed in a way that allow us to speak of
a ``classical limit''.
\end{itemize}
The latter observation will be discussed in Chapter~\ref{Chap2}, in
which we will 
provide our quantization procedure and a suitable classical limit,
whereas the CCR problem will be the core of the next Section.
\subsection{Finite dimensional Quantization on the
torus}\label{FDQoT}\vspace{3mm}  
We now consider the (non commutative) finite dimensional algebra
${\cal M}_N$ of $N\times N$ matrices acting on
a $N$--dimensional Hilbert space ${\cal H}_N={\IC}^N$. Let us give a
Definition of a state $\tn$ on matrix algebras, that will be used in
the following. \newpage
\begin{quote}
\begin{DDD}{}\ \\[-5.5ex]
\begin{Ventry}{}\label{tauenne}
	\item[] We will denote by $\tn$ the state given by the following
positive, linear and normalized functional over ${\cal M}_N$:
\begin{equation*}
\tn:{\cal M}_N\ni
M\longmapsto\tau_N(M)\coleq\frac{1}{N}\Tr\pt{M}\in \IR^{+}\ \cdot 
\notag
\end{equation*}
\end{Ventry}
\end{DDD}
\end{quote}
\noindent
Due to the finiteness dimension, it is impossible to find in ${\cal
M}_N$ two operators $\hat{Q}$, $\hat{P}$ playing the role of position,
respectively momentum, satisfying CCR~\cite{Thi83:1}.
Indeed, taking the trace
of the basic equation
\begin{equation}
\pq{\hat{Q},\hat{P}} = i\,\hbar\,\Id\ ,
\label{QPcomm}
\end{equation}
the ciclicity property of the trace gives us $0=i\,\hbar\,N$.

Nevertheless, as in the Schr\"odinger representation, $\hat{P}$ is the
generator of the 
(compact) Lie group of space translations, while $\hat{Q}$
acts as the generator of the group of momentum translations.
The form and the action of the shift operator $\hat{U}$, $\hat{V}$,
in position $(q)$, respectively momentum $(p)$, coordinates are given by:
\begin{alignat}{2}
\hat{U}\pt{\ud q}\ket{q}
&\coleq e^{-\frac{i\,\hat{P}\,\ud q}{\hbar}}\ket{q}=\ket{q+\ud q}
\quad&\text{,}\quad
\hat{U}\pt{\ud q}\ket{p}
&\coleq e^{-\frac{i\,\hat{P}\,\ud
q}{\hbar}}\ket{p}=e^{-\frac{i\,p\,\ud q}{\hbar}}\ket{p}\ ,
\label{UVcomm2}\\ 
\hat{V}\pt{\ud p}\ket{p}&\coleq e^{\phantom{-}\frac{i\,\hat{Q}\,\ud p}{\hbar}}\ket{p} 
=\ket{p+\ud p}
\quad&\text{,}\quad
\hat{V}\pt{\ud p}\ket{q}&\coleq e^{\phantom{-}\frac{i\,\hat{Q}\,\ud p}{\hbar}}\ket{q}
=e^{\phantom{-}\frac{i\,q\,\ud p}{\hbar}}\ket{q}\ .
\label{UVcomm3}
\end{alignat}
Using~\eqref{QPcomm} and the Baker--Hausdorff's Lemma, we get
from~(\ref{UVcomm2}--\ref{UVcomm3}):
\begin{equation}
\hat{U}\pt{\ud q}\hat{V}\pt{\ud p}=\hat{V}\pt{\ud p}\hat{U}\pt{\ud q}
e^{-\frac{i\,\ud q\,\ud p}{\hbar}}\ .
\label{CCRUV}
\end{equation}
Of course~\eqref{CCRUV} is unchanged if we define $\hat{U}$ and
$\hat{V}$ up to phases.

The latter relation can be a good starting point for a
quantization procedure~\cite{DeB98:1,DeB01:1}. Given a
$N$--dimensional Hilbert space ${\cal  
H}_N={\IC}^N$, its basis can be labeled by
${\pg{\ket{q_\ell}}}_{\ell=0\,\ldots,N-1}$. If we want to interpreter
this basis 
as a ``$\IT$: $q$--coordinates'' basis, we have to respect the toral
topology and to add the folding
condition, namely $\ket{q_{\ell + N}}=\ket{q_\ell}$ for all $\ell$
belonging to 
$\ZNZn$, the residual class$\pmod{N}$. In a similar way we could
choose a ``$\IT$: $p$--coordinates'' representation, by choosing a basis
${\pg{\ket{p_m}}}_{m=0\,\ldots,N-1}$ endowed with the same folding
condition $\ket{p_{m + N}}=\ket{p_m},\ \forall m\in\ZNZn$.

\noindent The coordinates
$\pt{q_\ell,p_m}=\pt{\frac{\ell}{N},\frac{m}{N}}$ will label the
points of a square grid of lattice spacing $\frac{1}{N}$ 
lying on the torus $\IT$.

On this grid, we can construct two unitary shift operators $U_N$ and
$V_N$ 
mimicking equations~(\ref{UVcomm2}--\ref{UVcomm3}); we will
explicitly indicate the dependence of the representation on two
arbitrary phases
$(\alpha,\beta)$:
\begin{alignat}{2}
U_N\pt{\ud q}\ket{q_j}
&\coleq e^{i\,\alpha}\ket{q_j+\ud q}
\quad&\text{,}\quad
U_N\pt{\ud q}\ket{p_j}
&\coleq e^{i\,\alpha}e^{-i\frac{j\,\ud q}{N\hbar}}\ket{p_j}\ ,
\label{UVcomm4}\\ 
V_N\pt{\ud p}\ket{p_j}
&\coleq e^{i\,\beta}\ket{p_j+\ud p}
\quad&\text{,}\quad
V_N\pt{\ud p}\ket{q_j}&\coleq 
e^{i\,\beta}e^{\phantom{-}i\frac{j\,\ud p}{N\hbar}}\ket{q_j}\ .
\label{UVcomm5}
\end{alignat}
If we want that these operator act ``infinitesimally'', we have to
tune them according to the minimal distance (in $q$ and $p$) coordinates
permitted by the granularity of the phase--space, that is we have to fix 
$U_N\coleq \hat{U}\pt{\ud q=q_1=\frac{1}{N}}$ and 
$V_N\coleq \hat{V}\pt{\ud p=p_1=\frac{1}{N}}$.
Thus the action of $U_N$ and $V_N$ on the $q$--basis can be
rewritten as
\begin{equation}
U_N\ket{q_j}
\coleq e^{i\,\alpha}\ket{q_{j+1}}
\quad\text{,}\quad
V_N\ket{q_j}\coleq 
e^{i\,\beta}e^{\phantom{-}\frac{i}{N}\frac{1}{N\hbar}j}\ket{q_j}\ .
\label{UVcomm6}
\end{equation}

Now it remains to impose the folding condition on the operators $U_N$ and
$V_N$, that is
\begin{equation}
  U_N^N = e^{2i\pi u}\, \Id_N,\quad
  V_N^N = e^{2i\pi v}\, \Id_N\  
\label{FoldUV}
\end{equation}
where $u=\alpha\,\frac{N}{2\pi}$ and $v=\beta\,\frac{N}{2\pi}$ can be
chosen to belong to $[0,1)$ and are parameters labeling the representations.

If we want $V_N^N = e^{2i\pi v}\, \Id_N$ to hold 
we have o fix $\frac{1}{N\hbar}=\frac{2\pi}{Nh}=2\pi s\in\IZ$;
without loss of generality~\cite{Deg93:1}, we choose $s=-1$.\\ 
Then, from identity $h=-\frac{1}{N}$, it turns out that our
quantization parameter is given by $N$, the dimension of Hilbert
space, and we expect to recover the classical behaviour (namely
commutativity) when $N\rightarrow\infty$. This is evident
from~\eqref{CCRUV},
that now reads
\begin{equation}
 U_NV_N=e^{\frac{2\pi i}{N}}\, V_NU_N\ \cdot
\label{algUV}
\end{equation} 
Upon changing the labels of the o.n.b.\footnote{orthonormal basis} of
the ${\cal H}_N$ by letting $\ket{q_j}\longmapsto\ket{j}$,
equation~\eqref{UVcomm6} can more conveniently be written as 
\begin{equation}
U_N\ket{j}
\coleq e^{\frac{2\pi i}{N}u}\ket{j+1}
\quad\text{and}\quad
V_N\ket{j}\coleq 
e^{\frac{2\pi i}{N}\pt{v-j}}\ket{j}\ ,
\label{UVcomm7}
\end{equation}
By mimicking the usual algebraic approach to CCR in the continuous
case, we introduce Weyl operators labeled by $\bs{n} =
(n_1,n_2)\in\IZ^2$ 
\begin{align}
  W_N(\bs{n})& \coleq e^{\frac{i\pi}{N}n_1n_2}\, V_N^{n_2}U_N^{n_1}\
,\label{Weyl1}\\  
W_N^*(\bs{n})  & =  W_N^{\phantom{*}}(-\bs{n}) \label{Weyl1bis}\ \cdot
\end{align}
Their explicit action on the
o.n.b. ${\pg{\ket{j}}}_{j=0,1,\cdots,N-1}$ is given by 
\begin{equation}
 W_N(\bs{n})\, \ket{j} = \exp\pt{\frac{i\pi}{N}(-n_1n_2+ 2n_1 u+ 2n_2
 v)}\, \exp\pt{-\frac{2i\pi}{N}jn_2}\, \ket{j+n_1}\ ,
\label{Weyl4}
\end{equation}
whence
\begin{align}
 W_N(N\bs{n})
 &= e^{i\pi(N n_1n_2+ 2n_1 u+ 2n_2 v)}\ ,
\label{Weyl2} \\
 W_N(\bs{n})W_N(\bs{m})
 &= e^{\frac{i\pi}{N}\sigma(\bs{n},\bs{m})}\, W_N(\bs{n}+\bs{m}),
\label{Weyl3}
\end{align}
where $\sigma(\bs{n},\bs{m})=n_1m_2-n_2m_1$ is the so--called simplectic
form. From equation~(\ref{Weyl3}) one derives
\begin{equation*}
 \pq{W_N(\bs{n}), W_N(\bs{m})} =
 2i\sin\pt{\frac{\pi}{N}\:\sigma(\bs{n},\bs{m})}\, W_N(\bs{n}+\bs{m}), 
\end{equation*}
which shows once more how recovering Abelianness is related to
$N\longrightarrow\infty$.\\[-2ex] 
\begin{quote}
\begin{DDD}{}\ \label{weylgroup}\\
The \enfasi{Weyl Algebra} is the C*-algebra over
$\IC$ generated by the (discrete) group of Weyl operators
\begin{equation*}
 {\pg{W_N(\bs{n})}}_{\bs{n}\in{\IZ}^2}\ \cdot
\end{equation*}
\end{DDD}
\ \\[-5.5ex]
\begin{NNS}[The Weyl group]{}\ \\[-5.5ex]
\begin{Ventry}{ii)}\label{unitarity}
	\item[i)]
Let us comment now on the role played by the two parameter $(u,v)$
introduced in~\eqref{FoldUV}: until now they are
arbitrary parameters and we will fix them by inserting the
dynamics into our scheme of quantization. Actually, although the Weyl
group introduced in Definition~\ref{weylgroup} is just supposed to
fulfill relations~\eqref{Weyl1bis} and~\eqref{Weyl3}, 
choosing a couple of parameters $(u,v)$ we choose a definite
representation $\pi_{(u,v)}$ of the (abstract) Weyl
group ${\pg{W_N(\bs{n})}}_{\bs{n}\in{\IZ}^2}$. In order to
classify all possible representations of the Weyl group, we cite
now~\cite{Deg93:1} a useful\\[-3.5ex]
\begin{TT}{:} \label{classification}
\begin{Ventry}{a)}
	\item[a)] $\pi_{(u,v)}$ is an irreducible *-representation of
	${\pg{W_N(\bs{n})}}_{\bs{n}\in{\IZ}^2}$ 
	\item[b)] $\pi_{(u,v)}$ is unitarily equivalent%
\footnote{It means that exists an unitary operator $U$ such that 
$U \pi_{(u,v)} U^{\dagger} = \pi_{(\widetilde{u},\widetilde{v})}$}
	to $\pi_{(\widetilde{u},\widetilde{v})}$ iff
	$(u,v)=(\widetilde{u},\widetilde{v})$ 
\end{Ventry}
\end{TT}
	\item[ii)] Once the generators of the group%
\footnote{Here $\bs{\hat{e}}_1\coleq\pt{\begin{smallmatrix} 1\\ 0
	\end{smallmatrix}}$ and
	$\bs{\hat{e}}_2\coleq\pt{\begin{smallmatrix} 0\\ 1 
	\end{smallmatrix}}$, the two basis vector of $\IR^2$.} 
	$w_i\coleq W_N(\bs{\hat{e}}_i)$ are assigned, a representation
	$\pi$ is chosen; the whole Weyl
	group ${\pg{W_N(\bs{n})}}_{\bs{n}\in{\IZ}^2}$ can be
	constructed just by using relations~\eqref{Weyl1bis} 
	and~\eqref{Weyl3}.
\end{Ventry}
\end{NNS}
\end{quote}

Finally, by manipulating the matrix element of $W_N(\bs{n})$ given
in~\eqref{Weyl4}, one easily derives the following\\[-2ex]
\begin{quote}
\begin{PRS}{}\ \label{proper121}\\
Let $\tau_N$ of Definition~\ref{tauenne} be our quantum reference
state; then it holds  
\begin{align}
 &\tau_N(W_N(\bs{n})) = e^{\frac{i\pi}{N}(-n_1n_2+ 2n_1 u+ 2n_2 v)}\,
 \delta^{(N)}_{\bs{n},\bs{0}}\ ,
\label{Weyl5} \\[6pt]
 &\frac{1}{N} \sum_{p_1,p_2=0}^{N-1} W_N(-\bs{p})\, W_N(\bs{n})\,
 W_N(\bs{p}) = \Tr \pt{W_N(\bs{n})}\,\Id_N\ ,
\label{Weyl8} \\
 &{\cal M}_N \ni X = \sum_{p_1,p_2=0}^{N-1}
 \tau_N \Bigl(X\,W_N(-\bs{p})\Bigr)\, W_N(\bs{p})\ ,
\label{Weyl9}
\end{align}
where in~(\ref{Weyl5}) we have introduced the periodic Kronecker delta, that is
\mbox{$\delta^{(N)}_{\bs{n},\bs{0}}=1$} if and only if
$\bs{n}=\bs{0}\,\pmod{N}$.

Notice that, according to~(\ref{Weyl9}), the Weyl algebra coincides
with the $N\times N$ matrix algebra ${\cal M}_N$.
\end{PRS}
\end{quote}
\noindent	
\subsection{Weyl Quantization on the torus}\label{WQ}\vspace{3mm}
Weyl operators have a nice interpretation in terms of the group of
translations generated by $\hat{Q}$ and
$\hat{P}$. The two operators $U_N$ and $V_N$ are given
in~(\ref{UVcomm4}--\ref{UVcomm5}) by 
mimicking the action of 
$\hat{U}(\ud q)$ and $\hat{V}(\ud p)$ in~(\ref{UVcomm2}--\ref{UVcomm3})
(up to two phases $u$
and $v$), with $\ud q = \ud p = - h =
\frac{1}{N}$. Explicitly, they are formally related to $\hat{Q}$
and $\hat{P}$ by
\begin{equation}
U_N
=e^{\,2 \pi i\,\hat{P}}       
\quad\text{and}\quad
V_N
=e^{-2 \pi i\,\hat{Q}}
\label{UVcomm8}
\end{equation}
Using Baker--Hausdorff's Lemma, together with~\eqref{Weyl1}, we obtain
\begin{equation}
W_N(\bs{n})=
e^{\,2 \pi i\pt{n_1\hat{P}-n_2\hat{Q}}}\ .
\label{WeylPQ}
\end{equation}
If now we restrict to a subalgebra ${\cal A}_{\cal X}$ consisting of
functions sufficiently 
smooth and regular to be Fourier decomposed, denoting with $\bs{x}
=\pt{x_1 , x_2}$ the canonical coordinates $\pt{q,p}$, then the
exponential functions
${\pg{e^{\,2 \pi i\,\sigma(\bs{n},\bs{x})}}}_{\bs{n} \in
{\IZ}^2}$ generate ${\cal A}_{\cal X}$, in the sense that
\begin{align}
f(\bs{x}) & = \sum_{\bs{n} \in {\IZ}^2} \hat{f}_{\bs{n}} \;
e^{\,2 \pi i\,\sigma(\bs{n},\bs{x})}
\label{RoeiW_1}\\
\text{where} \quad \hat{f}_{\bs{n}} & = \int\!\!\!\!\!\int_{\IT} \ud \mu
(\bs{x}) \; f (\bs{x}) \; 
e^{-2 \pi i\,\sigma(\bs{n},\bs{x})}
\label{RoeiW_2}
\end{align}
The Weyl Quantization procedure associates functions $f$ to operators
$W_{N,\infty}\pt{f}\in {\cal M}_N$ via the following
*-morphism~\cite{DeB98:1,DeB01:1} 
\begin{equation}
W_{N,\infty} : {\cal A}_{\IT} \ni f \longmapsto 
W_{N,\infty}\pt{f} = \sum_{\bs{n} \in {\IZ}^2} \hat{f}_{\bs{n}}
\;W_N(\bs{n}) \label{WeylQ}\in{\cal M}_N\ .
\end{equation}
We will postpone the construction of the de--quantizing
*-morphism $W_{\infty,N}$, because it involves coherent states that will be
introduced in Section~\eqref{CST}, moreover this construction is
completely analogue to the Anti--Wick way of de--quantizing, presented in
Section~\eqref{AWQ}. In Section~\ref{WDGriglIA} we will construct a
concrete example of a Weyl ``quantization'' procedure, and in
Section~\ref{DDW} we will invert such a scheme.
\section{Discretization of the torus over a $N\times N$
square grid}\label{DPR}\vspace{6mm} 
In the following we proceed to a discretization of classical
dynamical systems on the torus $\IT$ that, according
to Section~\ref{Aset}, will be identified with $\pt{{\cal
A}_{\IT},\omega_\mu,\Theta}$. As in the 
introduction to the quantization methods, we postpone the role played
by the dynamics to Chapter~\ref{Chap2} and we start with
phase--space discretization. 

Roughly speaking, given an integer $N$, we shall force the continuous
classical systems $\pt{{\cal A}_{\IT},\omega_\mu,\Theta}$
to live on a lattice $L_N\subset \IT$, of lattice--spacing $\frac{1}{N}$
\begin{equation}
L_N \coleq \pg{\frac{\bs{p}}{N} \ \Big|\  \bs{p}\in {\pt{\IZ / N \IZ}}^2}\ ,
\label{llnn}
\end{equation}
where $\pt{\IZ / N \IZ}$ denotes the residual
class$\pmod{N}$.
In order to set an algebraic structure for the discretization scheme,
we give now some  
\\[-2ex]
\begin{quote}
\begin{DDS}{}\ \\[-5.5ex]
\begin{Ventry}{iii.}\label{diagonali}
	\item[i.] ${\cal H}_N^D$ will denote an $N^2$--dimensional
	Hilbert space; 
	\item[ii.] ${\cal D}_N$ will denote
	the abelian algebra $D_{N^2}\pt{\IC}$ of $N^2 \times 
	N^2$ matrices (${\cal D}$ standing for diagonal with respect
	to a chosen  
	o.n.b. ${\pg{\ket{\bs{\ell}}}}_{\bs{\ell}\in 
	{\pt{\IZ / N \IZ}}^2}\in{\cal H}_N^D$);
	\item[iii.] To avoid difficulties due to the fact that the
	``quantum'' algebra ${\cal
	M}_N$ and the ``discretized'' algebra ${\cal D}_N$ are indexed
	by the same ``$N$'' but 
	their dimension is different ($N\times N$, $N^2\times N^2$
	respectively), when it will be important to refer to the
	dimensionality of Hilbert spaces (${\cal H}_N$, ${\cal H}_N^D$
	respectively) we will use the symbol: 
	$\nh\coleq\dim\pt{\cal H}$.
\end{Ventry}
\end{DDS}
\end{quote}
We can compare discretization of classical continuous
systems with quantization; 
to this aim, we define in the next Section
a discretization procedure resembling the Weyl quantization of
Sections~\ref{FDQoT} and~\ref{WQ};  
in practice, we will construct a *-morphism ${\cal J}_{N , \infty}$
from 
${\cal A}_{\IT}$ into the \underline{abelian} algebra ${\cal D}_N$.
The basis vectors
will be labeled by the points of $L_N$, defined in~\eqref{llnn}

The main point is that, although ${\cal J}_{N , \infty}$ maps
${\cal A}_{\IT}$ into a finite dimensional algebra ${\cal D}_N$
(and this will be very useful for our purpose), ${\cal D}_N$ is
abelian, and so endowed with very nice properties. In this scheme
discretization can be considered a very useful ``toy model'' for
testing the similarities with quantization and quantum systems as
source of granular description, leaving inside non--commutativity.
\subsection{Weyl Discretization: from $\Cspace{0}{\IT}$ to
${\cal D}_N$}\label{WDGriglIA}\vspace{3mm}
In order to define the discretization morphism ${\cal J}_{N ,
\infty}$, we use Fourier analysis and restrict ourselves to the
*-subalgebra ${\cal W}_{\text{exp}}\in{\cal A}_{\IT}$ generated
by the exponential functions
\begin{equation}
W(\bs{n})(\bs{x})=\exp(2\pi i\:\bs{n}\cdot\bs{x})\ , \label{equ_22}
\end{equation}
where 
$\bs{n}\cdot\bs{x}=n_1\,x_1+n_2\,x_2$.
The generic element of ${\cal
W}_{\text{exp}}$ is: 
\begin{equation}
f(\bs{x}) = \sum_{\bs{n} \in {\IZ}^2} \hat{f}_{\bs{n}}
W(\bs{n})(\bs{x})\label{RoeiW_1}
\end{equation} with finitely many coefficients $\displaystyle
\hat{f}_{\bs{n}} = \int\!\!\!\!\!\int_{\IT} \ud
\bs{x} \; f (\bs{x}) \; e^{- 2\pi i \bs{n}\bs{x}}$ different
from zero.\\
On ${\cal W}_{\text{exp}}$, formula~\eqref{omegamu} defines a state such that
\begin{equation}
	\omega_\mu\left(W(\bs{n})\right)=\delta_{\bs{n},\bs{0}} 
\label{fabio1}\cdot    
\end{equation}
Following Weyl quantization, we get elements of
${\cal D}_N$ out of elements  of ${\cal W}_{\text{exp}}$ by replacing,
in~\eqref{RoeiW_1}, exponentials with 
diagonal matrices:
\begin{equation}
W(\bs{n}) \longmapsto \widetilde{W}(\bs{n}) \coleq \sum_{\bs{\ell} \in {(\ZNZ{N})^2}} 
e^\frac{\:2\pi i \bs{n}\bs{\ell}}{N}
\ket{\bs{\ell}}\bra{\bs{\ell}} \ ,\qquad
\bs{\ell} =\pt{\ell_1 , \ell_2}\cdot\label{RoeiW_4}
\end{equation}
We will denote by ${\cal J}_{N ,
\infty}^{\cal W}$, the *-morphism from the 
\mbox{*-algebra} ${\cal W}_{\text{exp}}$ into the diagonal matrix algebra
${\cal D}_N$, given by:
\begin{align}
{\cal W}_{\text{exp}}\ni f \longmapsto {\cal J}_{N , \infty}^{\cal W}(f) &
\coleq \sum_{\bs{n} \in {\IZ}^2} \hat{f}_{\bs{n}}
\;\widetilde{W}(\bs{n})\label{RoeiW_50} \\
& = \sum_{\bs{\ell} \in {(\ZNZ{N})^2}} 
f\pt{\frac{\bs{\ell}}{N}}
\ket{\bs{\ell}}\bra{\bs{\ell}}\cdot\label{RoeiW_5}
\end{align} 
\\[-7.5ex]
\begin{quote}
\begin{NNS}{}\ \\[-5.5ex]
\begin{Ventry}{iii)}
	\item[\mdseries i)] The completion of the subalgebra ${\cal
	W}_{\text{exp}}$ with respect to the uniform norm
	given in equation~\eqref{uninorm} is the C*-algebra
	$\Cspace{0}{\IT}$~\cite{Ree72:1}.  
	\item[\mdseries ii)] With the usual \enfasi{operator norm}
	$\norm{\cdot}{N^2}$ of ${\cal B}\pt{{\cal H}_N^D}$,
	$\pt{{\cal D}_N 
	,\norm{\cdot}{N^2}}$ is the C* algebra of $N^2 \times N^2$
	diagonal matrices.
	\item[\mdseries iii)] 
	The *-morphism ${\cal J}_{N ,
	\infty}^{\cal W}:\pt{{\cal
	W}_{\text{exp}},\norm{\cdot}{0}}\longmapsto 
	\pt{{\cal D}_N,\norm{\cdot}{N^2}}$ is 
	bounded by \mbox{$||{\cal J}_{N ,
	\infty}^{\cal W}|| = 1$.} Using the Bounded Limit
	Theorem~\cite{Ree72:1}, ${\cal J}_{N
	,\infty}^{\cal W}$ can be uniquely extended to a bounded linear
	transformation (with the same bound)\\
	${\cal J}_{N
	,\infty}: \pt{\Cspace{0}{\IT},\norm{\cdot}{0}}\longmapsto 
	\pt{{\cal D}_N,\norm{\cdot}{N^2}}$.
	\item[\mdseries iv)] Using Remark {\mdseries iii},
	equation~\eqref{RoeiW_5} can be taken as a definition of
	${\cal J}_{N 
	,\infty}$, as in the following
\end{Ventry}
\end{NNS}
\end{quote}
\begin{quote}
\begin{DDD}{}\ \\[-5.5ex]
\begin{Ventry}{}\label{RoeiW_51}
	\item[] We will denote by ${\cal J}_{N ,
	\infty}$, the *-morphism from the C*--algebra
	$\Cspace{0}{\IT}$ into the diagonal matrix algebra
	${\cal D}_N$, given by:
\begin{equation*}
{\cal J}_{N , \infty}:\Cspace{0}{\IT}\ni f \longmapsto {\cal J}_{N
, \infty}(f)  = \sum_{\bs{\ell} \in {(\ZNZ{N})^2}} 
f\pt{\frac{\bs{\ell}}{N}}
\ket{\bs{\ell}}\bra{\bs{\ell}}\in {\cal D}_N
\ \cdot\notag
\end{equation*} 
\end{Ventry}
\end{DDD}
\begin{NNN}{}\ \\[-5.5ex]\label{tauenneonD}
\begin{Ventry}{}
\item[\mdseries i.] The expectation $\tn\pt{{\cal
J}_{N,\infty}\pt{f}}$ ($\tn$ given in Definition~\ref{tauenne})
corresponds to the numerical calculation of the integral of $f$
 realized on the grid $L_N$ of~\eqref{llnn}.
\end{Ventry}
\end{NNN}
\end{quote}
\noindent
\section{Coherent States}\label{CST}\vspace{6mm} 
The next Quantization procedure we are going to consider, the
Anti--Wick quantization, makes use of coherent states (CS, for short). 
Moreover, when we use Weyl quantization, the dequantizing
operator is constructed by means of CS; actually the most successful
semi-classical tools used to study the classical limit, are based
on the use of CS. For this reason, in this section we will give a
suitable definitions of CS, in the abelian case 
$\tripAAoT$ and in the non--abelian one $\tripQT$, that will be of use
in quantization schemes. 

We remind the reader that in the following, in particular in
Definition~\ref{coh}, $\nh$ introduced in
Definition~\ref{diagonali} (iii.), denotes the 
dimension of the Hilbert space ${\cal H}_N={\IC}^{\nh}$ associated to the
algebra ${\cal M}_N$ (of $\nh\times \nh$ matrices). As it has already 
been seen in Section~\ref{FDQoT}, $\nh$ play also the role of
quantization 
parameter,  i.e. we use $1/\nh$ as an $h$-like parameter. 
The quantum reference state is $\tau_N$ of Definition~\ref{tauenne}
and the dynamics is given in terms of a unitary 
operator $U$ on ${\cal H}_N$ in the standard way: $\Theta_N(X) :=
U^*X\,U$. 

In full generality, coherent states  will be identified as follows.\\[-2ex]
\begin{quote}
\begin{DDD}{}\ \\
\label{coh}
 A family $\{\vert C_N(\bs{x})\rangle \mid \bs{x}\in\c X\}\in {\cal H}_N$ of vectors, 
 indexed by points
 $\bs{x}\in\c X$, constitutes a set of coherent states if it satisfies the
 following requirements:
 \begin{enumerate}
 \item
  \co{Measurability}: $\bs{x} \mapsto \vert C_N(\bs{x})\rangle$ is measurable on
 $\c X$;\\[-1ex]
 \item
  \co{Normalization}: $\|C_N(\bs{x})\|^2 = 1$, $\bs{x}\in\c X$;\\[-1ex]
 \item
  \co{Overcompleteness}: $\nh \int_{\c X}\mu(\ud\bs{x})\, \coh{N}{\bs{x}}
  \lcoh{N}{\bs{x}} = \idty$;\\[-1ex]
 \item
  \co{Localization}: given $\varepsilon>0$ and $d_0>0$, there exists 
  $N_0(\epsilon,d_0)$ such that for $N\ge N_0$ and
 $d_{\cal X}(\bs{x},\bs{y})\ge d_0$ one has  
  \begin{equation*} 
   \nh |\< C_N(\bs{x}), C_N(\bs{y}) \>|^2 \le \varepsilon.
  \end{equation*}
 \end{enumerate}
\end{DDD}
\end{quote}
\noindent
The symbol $d_{\cal X}(\bs{x},\bs{y})$ used in the \co{localization} property 
stands for the length of the shorter segment
connecting the two points on ${\cal X}$. Of course the latter quantity
does depend on the topological properties of ${\cal X}$ so, with the
aim of using it when ${\cal X}=\IT$, we give now the following\\[-2ex]
\begin{quote}
\begin{DDD}{}\ \label{dont}\\
We shall denote by
\begin{equation}
d_{\IT}\pt{\bs{x},\bs{y}}  \coleq
\min_{\bs{n}\in{\IZ}^2} \norm{\bs{x}-\bs{y}+\bs{n}}{{\IR}^2}
\label{Gnbar_m2}
\end{equation}
the distance on ${\IT}$.
\end{DDD}
\end{quote}
\noindent
The \co{overcompleteness} condition may be written in dual form as
\begin{equation*}
 \nh \int_{\c X}\mu(\ud\bs{x})\, \<C_N(\bs{x}), X\, C_N(\bs{x})\> = \tr X, \quad X\in\c 
 M_N.
\end{equation*}
Indeed,
\begin{equation*}
 \nh \int_{\c X}\mu(\ud\bs{x})\, \<C_N(\bs{x}), X\, C_N(\bs{x})\> = \nh
 \tr \pt{\int_{\c 
 X}\mu(\ud\bs{x})\, \coh{N}{\bs{x}} \lcoh{N}{\bs{x}}\, X} = \tr X. 
\end{equation*}
In the next three Sections, we define three different sets of states,
two of them satisfying Properties in Definition~\ref{coh} and then
rightly named Coherent States (CS).
These sets belong to the two different
Hilbert spaces ${\cal H_N}$ and ${\cal H_N^D}$ defined up to now, and
are indexed by points of the torus ${\IT}$.
\subsection{First set of C.S.: $\{\vert C_N^1(\bs{x})\rangle \mid \bs{x}\in\IT\}\in {\cal H}_N$}\label{CST1}\vspace{3mm} 
We shall construct a family $\{\vert C_N^1(\bs{x})\rangle \mid
\bs{x}\in\IT\}$ of  
coherent states on the 2-torus $\IT$ by means of the discrete Weyl
group introduced in Definition~\ref{weylgroup}.
We define
\begin{equation}
 \vert C_N^1(\bi x)\rangle := W_N(\floor{N \bi x})\, \vert C_N\rangle,
\label{coh1}
\end{equation}
where $\floor{N \bi x} = (\floor{Nx_1},\floor{Nx_1})$,
$0\leq\floor{Nx_i}\leq N-1$ is the largest integer smaller than $Nx_i$ and
the fundamental vector $\vert C_N\rangle$ is chosen to be
\begin{equation}
 \vert C_N\rangle = \sum_{j=0}^{N-1} C_N(j)\vert j\>,\qquad
 C_N(j)\coleq \frac{1}{2^{(N-1)/2}}\sqrt{\binom{N-1}{j}}.
\label{coh2}
\end{equation}
\co{Measurability} and \co{normalization} are immediate,
 \co{overcompleteness} comes 
as follows. 
Let $Y$ be the operator 
in Definition~\ref{coh}
on the left hand side of
property $3$ . If $\tau_N(Y\, W_N(\bi n)) =
\tau_N(W_N(\bi n))$ for 
all $\bi n = (n_1,n_2)$ with $0\leq n_i\leq N-1$, then according
to~(\ref{Weyl9}) applied to $Y$ it follows that $Y=\idty$. This is
indeed the case as, using~\eqref{Weyl2},~\eqref{Weyl5} and $N$-periodicity,
\begin{align}
 \tau_N(Y\, W_N(\bi n))
 &= \int_{\Ts} d\bi x\, \<C_N^1(\bi x), W_N(\bi n)\, C_N^1(\bi x) \>
\nonumber \\
 &= \int_{\Ts} d\bi x\, \exp{\Bigl(\frac{2\pi i}{N}\sigma(\bi
 n,\floor{N\bi x})\Bigr)}\, \< C_N, W_N(\bi n)\, C_N\>
\nonumber \\
 &= \frac{1}{N^2} \sum_{p_1,p_2=0}^{N-1} \exp{\Bigl( \frac{2\pi
 i}{N}\sigma(\bi n,\bi p)\Bigr)}\, \< C_N, W_N(\bi n)\, C_N\>
\nonumber \\
 &= \tau_N (W_N(\bi n)).
\label{coh3}
\end{align}
In the last line we used that when $\bi x$ runs over $[0,1)$,
$\floor{Nx_i}$,  $i=1,2$ runs over the set of integers $\pg{0,1,\cdots, N-1}$.

The proof the \co{localization} property in Definition~\ref{coh}
requires several steps. 
First, we observe that, due to~(\ref{FoldUV}),
\begin{align}
 E(n)
 &\coleq \Bigl| \< C_N, W_N(\bi n)\, C_N \> \Bigr|
\nonumber \\
 &= \frac{1}{2^{N-1}} \left| \sum_{\ell=0}^{N-n_1-1}
 \exp\Bigl( -\frac{2\pi i}{N}\ell n_2 \Bigr)
 \sqrt{\binom{N-1}{\ell} \binom{N-1}{\ell+n_1}} \right.
\nonumber \\
 &\quad + \left.\sum_{\ell=N-n_1}^{N-1}
 \exp\Bigl( -\frac{2\pi i}{N}\ell n_2 \Bigr)
 \sqrt{\binom{N-1}{\ell} \binom{N-1}{\ell+n_1-N}} \right|
\label{loc2} \\
 &\leq \frac{1}{2^{N-1}} \left[ \sum_{\ell=0}^{N-n_1-1}
 \sqrt{\binom{N-1}{\ell} \binom{N-1}{\ell+n_1}} \right.
\nonumber \\
 &\quad + \left. \sum_{\ell=N-n_1}^{N-1} 
 \sqrt{\binom{N-1}{\ell} \binom{N-1}{\ell+n_1-N}} \right].
\label{loc3}
\end{align}
Second, using the entropic bound of the binomial coefficients
\begin{equation}
 \binom{N-1}{\ell} \leq  2^{(N-1) \:\eta(\frac{\ell}{N-1})}\quad ,
\label{loc4}
\end{equation}
where
\begin{equation}
 \eta(t) \coleq \begin{cases} -t\log_2 t-(1-t)\log_2(1-t) & \text{if }
 0<t\leq 1 \\ 
 0 & \text{if } t=0 \end{cases}\quad ,
\label{loc5}
\end{equation}
we estimate
\begin{align}
 E(n) &\leq \frac{1}{2^{N-1}}  \left[\sum_{
\phantom{N n_1}\ell=0\phantom{N n_1}
}^{N-1-n_1}
 2^{\frac{N-1}{2} \Big[ \eta\big(\frac{\ell}{N-1}\big) +
 \eta\big(\frac{\ell+\,n_1}{N-1}\big) \Big]} \right.
\nonumber \\
 &+ \left. \sum_{\ell=N-n_1}^{N-1} 2^{\frac{N-1}{2}\Big[
 \eta\big(\frac{\ell}{N-1}\big) +
 \eta\big(\frac{\ell+\,n_1-N}{N-1}\big)\Big]} \right]\quad .
\label{loc6}
\end{align}

The exponents in the two sums are bounded by their maxima
\begin{align}
 \eta\left(\frac{\ell}{N-1}\right) +
 \eta\left(\frac{\ell+n_1}{N-1}\right) 
 &\leq 2\,\eta_1(n_1),\quad \left(0\leq\ell\leq N-n_1-1\right)
\label{loc6a} \\
 \eta\left(\frac{\ell}{N-1}\right) +
 \eta\left(\frac{\ell+n_1-N}{N-1}\right) 
 &\leq 2\,\eta_2(n_1),\quad \left(N-n_1\leq\ell\leq N-1\right)
\label{loc6b}
\end{align}
where
\begin{align}
 \eta_1(n_1)
 &:= \eta\left(\frac{1}{2} - \frac{n_1}{2(N-1)}\right) \leq 1
\label{loc6c} \\
 \eta_2(n_1)
 &:= \eta\left(\frac{1}{2} + \frac{N-n_1}{2(N-1)}\right) \leq \eta_2<1.
\label{loc6d} 
\end{align}
Notice that $\eta_2$ is automatically $<1$, while $\eta_1(n_1)<1$ if 
${\displaystyle \lim_{N\rightarrow\infty}}\frac{n_1}{N} \ne 0$.
If so, the upper bound
\begin{equation}
 E(n) \leq N\Bigl( 2^{-(N-1)(1-\eta_1(n_1))} \,+\, 2^{-(N-1)(1-\eta_2)} \Bigr)
\label{loc7a}
\end{equation}
implies $N\bigl| \< C_N, W_N(\bi n)\, C_N \> \bigr|^2\longmapsto 0$
exponentially with $N\to\infty$.

\noindent The condition for which $\eta_1(n_1)<1$ is fulfilled when 
$|x_1-y_1|>\delta$; in fact, $\bi n=\floor{N\bi y}-\floor{N\bi x}$ and
\begin{equation*}
\lim_{N\rightarrow\infty}\frac{\floor{Nx_1}-\floor{Ny_1}}{N} =
x_1-y_1\ \cdot 
\end{equation*}
On the other hand, if
$x_1=y_1$ and $n_2 = \floor{Nx_2}-\floor{Ny_2} \ne 0$, one explicitly computes
\begin{equation}
 N \bigl| \< C_N, W_N((0,n_2))\, C_N \> \bigr|^2 =
 N{\pt{\cos^2\pt{\frac{\pi n_2}{N}}}}^{N-1}.
\label{loc8}
\end{equation}
Again, the above expression goes exponentially fast to zero, if
${\displaystyle \lim_{N\rightarrow\infty}} \frac{n_2}{N} \ne 0$ which is the
case if $x_2 \ne y_2$. 
\subsection{A second set of states, not overcomplete: $\{\vert
\beta_N(\bs{x})\rangle \mid \bs{x}\in\IT\}\in {\cal
H}_N^D$}\label{CST2}\vspace{3mm}  

As in the previous Section~\ref{CST2}
$\floor{\cdot}$ denotes the integer part of a real number; moreover we
introduce the notation 
$\bk{\cdot}$ to denote
fractional parts, namely $\bk{x}\coleq x - \floor{x}$, so that
we can express each $\bs{x}\in{\IT}$
as 
\begin{equation*}
\bs{x} =\pt{\frac{\floor{N
x_1}}{N},\frac{\floor{N x_2}}{N}} + 
\pt{\frac{\bk{N x_1}}{N},\frac{\bk{N x_2}}{N}}\ \cdot
\end{equation*}
Then we associate
$\bs{x}\in\IT$ with vectors of $\vert \beta_N(\bs{x})\rangle\in{\cal
H}_N^D$ as follows:   
\begin{multline}
\label{RoeiW_6}
\IT\ni\bs{x} \mapsto  \vert \beta_N(\bs{x})\rangle= 
\lambda_{00}\pt{\bs{x}}\ket{\floor{N x_1},\floor{N
x_2}} +\\ 
+ \lambda_{01}\pt{\bs{x}}\ket{\floor{N x_1},\floor{N
x_2}+1} +
\lambda_{10}\pt{\bs{x}}\ket{\floor{N x_1}+1,\floor{N
x_2}} +\\ 
+ \lambda_{11}\pt{\bs{x}}\ket{\floor{N x_1}+1,\floor{N
x_2}+1}\in {\cal H}_N^D\cdot
\end{multline}
We choose the coefficients $\lambda_{\mu\nu}$ in order to have
\co{Measurability}, \co{normalization}, and invertibility of the
mapping in~\eqref{RoeiW_6}:
\begin{equation}
\begin{cases}
\lambda_{00}\pt{\bs{x}}= \cos\pt{\frac{\pi}{2}\bk{N x_1}} 
\cos\pt{\frac{\pi}{2}\bk{N x_2}}\\
\lambda_{01}\pt{\bs{x}}= \cos\pt{\frac{\pi}{2}\bk{N x_1}} 
\sin\pt{\frac{\pi}{2}\bk{N x_2}}\\
\lambda_{10}\pt{\bs{x}}= \sin\pt{\frac{\pi}{2}\bk{N x_1}} 
\cos\pt{\frac{\pi}{2}\bk{N x_2}}\\
\lambda_{11}\pt{\bs{x}}= \sin\pt{\frac{\pi}{2}\bk{N x_1}} 
\sin\pt{\frac{\pi}{2}\bk{N x_2}}
\end{cases}
\label{RoeiW_65}
\end{equation}
Before going in the proof of other properties, let us remind to the
reader that in this case $\nh$ in Definition~\ref{coh} stands for
$N^2$, the
dimension of the Hilbert space.

\co{Overcompleteness}  
fails and we refer to Appendix~\ref{app_A} for a proof. Since we will
not use them in the 
Anti--Wick quantization, in which that property is required, this is
no trouble.
On the other hand, the states $\vert
\beta_N(\bs{x})\rangle$ are useful to invert the Weyl discretization
developed in Section~\ref{WDGriglIA}, as we shall see in
Section~\ref{DDW}. Here we simply note that, although \co{overcompleteness}
is not satisfied by this family of states, nevertheless it is
``not too far from being true'', in the sense that they provide via 
Definition~\ref{coh} property $3$ 
an operator
$I_{\bs{\ell},\bs{m}}$ actually very near to the identity operator $\Id$.

We now prove \co{localization}.

The states $|\beta_N(\bs{x})\rangle$ are constructed by choosing, among
the elements of the basis of ${\cal H}_N^D$, the four ones labeled by
elements of $L_N$ that are
neighbors of $\bs{x}$; it follows that $|\beta_N(\bs{x})\rangle$ is
orthogonal to 
every basis element labeled by a point of $L_N$ whose toral
distance $d_{\IT}$ (see Definition~\eqref{dont}) from $\bs{x}$ is
greater than $\sqrt{2}/N$. 
 
As a consequence, the quantity $N^2\langle
\beta_N(\bs{x}),\beta_N(\bs{y})\rangle=0$ 
if the distance on
the torus between $\bs{x}$ and $\bs{y}$ is greater than $2\sqrt{2}/N$.

\noindent Thus, given $d_0>0$, it is sufficient that $d_0>2\sqrt{2}/N$, that is
$N_0(\epsilon,d_0)>2\sqrt{2}/d_0$, to have 
\begin{equation*}
N>N_0(\epsilon,d_0) \Longrightarrow N^2\langle
\beta_N(\bs{x}),\beta_N(\bs{y})\rangle=0
\end{equation*}
\subsection{A third set of C.S.: $\{\vert C_N^3(\bs{x})\rangle
\mid \bs{x}\in\IT\}\in {\cal H}_N^D$}\label{CST3}\vspace{3mm}  
The new family of CS we are going to introduce in this Section, is
not too different from the one introduced in the previous Section, 
as it will also consist of states in the same Hilbert space 
and constructed by grouping a small cluster of nearest neighbors in
the basis of ${\cal H}_N^D$. 

Nevertheless there is one big difference between the two
example: in the present case, the mapping from $\IT$ into ${\cal
H}_N^D$ defining the family of coherent states is as follows:
\begin{equation}
\label{CSforL1}
\IT\ni\bs{x} \mapsto  \vert C_N^3(\bs{x})\rangle= 
\ket{\floor{N x_1 + \pum},\floor{N x_2 + \pum}}
\in {\cal H}_N^D
\end{equation}
and is not invertible. \co{Measurability} and \co{normalization} are
clearly satisfied and \co{localization} can be proved in the same way
as in the previous Section. Now we shall give a direct proof
of \co{overcompleteness}.

Overcompleteness property of Definition~\ref{coh} can be expressed
as\footnote{For the definition of $\delta^{(N)}_{\bs{\ell},\bs{m}}$,
see in Properties~\ref{proper121}.}
\begin{equation}
N^2 \int_{\c X}\mu(\ud\bs{x})\, 
\langle\bs{\ell}\ 
\vert C_N^3(\bs{x})\rangle\langle C_N^3(\bs{x})\vert
\ \bs{m}\rangle
= \delta^{(N)}_{\bs{\ell},\bs{m}},\qquad\forall\bs{\ell},\bs{m}\in\ZNZD
\label{app3}
\end{equation}
and this is exactly what we are going to prove; let us take the quantity
\begin{align}
I_{\bs{\ell},\bs{m}}&\coleq
N^2 \int_{\c X}\mu(\ud\bs{x})\,  
\langle\bs{\ell}\ 
\vert C_N^3(\bs{x})\rangle\langle C_N^3(\bs{x})\vert
\bs{m}\rangle\notag\\
&= 
N^2 
\int_{0}^{1}\ud x_1\,
\int_{0}^{1}\ud x_2\;
\Bigg\langle \ell_1,\ell_2\ \Bigg|\ 
\floor{N x_1+\pum},\floor{N x_2+\pum}
\Bigg\rangle \ \times\notag\\
&\ \ \ \ \ \times\ \Bigg\langle 
\floor{N x_1+\pum},\floor{N x_2+\pum}
\ \Bigg|\ m_1,m_2\displaybreak
\Bigg\rangle\ =\notag\\
&= 
N^2 
\int_{0}^{1}\ud x_1\,
\int_{0}^{1}\ud x_2\;\ 
\delta^{(N)}_{\ell_1\:,\:\floor{N x_1+\pum}}\;
\delta^{(N)}_{\ell_2\:,\:\floor{N x_2+\pum}}\;
\delta^{(N)}_{m_1\:,\:\floor{N x_1+\pum}}\;
\delta^{(N)}_{m_2\:,\:\floor{N x_2+\pum}}\;\ =\notag\\
&= 
N^2 \ 
\delta^{(N)}_{\ell_1\:,\:m_1}\;
\delta^{(N)}_{\ell_2\:,\:m_2}\;\pq{
\int_{0}^{1}\ud x_1\,
\delta^{(N)}_{\ell_1\:,\:\floor{N x_1+\pum}}}\;\pq{
\int_{0}^{1}\ud x_2\;\ 
\delta^{(N)}_{\ell_2\:,\:\floor{N x_2+\pum}}}\;\label{CS3f1}\ \cdot
\end{align}
Note that,
in order to have
the integrand of~\eqref{CS3f1} different from zero we must have
$\ell_i\leq N x_i+\pum<\ell_i+1$ for $i=1,2$, that is
$\frac{\ell_i-\pum}{N}\leq
x_i<\frac{\ell_i+\pum}{N}$. Then~\eqref{CS3f1} reads: 
\begin{align}
I_{\bs{\ell},\bs{m}}&
 = N^2 \ 
\pt{\delta^{(N)}_{\ell_1\:,\:m_1}\;
\delta^{(N)}_{\ell_2\:,\:m_2}}\;\pq{
\int_{\frac{\ell_1-\pum}{N}}^{\frac{\ell_1+\pum}{N}}\ud x_1\,
}\;\pq{
\int_{\frac{\ell_2-\pum}{N}}^{\frac{\ell_2+\pum}{N}}\ud x_2}\;\ 
\; \cdot\notag\\
&
 = N^2 \ 
\delta^{(N)}_{\bs{\ell},\bs{m}}\;\times\;\frac{1}{N}\;\times\;\frac{1}{N}
 = \delta^{(N)}_{\bs{\ell},\bs{m}}
\;\label{CS3f2}\ 
\end{align}
and hence \co{overcompleteness} is proved.
\section{Anti--Wick Quantization}\label{AWQ}\vspace{6mm}	
In order to study the classical limit and, more generally,
the semi--classical behaviour of $(\c M_N ,\Theta_N,\tau_N)$
when $N\to\infty$, we introduce two linear maps. The first,
${\cal J}_{N\infty}$, (anti-Wick quantization) associates $N\times N$
matrices to functions in ${\cal A}_{\cal X}$, 
the second one, ${\cal J}_{\infty N}$, maps $N\times N$ matrices into
functions  
in ${\cal A}_{\cal X}$.\\[-2ex]
\begin{quote}
\begin{DDS}{}\label{qWick}\ \\
 Given a family $\{\vert C_N(\bs{x})\rangle \mid \bs{x}\in\c X\}$ of
 coherent states in 
 ${\cal H}_N$, the Hilbert space of dimension $\nh$, the anti-Wick
 quantization scheme will be described by a 
 (completely) positive unital map ${\cal J}_{N\infty}: {\cal A}_{\cal X}\to\c
 M_N$ 
 \begin{equation*}
 {{\cal A}_{\cal X}\ni} f \mapsto   
 \nh \int_{\c X}\mu(\ud \bs{x})\, f(\bs{x})\,
  \coh{N}{\bs{x}} \lcoh{N}{\bs{x}}=:{\cal J}_{N\infty}(f)\in \c M_N\quad .
 \end{equation*}
 The corresponding dequantizing map ${\cal J}_{\infty N}: \c
 M_N\to{\cal A}_{\cal X}$ 
 will correspond to the (completely) positive unital map
 \begin{equation*}
  \c M_N \ni X \mapsto 
  \<C_N(\bs{x}), X\,C_N(\bs{x})\>=:{\cal J}_{\infty N}(X)(\bs{x})
 \in{\cal A}_{\cal X} 
  \quad .
 \end{equation*}
\end{DDS}
\end{quote}
\noindent Both maps are identity preserving because of the conditions
imposed 
on the family of coherent states and are also completely positive
since the domain of ${\cal J}_{N\infty}$ is a commutative algebra as
well as the range of ${\cal J}_{\infty N}$. Moreover, 
\begin{equation}
 \norm{{\cal J}_{\infty N} \circ {\cal J}_{N\infty}(g)}{} \le
 \norm{g}{},\quad g\in {\cal A}_{\cal X},
\label{bound}
\end{equation}
where $\norm{\cdot}{}$ denotes the norm with respect to which 
the C*-algebra 
${\cal A}_{\cal X}$ is complete ($\norm{\cdot}{0}$ for
$\Cspace{0}{\cal X}$ and $\norm{\cdot}{\infty}$ for
$\Lspace{\infty}{\cal X}$).
\subsection{Classical limit in the anti--Wick quantization
scheme}\label{CLAW}\vspace{3mm} 
Performing the classical limit or a semi-classical analysis consists 
in studying how a family of
algebraic triples $({\cal M}_N,\tn,\Theta_N)$ 
depending on a quantization $\hbar$-like parameter 
is mapped onto $({\cal A}_{\cal X},\om,\Theta)$
when the parameter goes to zero. 

We shall give now two equivalent properties that can be taken as
requests on any  
well--defined quantization--dequantization scheme for observables.
In the sequel, we shall need the notion of quantum dynamical systems
$({\cal M}_N,\tn,\Theta_N)$ tending to the classical limit $(\c
X,\mu,T)$. Indeed, a request upon any sensible quantization procedure is
to recover 
the classical description in the limit $\hbar \to 0$; in a
similar way, our quantization (or discretization) should recover the
classical (or continuous)
system in the $\frac{1}{N}\to 0$ limit. Moreover we not only need
convergence of observables but also 
of the dynamics: this aspect will be considered in Section~\ref{CCLD}.

Here ${\cal M}_N$ will denote a general $\nh\times\nh$ matrix algebra
and the following two proposition will be proved for both 
${\cal A}_{\cal X}=\Cspace{0}{\cal X}$ and 
${\cal A}_{\cal X}=\Lspace{\infty}{\cal X}$, the two functional spaces
introduced in 
Section~\ref{TuAt}.\\[-2ex]
\begin{quote}
\begin{PPP}{}\label{prop1}\ \\
 For all $f\in{\cal A}_{\cal X}$
 \begin{equation*}
  \lim_{N\to\infty} {\cal J}_{\infty N} \circ {\cal J}_{N\infty}(f) =
  f\quad \mu\text{ -- a.e.}
 \end{equation*}
\end{PPP}
\end{quote}
\noindent\ \\[-2ex]
\begin{quote}
\begin{PPP}{}\label{prop2}\ \\
 For all $f,g\in{\cal A}_{\cal X}$
 \begin{equation*}
  \lim_{N\to\infty} \tau_N \bigl( {\cal J}_{N\infty}(f)^*
  {\cal J}_{N\infty}(g) \bigr) = \omega_\mu(\overline f g) = \int_{\c X}
  \mu(\ud\bs{x})\, \overline{f(\bs{x})}g(\bs{x}).
 \end{equation*}
\end{PPP}
\end{quote}
\noindent  

{\setlength{\parindent}{0pt}  
\textbf{Proof of Proposition~\ref{prop1}:}\\[2.5ex]
We first prove the assertion when ${\cal A}_{\cal X}=\Cspace{0}{\cal
X}$ and 
then we extend to ${\cal A}_{\cal X}=\Lspace{\infty}{\cal X}$. We show
that the quantity 
\begin{align*}
 F_N(\bs{x})
 &\coleq \Bigl|f(\bs{x}) - {\cal J}_{\infty N} \circ {\cal J}_{N\infty}(f)(\bs{x})
 \Bigr| 
 \\
 &= \left|f(\bs{x}) - N \int_{\c X}\mu(\ud\bs{y})\, f(\bs{y})\, |\< C_N(\bs{x}), C_N(\bs{y})\>|^2
 \right| \\
 &= N\left| \int_{\c X }\mu(\ud\bs{y})\, (f(\bs{y}) - f(\bs{x}))\, |\< C_N(\bs{x}), C_N(\bs{y})\>|^2 
 \right|
\end{align*}
becomes arbitrarily small for $N$ large enough, uniformly in $\bs{x}$.
Selecting a ball $B(\bs{x},d_0)$ of radius $d_0$, using the mean-value
theorem and property~(\ref{coh}.3), we derive the upper bound
\begin{align}
 F_N(\bs{x})
 &\le N\left|\int_{B(\bs{x},d_0)} \mu(\ud\bs{y})\, (f(\bs{y}) - f(\bs{x}))\, |\< C_N(\bs{x}),
 C_N(\bs{y})\>|^2 \right|
\nonumber \\
 &\quad + N\left|\int_{\c X\setminus B(\bs{x},d_0)} \mu(\ud\bs{y})\, (f(\bs{y}) -
 f(\bs{x}))\, |\< C_N(\bs{x}), C_N(\bs{y})\>^2 \right| \\
 &\le |f(\bs{c}) - f(\bs{x})|
 + \int_{\c X\setminus B(\bs{x},d_0)} \mu(\ud\bs{y})\, |f(\bs{y}) - f(\bs{x})|\, N\,
 |\< C_N(\bs{x}), C_N(\bs{y})\>|^2,
\label{prop1.2}
\end{align}
where $c\in B(\bs{x},d_0)$.

Because $\c X$ is compact, $f$ is uniformly continuous. Therefore, we
can choose $d_0$ in such a way that $|f(\bs{c}) - f(\bs{x})| < \varepsilon$
uniformly in $\bs{x}\in\c X$. On the other hand, from the localization
property~(\ref{coh}.4), given $\varepsilon'>0$, there exists an
integer $N_0(\varepsilon', d_0)$ such that $N|\< C_N(\bs{x}), C_N(\bs{y})\>|^2 
< \varepsilon'$ whenever $N>N_0(\varepsilon',d_0)$. This choice
leads to the upper bound
\begin{align}
 F_N(\bs{x})
 &\le \varepsilon + \varepsilon'  \int_{\c X\setminus B(\bs{x},d_0)}
 \mu(\ud\bs{y})\, |f(\bs{y}) - f(\bs{x})|
\nonumber\\
 &\le \varepsilon + \varepsilon'  \int_{\c X} \mu(\ud\bs{y})\, |f(\bs{y}) - f(\bs{x})| \le
 \varepsilon + 2 \varepsilon' \|f\|_\infty.
\end{align}
To get rid of the continuity of $f$, that is when
$f\in\Lspace{\infty}{\cal X}$,
we use a corollary~\ref{LusCOR} of Lusin's
theorem~\cite{Hew69:1,Rud87:1,Rie55:1}. For later use, we write down 
both statements,
in a form slightly adapted to our
case (for instance our compact space ${\cal X}=\IT$ is a ``locally compact
Hausdorff space'', but we do not need these generic settings, and the
same is true for the class of $f$ we will refer to):\newpage
\begin{quote}
\begin{TT}[Lusin's]{:} \label{LusTEO}
Every measurable function
$f\pt{\bs{x}}$ on a measurable set ${\cal X}$ can be made
continuous by removing from ${\cal X}$ the points contained in
suitably chosen open intervals whose total measure is arbitrarily small.
\end{TT}
\begin{CC}[of Lusin's Theorem]{:}\label{LusCOR}
Given $f\in\Lspace{\infty}{\cal X}$,
with $\c X$ compact,
there exists a sequence $\{f_n\}$ of continuous functions on $\c X$
such that $|f_n| \le \|f\|_\infty$ and converging to $f$ $\mu$ -- almost
everywhere. 
\end{CC}
\end{quote}
\noindent
Thus, for $f \in \Lspace{\infty}{\cal X}$, we pick such a
sequence and estimate
\begin{align*}
 F_N(\bs{x})
 &\le \Bigl|f(\bs{x})-f_n(\bs{x})\Bigr| + \Bigl|f_n(\bs{x}) - {\cal J}_{\infty
 N}\circ{\cal J}_{N\infty}(f_n)(\bs{x}) \Bigr| \\
 &\quad+ \Bigl|{\cal J}_{\infty N} \circ {\cal J}_{N\infty}(f_n- f)(\bs{x})\Bigr|.
\end{align*}
The first term can be made arbitrarily small ($\mu$ -- a.e) by choosing $n$ large
enough because of Lusin's theorem, while the second one goes to $0$
when  $N\to\infty$ since $f_n$ is continuous.
Finally, the third term becomes as well vanishingly small with $n\to\infty$
as one can deduce from
\begin{align*}
 &\int_{\c X} \mu(\ud\bs{x})\,  \Bigl|{\cal J}_{\infty N} \circ {\cal J}_{N\infty}
 (f - f_n)(\bs{x})\Bigr| \\
 &\quad= \int_{\c X} \mu(\ud\bs{x})\, \left| \int_{\c X} \mu(\ud\bs{y})\, (f(\bs{y}) -
 f_n(\bs{y}))\, N\, |\< C_N(\bs{x}), C_N(\bs{y})\>|^2 \right| \\
 &\quad\le \int_{\c X} \mu(\ud\bs{y})\, |f(\bs{y}) - f_n(\bs{y})|\, \int_{\c X}
 \mu(\ud\bs{x})\, N\, |\< C_N(\bs{x}), C_N(\bs{y})\>|^2 \\
 &\quad= \int_{\c X} \mu(\ud\bs{y})\, |f(\bs{y}) - f_n(\bs{y})|,
\end{align*}
where exchange of integration order is harmless
because of the existence of the integral~(\ref{bound}).
The last integral goes to zero with $n$ by dominated convergence and thus
the result follows.\hfill$\qed$\\[3ex] 
{\setlength{\parindent}{0pt}  
\textbf{Proof of Proposition~\ref{prop2}:}\\[3ex] 
Consider
\begin{align*}
 \Omega_N
 &\coleq \Bigl| \tau_N \bigl({\cal J}_{N\infty}(f)^*
 {\cal J}_{N\infty}(g)\bigr) - \omega_\mu(\overline f g)\Bigr| \\
 &=N \left| \int_{\c X} \mu(\ud\bs{x})\, \overline{f(\bs{x})} \int_{\c X}
 \mu(\ud\bs{y})\, (g(\bs{y})-g(\bs{x}))\, |\< C_N(\bs{x}), C_N(\bs{y})\>|^2 \right| \\
 &\le \int_{\c X} \mu(\ud\bs{x})\, |f(\bs{x})|\, \left| \int_{\c X} \mu(\ud\bs{y})\,
 (g(\bs{y})-g(\bs{x}))\, N\, |\< C_N(\bs{x}), C_N(\bs{y})\>|^2 \right|.
\end{align*}
By choosing a sequence of continuous $g_n$ approximating {\bf $g\in
\Lspace{\infty}{\cal X}$}, and arguing as in the previous
proof, we get the following upper bound:
\begin{align*}
 \Omega_N
 &\le N \int_{\c X} \mu(\ud\bs{x})\, |f(\bs{x})|\,
 \left| \int_{\c X} \mu(\ud\bs{y})\, (g(\bs{y})-g_n(\bs{y}))\, |\< C_N(\bs{x}), C_N(\bs{y})\>|^2
 \right| \\
 &+ N \int_{\c X} \mu(\ud\bs{x})\, |f(\bs{x})|\, \left|
 \int_{\c X} \mu(\ud\bs{y})\, (g_n(\bs{y})-g_n(\bs{x}))\, |\< C_N(\bs{x}), C_N(\bs{y})\>|^2
 \right| \\
 &+ N \int_{\c X} \mu(\ud\bs{x})\, |f(\bs{x})|\, \left|
 \int_{\c X} \mu(\ud\bs{y})\, (g(\bs{x})-g_n(\bs{x}))\, |\< C_N(\bs{x}), C_N(\bs{y})\>|^2 \right|.
\end{align*}
The integrals in the first and third lines go to zero by dominated 
convergence and Lusin's theorem. As regards the middle line, one can
apply the argument used for the quantity $F_N(\bs{x})$ in the proof of
Proposition~\ref{prop1}.\hfill$\qed$\\[-3ex]
\subsection{Discretization/Dediscretization of $\Lspace{\infty}{\cal
X}$ by means of ${\{\vert
C_N^3(\bs{x})\rangle\}}_{\bs{x}\in\IT}$}\label{AWSM}\vspace{3mm}   

Now that we have proved the so called classical limit for the
anti--Wick quantization in the general case, we have all ingredient to
build a concrete example of such a quantization procedure. In
particular we will apply Definitions~\ref{qWick} and
discretize $\Lspace{\infty}{\cal X}$ by means of the CS set 
$\{\vert C_N^3(\bs{x})\rangle
\mid \bs{x}\in\IT\}\in {\cal H}_N^D$ introduced in Section~\ref{CST3}.

In this framework, the discretizing/dediscretizing operators
of Definitions~\ref{qWick} now reads:
 \begin{align}
 {\Lspace{\infty}{\IT}\ni} f & \mapsto   
 N^2 \int_{\IT}\mu(\ud \bs{x})\, f(\bs{x})\,
 \vert C_N^3(\bs{x})\rangle\langle C_N^3(\bs{x})\vert
=:{\cal J}_{N\infty}(f)\in \c D_N\quad \cdot\label{AWSM_1}\\
  \c D_N \ni X & \mapsto 
  \langle C_N^3(\bs{x})\,,\, X\, C_N^3(\bs{x})\rangle
  =:{\cal J}_{\infty N}(X)(\bs{x})
 \in{\cal S}\pt{\IT}\protect\footnotemark\subset
\Lspace{\infty}{\IT}
  \quad \cdot\label{AWSM_2}
 \end{align}
\protect\footnotetext{
The symbol ${\cal S}\pt{\IT}$ denotes the set of {\it simple functions}
on the torus. A function $f$ belong to that set if it holds that
$f$ assumes on $\IT$ only a finite number of
values~\protect\cite{Hew69:1}. The relation $\mathrm{Ran}\pt{{\cal
J}_{\infty N}}={\cal S}\pt{\IT}$ will be shown in~\protect\eqref{AWJIN1}.}
In this Section we will give an interpretation of these two operators
 that will be useful in the following. Let us start by computing the
 matrix elements of ${\cal J}_{N\infty}(f)$ in~\eqref{AWSM_1}.
\begin{align}
M_{\bs{\ell},\bs{m}}^{(f)}&\coleq
\langle \bs{\ell}\,,\, {\cal J}_{N\infty}(f)\,
\bs{m}\rangle\label{AWSM_3}\\ 
&=N^2 \int_{\c X}\mu(\ud\bs{x})\, f(\bs{x})\,
\langle\bs{\ell}\ 
\vert C_N^3(\bs{x})\rangle\langle C_N^3(\bs{x})\vert
\bs{m}\rangle\label{AWSM_4}\\
&= 
N^2 
\int_{0}^{1}\ud x_1\,
\int_{0}^{1}\ud x_2\; f(\bs{x})\;
\Bigg\langle \ell_1,\ell_2\ \Bigg|\ 
\floor{N x_1+\pum},\floor{N x_2+\pum}
\Bigg\rangle \ \times\notag\\
&\ \ \ \ \ \times\ \Bigg\langle 
\floor{N x_1+\pum},\floor{N x_2+\pum}
\ \Bigg|\ m_1,m_2\displaybreak
\Bigg\rangle\ \label{AWSM_5}\\
&= 
N^2 
\int_{0}^{1}\ud x_1\,
\int_{0}^{1}\ud x_2\, f(\bs{x})\,
\delta^{(N)}_{\ell_1\:,\:\floor{N x_1+\pum}}\;
\delta^{(N)}_{\ell_2\:,\:\floor{N x_2+\pum}}\;
\delta^{(N)}_{m_1\:,\:\floor{N x_1+\pum}}\;
\delta^{(N)}_{m_2\:,\:\floor{N x_2+\pum}}\;\notag\\
&= 
N^2 \ 
\delta^{(N)}_{\ell_1\:,\:m_1}\;
\delta^{(N)}_{\ell_2\:,\:m_2}\;
\int_{0}^{1}\ud x_1\,
\int_{0}^{1}\ud x_2\; f(\bs{x})\;
\delta^{(N)}_{\ell_1\:,\:\floor{N x_1+\pum}}\;
\delta^{(N)}_{\ell_2\:,\:\floor{N x_2+\pum}}\;\ \cdot\label{AWSM_6}
\end{align}
As already observed (between
eqs.~\eqref{CS3f1} and~\eqref{CS3f2}),
in order to have
the integrand of~\eqref{AWSM_6} different from zero we must have
$\ell_i\leq N x_i+\pum<\ell_i+1$ for $i=1,2$, that is
$\frac{\ell_i-\pum}{N}\leq x_i<\frac{\ell_i+\pum}{N}$
and~\eqref{AWSM_6} reads:
\begin{equation}
M_{\bs{\ell},\bs{m}}^{(f)}
 = N^2 \;\delta^{(N)}_{\bs{\ell},\bs{m}}
\;
\int_{\frac{\ell_1-\pum}{N}}^{\frac{\ell_1+\pum}{N}}\ud x_1\,
\;
\int_{\frac{\ell_2-\pum}{N}}^{\frac{\ell_2+\pum}{N}}\ud x_2\; f(\bs{x})\;
\ \cdot\label{AWSM_7}
\end{equation}
From the latter equation we see that $\mathrm{Ran}\pt{{\cal J}_{N\infty}}={\cal
D}_N$. We will reduce~\eqref{AWSM_7} to a nicer expression, but to
this aim we introduce now a new\\[-2ex]
\begin{quote}
\begin{DDD}[Running Average Operator (RAO)]{}\ \label{runave}\\[-5.5ex]
\begin{Ventry}{}
	\item[--] We will denote by $Q_N\pt{\bs{x}}$ the small square of
	side $1/N$, oriented with sides parallel to the axis
	of the torus, centered on $\bs{x}$. 
	\item[--] By means of the latter, we introduce now the Running
	Average Operator $\Gamma_N :  \Lspace{\infty}{\cal
	X}\longmapsto \Cspace{0}{\IT}$ defined by:
\begin{equation}
\Lspace{\infty}{\cal X}
\ni f(\bs{x})\longmapsto\Gamma_N\pt{f}(\bs{x})=:
N^2 
\;
\int_{Q_N\pt{\bs{x}}} \mu(\ud\bs{x})\;f(\bs{y})\;
\in\Cspace{0}{\IT}
\ \cdot\notag
\end{equation}
\end{Ventry}
\end{DDD}
\begin{PPS}{}\ \label{runaveprps}\\[-5.5ex]
\begin{Ventry}{}
	\item[$1$)] Given $f\in \Lspace{\infty}{\IT}$, the function
$f_N^{(Q)}\coleq\Gamma_N\pt{f}$ is uniformly \mbox{continuous on $\IT$}
	\item[$2$)] Denoting by
\begin{equation}
\norm{\Gamma_N}{\cal B}\coleq\sup_{f\in \Lspace{\infty}{\IT}}
\frac{\norm{\Gamma_N\pt{f}}{0}}{\norm{f}{\infty}}\label{AWSM_8}
\end{equation}
we have that $\norm{\Gamma_N}{\cal B}=1$. 
\end{Ventry}
\end{PPS}
\end{quote}
\noindent
\textbf{Proof of Propositions~\ref{runaveprps}:}\\[3ex]
Given a two functions, $f\in \Lspace{\infty}{\IT}$ and $g\in
\Lspace{1}{\IT}$, and denoting by 
$\norm{\cdot}{1}$ the $\Lspace{1}{\IT}$-norm, that is 
\begin{align}
\norm{g}{1}&\coleq\int_{\IT} \mu(\ud\bs{x})\, \abs{g(\bs{x})}\
,\label{runaveprps1}
\intertext{one has:}
\norm{fg}{1}&\leq\norm{f}{\infty}\norm{g}{1}\label{runaveprps2}
\end{align} 
Equation~\eqref{runaveprps2} can be seen as an extension of the
H\"older's inequality, but its proof is more easily deduced by
integrating the obvious relation
$\abs{fg}\leq\norm{f}{\infty}\abs{g}$; using~\eqref{runaveprps2} we
are going to prove the two statement of Proposition~\ref{runaveprps}.

\textbf{1)}\quad Let's take two points $\bs{x}_0\in\IT$ and $\bs{x}\in
Q_N\pt{\bs{x}_0}$, and let ${\cal X}_E$ denote the characteristic function of 
$E\subset\IT$. By Definition~\eqref{runave}:
\begin{align}
\abs{f_N^{(Q)}\pt{\bs{x}_0}-f_N^{(Q)}\pt{\bs{x}}}
&=N^2\;
\abs{\int_{\IT}\mu(\ud\bs{y})\;f(\bs{y})\;
\pt{{\cal X}_{Q_N\pt{\bs{x}_0}}(\bs{y})-{\cal
X}_{Q_N\pt{\bs{x}}}(\bs{y})} 
}\notag
\intertext{therefore, carrying $\abs{\cdot}$ inside the integral and using~\eqref{runaveprps2}}
&\leq N^2\;\norm{f}{\infty}
\int_{\IT}\mu(\ud\bs{y})\;
\abs{{\cal X}_{Q_N\pt{\bs{x}_0}}(\bs{y})-{\cal
X}_{Q_N\pt{\bs{x}}}(\bs{y})}\notag\\
&\leq N^2\;\norm{f}{\infty}\bigg[
\mu\Big(Q_N\pt{\bs{x}_0}\cup Q_N\pt{\bs{x}}\Big)\;-\;
\mu\Big(Q_N\pt{\bs{x}_0}\cap Q_N\pt{\bs{x}}\Big)
\bigg]\ \cdot\notag
\end{align}
According to our hypothesis, $\bs{x}\in
Q_N\pt{\bs{x}_0}$, thus geometrical considerations lead to:
\begin{align}
\mu\Big(Q_N\pt{\bs{x}_0}\cup Q_N\pt{\bs{x}}\Big)&<
\Big(\frac{1}{N}+\abs{x_1-x_{01}}\Big)
\Big(\frac{1}{N}+\abs{x_2-x_{02}}\Big)\notag\\
\mu\Big(Q_N\pt{\bs{x}_0}\cap Q_N\pt{\bs{x}}\Big)&=
\Big(\frac{1}{N}-\abs{x_1-x_{01}}\Big)
\Big(\frac{1}{N}-\abs{x_2-x_{02}}\Big)\notag\\
\mu\Big(Q_N\pt{\bs{x}_0}\cup Q_N\pt{\bs{x}}\Big)\;& -\;
\mu\Big(Q_N\pt{\bs{x}_0}\cap Q_N\pt{\bs{x}}\Big)<
\frac{2}{N}\Big(\abs{x_1-x_{01}}+\abs{x_2-x_{02}}\Big)\notag\\
& \phantom{-\;\ \,
\mu\Big(Q_N\pt{\bs{x}_0}\cap Q_N\pt{\bs{x}}\Big)}<\;
\frac{2\sqrt{2}}{N}\;
\norm{\bs{x}_0-\bs{x}}{}\notag
\end{align}
Finally we can write
\begin{equation}
\abs{f_N^{(Q)}\pt{\bs{x}_0}-f_N^{(Q)}\pt{\bs{x}}}\leq
2\sqrt{2}\;N\;\norm{f}{\infty}\;\norm{\bs{x}_0-\bs{x}}{}\notag
\end{equation}
and this prove continuity of $f_N^{(Q)}$ (and so uniform continuity
too, being $\IT$ a compact space).\hfill$\qed$

\textbf{2)}\quad A straightforward application of~\eqref{runaveprps2}
give us that $\norm{\Gamma_N}{\cal B}\leq 1$. We reach the maximum by
noting that it is attained when we choose $f$ constant.\hfill$\qed$

We can now rewrite equation~\eqref{AWSM_1} by using RAO together
with~\eqref{AWSM_7}; as a result:  
\begin{equation}
{\cal J}_{N\infty}(f)
=\sum_{\bs{\ell} \in {(\ZNZ{N})^2}} 
\Gamma_N\pt{f}\pt{\frac{\bs{\ell}}{N}}
\ket{\bs{\ell}}\bra{\bs{\ell}}
=\sum_{\bs{\ell} \in {(\ZNZ{N})^2}} 
f_N^{(Q)}\pt{\frac{\bs{\ell}}{N}}
\ket{\bs{\ell}}\bra{\bs{\ell}}
\label{AWJNI1}\ \cdot
\end{equation}
We now compute explicitly~\eqref{AWSM_2}, 
considering the matrix 
elements ${\pg{X_{\bs{\ell},\bs{\ell}}}}_{\bs{\ell} \in
{(\ZNZ{N})^2}}$ of $X$; namely
\begin{equation}
X =\sum_{\bs{\ell} \in {(\ZNZ{N})^2}} 
X_{\bs{\ell},\bs{\ell}}
\ket{\bs{\ell}}\bra{\bs{\ell}}\label{AWJIN1}\ \cdot
\end{equation}
By putting~\eqref{AWJIN1} in~\eqref{AWSM_2} we get:
\begin{align}
{\cal J}_{\infty N}(X)(\bs{x})&=\sum_{\bs{\ell} \in {(\ZNZ{N})^2}} 
X_{\bs{\ell},\bs{\ell}}\;
\langle
C_N^3(\bs{x})\ket{\bs{\ell}}\bra{\bs{\ell}}C_N^3(\bs{x})\rangle\notag\\
&=\sum_{\bs{\ell} \in {(\ZNZ{N})^2}} 
X_{\bs{\ell},\bs{\ell}}\;
 \Bigg\langle 
\floor{N x_1+\pum},\floor{N x_2+\pum}
\ \Bigg|\ \ell_1,\ell_2
\Bigg\rangle\ \times\notag\\
&\ \ \ \ \ \ \ \ \ \ \times\ \Bigg\langle \ell_1,\ell_2\ \Bigg|\ 
\floor{N x_1+\pum},\floor{N x_2+\pum}
\Bigg\rangle=\notag\\
&=\sum_{\bs{\ell} \in {(\ZNZ{N})^2}} 
X_{\bs{\ell},\bs{\ell}}\;{\pt{
\delta^{(N)}_{\ell_1\:,\:\floor{N x_1+\pum}}}}^2\;
{\pt{\delta^{(N)}_{\ell_2\:,\:\floor{N x_2+\pum}}}}^2\;\notag\\
&=\sum_{\bs{\ell} \in {(\ZNZ{N})^2}} 
X_{\bs{\ell},\bs{\ell}}\;{\cal
X}_{Q_N\pt{\frac{\bs{\ell}}{N}}}(\bs{x})
\label{AWJIN1}\ ,
\end{align}
and it proves that $\mathrm{Ran}\pt{{\cal
J}_{\infty N}}={\cal S}\pt{\IT}$.

Moreover, equations~\eqref{AWJNI1} and ~\eqref{AWJIN1} can be combined
and we get the form of the (simple) function that arises from one in
$\Lspace{\infty}{\cal X}$, by performing the anti--Wick
quantization/dequantization:
\begin{equation}
\pt{{\cal J}_{\infty N}\circ{\cal J}_{N\infty}}(f)(\bs{x})
=\sum_{\bs{\ell} \in {(\ZNZ{N})^2}} 
\Gamma_N\pt{f}\pt{\frac{\bs{\ell}}{N}}
\;{\cal X}_{Q_N\pt{\frac{\bs{\ell}}{N}}}(\bs{x})
\label{AWJINJNI}\ \cdot
\end{equation}
\section{%
Inverting the Weyl discretization
by means of ${\{\vert
\beta_N(\bs{x})\rangle\}}_{\bs{x}\in \IT}$ states}
\label{DDW}\vspace{6mm}    

In Sections~\ref{WQ} and~\ref{WDGriglIA} we developed the procedures
of Weyl quantization, Weyl discretization respectively. In this
Section, we will refer to the second scheme, the  
Weyl discretization of $\Cspace{0}{\IT}$ into ${\cal D}_N$
described in Section~\ref{WDGriglIA}.

In the algebraic discretization, our goals was to find
the operator ${\cal J}_{N ,\infty}$ in
Definition~\ref{RoeiW_51}; actually this is not complete. As in the
anti--Wick scheme of quantization/de--quantization (equivalently
discretization/de--discretization) developed in Section~\ref{AWQ}, described
in Definitions~\ref{qWick} and based on a couple of *-automorphisms
${\cal J}_{N\infty}$ and ${\cal J}_{\infty N}$, also here we have to go
back from ${\cal D}_N$ to $\Cspace{0}{\IT}$ by defining a *-morphism
${\cal J}_{\infty , N}$ that
``inverts'' the ${\cal J}_{N\infty}$ introduced in Definition~\ref{RoeiW_51},
at least in the $N \to \infty$ limit. We construct this operator by
means of the family of states $\{\vert \beta_N(\bs{x})\rangle
\mid \bs{x}\in\IT\}\in {\cal H}_N^D$ introduced in Section~\ref{CST2}, as
follows\\[-2ex]
\begin{quote}
\begin{DDD}{}\ \\[-5.5ex]
\begin{Ventry}{}\label{RoeiW_51b}
	\item[] We will denote by ${\cal J}_{\infty
	,N}: {\cal D}_N \longmapsto  $ the *-morphism 
	defined by:
\begin{equation}
{\cal J}_{\infty,N} : {\cal D}_N \ni M \longmapsto{\cal J}_{\infty ,
N}(M)\pt{\bs{x}} \coleq \bra{\beta_N\pt{\bs{x}}} M
\ket{\beta_N\pt{\bs{x}}}\in\Cspace{0}{\IT}\ \cdot\notag
\end{equation}
\end{Ventry}
\end{DDD}
\end{quote}
\noindent

Therefore, from definitions~\eqref{RoeiW_51} and \eqref{RoeiW_51b} it
follows that, when mapping $\Cspace{0}{\IT}$ onto
${\cal D}_N$ and the latter back into $\Cspace{0}{\IT}$, we get%
\footnote{we omit here the details of the cumbersome calculation:
equation~\eqref{RoeiW_8} is derived by using the same technique as
the one showed in Appendix~\ref{app_A}.}:
\begin{multline}
\widetilde{f}_N\pt{\bs{x}}\coleq
\pt{{\cal J}_{\infty , N}\circ {\cal J}_{N , \infty}}
\pt{f}\pt{\bs{x}} =
\sum_{\bs{\ell} \in {(\ZNZ{N})^2}} 
f\pt{\frac{\bs{\ell}}{N}}
{\abs{\bk{ \beta_N\pt{\bs{x}}\big| \bs{\ell}}}}^2 =\\
= \frac{1}{4}\sum_{\pt{\mu,\nu,\rho,\sigma}\in{\pg{0,1}}^4} 
\cos\pt{\pi \mu \bk{N x_1}} \cos\pt{\pi \nu \bk{N x_2}}
{\pt{-1}}^{\mu \rho + \nu \sigma} \times\\
\times f\pt{\frac{\floor{N x_1}+\rho}{N},\frac{\floor{N x_2}+\sigma}{N}}
\label{RoeiW_8}
\end{multline}
\begin{quote}
\begin{NNS}{}\ \\[-5.5ex]
\label{rem_24}
\begin{Ventry}{iii)}
	\item[\mdseries i)] As it was pointed out in
	Section~\ref{CST2}, the family of states $\{\vert
	\beta_N(\bs{x})\rangle \mid \bs{x}\in\IT\}$ cannot be
	considered as a set of C.S., for they does not fulfill
	\co{overcompleteness}. The latter property
	is a necessary condition in order to prove the classical
	limit in the anti--Wick scheme of quantization/dequantization,
	and to this aim it has been profitably used in the proofs of
	property~(\ref{prop1}--\ref{prop2}). Nevertheless, in the Weyl
	scheme, we will provide in Theorem~\ref{convWeyl} an equivalent
	proof of the classical 
	limit that does not depend on \co{overcompleteness}; conversely
	we can use other nice properties of our family of states, like
	invertibility. 
	\item[\mdseries ii)] From~\eqref{RoeiW_8},
	$f = \widetilde{f}_N$ on
	lattice points. Moreover, although the first derivative
	of~\eqref{RoeiW_8} is not defined on the lattice, its
	limit exists there and it is zero; thus, we can extend by continuity 
	$\widetilde{f}_N$ to a function in $\Cspace{1}{\IT}$ that we
	will denote again as $\widetilde{f}_N$.
	\item[\mdseries iii)] We note that $\mathrm{Ran}\pt{{\cal
	J}_{\infty , N}}$ is a subalgebra 
 	strictly contained in ${\cal A}_{\cal X}$; this is not
	surprising and comes as consequence of Weyl quantization,
	where this phenomenon is quite
	typical~\cite{DeB98:1,DeB01:1}.
\end{Ventry}
\end{NNS}
\end{quote}
\noindent
We show below that \mbox{${\cal J}_{\infty , N} \circ {\cal J}_{N ,
\infty}$} approaches ${\Id}_{\Cspace{0}{\IT}}$ (the identity function
in ${\Cspace{0}{\IT}}$) when $N \to \infty$.  
Indeed, a request upon any sensible quantization procedure is
to recover 
the classical description in the limit $\hbar \to 0$; in a
similar way, our discretization should recover the continuous
system in the $N\to \infty$ limit. 
\begin{quote}
\begin{TT}{:}\label{convWeyl}
Given $f \in \Cspace{0}{\IT},$
\mbox{$\displaystyle
\  \lim_{N\to\infty}{\bigg\|\pt{{\cal J}_{\infty , N} \circ {\cal
J}_{N , \infty}-{\Id}_{\Cspace{0}{\IT}}}\pt{f}\bigg\|_0} = 0
\; \cdot
$}
\end{TT}
\end{quote}
\noindent
\textbf{Proof of Theorem~\ref{convWeyl}:}\\[2.5ex]

{\bf i)} Since ${\cal X}= \IT$ is compact, $f$ is uniformly continuous on it,
that is%
\footnote{Notation: we use $\displaystyle \frac{\floor{N\bs{x}}}{N}$ to denote 
$\displaystyle \pt{\frac{\floor{N x_1}}{N},\frac{\floor{N x_2}}{N}}\ \cdot$}
\begin{align}
\forall \; \varepsilon
> 0, \; \exists\;
\delta_{f,\varepsilon} > 0 \; \text{ s.t. } 
&\bigg|\bs{x} - \frac{\floor{N\bs{x}}}{N}\bigg| <
\delta_{f,\varepsilon} \Longrightarrow \nonumber\\ 
\Longrightarrow \bigg|f\pt{\bs{x}} - f \pt{\frac{\floor{N\bs{x}}}{N}} 
&\bigg| < \frac{\varepsilon}{2} \; ,\; \forall \bs{x} \in \IT,\;
\forall N\in {\IN}^+ 
\end{align}
Further, we can choose an $\bar{N}_{f,\varepsilon} =
\bar{N}_{f,\varepsilon}(\delta_{f,\varepsilon})$ 
s.t. $\displaystyle \abs{\bs{x} -  
\frac{\floor{\bar{N}_{f,\varepsilon}\bs{x}}}{\bar{N}_{f,\varepsilon}}}
< \delta_{f,\varepsilon}$, that is%
\footnote{$\displaystyle \forall \,i\in\pg{1,2}, \
\bigg|N x_i -   
\floor{N x_i}\bigg|<1 \Longrightarrow
\bigg|N \bs{x} -   
\floor{N\bs{x}}\bigg|  < \sqrt{2}
\Longrightarrow \bigg|\bs{x} -   
\frac{\floor{N\bs{x}}}{N}\bigg|
< \frac{\sqrt{2}}{N}\Longrightarrow$\\
$\phantom{.}\;\!\displaystyle\Longrightarrow \bigg|\bs{x} -   
\frac{\floor{N\bs{x}}}{N}\bigg|
< \frac{\sqrt{2}}{\bar{N}_{f,\varepsilon}}, \ \forall\
N>\bar{N}_{f,\varepsilon}\cdot$} 
$\displaystyle \bar{N}_{f,\varepsilon} >
\frac{\sqrt{2}}{\delta_{f,\varepsilon}}$; therefore
\begin{align}
 \forall \varepsilon
> 0, \; \exists \bar{N}_{f,\varepsilon}\in {\IN}^+
\; \text{ s.t. } N>\bar{N}_{f,\varepsilon} & \Longrightarrow 
\:\abs{f\pt{\bs{x}} - f\pt{\frac{\floor{N\bs{x}}}{N}}\:} <
 \frac{\varepsilon}{2} \; ,\  \forall\,
 \bs{x} \in \IT\ \cdot\nonumber\\
\intertext{{\bf ii)} $\widetilde{f}_N\in\Cspace{1}\IT\subset\Cspace{0}\IT$ and the
previous point {\bf i)} let us write}
 \forall \varepsilon
> 0, \; \exists \bar{N}_{f,\varepsilon}^{\prime}\in {\IN}^+
\; \text{ s.t. } N>\bar{N}_{f,\varepsilon}^{\prime}  & \Longrightarrow 
\:\abs{\widetilde{f}_N\pt{\bs{x}} - \widetilde{f}_N\pt{\frac{\floor{N\bs{x}}}{N}}\:} < \frac{\varepsilon}{2} \; ,\  \forall\,
 \bs{x} \in \IT\ \cdot\nonumber
\intertext{{\bf iii)} $\displaystyle \widetilde{f}_N\pt{\frac{\floor{N\bs{x}}}{N}} =
f\pt{\frac{\floor{N\bs{x}}}{N}}\;\;\forall \bs{x} \in \IT,\;
\forall N\in {\IN}^+$. (see Remark~\ref{rem_24}.ii)}
\intertext{Then, using the triangle inequality, with
$\bar{N}_{f,\varepsilon}^{\prime\prime}=\max\pg{\bar{N}_{f,\varepsilon}
^{\prime},\bar{N}_{f,\varepsilon}}$,  
we get}
 \forall \varepsilon
> 0, \; \exists \bar{N}_{f,\varepsilon}^{\prime\prime}\in {\IN}^+
\; \text{ s.t. } N>\bar{N}_{f,\varepsilon}^{\prime\prime}  &
\Longrightarrow \nonumber\\
\Longrightarrow 
 \bigg|f\pt{\bs{x}}-\widetilde{f}_N \pt{\bs{x}}\bigg|\leq
\bigg|f\pt{\bs{x}} - f\bigg(&\frac{\floor{N\bs{x}}}{N}\bigg)\bigg| + 
\bigg|\widetilde{f}_{N}\pt{\frac{\floor{N\bs{x}}}{N}} -
\widetilde{f}_{N}\pt{\bs{x}} \bigg|
< \varepsilon
\; ,\  \forall\,
 \bs{x} \in \IT\nonumber
\end{align}
that is:
\begin{equation}
\lim_{N\to\infty}{\Big\| 
\;\widetilde{f}_N - f\;
\Big\|}_0 = 0
\; \cdot\hfill\tag*{\qed}
\end{equation}
\pagestyle{fancyplain}
\chapter{Quantization of the Dynamics and its classical limit}\label{Chap2}\vspace{9mm}
\section{Classical Automorphisms on the Torus}\vspace{6mm}
\subsection{Classical description of Sawtooth Maps and Cat
Maps}\label{AATT}\vspace{3mm} 
The special kind of automorphisms of the torus that we are going to
consider in this Section,   
namely the Sawtooth Maps~\cite{Che92:1,Vai92:1}, are a big family
including the well known Cat Maps as a subset.
From a classical point of view, in the spirit of Section~\ref{CDS}, we
describe these systems by means of triples $({\cal X},\mu,S_\alpha)$
where 
\begin{subequations}
\label{AoDC_1}
\begin{align}
{\cal X}&=\IT
\label{AoDC_1a}\\
S_\alpha 
\begin{pmatrix}
x_1\\
x_2
\end{pmatrix}& =
\begin{pmatrix}
1+\alpha & 1\\
\alpha & 1
\end{pmatrix}
\begin{pmatrix}
\bk{x_1}\\
x_2
\end{pmatrix}
\ \pmod{1}\ ,\quad
\alpha\in\IR
\label{AoDC_1b}\\
\mu(\ud\bs{x})&=\ud x_1\; \ud x_2 \ ,
\label{AoDC_1c}
\end{align} 
\end{subequations}
where $\bk{\cdot}$ denotes
the fractional part of a real number. Without
$\bk{\cdot}$,~\eqref{AoDC_1b} is not well defined on $\IT$ for
not--integer $\alpha$; in fact, without taking the fractional part,
the same point $\bs{x} = \bs{x} + \bs{n} \in \IT , \bs{n} \in
{\IZ}^2$, would have (in general) $S_\alpha \pt{\bs{x}} \neq S_\alpha
\pt{\bs{x}+\bs{n}}$. Of course, $\bk{\cdot}$ is not necessary when
$\alpha\in{\IZ}$.\\
The Lebesgue measure defined in~\eqref{AoDC_1c} is
$S_\alpha$--\enfasi{invariant} for all $\alpha\in\IR$.\\
After identifying $\bs{x}$ with canonical coordinates
$(q,p)$ and imposing the$\pmod{1}$ condition on
both of them, the above dynamics can be rewritten as:
\begin{equation}
\begin{cases}
q^\prime &= q + p^\prime\\
p^\prime &= p + \alpha  \bk{q}
\end{cases}
\pmod{1},
\label{AoDC_11}
\end{equation}
This is nothing but the Chirikov Standard Map~\cite{Cas95:1} in which
$-\frac{1}{2\pi}\sin(2\pi q)$ is replaced by 
$\bk{q}$.
The dynamics in~\eqref{AoDC_11} can also be thought of as generated
by the (singular) Hamiltonian 
\begin{equation}
H(q,p,t)=\frac{p^2}{2}-\alpha\, 
\frac{{\bk{q}}^2}{2}\,\delta_p(t),
\end{equation}
where $\delta_p(t)$ is the periodic Dirac delta which makes the
potential act through
periodic kicks with period 
$1$~\cite{For91:1}.

\noindent Sawtooth Maps are invertible and the inverse is given by the
expression   
\begin{gather}
S_\alpha^{-1}
\begin{pmatrix}
x_1\\
x_2
\end{pmatrix} =
\begin{pmatrix}
\phantom{-}1 & 0\\
-\alpha & 1
\end{pmatrix}
\bk{\begin{pmatrix}
1 & -1\\
0 & \phantom{-}1
\end{pmatrix}
\begin{pmatrix}
x_1\\
x_2
\end{pmatrix}}
\ \pmod{1}
\label{AoDC_1d}\\
\intertext{or, in other words,}
\begin{cases}
q &= \phantom{-\alpha\,}q^\prime - p^\prime\\
p &= -\alpha\,q^{\phantom{\prime}} + p^\prime
\end{cases}
\pmod{1}\ .
\label{AoDC_1e}
\end{gather}
It can indeed be checked that
$S_\alpha\pt{S_\alpha^{-1}\pt{\bs{x}}} =
S_\alpha^{-1}\pt{S_\alpha\pt{\bs{x}}}=\bs{x}$\footnote{%
of course $\bs{x}$ has to be intended as an element of the torus, that
is an equivalent class of $\IR^2$ points whose coordinates differ for
integer value, indeed $\IT={\IR}^2\setminus{\IZ}^2$.}$,\ \forall
\bs{x}\in\IT$.

Another dynamics we are going to consider is the one described
by~\eqref{AoDC_1}, but with~\eqref{AoDC_1b} replaced by
\begin{equation}
T\pt{\bs x} =T\cdot\bs x\ \pmod{1}\ ,\tag{\ref{AoDC_1b}$'$}
\end{equation}
where 
$T\in{\text{SL}}_2\left(\IT\right)$, namely the $2\times2$
matrices with integer entries and determinant equal to one.
The Lebesgue measure defined in~\eqref{AoDC_1c} is
$T$--\enfasi{invariant} 
for all $T\in{\text{SL}}_2\left(\IT\right)$.\\[-2ex] 
\begin{quote}
\begin{DDD}{}\ \\[-5.5ex]
\begin{Ventry}{$\tripCTa$}\label{def_SeT}
	\item[] When $\alpha\in\IZ$,
	we shall write $T_\alpha$ instead of
	$S_\alpha$.  In particular:
	\item[$\tripCTa$] will be the classical dynamical system
	representing a generic $T_{\alpha}$ automorphism;
	\item[$\tripCSa$] will represent Sawtooth Maps;
	\item[$\tripCT$] with $T\in{\text{SL}}_2\left(\IT\right)$,
	will represent the 
	so--called \enfasi{Unitary Modular Group}~\cite{Kat99:1} (UMG
	for short).
\end{Ventry}
\end{DDD}
\end{quote}
\noindent\\[-3ex]
$T_1=\pt{\begin{smallmatrix} 2 & 1\\ 1 & 1
\end{smallmatrix}}$ is the Arnold Cat Map~\cite{Kat99:1}.

\noindent Then, $\displaystyle T_1 \in 
{\left\{T_{\alpha}\right\}}_{\alpha\in\IZ} \subset
{\text{SL}}_2\left(\IT\right)\subset
{\text{GL}}_2\left(\IT\right)\subset
{\text{ML}}_2\left(\IT\right)$ where
${\text{ML}}_2\left(\IT\right)$ is the set of $2\times2$
matrices with integer entries and
${\text{GL}}_2\left(\IT\right)$ is the subset of
invertible matrices.\\[-2ex]
\begin{quote}
\begin{NNS}{}\ \\[-5.5ex]\label{Rem_21}
\begin{Ventry}{\mdseries vii.}
\item[\mdseries i.] Sawtooth Maps $\{S_\alpha\}$ are
	\enfasi{discontinuous} on
	the subset \\$\gamma_0\coleq\pg{\bs{x} = \pt{0,p},\
	p\in{\ITu}}\in \IT$: 
	two points close to this border,
	\mbox{$A\coleq\pt{\varepsilon,p}$} and
	$B\coleq\pt{1-\varepsilon,p}$,
	have images that differ, in the $\varepsilon \rightarrow 0$ limit, by a
	vector $d^{(1)}_{S_{\alpha}^{\phantom{-1}}}(A,B)=\pt{\alpha,\alpha} \pmod{1}$.
\item[\mdseries ii.] Inverse Sawtooth Maps $\{S_\alpha^{-1}\}$ are
	\enfasi{discontinuous} on
	the subset \\$\gamma_{-1}\coleq S_\alpha\pt{\gamma_0} =
	\pg{\bs{x} = \pt{p,p},\
	p\in{\ITu}}\in \IT$:
	two points close to this border,
	\mbox{$A\coleq\pt{p+\varepsilon,p-\varepsilon}$} and
	$B\coleq\pt{p-\varepsilon,p+\varepsilon}$,
	have images that differ, in the $\varepsilon \rightarrow 0$ limit, by a
	vector $d^{(1)}_{S_{\alpha}^{-1}}(A,B)=\pt{0,\alpha} \pmod{1}$.
\item[\mdseries iii.] The maps $T_\alpha$ and $T_\alpha^{-1}$ are
	\enfasi{continuous}: \\
	$\alpha\in\IZ \Longrightarrow
	d^{(1)}_{T_{\alpha}^{\phantom{-1}}}(A,B) =
	d^{(1)}_{T_{\alpha}^{-1}}(A,B) = \pt{0,0} \pmod{1}$.  \\
	Also, all $T\in{\text{SL}}_2\left(\IT\right)$ are continuous.
\item[\mdseries iv.] The eigenvalues of 
	$\pt{\begin{smallmatrix} 1+ \alpha  & 1\\ \alpha & 1 
	\end{smallmatrix}}$ are  
 	$\pt{\alpha+2\pm\sqrt{(\alpha+2)^2-4}}\Big/2$.
	They are complex conjugates 
	if
	$\alpha\in\pq{-4,0}$, while one eigenvalue $\lambda>1$ and the
	other $\lambda^{-1}<1$ if 
	$\alpha\not\in\pq{-4,0}$.	
\item[\mdseries v.] For a generic 
	matrix $T\in{\text{SL}}_2\left(\IT\right)$, denoting
	$t=\Tr\pt{T}/2$, the eigenvalues 
	are $(t\pm\sqrt{t^2-1})$.
	They are
	conjugate complex numbers if
	\mbox{$\abs{t}<1$}, while one eigenvalue $\lambda>1$ if 
	$\abs{t}>1$. The latter is our case of interest, indeed we
	will use only \enfasi{hyperbolic}
	$T\in{\text{SL}}_2\left(\IT\right)$, that is $\abs{t}>1$.
\item[\mdseries vi.] When a positive eigenvalue is present, that is
	$\lambda>1$, 
	distances are stretched along the direction of the
	eigenvector $|\bs{e}_+\rangle$,
	$S_\alpha|\bs{e}_+\rangle=\lambda|\bs{e}_+\rangle$,
	contracted along 
	that of $|\bs{e}_-\rangle$, $S_\alpha|\bs{e}_-\rangle=
	\lambda^{-1}|\bs{e}_-\rangle$. In this case, we can see
	in $\log\lambda$, the (positive) Lyapounov exponent
	(compare~\eqref{Lyap1}). 	
\item[\mdseries vii.] The Lebesgue measure in~\eqref{AoDC_1c} is
	$S_\alpha^{-1}$--invariant. 
\end{Ventry}
\end{NNS}\newpage
\begin{NON}{}\ \label{STdot}\\
Let $S_\alpha$ be the matrix $\pt{\begin{smallmatrix} 1+ \alpha  & 1\\
\alpha & 1 \end{smallmatrix}}$. Then the expression
$S_\alpha\pt{\bs{x}}$ will denote the action represented
by~\eqref{AoDC_1b}, whereas $S_\alpha\cdot \bs{x}$ will denote the
matrix action of $S_\alpha$ on the vector $\bs{x}$.

When the dynamics arises from the action of a UMG map (so, in
particular, when ${\left\{T_{\alpha}\right\}}_{\alpha\in\IZ}$
is the family of toral automorphisms), the equation~\eqref{AoDC_1b} assumes
the simpler form 
$T_\alpha \pt{\bs{x}}= T_\alpha\cdot \bs{x}
\ \pmod{1}$.

Analogously, expression like $T_\alpha \cdot\bs{x},\
T_\alpha^{\text{tr}} \cdot\bs{x},\ T_\alpha^{-1} \cdot\bs{x}$
and $\pt{T_\alpha^{\text{tr}}}^{-1}\cdot\bs{x}$, will denote action by
$T_\alpha$ itself, its transposed, its
inverse and the inverse of the transposed, respectively.
\end{NON}
\end{quote}
\noindent
\subsection{Algebraic description for the classical dynamical systems
of toral automorphisms}\label{AdoT}\vspace{3mm} 
In this Section we make use of the two commutative algebras introduced in
Section~\ref{TuAt} in order to describe the two family
of toral automorphisms defined up to now.

For the (continuous) automorphisms in the $\pg{T_\alpha}$ family, a
convenient 
algebra~\cite{DeB98:1,DeB01:1} of observables is the C*--algebra
$\Cspace{0}{\cal X}$, equipped with the uniform norm given  
in~\eqref{uninorm}. 

The discrete--time dynamics generates automorphisms $\Theta_{\alpha}$
and its iterates
$\Theta_{\alpha}^{j}: \Cspace{0}{\cal X}\mapsto\Cspace{0}{\cal X} $ as
follows: 
\begin{equation}
\Theta_{\alpha}^j\pt{f}\pt{\bs{x}}\coleq f(T_{\alpha}^j \cdot \bs{x})\
, \quad j\in \IZ\ ,\ \alpha\in\IZ\ \cdot
\label{AoDCT_3}
\end{equation}
They preserve the state
$\omega_\mu\circ\Theta_{\alpha}^j=\omega_\mu$. 

Due to the discontinuity of Sawtooth Maps, the maps $\Theta_{\alpha}^j$ in
equation~\eqref{AoDCT_3}, with
$\alpha\in\IR\setminus\IZ$,  are no longer automorphisms of
$\Cspace{0}{\cal X}$. 

For this reason, in order to deal with
Sawtooth maps, we will make use of the (Von~Neumann) algebra
$\Lspace{\infty}{\cal X}$ of essentially bounded functions defined in 
Section~\ref{TuAt}.

We define
$\Theta_{\alpha}^{j}: \Lspace{\infty}{\cal X}\mapsto
\Lspace{\infty}{\cal X}$ by  
\begin{equation}
\Theta_{\alpha}^j\pt{f}\pt{\bs{x}}\coleq f(S_{\alpha}^j \pt{\bs{x}})\ , \quad j\in \IZ\ ,\ \alpha\in\IR\ ,
\label{AoDCT_3mod}
\end{equation}
These maps are now automorphisms of $\Lspace{\infty}{\cal X}$ and leave the
state $\omega_\mu$ invariant. 

Even if the maps $T$ belonging to the UMG are continuous, 
when dealing with quantized UMG
it is preferable to make use of the Von~Neumann algebra
$\Lspace{\infty}{\cal X}$, 
in conjunction with the state $\om$ and another automorphism, different
from~(\ref{AoDCT_3}--\ref{AoDCT_3mod}), given by: 
\begin{equation}
\Pi : \Lspace{\infty}{\cal X}\ni f\pt{\bs{x}}\mapsto
\Pi\pt{f}\pt{\bs{x}}\coleq f(T^{-1} \cdot \bs{x})\in
\Lspace{\infty}{\cal X}\ \cdot
\label{AoDCT_3mod2}
\end{equation}
The maps $\Pi^j$ are measure preserving automorphisms on
$\Lspace{\infty}{\cal X}$. 
The following definitions are thus justified:\\[-2ex]
\begin{quote}
\begin{DDS}{}\ \\[-5.5ex]
\begin{Ventry}{}\label{defalgST}
	\item[--] The triplets describing UMG automorphisms will
	be chosen between either $\tripAT$ or $\tripATa$.
	\item[--] Sawtooth Maps will
	be identified by triplets
	$\tripASa$. 
\end{Ventry}
\end{DDS}
\end{quote}
\section{Quantum Automorphisms on the Torus}\label{}\vspace{6mm}
\subsection{Dynamical evolution of the Weyl operators}\label{DEoWO}\vspace{3mm} 
In the quantization of dynamical systems $\tripAT$ and the study of
their classical limit, the main
role is played by the evolution of the Weyl operators. Indeed, from
the time evolution in the Weyl pictures, one easily goes to the 
anti--Wick scheme realized by means of C.S.
of the form~\eqref{coh1}.

We will derive
this evolution basing our analysis on the Weyl scheme, described in
Section~\ref{WQ}, and then we will extend such kind of evolution to
the Anti--Wick scheme, that will be used in the rest of our work. Of
course in Section~\ref{CLfQCM} we will provide a proof that our definition of
quantum dynamics is ``well posed'', in the sense that it leads to a
well defined classical limit.   

We introduce our evolution operator $\Theta_N^j : {\cal M}_N \to {\cal
M}_N$ by giving this requirement: 
\begin{equation}
\forall \,f\in {\cal A}_{\IT},\qquad \Theta_N^j \pt{W_{N,\infty}\pt{f}} =
W_{N,\infty}\pt{\Pi^j \pt{f}} \label{Weylevol}\ ,
\end{equation} 
where once more we suppose ${\cal A}_{\IT}$  to consist of functions
sufficiently 
smooth and regular, namely to be Fourier decomposed, $\Pi$ is the
(measure preserving) automorphism on
$\Lspace{\infty}{\cal X}$ given in~\eqref{AoDCT_3mod2}, 
and $W_{N,\infty}$ is the Weyl quantization operator,
whose definition~\eqref{WeylQ} leads to:
\begin{align}
\Theta_N^j \pt{W_{N,\infty}\pt{f}}
& = \sum_{\bs{n} \in {\IZ}^2} \hat{f}_{\bs{n}}
\;\Theta_N^j \pt{W_N(\bs{n})}\label{thetaNderiv1}\ \cdot\displaybreak
\intertext{Conversely, from~\eqref{Weylevol} and~\eqref{AoDCT_3mod2}, we have}
\Theta_N^j \pt{W_{N,\infty}\pt{f}} 
& = W_{N,\infty}\pt{\Pi^j \pt{f}} \label{thetaNderiv2}\\
& = W_{N,\infty}\pt{f\circ T^{-j}}\label{thetaNderiv3}\\
& =\sum_{\bs{n} \in {\IZ}^2} 
W_N(\bs{n})\;\widehat{\pt{f\circ T^{-j}}}_{\bs{n}}\label{thetaNderiv4}\\
& =\sum_{\bs{n} \in {\IZ}^2} 
W_N(\bs{n})\;\int\!\!\!\!\!\int_{\IT} \ud \mu
(\bs{x}) \; f (T^{-j}\cdot\bs{x}) \; 
e^{-2 \pi i\,\sigma(\bs{n}\:,\:\bs{x})}\label{thetaNderiv5}
\intertext{and changing variable $\bs{y}\coleq T^{-j}\cdot\bs{x}$
(note that $T\in{\text{SL}}_2\left(\IT\right)$, so
$\det\pt{T^{-j}}=1$)} 
& =\sum_{\bs{n} \in {\IZ}^2} 
W_N(\bs{n})\;\int\!\!\!\!\!\int_{\IT} \ud \mu
(\bs{y}) \; f (\bs{y}) \; 
e^{-2 \pi i\,\sigma(\bs{n}\:,\:T^{j}\cdot\bs{y})}\label{thetaNderiv6}\\
& =\sum_{\bs{n} \in {\IZ}^2} 
W_N(\bs{n})\;\int\!\!\!\!\!\int_{\IT} \ud \mu
(\bs{y}) \; f (\bs{y}) \; 
e^{-2 \pi i\,\sigma(T^{-j}\cdot\bs{n}\:,\:\bs{y})}\label{thetaNderiv7}
\intertext{(indeed the simplectic form $\sigma(\:\cdot\:,\:\cdot\:)$
is ${\text{SL}}_2\left(\IT\right)$-invariant,
i.e. $\sigma(T\bs{p},T\bs{q})=\sigma(\bs{p},\bs{p})$)}    
& =\sum_{\bs{n} \in {\IZ}^2} 
\widehat{f}_{T^{-j}\cdot\bs{n}}
\;W_N(\bs{n})
\label{thetaNderiv8}\\
& =\sum_{T^{j}\cdot \bs{m} \in {\IZ}^2} 
\widehat{f}_{\bs{m}}
\;W_N(T^{j}\cdot\bs{m})
\label{thetaNderiv9}\\
& =\sum_{\bs{m} \in {\IZ}^2} 
\widehat{f}_{\bs{m}}
\;W_N(T^{j}\cdot\bs{m})
\label{thetaNderiv10}
\end{align}
where in the latter equality we have used the fact that
matrices belonging to ${\text{SL}}_2\left(\IT\right)$ map ${\IZ}^2$
onto itself. Comparing~\eqref{thetaNderiv1} and~\eqref{thetaNderiv10}
we obtain the result
\begin{equation}
\Theta_N^j \pt{W_N(\bs{n})}=W_N\pt{T^{j}\cdot\bs{n}}\label{Weylevolsing}\ ,
\end{equation} 
that will be taken as a definition also in the anti--Wick scheme.

In order to describe the quantum dynamical system during its
(discrete) temporal evolution, we need to have the evolution unitarily
implemented on the Weyl algebra, that is $\Theta_N^j
\pt{W_N(\bs{n})}=U_T\,W_N(\bs{n})\,U_T^*$, 
with $U_T$ unitary operator on ${\cal H}_N$. In other words, the
representation generated by the two generators
$W_N(\bs{\hat{e}}_1)$ and $W_N(\bs{\hat{e}}_2)$, and the
one generated by $\Theta_N
\pt{W_N(\bs{\hat{e}}_1)}={\pq{W_N(T\cdot\bs{\hat{e}}_1)}}$ and $\Theta_N
\pt{W_N(\bs{\hat{e}}_2)}={\pq{W_N(T\cdot\bs{\hat{e}}_2)}}$, has to be
unitarily equivalent.
If this is obtained, by
point (b) of Theorem~\ref{classification} in Remark~\ref{unitarity},
we see that the two representations have to be labeled by the same
$u$ and $v$. Therefore:
\begin{equation}
{\pq{W_N(\bs{\hat{e}}_1)}}^N={\pq{W_N(T\cdot\bs{\hat{e}}_1)}}^N
\qquad\text{and}\qquad
{\pq{W_N(\bs{\hat{e}}_2)}}^N={\pq{W_N(T\cdot\bs{\hat{e}}_2)}}^N
\label{UVcond}\ \cdot 
\end{equation} 
The latter request restrict the possible couples $(u,v)$ available, as
it is showed in the next~\cite{Deg93:1}\\[-2ex]
\begin{quote}
\begin{TT}[Degli Esposti]{:} \label{rightUV}
Let $T=\pt{\begin{smallmatrix} a & b\\ c & d
\end{smallmatrix}}\in{\text{SL}}_2\left(\IT\right)$ and $N$ be
positive integer. Then~\eqref{UVcond} can be fulfilled, more
precisely:

for any given automorphism $T$, all admissible representations are
labeled by all $(u,v)\in\IT$ such that
\begin{equation}
\pt{T^{\text{tr}} - \Id}
\begin{pmatrix}
u\\
v
\end{pmatrix} =
\frac{1}{2}N
\begin{pmatrix}
ac\\
bd
\end{pmatrix}
+
\begin{pmatrix}
m_1\\
m_2
\end{pmatrix}
,\quad m_1,m_2\in\IZ
\label{UVcond1}
\end{equation}
\end{TT}
\end{quote}
Equation~\eqref{UVcond1} can be trivially solved\footnote{
Note that, in equation~\eqref{UVcond2}, $\Tr\pt{T}\neq 2$
because of the choice in Remark~\ref{Rem_21}.v.}, and we get for the
couples $(u,v)$:
\begin{equation}
\begin{pmatrix}
u\\
v
\end{pmatrix} =\frac{1}{\Tr\pt{T}-2}
\begin{pmatrix}
1-d & c\\
b & 1-a
\end{pmatrix} 
\pq{\bs{p}+
\begin{pmatrix}
m_1\\
m_2
\end{pmatrix}}\pmod{1},\quad m_1,m_2\in\IZ
\label{UVcond2}
\end{equation}
where $\bs{p}=\bs{0}$ for even $N$, whereas for odd $N$ it depends from the
parity of the elements of the matrix $T$. In particular, in agreement
with the condition $\det\pt{T}=1$, we have three allowed sets of
matrices $T$, that are:
\begin{gather}
{\cal N}_1\coleq\pg{\begin{pmatrix}
\text{\scriptsize even} & \text{\scriptsize odd} \\
\text{\scriptsize odd}  & \text{\scriptsize even}
\end{pmatrix},\begin{pmatrix}
\text{\scriptsize odd}  & \text{\scriptsize even}\\
\text{\scriptsize even} & \text{\scriptsize odd} 
\end{pmatrix}}\qquad,\qquad
{\cal N}_2\coleq\pg{\begin{pmatrix}
\text{\scriptsize even} & \text{\scriptsize odd} \\
\text{\scriptsize odd}  & \text{\scriptsize odd}
\end{pmatrix},\begin{pmatrix}
\text{\scriptsize odd}  & \text{\scriptsize even}\\
\text{\scriptsize odd} & \text{\scriptsize odd} 
\end{pmatrix}}\qquad,\notag\\
{\cal N}_3\coleq\pg{\begin{pmatrix}
\text{\scriptsize odd} & \text{\scriptsize odd} \\
\text{\scriptsize odd}  & \text{\scriptsize even}
\end{pmatrix},\begin{pmatrix}
\text{\scriptsize odd}  & \text{\scriptsize odd}\\
\text{\scriptsize even} & \text{\scriptsize odd} 
\end{pmatrix}}\notag
\end{gather}
whose corresponding vector $\bs{p}$ are $\bs{p}\pt{{\cal N}_1}=\bs{0}$,
$\bs{p}\pt{{\cal N}_2}=\frac{1}{2}\,\bs{\hat{e}}_2$ and
$\bs{p}\pt{{\cal N}_3}=\frac{1}{2}\,\bs{\hat{e}}_1$.

The set ${\cal N}_1$, whose corresponding (unique) couple $\pt{u,v}$
is $\pt{0,0}$, is also important for historical reasons: indeed this
set of matrices was used to develop the first quantization of Cat Maps 
obtained by Berry and Hannay~\cite{Ber80:1}\footnote{For recent
developments of the Berry's approach to the quantization of Cat Maps,
see~\cite{Kea91:1,Kea00:1,Mez02:1}.}. 

We end up this Section by noting that unitarity of dynamics guarantees:
\begin{equation}
\tau_N(W_N(T\cdot\bs{n})) = \tau_N(W_N(\bi n))\ \cdot
\label{Weyl7} 
\end{equation}
\section{Discretization of Sawtooth Map families}\label{}\vspace{6mm}
\subsection{Dynamical evolution on ${\cal D}_N$ arising from the
$\pg{T_\alpha}$ subfamily of UMG}\label{DETa}\vspace{3mm}   
In this Section we will parallel what we have done in
Section~\ref{DEoWO} in the framework where the dynamical
evolution is
dictated by the $\pg{T_\alpha}$ subfamily of UMG (see Definitions
~\ref{def_SeT} and ~\ref{defalgST}), namely we will use
the C*--algebra $\Cspace{0}{\IT}$
to describe the continuous system
and we
develop a technique of discretization/dediscretization presented in
Section~\ref{WDGriglIA}, respectively Section~\ref{DDW}.

The calculations follow (\ref{Weylevol}--\ref{Weylevolsing}): starting from the request
\begin{equation}
\forall \,f\in \Cspace{0}{\IT},\qquad \widetilde{\Theta}_{N,\alpha}^j \pt{{\cal
J}_{N,\infty}\pt{f}} = 
{\cal J}_{N,\infty}\pt{\Theta_\alpha^j \pt{f}} \label{WeylevolTa}\ ,
\end{equation} 
we get
\begin{equation}
\widetilde{\Theta}_{N,\alpha}^j \pt{W_N(\bs{n})}=W_N\pt{{\pt{T_\alpha^{\text{tr}}
}}^j\cdot\bs{n}}\label{WeylevolsingTa}\ \cdot
\end{equation} 
Few comment are now in order: in Section~\ref{DEoWO}
we used Weyl 
operators defined by means of the symplectic form
$\sigma\pt{\bs{n}\,,\,\bs{x}}$
(see~(\ref{WeylPQ}--\ref{WeylQ})), which is invariant under the
action of $T\in{\text{SL}}_2\left(\IT\right)$, that is
$\sigma\pt{T\cdot\bs{n}\,,\,T\cdot\bs{x}}
=\sigma\pt{\bs{n}\,,\,\bs{x}}$. 
Therefore the dynamics $\bs{x}\mapsto
T^{-j} \cdot \bs{x}$ defined in~\eqref{AoDCT_3mod2} actually affects
the indices of Weyl operators in the
in the sense that $\bs{n}\mapsto
T^{j} \cdot \bs{n}$ (indeed
$\sigma\pt{\bs{n}\,,\,T^{-j}\cdot\bs{x}} =
\sigma\pt{T^{j}\cdot\bs{n}\,,\,\cdot\bs{x}})$.

In the present case, Weyl 
operators are defined using not the symplectic form but 
scalar
product $\bkk{\bs{n}}{\bs{x}}=\bs{n}\cdot\bs{x}$ instead, so that
$\Big\langle\bs{n}\Big\vert T_\alpha^j\cdot\bs{x}\Big\rangle=
\bkk{{\pt{T_\alpha^{\text{tr}}
}}^j\cdot\bs{n}}{\bs{x}}$, whence~\eqref{WeylevolsingTa}.

Although we defined the dynamics on Weyl operators, neither
the discretizing operator ${\cal J}_{N ,\infty}$ in Definition~\ref{RoeiW_51}
nor the dediscretizing one ${\cal J}_{\infty,N}$ in
Definition~\ref{RoeiW_51b} do depend 
(explicitly) on Weyl operators. For this reason we introduce now
the operator $\widetilde{\Theta}_{N,\alpha}$ on ${\cal D}_N$ in a way
compatible 
with~\eqref{WeylevolsingTa}. To this aim we introduce first a new
family of maps on the torus $\IT\pt{{[0,N)}^2}$, namely ${[0,N)}^2 \pmod{N}$,
given by\footnote{%
Although we are now dealing with the $\pg{T_\alpha}$ family of maps,
definition~\eqref{Ualpha} is formulated for Sawtooth maps; when
$\alpha\in\IZ$,~\eqref{Ualpha} obviously simplify to direct matrix
action, which is not true when $\alpha\in\IR\setminus\IZ$, which will
be the case later on.}\newpage
\begin{subequations}\label{Ualpha}
\begin{alignat}{2}
\IT\pt{{[0,N)}^2} \ni\bs{x}&&\ \mapsto 
U_\alpha^0\pt{\bs{x}} &\coleq 
\bs{x}
\notag\\
&&&= N\,S_\alpha^0\pt{\frac{\bs{x}}{N}}
\in\IT\pt{{[0,N)}^2}\ ,\label{Ualphaa}\\
\IT\pt{{[0,N)}^2} \ni\bs{x}&&\ \mapsto 
U_\alpha^{\pm 1}\pt{\bs{x}} &\coleq N\,
S_\alpha^{\pm 1}\pt{\frac{\bs{x}}{N}}
\in\IT\pt{{[0,N)}^2}\ ,\label{Ualphab}\\
\IT\pt{{[0,N)}^2} \ni\bs{x}&&\ \mapsto 
U_{\alpha}^{\pm j}\pt{\bs{x}}&\coleq
\underbrace{
U_{\alpha}^{\pm 1}(\,
U_{\alpha}^{\pm 1}(\,
\cdots
U_{\alpha}^{\pm 1}(\,
U_{\alpha}^{\pm 1}(
}_{j\ \text{times}}
\bs{x}\,)\,)\cdots \,)\,)\ ,
\quad j\in\IN\ ,\notag\\
&&&= N\,
S_\alpha^{\pm j}\pt{\frac{\bs{x}}{N}}
\in\IT\pt{{[0,N)}^2}\ \cdot\label{Ualphac}
\end{alignat}
\end{subequations}
Using the latter set of maps, we can give the following\\[-2ex]
\begin{quote}
\begin{DDD}{}\ \\[-5.5ex]\label{ThetaNTa}
	\item[$\widetilde{\Theta}_{N,\alpha}$] is the *automorphism of
	${\cal D}_N$ defined by:
	\begin{equation}
	{\cal D}_N\ni X \mapsto
	\widetilde{\Theta}_{N,\alpha}^{\phantom{t}}\pt{X}  
	\coleq\sum_{\bs{\ell} \in {(\ZNZ{N})^2}} 
	X_{U_\alpha\pt{\bs{\ell}},U_\alpha\pt{\bs{\ell}}}
	\ket{\bs{\ell}}\bra{\bs{\ell}}\in{\cal D}_N
	\ \cdot \label{CoAFE_11}
	\end{equation}
\end{DDD}
\begin{NNS}{}\ \\[-5.5ex]\label{def_Uj}
\begin{Ventry}{\mdseries iii.}
\item[\mdseries i.] $\widetilde{\Theta}_{N,\alpha}^{\phantom{t}}$ is a
*automorphism because the map\mbox{
${\pt{\IZ / N \IZ}}^2\ni\bs{\ell}\longmapsto
U_\alpha\pt{\bs{\ell}}\in{\pt{\IZ / N \IZ}}^2$} is a
bijection. For the same reason the state $\tn$ is $\widetilde{\Theta}_{N,\alpha}^{\phantom{t}}$--invariant.
\item[\mdseries ii.] One can check that, given $f\in\Cspace{0}{\IT}$, 
\begin{equation}
\widetilde{\Theta}_{N,\alpha}^{\phantom{t}}\pt{{\cal J}_{N ,
\infty}\pt{f}} \coleq \sum_{\bs{\ell} \in {(\ZNZ{N})^2}} 
f\pt{\frac{U_\alpha\pt{\bs{\ell}}}{N}}
\ket{\bs{\ell}}\bra{\bs{\ell}}\ \cdot
\label{CoAFE_011}
\end{equation}
\item[\mdseries iii.]
Also, $\widetilde{\Theta}_{N,\alpha}^{j}\circ{\cal J}_{N ,
\infty}	= {\cal J}_{N ,\infty}\circ\Theta_\alpha^{j}$ for all $j\in\IZ$.
\end{Ventry}
\end{NNS}
\end{quote}
The automorphism $\widetilde{\Theta}_{N,\alpha}^{\phantom{t}}$ can be
rewritten in the more familiar form
\begin{align}
\widetilde{\Theta}_{N,\alpha}^{\phantom{t}}\pt{X} & 
= \sum_{\bs{\ell} \in {(\ZNZ{N})^2}} 
X_{U_\alpha\pt{\bs{\ell}},U_\alpha\pt{\bs{\ell}}}
\ket{\bs{\ell}}\bra{\bs{\ell}}=\label{UXUT}\\
& = \sum_{U_\alpha^{-1}\pt{\bs{s}} \in {(\ZNZ{N})^2}} 
X_{\bs{s},\bs{s}}
\ket{U_\alpha^{-1}\pt{\bs{s}}}\bra{U_\alpha^{-1}\pt{\bs{s}}}=\displaybreak
\notag\\
\!\!\!\!\!\!\!\!\!{\textstyle\text{(see Remark~\ref{rem_44} i and ii)}}\qquad
& = U_{\alpha,N}^{\phantom{*}}\pt{\sum_{\newatop{\text{all equiv.}}{\text{classes}}} 
X_{\bs{s},\bs{s}}
\ket{\bs{s}}\bra{\bs{s}}}U_{\alpha,N}^{*}= \label{CoAFE_121}\\
& = U_{\alpha,N}^{\phantom{*}}
\;X\;\;
U_{\alpha,N}^{*}\ ,\label{CoAFE_131}
\end{align}
where the operators $U_{\alpha,N}$ defined by
\begin{equation}
{\cal H}_{N^2}\ni\big|\bs{\ell}\big\rangle\longmapsto
U_{\alpha,N}^{\phantom{*}}\big|\bs{\ell}\big\rangle\coleq\ket{
U_\alpha^{-1}\pt{\bs{\ell}}}\ \cdot\label{aggiunta}
\end{equation} 
are unitary (see below).
\\[-2ex]
\begin{quote}	
\begin{NNS}{}\ \label{rem_44}\\[-5.5ex]
\begin{Ventry}{ii)}
	\item[\mdseries i)] All of $T_\alpha$, $T_\alpha^{-1}$,
	$T_\alpha^{\text{tr}}$ and ${\pt{T_\alpha^{-1}}}^{\text{tr}}$ belong to
	$SL_2\pt{\IZ/N\IZ}$; in particular these matrices are
	automorphisms on ${\pt{\IZ/N\IZ}}^2$ (and so $U_\alpha$ and
	$U_\alpha^{-1}$ as well) so that,
	in~\eqref{CoAFE_121}, one can sum over the equivalence classes.
	\item[\mdseries ii)] The same argument as before proves that
	the operators in~\eqref{aggiunta} are unitary, whence
	$\widetilde{\Theta}_{N,\alpha}$ is a 
	*-automorphism of ${\cal D}_N$.
\end{Ventry} 
\end{NNS}
\end{quote}
\noindent
\subsection{Dynamical evolution on ${\cal D}_N$ arising from the
action of Sawtooth maps}\label{DynSMf}\vspace{3mm}   

From a the measure theoretical point of view, the dynamical systems
$\tripCSa$ can be thought as extensions of $\tripCTa$. Both kind of systems
are defined on the same space $\IT$, endowed with the same
measure $\mu$ and the $S_\alpha$ maps reduce to $T_\alpha$ when we restrict
the domain of $\alpha$ from $\IR$ to $\IZ$.

From an algebraic point of view, we note that the algebra describing
classical dynamical systems given by $\tripASa$
is larger than the one of $\tripATa$ as $\Cspace{0}{\cal
X}\subset\Lspace{\infty}{\cal X}$, while the
state $\om$ is the same and 
for the dynamics the same consideration of above holds.

For what concern discretization, while the Weyl scheme can be
used for $\tripATa$, we note that a straightforward
application of the same procedure for $\tripASa$ would
not work. Indeed,
$f\in\Lspace{\infty}{\cal X}$ could be unbounded
on the grid $L_N$, because the grid has null
$\mu$-measure and thus not felt by
the essential norm
$\norm{\cdot}{\infty}$ in~\eqref{AdoUMG_1}.

Moreover, while in~\eqref{WeylevolsingTa} one sees that the (matrix)
action of $T_\alpha$ on
$\bs{x}$ transfers to the index of the Weyl operators, this
property no longer holds in the case of the action of
$S_\alpha$, $\alpha\not\in\IZ$, which is non--matricial.

In the (algebraic) discrete description, we deal with two 
families of ``quantum'' dynamical systems, namely $\tripQTa$ for
$\pg{T_\alpha}$  and
$\big({\cal D}_N,\tau_N,\;\text{{\bf ?}\footnotemark}\;\big)$ for
Sawtooth maps,
\footnotetext{Here the dynamics is not yet set.}
 in which the Abelian
finite dimensional algebra ${\cal 
D}_N$ and the tracial state $\tn$ are the same, for both.

Thus it proves convenient to extend the automorphisms
$\widetilde{\Theta}_{N,\alpha}$ in a way more suited to 
Sawtooth maps, which coincide with Definition~\ref{ThetaNTa} when
$\alpha\in\IZ$. 
To this aim we introduce first new family of maps $V_\alpha$ from the torus
$\IT\pt{{[0,N)}^2}\coleq[0,N)\times[0,N) \pmod{N}$, to $\ZNZn\times\ZNZn$
(where $\ZNZn$ denotes the residual class$\pmod{N}$). These maps are
defined as follows:
\begin{subequations}\label{Valpha}
\begin{alignat}{2}
\IT\pt{{[0,N)}^2} \ni\bs{x}&&\ \mapsto 
V_\alpha^0\pt{\bs{x}} &\coleq 
\floor{\bs{x}}
\notag\\
&&&= \floor{U_\alpha^{0}\pt{\bs{x}}}\in
\ZNZD
\ ,\label{Valphaa}\\
\IT\pt{{[0,N)}^2} \ni\bs{x}&&\ \mapsto 
V_\alpha\pt{\bs{x}} &\coleq 
\floor{U_\alpha\pt{\bs{x}}}\in
\ZNZD
\ ,\label{Valphab}\\
\IT\pt{{[0,N)}^2} \ni\bs{x}&&\ \mapsto 
\:V_{\alpha}^{j}\pt{\bs{x}}&\coleq
\underbrace{
V_{\alpha}(\,
V_{\alpha}(\,
\cdots
V_{\alpha}(\,
V_{\alpha}(
}_{j\ \text{times}}
\bs{x}\,)\,)\cdots \,)\,)\in
\ZNZD\ ,
\quad j\in\IN^{+}\ ,\notag\\
&&&= 
\underbrace{
\lfloor U_{\alpha}(\,
\lfloor U_{\alpha}(\,
\cdots
\lfloor U_{\alpha}(\,
\lfloor U_{\alpha}(
}_{j\ \text{times}}
\bs{x}\,)\rfloor\,)\rfloor\cdots \,)\rfloor\,)\rfloor
\ \cdot\label{Valphac}
\end{alignat}
\end{subequations}
By means of the latter set ${\pg{V_\alpha^j}}_{j\in\IZ}$, we can proceed
to\\[-2ex] 
\begin{quote}
\begin{DDD}{}\ \\[-5.5ex]\label{ThetaNSa}
	\item[$\widetilde{\Theta}_{N,\alpha}$] is the *-automorphism of
	${\cal D}_N$ defined by:
	\begin{equation}
	{\cal D}_N\ni X \mapsto
	\widetilde{\Theta}_{N,\alpha}^{\phantom{t}}\pt{X}  
	\coleq\sum_{\bs{\ell} \in {(\ZNZ{N})^2}} 
	X_{V_\alpha\pt{\bs{\ell}},V_\alpha\pt{\bs{\ell}}}
	\ket{\bs{\ell}}\bra{\bs{\ell}}\in{\cal D}_N
	\ \cdot \label{CoAFE_11}
	\end{equation}
\end{DDD}
\begin{NNS}{}\ \\[-5.5ex]\label{def_Vj}
\begin{Ventry}{\mdseries iii.}
\item[\mdseries i.] Note that $\widetilde{\Theta}_\alpha^{j}\coleq
\underbrace{\widetilde{\Theta}_\alpha^{\phantom{t}}\circ  
\widetilde{\Theta}_\alpha^{\phantom{t}}\circ\cdots\circ
\widetilde{\Theta}_\alpha^{\phantom{t}}}_{j\ \text{times}}$ is implemented by
$V_{\alpha}^{j}\pt{\bs{\ell}}$ given
in~\eqref{Valphac}. 
\item[\mdseries ii.] $\widetilde{\Theta}_{N,\alpha}^{\phantom{t}}$ is a
*-automorphism because the map\mbox{
${\pt{\IZ / N \IZ}}^2\ni\bs{\ell}\longmapsto
V_\alpha\pt{\bs{\ell}}\in{\pt{\IZ / N \IZ}}^2$} is a
bijection. For the same reason the state $\tn$ is $\widetilde{\Theta}_{N,\alpha}^{\phantom{t}}$--invariant and $V_\alpha$ is invertible too.
\item[\mdseries iii.] When $\alpha\in\IZ$,
${\pt{\IZ / N \IZ}}^2\ni\bs{\ell}\longmapsto V_\alpha\pt{\bs{\ell}} =
T_\alpha \bs{\ell}\in{\pt{\IZ / N \IZ}}^2$, namely the action of the
map $V_{\alpha}$ becomes 
that of a matrix$\pmod{N}$. Moreover, in that case, $U_\alpha$ and
$V_\alpha$ coincide and Definitions~\ref{ThetaNTa}
and~\ref{ThetaNSa} so do.
\end{Ventry}
\end{NNS}
\end{quote}
A the same argument as in~(\ref{UXUT}--\ref{CoAFE_131}), 
proves that the automorphism
$\widetilde{\Theta}_{N,\alpha}^{\phantom{t}}$ is unitarily implemented,
\begin{align}
\widetilde{\Theta}_{N,\alpha}^{\phantom{t}} & =
U_{\alpha,N}^{{\prime}\phantom{*}} 
\;X\;\;
U_{\alpha,N}^{{\prime}*}\ ,\label{CoAFE_131Sa}
\intertext{with $U_{\alpha,N}^{\prime}$ unitary and given by}
{\cal H}_{N^2}\ni\big|\bs{\ell}\big\rangle&\longmapsto
U_{\alpha,N}^{{\prime}\phantom{*}}\big|\bs{\ell}\big\rangle\coleq\ket{
V_\alpha^{-1}\pt{\bs{\ell}}}\ \label{aggiuntaSa}
\end{align}
(note that Remark~\ref{def_Vj}.ii. allows us to use $V_\alpha^{-1}$).
\section{Classical/Continuous limit of the
dynamics}\label{CCLD}\vspace{6mm}
One of the main issues in the semi-classical analysis is to compare
if and how the quantum and classical time evolutions mimic each other
when a quantization parameter goes to zero.

In the case of classically chaotic quantum systems, the situation is
strikingly different from the case of classically integrable quantum
systems. In the former case, classical and quantum mechanics agree on
the level of coherent states only over times which scale as
$-\!\log\hbar$. 

As we shall see, such kind of scaling is not related with non
commutativity. The quantization--like procedure we developed until
now, depending on the lattice spacing $\frac{1}{N}$, has been set with
the purpose to exhibit such a behavior also in the classical limit
(actually continuous limit) connecting functional Abelian algebras
${\cal A}_{\cal X}$ with Abelian (finite dimensional) ones, consisting
of diagonal matrices. 

\subsection{A useful localization property leading to a well defined
classical limit}\label{}\vspace{3mm} 
As before, let $T$ denote the evolution on the
classical phase space $\c X$ and $U_T$ the unitary single step
evolution on $\Cx^N$ which represent its ``quantization''. We formally 
state the semi--classical correspondence of
classical and quantum evolution using coherent states:\ \\[-2ex] 
\newpage
\begin{quote}
\begin{CON}[Dynamical localization]{}\ \\\label{dynloc}
\noindent \!\!There exists an $\alpha>0$ such that for all choices of
 $\varepsilon>0$ and $d_0>0$ there exists an $N_0\in\Nl$ with the
 following property: if  $N > N_0$ and $k\le \alpha \log N$, then $N
 |\< C_N(\bs{x}), U_T^k\,C_N(\bs{y})\>|^2 \le \varepsilon$ whenever $d(T^k\bs{x},\bs{y})
 \ge d_0$.
\end{CON}\ \\[-5ex]
\begin{NNN}{}\   \\
The condition of dynamical localization is what is expected of
a good choice of coherent states, namely, on a time scale
logarithmic in the inverse of the
semi-classical parameter, evolving coherent states should stay localized around the
classical trajectories.
Informally, when $N\to\infty$, the quantities
\begin{equation}
\label{dyn-loc}
K_{k}(\bs{x},\bs{y}) := \< C_N(\bs{x}), U_T^k C_N(\bs{y})\>
\end{equation}
should behave as if
$N|K_k(\bs{x},\bs{y})|^2\simeq\delta(T^k\bs{x}-\bs{y})$ (note that
this hypothesis makes our quantization consistent with 
the notion of \enfasi{regular
quantization} described in Section V of~\cite{Slo94:1}).
The constraint $k\le \alpha\log N$ is typical of hyperbolic classical
behaviour  
and comes heuristically as follows.
The maximal localization of coherent states cannot exceed the minimal
coarse-graining dictated by $1/N$; if, while evolving,
coherent states stayed localized forever around the classical trajectories, 
they would get more and more localized along the contracting direction.
Since for hyperbolic systems the increase of localization is 
exponential with Lyapounov exponent $\log\lambda>0$, this sets 
the upper bound and indicates that $\alpha\simeq1/\log\lambda$.
\end{NNN}
\begin{PPP}{}\ \\
\label{prop3}
 Let $(\c M_N ,\Theta_N,\tau_N)$ be a general quantum
 dynamical system as defined in Section~\ref{QUAPRO} and suppose
 that it satisfies Condition~\ref{dynloc}. Let $\|X\|_2 := \sqrt{
 \tau_N(X^*X)}$, $X\in \c M_N$ denote the normalized
 Hilbert-Schmidt norm. In the ensuing topology 
 \begin{equation}
  \lim_{\substack{k,\ N\to\infty \\k< \alpha\log N}} \| \Theta_N^k
  \circ {\cal J}_{N \infty}(f) - {\cal J}_{N \infty} \circ \Theta^k(f)
  \|_2 = 0.\label{added}
 \end{equation}
\end{PPP}
\end{quote}
\noindent\newpage
\textbf{Proof of Proposition~\ref{prop3}:}\\[2ex]
One computes
\begin{align}
 &\|\Theta_N^k \circ {\cal J}_{N\infty}(f) - {\cal J}_{N\infty} \circ
 \Theta^k(f)\|_2^2 
\nonumber \\
 &= 2 N \int_{\c X} \mu(\ud\bs{x})\, \int_{\c X} \mu(\ud\bs{y})\, \overline{f(\bs{x})}\, f(\bs{y})\, 
 |\< C_N(\bs{x}), C_N(\bs{y}) \>|^2 
\nonumber \\
 &\quad - 2 N \;{\Re}\left[\int_{\c X} \mu(\ud\bs{x})\, \int_{\c X}
 \mu(\ud\bs{y})\,  \overline{f(\bs{y})}\, f(T^k \bs{x})  |\< C_N(\bs{x}), U_T^k C_N(\bs{y}) \>
 |^2 \right].
\end{align}

The double integral in the first term goes to $\int \mu(\ud\bs{x}) |f(\bs{x})|^2$.
So, we need to show that the second integral, which we shall denote by
$I_N(k)$, does the same. We will concentrate on the case of
continuous $f$, the extension to essentially bounded $f$ is straightforward.
Explicitly, selecting a ball $B(T^k \bs{x},d_0)$, one derives
\begin{align*}
 &\left| I_N(k) -
 \int_{\c X} \mu(\ud\bs{y})\, |f(\bs{y})|^2 \right| \\
 &=  \left| \int_{\c X} \mu(\ud\bs{x})\, \int \mu(\ud\bs{y})\, \overline{f(\bs{y})}\,
 \bigl( f(T^k \bs{x}) - f(\bs{y}) \bigr)\, N|\< C_N(\bs{x}), U_T^k C_N(\bs{y})\>|^2
 \right| \\
 & \le  \left| \int_{\c X} \mu(\ud\bs{x})\, \int_{B(T^k \bs{x},d_0)} \mu(\ud\bs{y})\,
 \overline{f(\bs{y})} \bigl(f(T^k \bs{x}) - f(\bs{y})\bigr) N|\< C_N(\bs{x}),
 U_T^k C_N(\bs{y})\>|^2 \right| \\
 &+ \left| \int_{\c X} \mu(\ud\bs{x})
 \int_{\c X\setminus B(T^k \bs{x},d_0)} \mu(\ud\bs{y})
 \overline{f(\bs{y})}\bigl(f(T^k \bs{x}) - f(\bs{y})\bigr) N|\< C_N(\bs{x}),
 U_T^k C_N(\bs{y})\> |^2 \right|.
\end{align*}
Applying the mean value theorem and approximating the integral of the
kernel as in the proof of Proposition~\ref{prop1}, we get that 
$\exists \:\bs{c} \in B(T^k \bs{x}, d_0)$ such that
\begin{align*}
 &\left| I_N(k) -
 \int_{\c X} \mu(\ud\bs{y})\, |f(\bs{y})|^2 \right|\\
 &\le \left| \int_{\c X} \mu(\ud\bs{x})\, 
 \overline{f(\bs{c})}\, \bigl(f(T^k \bs{x}) -
 f(\bs{c})\bigr)\, \int_{B(T^k \bs{x},d_0)} \mu(\ud\bs{y})\,N|\< C_N(\bs{x}), U_T^k C_N(\bs{y})\>|^2 \right| \\
 &+ \left| \int_{\c X} \mu(\ud\bs{x}) \int_{\c X\setminus B(T^k \bs{x},d_0)} 
 \mu(\ud\bs{y})
 \overline{f(\bs{y})} \bigl(f(T^k \bs{x}) - f(\bs{y})\bigr) N|\<  C_N(\bs{x}),
 U_T^k C_N(\bs{y})\>|^2 \right|
\intertext{and using property~(\ref{coh}.3)}
 &\le \left| \int_{\c X} \mu(\ud\bs{x})\, 
 \overline{f(\bs{c})}\, \bigl(f(T^k \bs{x}) -
 f(\bs{c})\bigr)\right| \\
 &+ \left| \int_{\c X} \mu(\ud\bs{x}) \int_{\c X\setminus B(T^k \bs{x},d_0)} 
 \mu(\ud\bs{y})
 \overline{f(\bs{y})} \bigl(f(T^k \bs{x}) - f(\bs{y})\bigr) N|\<  C_N(\bs{x}),
 U_T^k C_N(\bs{y})\>|^2 \right|\ \cdot
\end{align*}
By uniform continuity we can bound the first term by some arbitrary
small $\varepsilon$, provided  we choose $d_0$ small enough. Now, for
the second integral we use our localization condition~\ref{dynloc}. 
As the constraint $k \le \alpha \log N$ has to be enforced, we
have to take a joint  limit of time and size of the system with this
constraint. In that case the second integral can  also be bounded by
an arbitrarily small $\varepsilon'$, provided $N$ is large
enough. \hfill$\qed$
\subsection{Classical Limit for Quantum Cat Maps}\label{CLfQCM}\vspace{3mm}   
We shall not prove the dynamical localization condition~\ref{dynloc} for 
the quantum cat maps but instead provide a direct derivation of 
formula~\eqref{added} based on the simple expression~(\ref{Weylevolsing})
of the dynamics when acting on Weyl operators.  
For this reason, we remind the reader the Weyl quantization
operator $W_{N,\infty}$, already introduced in~\eqref{WeylQ}, together
with some other useful tools.\\[-2ex]
\begin{quote}
\begin{DDS}{}\ \\\label{Weyl-q}
 Let $f$ be a function in ${\cal A}_{\IT}\subset\Lspace{\infty}{\IT}$,
 ${\cal A}_{\IT}$ denoting 
 the subset of $\Lspace{\infty}{\IT}$ characterized by functions whose
 Fourier series   
 $\hat f$ have only finitely many non-zero terms. We shall denote by
 $\r{Supp}(\hat f)$ the support of $\hat f$ in $\IZ^2$. Then, in the
 Weyl quantization scheme, one associates to $f$ the $N\times N$
 matrix
\begin{equation*}
W_{N,\infty} : {\cal A}_{\IT} \ni f \longmapsto 
W_{N,\infty}\pt{f} = \sum_{\bs{k} \in {\r{Supp}(\hat f)}} \hat{f}_{\bs{k}}
\;W_N(\bs{k}) \in{\cal M}_N\ .
\end{equation*}
\end{DDS}
\end{quote}
\noindent
Our aim is to prove: \\[-2ex]
\begin{quote}
\begin{PPP}{}\ \\ \label{prop3cat}
 Let $\tripQT$ be a sequence of quantum cat maps
 tending with $N\to\infty$ to a classical cat map with Lyapounov exponent
 $\log\lambda$; then
 \begin{equation*}
  \lim_{\substack{k,\ N\to\infty \\k< \log N/(2\log\lambda)}} \|
  \Theta_N^k \circ {\cal J}_{N \infty}(f) - {\cal J}_{N \infty} \circ
  \Theta^k(f) \|_2 = 0\quad ,
  \end{equation*}
where $\|\,\cdot\,\|_2$ is the Hilbert-Schmidt norm of Proposition~\eqref{prop3}
\end{PPP}
\end{quote}
\noindent
First we prove an auxiliary result.\\[-2.5ex]\newpage
\begin{quote}
\begin{LLL}{}\ \label{Lemma0}\\[2ex] 
 If $\bi n=(n_1,n_2)\in\Ir^2$ is such that $0\le n_i\le N-1$ and
 $\lim_N  \frac{n_i}{\sqrt{N-1}}=0$, then the expectation of Weyl
 operators $W_N(\bi n)$ with respect to the state $| C_N \>$ given
 in~\eqref{coh2} is such that
 \begin{equation*}
  \lim_{N\to\infty} \< C_N, W_N(\bi n)\, C_N \> = 1\ \cdot
 \end{equation*}
\end{LLL}
\end{quote}
\noindent
\textbf{Proof of Lemma~\ref{Lemma0}:}\\[2ex]
The idea of the proof is to use the fact that, for large $N$, the
binomial coefficients $\binom{N-1}{j}$ contribute to the binomial sum
only when $j$ stays within a neighborhood of $(N-1)/2$ of width
$\simeq \sqrt{N}$, in which case they can be approximated by a
normalized Gaussian function. We also notice that, by expanding the
exponents in the bounds~(\ref{loc7a})  and~(\ref{loc8}), the
exponential decay fails only if $n_{1,2}$ grow with $N$ slower 
than $\sqrt{N}$, which is surely the case for fixed finite $n$, whereby
it also follows that we can disregard the second term
in the sum comprising the contributions~(\ref{loc2}). We then write
the $j$'s in the binomial coefficients as
\begin{equation*}
 j = \floor{ \frac{N-1}{2}} + k = \frac{N-1}{2} + k - \alpha, \quad
 \alpha\in\pg{0,{\frac{1}{2}}},
\end{equation*}
and consider only $k = \r O(\sqrt N)$. Stirling's formula~\cite{Fel68:1}
\begin{equation*}
 L! = L^{L+1/2}\, \r e^{-L}\, \sqrt{2\pi}\,  \Bigl(1 + \r
 O(L^{-1})\Bigr),
\end{equation*}
allows us to rewrite the first term in the r.h.s.\ of~(\ref{loc2}) as 
\begin{align}
 &\exp{\left(-\frac{1}{2}\frac{n_1^2}{N-1}\right)}\
 \sum_{k=-\left[\frac{N-1}{2}\right]}^{N-1-\left[\frac{N-1}{2}\right]+n_1}
 \frac{2\,\r e^{\frac{2\pi i}{N}n_2
 \left(k+\left[\frac{N-1}{2}\right]\right)}}{\sqrt{2\pi(N-1)}}
\nonumber \\
 &\quad\times \exp{\left(-\frac{2(k-\alpha+\frac{n_1}{2})^2}{N-1}\right)}
 \Bigl(1 + \r O(N^{-1}) + \r O((k+n_1)^3\, N^{-2}) \Bigr).
\label{lemma1.1}
\end{align}
For any fixed, finite $\bi n$, both the sum and the factor in front tend to
1, the sum becoming the integral of a normalized Gaussian.\hfill$\qed$\\[3ex]
{\setlength{\parindent}{0pt}  
\textbf{Proof of Proposition~\ref{prop3cat}:}\\[2ex] 
Given $f\in\Lspace{\infty}{\IT}$ and $\varepsilon>0$, we choose $N_0$
such that the Fourier approximation $f_\varepsilon$ of $f$ with
$\#(\r{Supp}(\hat f)) = N_0$ is such that $\|f-f_\varepsilon\|
\le\varepsilon$, where $\|\cdot\|$ denotes the usual Hilbert space norm.
Next, we estimate
\begin{align*}
 I_N(f)
 &:= \bigl\| \Theta_N^k\circ{\cal J}_{N\infty}(f) -
 {\cal J}_{\infty N}\circ\Theta^k(f) \bigr\|_2 \\
 &\le \bigl\| \Theta_N^k\circ{\cal J}_{N\infty}(f-f_\varepsilon)
 \bigr\|_2 + \bigl\| {\cal J}_{N\infty}\circ\Theta^k(f-f_\varepsilon)
 \bigr\|_2 \\
 &\quad+ \bigl\| \Theta_N^k\circ{\cal J}_{N\infty}(f_\varepsilon) - 
 {\cal J}_{N\infty}\circ\Theta^k(f_\varepsilon) \bigr\|_2 \\
 &\le 2\|f-f_\varepsilon\| + I_N(f_\varepsilon).
\end{align*}

This follows from $\Theta_N$-invariance of the norm 
$\|\,\cdot\,\|_2$, from $T$-invariance of the measure $\mu$
and from the fact that the positivity inequality for unital
completely positive maps such as ${\cal J}_{N\infty}$ gives:
\begin{align*}
 \bigl\| {\cal J}_{N\infty}(g) \bigr\|_2^2
 &= \tau_N \bigl( {\cal J}_{N\infty}(g)^*{\cal J}_{N\infty}(g)
 \bigr) \le \tau_N\Bigl( {\cal J}_{N\infty}(|g|^2)\Bigr) \\
 &= \int_{\IT} d \bs{x}\, |g|^2(\bs{x}) = \|g\|^2\quad .
\end{align*}

We now use that $f_\varepsilon$ is a function with finitely 
supported Fourier transform and, inserting the Weyl quantization of
$f_\varepsilon$, we estimate 
\begin{equation}
 I_N(f_\varepsilon) \le \bigl\| {\cal J}_{N\infty}(f_\varepsilon)-
 W_{N,\infty}(f_\varepsilon) \bigr\|_2 + \bigl\| 
 {\cal J}_{N\infty} \circ \Theta^k(f_\varepsilon) - 
 \Theta^k_N(W_{N,\infty}(f_\varepsilon)) \bigr\|_2.
\label{prop3.1}
\end{equation}
Then, we concentrate on the square of the
second term, which we denote by $G_{N,k}(f_\varepsilon)$ and 
explicitly reads
\begin{align}
 G_{N,k}(f_\varepsilon)
 &= \tau_N \bigl( {\cal J}_{N\infty} \circ \Theta^k(f_\varepsilon)^*
 {\cal J}_{N\infty} \circ \Theta^k(f_\varepsilon) \bigr) + \tau_N
 \bigl( W_{N,\infty}(f_\varepsilon)^* W_{N,\infty}(f_\varepsilon)\bigr) 
\nonumber \\
 &\quad- 2\,\Re\;\Bigl( \tau_N \bigl(
 {\cal J}_{N\infty} \circ \Theta^k(f_\varepsilon)^*
 \Theta^k_N(W_{N,\infty}(f_\varepsilon))\bigr) \Bigr).
\label{prop3.2}
\end{align}

The first term tends to $\|f_\varepsilon\|^2$ as $N\to\infty$,
because of Proposition~\ref{prop2} and the same is true of the second term;
indeed,
\begin{equation*}
 \tau_N \bigl( W_{N,\infty}(f_\varepsilon)^*W_{N,\infty}(f_\varepsilon) \bigr) =
 \sum_{\bi{k,q}\in\r{Supp}(\hat f_\varepsilon)} \overline{\hat
 f_{\varepsilon\,\bs{k}}}\, \hat f_{\varepsilon\,\bs{q}}\,
 {\rm e}^{\frac{i\pi}{N}\sigma(\bi q,\bi k)}\,
 \tau_N \bigl( W_N(\bi{q-k})\bigr).
\end{equation*}
Now, since $\r{Supp}(\hat f_\varepsilon)$ is finite, the vector
$\bi{k-q}$ is uniformly bounded with respect to $N$. Therefore, with
$N$ large enough, (\ref{Weyl5}) forces $\bi k=\bi q$, whence the
claim. It remains to show that the same for the third term
in~(\ref{prop3.2}) which amounts to twice the real part of
\begin{align*}
 &\int_{\IT} \ud\bs{x}\, \overline{f_\varepsilon(T^{-k}\bs{x})}
 \< C_N(\bs{x}), \Theta^k_N(W_{N,\infty}(f_\varepsilon)) C_N(\bs{x})\> \\
 &\quad= \sum_{\bi p\in\r{Supp}(\hat f_\varepsilon)}
 \overline{\hat{f}_{\varepsilon\,\bs{q}}} \< C_N, W_N(T^k\bi p) C_N\>
 \int_{\IT} \ud\bs{x} \overline{f_\varepsilon(T^{-k}\bs{x})}
 \exp\Bigl(\frac{2\pi i}{N}\sigma(T^k\bi p,[N\bs{x}]) \Bigr).
\end{align*}
According to Lemma~\ref{Lemma0}, the matrix element 
$\< C_N,W_N(T^k\bi p) C_N\>$ tends to 1 as $N\to\infty$ whenever
the vectorial components $(T^k\bi p)_j$, $j=1,2$, satisfy
\begin{equation*}
 \lim_N \frac{(T^k\bi p)^2_j}{N} = C_{\bi u}(\bi p) ({\bi u})_j \lim_N
 \frac{\lambda^{2k}}{N} = 0,
\end{equation*}
where we expanded $\bi p = C_{\bi u}(\bi p){\bi u} +  C_{\bi v}(\bi p)
{\bi v}$ along the stretching and squeezing eigendirections of $T$
(see Remark~\ref{Rem_21}.v).  This fact sets the logarithmic time
scale $k < \frac{1}{2}\frac{\log N}{\log\lambda}$.  Notice that, when
$k=0$, $G_{N,k}(f_\varepsilon)$ equals the first term
in~(\ref{prop3.1}) and this concludes the proof.  \hfill
$\blacksquare$ \par}
\begin{quote}
\begin{NNN}{}\ \\
The previous result essentially points to the fact
that the time evolution and the classical limit do commute over time
scales that are logarithmic in the semi--classical parameter $N$. The
upper bound, which goes like
$\text{const.}\times\frac{\log N}{\log\lambda}$, is typical of quantum
chaos and is known as logarithmic {\it breaking--time}. Such a scaling
appear numerically in Section~\ref{sub_512}
also for discrete classical cat maps, converging in a suitable
classical limit to continuous cat maps.
\end{NNN}
\end{quote}
\subsection{Continuous limit for Sawtooth Maps}\label{ClfSM}\vspace{3mm}   

In Section~\ref{DETa} we provided a discretization procedure for
algebraic classical
dynamical systems by constructing two discretization/dediscretization
operators ${\cal J}_{N , \infty}:\Cspace{0}{\cal X}\mapsto{\cal D}_N$ and
${\cal J}_{\infty,N}:{\cal D}_N\mapsto\Cspace{0}{\cal X}$ such that
$\widetilde{\Theta}_{N,\alpha}^{j}\circ{\cal J}_{N , 
\infty}	= {\cal J}_{N ,\infty}\circ\Theta_\alpha^{j}$, where
$\widetilde{\Theta}_{N,\alpha}^{j}$, $\Theta_\alpha^{j}$ are the
quantized, respectively classical dynamical maps at timestep $j$.

However, the pictures drastically changes when we pass from
$T_\alpha$ to $S_\alpha$: then we are
forced to enlarge the algebra from $\Cspace{0}{\cal X}$ to
$\Lspace{\infty}{\IT}$, to define a different discretization scheme 
and a new *automorphism $\widetilde{\Theta}_{N,\alpha}$ on ${\cal
D}_N$. 

The origin of the inequality $\widetilde{\Theta}_\alpha^{j}\circ{\cal
J}_{N , \infty}\neq{\cal J}_{N ,\infty}\circ\Theta_\alpha^{j}$ (when
$\alpha\in\IR\setminus\IZ$) lies in the 
discontinuous character of the fractional part that appears
in~\eqref{AoDC_1b}.
Nevertheless, the equality is obtained 
when $N\longmapsto \infty$; more precisely, in
Proposition~\ref{propval}, we shall rigorously prove that
$\widetilde{\Theta}_\alpha^{j}\circ{\cal J}_{N , 
\infty}\pt{f}\longmapsto{\cal J}_{N
,\infty}\circ\Theta_\alpha^{j}\pt{f}$, for all $\Lspace{\infty}{\IT}$
with respect to the topology on
${\cal D}_N$ determined by the state $\tn$ through the Hilbert-Schmidt
norm $\norm{\cdot}{2}$ defined in Proposition~\ref{prop3}.

The above choice of topology is dictated by the fact that in the
continuous limit the set of discontinuities of the $j$--power of
Sawtooth Maps are a subset of zero measure. 
For later purposes, we now give a brief review of the discontinuities of
the maps
$S_\alpha$~\cite{Che92:1,Vai92:1,Per87:1}.

\noindent
As already noted in Remark~\ref{Rem_21}, $S_\alpha$ is discontinuous
on the circle $\gamma_0$;
therefore $S_\alpha^n$ will be discontinuous on the preimages 
\begin{equation}
\gamma_m\coleq S_\alpha^{-m}\pt{\gamma_0}\ \text{ for }\ 0\leq
m<n\label{gammapiu}\ ,  
\end{equation}
while the discontinuities of $S_\alpha^{-n}$ lie on the sets 
\begin{equation*}
\gamma_{-m}\coleq S_\alpha^{m}\pt{\gamma_0}\ \text{ for }\ 0< m\leq n\
\cdot 
\end{equation*}
\noindent
Apart from $\gamma_{-1}$, which is a closed curve on the torus
intersecting $\gamma_0$ transversally, each set of the type
$\gamma_m$ (for $\gamma_{-m}$ the argument is similar) is the
(disjoint) union of segments parallel to each other 
whose endpoints lie either on the same segment belonging to
$\gamma_p$, $p<m$, or on two different segments belonging to
$\gamma_p$ and $\gamma_{p^\prime}$, with $p^\prime\leq p <
m$~\cite{Vai92:1}.It proves convenient to introduce the
\enfasi{discontinuity set} of
$S_\alpha^n$,
\begin{equation}
\Gamma_n\coleq \bigcup_{p=0}^{n-1} \; \gamma_p \ ,\label{Gamman}
\end{equation}
which is a one dimensional sub--manifold of ${\cal
 X}=\IT$ and its complementary set, $G_n\coleq \IT \setminus
\Gamma_n$.

We now enlarge the above definition from 
continuous Sawtooth Maps, to discretized ones.\\[-2ex]
\begin{quote}
\begin{DDS}{}\ \\\label{Gnbar}
Given $\bs{x}\in\IT$, we shall denote by $\hat{\bs{x}}_N$ the element of
$\ZNZD$ given by: 
\begin{equation}
\hat{\bs{x}}_N\coleq\Big(\floor{N x_1 + \pum}\,,\,\floor{N x_2
+ \pum}\Big) \label{loc_c32}\ ,
\end{equation}
by segment $\pt{A,B}$, the shortest
curve joining $A,B\in\IT$, by $l\pt{{\gamma}_p}$ the length of the
curve ${\gamma}_p$ and by\footnote{%
The distance $d_{\IT}\pt{\cdot,\cdot}$ on the torus
has been introduced in Definition~\ref{dont}.}
\begin{align}
\overline{\gamma}_p\pt{\varepsilon} &\coleq 
\Big\{
\bs{x}\in\IT
\ \Big|\ 
d_{\IT}\pt{\bs{x},\gamma_p} \leq \varepsilon
\Big\}\label{Gnbar_0}
\intertext{the strip around $\gamma_p$ of width $\varepsilon$.}
\intertext{Further, we shall denote by}
\overline{\Gamma}_n\pt{\varepsilon} &\coleq 
\bigcup_{p=0}^{n-1}\;
\overline{\gamma}_p\pt{\varepsilon}
\label{Gnbar_1}
\intertext{the union of the strips up to $p=n-1$ and by
${G_n^N}\pt{\varepsilon}$
the subset of points}
{G_n^N}\pt{\varepsilon}&\coleq
\pg{\bs{x}\in\IT\ \Big\vert\
\frac{\hat{\bs{x}}_N}{N}\not\in\overline{\Gamma}_n\pt{\varepsilon}} \ \cdot
\label{Gnbar_2}
\end{align}
\end{DDS}
\end{quote}
\noindent
To prove that the discretized Sawtooth maps tend to Sawtooth maps in
the continuum when $N\longrightarrow\infty$, the main problem is to
control the discontinuities. In order to do that, we shall subdivide
the lattice points in a good and a bad set and prove that the images
of points in the good set, under the evolution $U_\alpha^q$,
$V_\alpha^q$ remain close to each other. This is not true for the bad
set, however we shall show that it tends with $N$ to a set of zero
Lebesgue measure and thus becomes ineffective.

To concretely implement the above strategy we need the next two Lemmas
whose proofs are 
given in Appendix~\ref{app_B}. 
\begin{quote}
\begin{LLL}{}\ \\[-1ex]
Using the notation of Definition~\ref{Gnbar}, we have:\ \\[-6ex]
\begin{Ventry}{3)}\label{Lemma2}
	\item[1)] Given the matrix
	$S_\alpha^{\phantom{-1}}=\pt{\begin{smallmatrix} 
	1+ \alpha  & 1\\ \alpha & 1 \end{smallmatrix}}$, 
	$\alpha\in\IR$, and its inverse
	$S_\alpha^{-1}=\pt{\begin{smallmatrix} 
	1 & - 1\\ -\alpha & 1 + \alpha \end{smallmatrix}}$, there
	exists a real number $\eta>1$ depending only on $\alpha$ such that,
	for any $A,\ B\in{\IR}^2$, we have, with $\bs{v}=A-B$,
\begin{subequations}
\label{lemma1_3} 
\begin{align}
\norm{S_\alpha^{\pm 1} \cdot \bs{v}}{{\IR}^2}\leq
\eta^{\phantom{-1}}\norm{\bs{v}}{{\IR}^2}\ 
,\label{lemma1_3a}\\
\norm{S_\alpha^{\pm 1} \cdot \bs{v}}{{\IR}^2}\geq
\eta^{-1}\norm{\bs{v}}{{\IR}^2}\ ,\label{lemma1_3b} 
\end{align}
\end{subequations}
where $S_\alpha^{\pm 1}\cdot \bs{v}$ denotes the matricial action of
	$S_\alpha^{\pm 1}$ on $\bs{v}$.
	\item[2)] Let $A,B\in\IT$ 
	and $d_{\IT}\pt{A,B}<\frac{1}{2}\;\eta^{-1}$:\\[2ex]
\begin{subequations}
	\noindent \label{t2r2tot}
	2a) If the segment $(A,B)$ does not cross $\gamma_{-1}$, then 
	\begin{align}
	d_{\IT}\pt{S_\alpha^{-1}\pt{A},S_\alpha^{-1}\pt{B}} & \leq
	\eta\;d_{\IT}\pt{A,B}\ \cdot\label{t2r2}
	\intertext{2b) If the $(A,B)$ does not cross
	$\gamma_{0\phantom{-}}$, then} 
	d_{\IT}\pt{S_\alpha^{\phantom{-1}}\pt{A},
	S_\alpha^{\phantom{-1}}\pt{B}} & \leq 
	\eta\;d_{\IT}\pt{A,B}\ \cdot\label{t2r2b}
	\end{align}
\end{subequations}
	\item[3)] For any given $\alpha\in\IR$, $p\in{\IN}^+$ and 
	$0\leq\varepsilon\leq\frac{1}{2}\,\eta^{-1}$, 
\begin{equation}
\bs{x}\in\overline{\gamma}_{p-1}\pt{\varepsilon}
\Longrightarrow S_\alpha^{-1}\pt{\bs{x}}\in\pt{
\overline{\gamma}_p\pt{\eta\,\varepsilon}\cup
\overline{\gamma}_0\pt{\eta\,\varepsilon}}\ \cdot
\label{Lemma21_0}
\end{equation}
	\item[4)] For any given $\alpha\in\IR$, $n\in{\IN}^+$ and
	$0\leq\varepsilon\leq\frac{1}{2}$, with $U_\alpha^q$ as
	in~\eqref{Ualpha},
\begin{equation}
\bs{x} \not\in \overline{\Gamma}_n\pt{\varepsilon} \Longrightarrow
d_{\IT}\pt{\frac{U_\alpha^q\pt{N\bs{x}}}{N},\gamma_0}>
\varepsilon\;\eta^{-q}   
\ ,\ \forall \ 0\leq q < n \ \cdot\label{nuovo_2}
\end{equation}
	\item[5)] For any given $\alpha\in\IR$ and $n\in{\IN}^+$, if
\begin{multline}
N>\widetilde{N}+3=2\sqrt{2}\,n\,\eta^{2n}+3\quad\text{and}\quad
\bs{x}\in {G_n^N}\pt{\frac{\widetilde{N}}{2N}}
\text{then}\\
d_{\IT}\pt{\frac{U_\alpha^p\pt{N\bs{x}}}{N},
\frac{V_\alpha^p\pt{\hat{\bs{x}}_N}}{N}}\leq
\frac{\sqrt{2}}{N}\pt{\frac{\eta^{p+1}-1}{\eta-1}}\quad,\quad\forall p\leq
n\ \cdot
\label{lemma1_1} 
\end{multline}
\end{Ventry}
\end{LLL}
\begin{LLL}{}\ \\ \label{Lemma1}
\noindent With the notation of Definition~\ref{Gnbar},  the following
relations hold for all
$p\in{\IN}$, $n\in{\IN}^{+}$ and 
$\varepsilon\in{\IR}^+$:
\begin{subequations} \label{lemma_1_0}
\begin{align} 
l\pt{\gamma_p}& \leq\eta^p
\ ,\label{lemma_1_1}\\
\mu\pt{\overline{\gamma}_p\pt{\varepsilon}}& \leq 2 \,\varepsilon\,
\eta^p + \pi\varepsilon^2 
\ ,\label{lemma_1_2}\\
\mu\pt{\overline{\Gamma}_n\pt{\varepsilon}} & \leq 
\varepsilon\,n\pt{2\,\eta^n + \pi\,\varepsilon}
\ \cdot\label{lemma_1_3}
\end{align} 
Denoting with $\pq{E}^{\circ}$ the complementary set of $E$ on the torus,
namely $\pq{E}^{\circ}\coleq\IT\setminus E$, it holds
\begin{equation}
\pq{\overline{\Gamma}_n\pt{\varepsilon+ 
\frac{1}{\sqrt{2}N}}}^{\circ}\subseteq 
{G_n^N}\pt{\varepsilon}
\ \cdot\label{lemma_1_5}
\end{equation}
Moreover, if $N\in\IN^{+}$ and
$\widetilde{N}=2\sqrt{2}\,n\,\eta^{2n}$
(cfr. Lemma~\ref{Lemma2}.5), we have:
\begin{equation}
N>\widetilde{N}+3\Longrightarrow
\mu
\pt{{\pq{G_n^N \pt{\frac{\widetilde{N}}{2N}}}}^{\circ}}  
\leq 
\frac{12\,\sqrt{2}\,n^2\,\eta^{3n}}{N}
\ \cdot\label{lemma_1_4}
\end{equation}
\end{subequations}
\end{LLL}
\end{quote}
By using Lemma~\ref{Lemma2}, we prove now a dynamical localization Property,
not too far from the one given in page~\pageref{dynloc}, which
involves the unitary single step
evolution operator $U_{\alpha,N}^{\prime}$ defined
in~\eqref{aggiuntaSa}. 
\begin{quote}
\begin{PPP}[Dynamical localization on $\bs{\{\vert
C_N^3(\bs{x})\rangle\}}$ states]{}\ \\[-2ex]\label{dynloc2} 
\!\!For any given $n\in\IN$ and $\alpha\in\IR$, for all 
$d_0>0$ there exists an $N_0\in\Nl$
with the following property: if $N > N_0$ and
$\bs{x}\in{G_n^N}\pt{\frac{\widetilde{N}}{2N}}$,
then 
$\big<C_N^3(\bs{x})\,\big\vert\, U_{\alpha,N}^{\prime \,n}\,C_N^3(\bs{y})\big>= 0$
 whenever $d_{\IT}\pt{S_\alpha^n\pt{\bs{x}},\bs{y}} \geq d_0$.
\end{PPP}
\end{quote}
\textbf{Proof of Proposition~\ref{dynloc2} :}\\[2ex]
Using the definitions of states $\{\vert
C_N^3(\bs{x})\rangle\}$ given in~\eqref{CSforL1}, together with
the unitary 
evolution operator $U_{\alpha,N}^{\prime}$ defined
in~\eqref{aggiuntaSa}, $\big<C_N^3(\bs{x})\,\big\vert\, U_{\alpha,N}^{\prime
\,n}\,C_N^3(\bs{y})\big>$ can be easily computed, as follows:
\begin{equation}
\big<C_N^3(\bs{x})\,\big\vert\, U_{\alpha,N}^{\prime
\,n}\,C_N^3(\bs{y})\big> =  
\Big\langle \hat{\bs{x}}_N\;\Big\vert\;
V_\alpha^{-n}
\pt{\hat{\bs{y}}_N}
\Big\rangle
=\delta^{(N)}_{\;
V_\alpha^n
\pt{\hat{\bs{x}}_N}
\:,\:\hat{\bs{y}}_N
}
\ \cdot\label{loc_c31}
\end{equation}
Using the triangular inequality, we get:
\begin{multline}
d_{\IT}\pt{\frac{U_\alpha^n\pt{N\bs{x}}}{N}\,,\,\bs{y}}\leq
d_{\IT}\pt{\frac{U_\alpha^n\pt{N\bs{x}}}{N}\,,
\,\frac{V_\alpha^n\pt{\hat{\bs{x}}_N}}{N}} +\\
+ d_{\IT}\pt{\frac{V_\alpha^n\pt{\hat{\bs{x}}_N}}{N}\,,
\,\frac{\hat{\bs{y}}_N}{N}} +
d_{\IT}\pt{\frac{\hat{\bs{y}}_N}{N}\,,
\,\bs{y}}
\label{loc_c33} 
\end{multline}
or equivalently,
\begin{multline}
d_{\IT}\pt{\frac{V_\alpha^n\pt{\hat{\bs{x}}_N}}{N}\,,
\,\frac{\hat{\bs{y}}_N}{N}}\geq 
d_{\IT}\pt{S_\alpha^n\pt{\bs{x}}\,,\,\bs{y}} -\\
d_{\IT}\pt{\frac{U_\alpha^n\pt{N\bs{x}}}{N}\,,
\,\frac{V_\alpha^n\pt{\hat{\bs{x}}_N}}{N}} - 
d_{\IT}\pt{\frac{\hat{\bs{y}}_N}{N}\,,
\,\bs{y}}
\label{loc_c34} 
\end{multline}
Now we will use these observations:

\begin{itemize}
\item From~\eqref{nuovopt_2} in Appendix~\ref{app_B} we
have
\begin{equation}
d_{\IT}\pt{\bs{y}\;,\;\frac{\hat{\bs{y}}_N}{N}}\leq
\frac{1}{\sqrt{2}N}\quad,\quad\forall\;\bs{y}\in\IT \ 
;\label{dYYN}
\end{equation}
\item The fact that $\bs{x}\in{G_n^N}\pt{\frac{\widetilde{N}}{2N}}$
permits us to use point 5 of Lemma~\ref{Lemma2}, 
that is
\begin{equation}
N>\widetilde{N}=2\sqrt{2}\,n\,\eta^{2n}\ \Longrightarrow\ 
d_{\IT}\pt{\frac{U_\alpha^n\pt{N\bs{x}}}{N}, 
\frac{V_\alpha^n\pt{\hat{\bs{x}}_N}}{N}}\leq
\frac{\sqrt{2}}{N}\pt{\frac{\eta^{n+1}-1}{\eta-1}}\ ;\label{dYYN2}
\end{equation}
\item $d_{\IT}\pt{S_\alpha^n\pt{\bs{x}}\,,\,\bs{y}}\geq d_0$ by hypothesis.
\end{itemize}
Then we can find a $N_0>\tilde{N}+3$ such that 
\begin{equation}
N>N_0\quad\Longrightarrow\quad
d_{\IT}\pt{\frac{V_\alpha^n\pt{\hat{\bs{x}}_N}}{N}\,, 
\,\frac{\hat{\bs{y}}_N}{N}}>\frac{1}{N}\ \label{dYYN4}
\end{equation}
Indeed, inserting~(\ref{dYYN}--\ref{dYYN2}) in
equation~\eqref{loc_c34}, we get
\begin{equation}
d_{\IT}\pt{\frac{V_\alpha^n\pt{\hat{\bs{x}}_N}}{N}\,,
\,\frac{\hat{\bs{y}}_N}{N}}\geq 
d_0 - \frac{\sqrt{2}}{N}\pt{\frac{\eta^{n+1}-1}{\eta-1}} -
\frac{1}{\sqrt{2}N} \ ;
\label{loc_c341} 
\end{equation}
therefore, we can choose $N_0$ such that for $N>N_0>\widetilde{N}+3$
the right hand side of the above inequality is larger than $\frac{1}{N}$.
Explicitly, it is sufficient to choose:
\begin{equation}
N_0 
=\max\pg{\frac{1}{d_0}\pq{1 + \sqrt{2}\pt{\frac{\eta^{n+1}-1}{\eta-1}} +
\frac{1}{\sqrt{2}}},\widetilde{N}+3}\ \cdot
\label{loc_c347} 
\end{equation}
Combining~\eqref{loc_c31} and~\eqref{dYYN4} the
proof is completed, by noting that if the toral distance
of two grid points exceeds $\frac{1}{N}$, then the distance between
the components of the integer vector labeling the two points is
different from zero and then the periodic Kronecker delta
in~\eqref{loc_c31} vanishes.\hfill$\qed$ \\[2ex] 
We now use the previous two Lemmas to prove the main result of this
section. 
\begin{quote}
\begin{PPP}{}\ \\
\label{propval}
 Let $\tripQSa$ be a quantum
 dynamical system as defined in Section~\ref{DynSMf} and suppose
 that it satisfies Condition~\ref{dynloc2}. For any fixed integer $k$,
 in the topology given by the
 Hilbert-Schmidt norm of Proposition~\eqref{prop3}, we have
 \begin{equation}
  \lim_{N\to\infty} \| \widetilde{\Theta}_{N,\alpha}^k
  \circ {\cal J}_{N \infty}(f) - {\cal J}_{N \infty} \circ \Theta_\alpha^k(f)
  \|_2 = 0.\label{added2}
 \end{equation}
\end{PPP}
\end{quote}
\noindent
\textbf{Proof of Proposition~\ref{propval}:}\\[2ex]
In this proof we will parallel the same strategy used in the proof of
Proposition~\ref{prop3}. As done there, we will consider
the case of $f$ 
continuous, being the extension to $f$--essentially bounded just an
application of Lusin's (Theorem~\ref{LusTEO}, Corollary~\ref{LusCOR},
page~\pageref{LusTEO}). Then we have to show that 
\begin{equation}
 I_N(k)\coleq N \int_{\c X} \mu(\ud\bs{x})\, \int_{\c X}
 \mu(\ud\bs{y})\,  \overline{f(\bs{y})}\, f(S_\alpha^k \bs{x})  |\< C_N^3(\bs{x}), U_{\alpha,N}^{\prime \,k} C_N^3(\bs{y}) \>
 |^2
\end{equation}
goes to $\int \mu(\ud\bs{z}) |f(\bs{z})|^2$. 
Resorting to ${G_n^N}\pt{\frac{\widetilde{N}}{2N}}$ in
Definition~\ref{Gnbar}, and to its complementary set
$\pq{G_n^N\pt{\frac{\widetilde{N}}{2N}}}^{\circ} =
{\cal X}\setminus{G_n^N}\pt{\frac{\widetilde{N}}{2N}}$, we can write
\begin{align}
 &\left| I_N(k) -
 \int_{\c X} \mu(\ud\bs{y})\, |f(\bs{y})|^2 \right| \nonumber\\
 &=  \left| \int_{\c X} \mu(\ud\bs{x})\, \int_{\c X} \mu(\ud\bs{y})\, \overline{f(\bs{y})}\,
 \bigl( f(S_\alpha^k \bs{x}) - f(\bs{y}) \bigr)\, N|\< C_N^3(\bs{x}), U_{\alpha,N}^{\prime \,k} C_N^3(\bs{y})\>|^2
 \right| \nonumber\\
 & \le  \left| \int_{\c X} \mu(\ud\bs{x})\, \int_{\pq{G_n^N\pt{\frac{\widetilde{N}}{2N}}}^{\circ}} \mu(\ud\bs{y})\,
 \overline{f(\bs{y})} \bigl(f(S_\alpha^k \bs{x}) - f(\bs{y})\bigr) N|\< C_N^3(\bs{x}),
 U_{\alpha,N}^{\prime \,k} C_N^3(\bs{y})\>|^2 \right| \nonumber\\
 &+ \left| \int_{\c X} \mu(\ud\bs{x})
 \int_{{G_n^N}\pt{\frac{\widetilde{N}}{2N}}} \mu(\ud\bs{y})
 \overline{f(\bs{y})}\bigl(f(S_\alpha^k \bs{x}) - f(\bs{y})\bigr) N|\< C_N^3(\bs{x}),
 U_{\alpha,N}^{\prime \,k} C_N^3(\bs{y})\> |^2 \right|.\label{secondterm}
\intertext{For the first integral in the r.h.s. of the previous
 expression we have:}
& \left| \int_{\c X} \mu(\ud\bs{x})\, \int_{\pq{G_n^N\pt{\frac{\widetilde{N}}{2N}}}^{\circ}} \mu(\ud\bs{y})\,
 \overline{f(\bs{y})} \bigl(f(S_\alpha^k \bs{x}) - f(\bs{y})\bigr) N|\< C_N^3(\bs{x}),
 U_{\alpha,N}^{\prime \,k} C_N^3(\bs{y})\>|^2 \right| \nonumber\\
& \le 2 {\pt{\norm{f}{\infty}}}^2
\left| \int_{\pq{G_n^N\pt{\frac{\widetilde{N}}{2N}}}^{\circ}}\mu(\ud\bs{y})\, \int_{\c X} \mu(\ud\bs{x})\, 
  N|\< C_N^3(\bs{x}),
 U_{\alpha,N}^{\prime \,k} C_N^3(\bs{y})\>|^2 \right| \nonumber
\intertext{(note that exchange of integration order is harmless
because of the existence of the integral)}
& \le 2 {\pt{\norm{f}{\infty}}}^2
\mu\pt{\pq{G_n^N\pt{\frac{\widetilde{N}}{2N}}}^{\circ}}\leq
\frac{24\,\sqrt{2}\,n^2\,\eta^{3n}}{N}{\pt{\norm{f}{\infty}}}^2\nonumber
\end{align}
where we have used Properties (2) and (3) of
Definition~\ref{coh} and equation~\eqref{lemma_1_4} from
Lemma~\ref{Lemma1}; this term becomes negligible for
large $N>\widetilde{N}$.
Now it remains to prove that the second term in~\eqref{secondterm} is also negligible for
large $N$:
 selecting a ball $B(S_\alpha^k \bs{x},d_0)$,
one derives 
\begin{align*}
 &\left| \int_{\c X} \mu(\ud\bs{x})\, \int_{{G_n^N}\pt{\frac{\widetilde{N}}{2N}}} \mu(\ud\bs{y})\, \overline{f(\bs{y})}\,
 \bigl( f(S_\alpha^k \bs{x}) - f(\bs{y}) \bigr)\, N|\< C_N^3(\bs{x}), U_{\alpha,N}^{\prime \,k} C_N^3(\bs{y})\>|^2
 \right| \\
 & \le  \left| \int_{\c X} \mu(\ud\bs{x})\,
 \int_{{G_n^N}\pt{\frac{\widetilde{N}}{2N}}\cap B(S_\alpha^k \bs{x},d_0)} \mu(\ud\bs{y})\,
 \overline{f(\bs{y})} \bigl(f(S_\alpha^k \bs{x}) - f(\bs{y})\bigr) N|\< C_N^3(\bs{x}),
 U_{\alpha,N}^{\prime \,k} C_N^3(\bs{y})\>|^2 \right| \\
 &+ \left| \int_{\c X} \mu(\ud\bs{x})
 \int_{{G_n^N}\pt{\frac{\widetilde{N}}{2N}}\cap\pt{{\c X}\setminus B(S_\alpha^k \bs{x},d_0)}} \mu(\ud\bs{y})
 \overline{f(\bs{y})}\bigl(f(S_\alpha^k \bs{x}) - f(\bs{y})\bigr) N|\< C_N^3(\bs{x}),
 U_{\alpha,N}^{\prime \,k} C_N^3(\bs{y})\> |^2 \right|.
\end{align*}
Applying the mean value theorem in the first double integral and
approximating the integral of the 
kernel as in the proof of Proposition~\ref{prop1}, we get that 
$\exists \:\bs{c} \in B(S_\alpha^k \bs{x}, d_0)$ such that
\begin{align*}
 &\left| \int_{\c X} \mu(\ud\bs{x})\, \int_{{G_n^N}\pt{\frac{\widetilde{N}}{2N}}} \mu(\ud\bs{y})\, \overline{f(\bs{y})}\,
 \bigl( f(S_\alpha^k \bs{x}) - f(\bs{y}) \bigr)\, N|\< C_N^3(\bs{x}), U_{\alpha,N}^{\prime \,k} C_N^3(\bs{y})\>|^2
 \right| \\
 &\le \left| \int_{\c X} \mu(\ud\bs{x})\, 
 \overline{f(\bs{c})}\, \bigl(f(S_\alpha^k \bs{x}) -
 f(\bs{c})\bigr)\, \int_{{G_n^N}\pt{\frac{\widetilde{N}}{2N}}\cap B(S_\alpha^k \bs{x},d_0)} \mu(\ud\bs{y})\,N|\< C_N^3(\bs{x}), U_{\alpha,N}^{\prime \,k} C_N^3(\bs{y})\>|^2 \right| \\
 &+ \left| \int_{\c X} \mu(\ud\bs{x})
 \int_{{G_n^N}\pt{\frac{\widetilde{N}}{2N}}\cap\pt{{\c X}\setminus B(S_\alpha^k \bs{x},d_0)}} \mu(\ud\bs{y})
 \overline{f(\bs{y})}\bigl(f(S_\alpha^k \bs{x}) - f(\bs{y})\bigr) N|\< C_N^3(\bs{x}),
 U_{\alpha,N}^{\prime \,k} C_N^3(\bs{y})\> |^2 \right|
\intertext{and using property~(\ref{coh}.3)}
 &\le \left| \int_{\c X} \mu(\ud\bs{x})\, 
 \overline{f(\bs{c})}\, \bigl(f(S_\alpha^k \bs{x}) -
 f(\bs{c})\bigr)\right| \\
 &+ \left| \int_{\c X} \mu(\ud\bs{x})
 \int_{{G_n^N}\pt{\frac{\widetilde{N}}{2N}}\cap\pt{{\c X}\setminus B(S_\alpha^k \bs{x},d_0)}} \mu(\ud\bs{y})
 \overline{f(\bs{y})}\bigl(f(S_\alpha^k \bs{x}) - f(\bs{y})\bigr) N|\< C_N^3(\bs{x}),
 U_{\alpha,N}^{\prime \,k} C_N^3(\bs{y})\> |^2 \right|\ \cdot
\end{align*}

By uniform continuity we can bound the first term by some arbitrary
small $\varepsilon$, provided  we choose $d_0$ small enough. For
the second integral, we use the localization condition~\ref{dynloc2}
that allow us to find $N_0=N_0(d_0,k)$ depending on the $d_0$ just
chosen to 
bound the first term and on the given fixed timestep $k$, such that the
second integral vanishes.\hfill$\qed$
\pagestyle{fancyplain}
\chapter{Entropies}\vspace{9mm}
\section{Classical Dynamical Entropy}\vspace{6mm}
Intuitively, one expects the instability proper to the presence of
a positive Lyapounov exponent to correspond to some degree of
unpredictability of the dynamics: classically, the metric entropy of  
Kolmogorov provides the link~\cite{For92:1}.
\subsection{Kolmogorov Metric Entropy}
\label{KSME}\vspace{3mm} 
For continuous classical systems $\tripCT$ such as
those introduced in 
Section~\ref{CDS}, the construction of the dynamical entropy of
Kolmogorov is based  
on 
subdividing $\cal X$ into measurable disjoint subsets
${\left\{E_\ell\right\}}_{\ell=1,2,\cdots, D}$ such that $\bigcup_\ell
E_\ell={\cal X}$ which form finite partitions (coarse graining) 
${\cal E}$.

Under the a dynamical maps $T:{\cal X}\to{\cal X}$ ,
any given ${\cal E}$ evolves into $T^{j}({\cal E})$ with atoms
$\displaystyle
T^{-j}(E_\ell)=\{x\in{\cal X}: T^jx\in E_\ell\}$;
one can then form finer partitions
\begin{alignat}{6}
{\cal E}_{[0,n-1]}& \coleq\bigvee_{j=0}^{n-1}T^{j}({\cal E})&\  & =\ 
 {\cal E}&&\bigvee \ \ \ T({\cal E})&&\bigvee&&\cdots&&\bigvee
 \ T^{n-1}({\cal E})\notag
\intertext{whose atoms}
 E_{i_0\,i_1\cdots i_{n-1}}&\coleq 
 \bigcap_{j=0}^{n-1}T^{-j}E_{i_j}&&=\ 
 E_{i_0}&&\bigcap T^{-1}(E_{i_1})&&\bigcap &&\cdots&&\bigcap 
 T^{-n+1}(E_{i_{n-1}})\notag
\intertext{have volumes}
\mu_{i_0\,i_1\cdots i_{n-1}}&\coleq\mu\left(
E_{i_0\,i_1\cdots i_{n-1}}
\right)&&\cdot&&&&&&&& \label{KSE_1}
\end{alignat}\\[-5.5ex]
\begin{quote}
\begin{DDD}{}\label{stringhe}\ \\
We shall set $\bs{i}=\pg{i_0\,i_1\cdots i_{n-1}}$ and
denote by $\Omega_D^n$ the set of $D^n$ n\_tuples with $i_j$ taking
values in $\pg{1, 2, \cdots, D}$.
\end{DDD}
\end{quote}
\noindent
The atoms of the partitions ${\cal E}_{[0,n-1]}$ describe segments 
of trajectories 
up to time $n$ encoded by the atoms of ${\cal E}$ that are traversed at 
successive times; the volumes $\mu_{\bs{i}}=\mu\pt{E_{\bs{i}}}$
corresponds to probabilities for the system to belong to the atoms 
$E_{i_0},E_{i_1},\cdots,E_{i_{n-1}}$ at successive times $0\leq j\leq
n-1$. The n\_tuples $\bs{i}$ by themselves provide a description of
the system in a symbolic dynamic. Of course, once the evolved
partition ${\cal E}_{[0,n-1]}$ is specified, not all strings (or words)
$\bs{i}\in\Omega_D^n$ would represent the possible trajectories of
the dynamical system. Therefore we can split the set $\Omega_D^n$ in
two: a set 
containing all admissible words, denoted by $k\pt{{\cal
E}_{[0,n-1]}}$, and its complementary set. Of course a word belongs to $k\pt{{\cal
E}_{[0,n-1]}}$, if the corresponding $E_{\bs{i}}\in{\cal E}_{[0,n-1]}$
contains (at least) one point. The study of the cardinality of the set
of admissible words $k\pt{{\cal E}_{[0,n-1]}}$, in the limit
$n\longmapsto\infty$, provide the simplest 
possibility of estimating the complexity of a dynamical
system~\cite{Ale81:1}. Moreover, we can expect that not all
trajectories would have the same weight for the system: in
particular there could be ``rare'' trajectories, encoded by $\bs{i}\in
k\pt{{\cal 
E}_{[0,n-1]}}$, whose weight (given by the measure of the corresponding
set $\mu_{\bs{i}}=\mu\pt{E_{\bs{i}}}$) is negligible.

The richness in diverse trajectories, that is the degree of
irregularity of the 
motion (as seen with the accuracy of the given coarse-graining)
correspond intuitively to our idea of ``complexity'' and 
can be measured better by the Shannon entropy~\cite{Ale81:1} 
\begin{equation}
S_\mu({\cal E}_{[0,n-1]})\coleq-\sum_{\bs{i}\in\Omega_D^n}\mu_{\bs{i}}
\log\mu_{\bs{i}}\ .
\label{KSE_2}
\end{equation}
In the long run, ${\cal E}$ attributes to the dynamics an entropy per 
unit time--step
\begin{equation}
h_\mu(T,{\cal E})\coleq\lim_{n\to\infty}\frac{1}{n}S_\mu({\cal E}_{[0,n-1]})\ .
\label{KSE_3}
\end{equation}
This limit is well defined~\cite{Kat99:1} and the ``average entropy
production'' $h_\mu(T,{\cal E})$ measure how predictable the dynamics
is on the coarse grained scale provided by the finite partition ${\cal
E}$. To remove the dependence on ${\cal
E}$, the Kolmogorov
entropy $h_\mu(T)$ of $\tripCT$ (or \co{KS} entropy)
is defined
as the supremum over all finite measurable
partitions~\cite{Kat99:1,Ale81:1}:
\begin{equation}
h_\mu(T)\coleq\sup_{{\cal E}}h_\mu(T,{\cal E})\qquad\
\cdot 
\label{KSE_4}
\end{equation}
If we go now to the problem of estiming a probability for 
the ``possible'' words in $k\pt{{\cal
E}_{[0,n-1]}}$, an important Theorem comes to our aid. Let ${\cal
E}_{[0,n-1]}\pt{\bs{x}}\in{\cal
E}_{[0,n-1]}$ denote the atom of containing $\bs{x}\in{\cal X}$. Then
it holds~\cite{Bil65:1}:\\[-3ex]
\begin{quote}
\begin{TT}[Shannon--Mc Millan--Breiman]{:} \label{SMB} If
$\tripCT$ is ergodic, then
\begin{equation}
\lim_{n\longrightarrow\infty}-\frac{1}{n}\log\mu\pt{{\cal
E}_{[0,n-1]}\pt{\bs{x}}} = h_\mu(T,{\cal E})\qquad \mu \text{ --
a.e.}\label{SMB1} 
\end{equation}
\end{TT}
\end{quote}
This Theorem implies that most of the words in $k\pt{{\cal
E}_{[0,n-1]}}$ have (asymptotically with $n$) the probability
$e^{-n h_\mu(T)}$; this is more precisely stated by the
next~\cite{Bil65:1}\\[-3ex] 
\begin{quote}
\begin{TT}[Asymptotic equipartition property]{:} \label{SMBEq} If
$\tripCT$ is ergodic, then for any $\varepsilon>0$ there exists a
positive integer $n_\varepsilon$ such that if $n\geq n_\varepsilon$
then $k\pt{{\cal E}_{[0,n-1]}}$ decomposes into two sets $H$ and $L$
such that
\begin{align*}
\sum_{\bs{i}\in L}&\mu_{\bs{i}}<\varepsilon
\intertext{and such that}
e^{-n\pt{h_\mu(T)+\varepsilon}}<&\mu_{\bs{i}}<
e^{-n\pt{h_\mu(T)-\varepsilon}}\label{SMB2} 
\end{align*}
for any $\bs{i}\in H$.
\end{TT}
\end{quote}
Using~\eqref{Lyap1} one can expect that the volumes~\eqref{KSE_1}
containing points with close--by trajectories decrease
as $\log\mu\pt{{\cal
E}_{[0,n-1]}\pt{\bs{x}}}\simeq - n \sum_j\log\lambda_j^+$, where
$\log\lambda_j^+$ are the positive Lyapounov exponents, and this would
fix a
relation between the Lyapounov exponents and (using Theorem~\ref{SMB}) the
\co{KS} entropy; this is indeed the
statement of the next important~\cite{Man87:1}\\[-3ex]
\begin{quote}
\begin{TT}[Pesin]{:} \label{Pesin} For smooth, ergodic
$\tripCT$:
\begin{equation}
h_\mu(T)=\sum_j\log\lambda_j^+\ \cdot
\label{Pesin1}
\end{equation}
\end{TT}
\end{quote}

\subsection{Symbolic Models as Classical Spin Chains}
\label{SMaCSC}\vspace{3mm} 
Finite partitions ${\cal E}$ of ${\cal X}$ provide symbolic models 
for the dynamical systems $\tripAAoT$ of Section~\ref{TuAt},
whereby the trajectories  
${\left\{T^jx\right\}}_{j\in\IZ}$ are encoded into sequences
${\left\{i_j\right\}}_{j\in\IZ}$ 
of indices relative to the atoms $E_{i_j}$ visited at successive times $j$;
the dynamics corresponds to the right--shift along the symbolic sequences. 
The encoding can be modelled as the shift along
a classical spin chain 
endowed with a shift--invariant state~\cite{Ali94:1}.
This will help to understand the quantum dynamical entropy which will be
introduced in the next Section.

Let $D$ be the number of atoms of a partition ${\cal E}$ of ${\cal
X}$, we shall denote by $\bs{A}_D$ the  
diagonal $D\times D$ matrix algebra generated by the characteristic
functions $e_{E_{\ell}}$ of the atoms $E_{\ell}$ and by
$\bs{A}_D^{[0,n-1]}$ the $n$-fold 
tensor product of $n$ copies of $(\bs{A}_D)$, that is the $D^n\times D^n$
diagonal matrix algebra
$\bs{A}_D^{[0,n-1]}\coleq(\bs{A}_D)_0\otimes(\bs{A}_D)_1\cdots
\otimes(\bs{A}_D)_{n-1}$.
Its typical elements are of the form $a_0\otimes a_1\cdots\otimes a_{n-1}$
each $a_j$ being a diagonal $D\times D$ matrix. 
Every $\bs{A}_D^{[p,q]}\coleq\otimes_{j=p}^q(\bs{A}_D)_j$ can be embedded into
the infinite tensor product 
$\displaystyle\bs{A}_D^\infty\coleq\otimes_{k=0}^\infty(\bs{A}_D)_k$ as
\begin{equation}
(\Id)_0\otimes\cdots\otimes(\Id)_{p-1}\otimes(\bs{A}_D)_p\otimes\cdots
\otimes(\bs{A}_D)_q\otimes(\Id)_{q+1}\otimes(\Id)_{q+2}
\otimes\cdots
\label{SMaCSC_1}
\end{equation}
The algebra $\bs{A}_D^\infty$ is a classical spin chain with a
classical $D$--spin at each site.

By means of the discrete probability measure
$\{\mu_{\bs{i}}\}_{\bs{i}\in\Omega_D^n}$, one can 
define a compatible family of states on the ``local'' algebras 
$\bs{A}_D^{[0,n-1]}$:
\begin{equation}
\rho_{{\cal E}}^{[0,n-1]}\left(a_0\otimes\cdots\otimes a_{n-1}\right)
=\sum_{\bs{i}\in\Omega_D^n}\mu_{\bs{i}}\,
(a_0)_{i_0i_0}\cdots
(a_{n-1})_{i_{n-1}i_{n-1}}\ .
\label{SMaCSC_2}
\end{equation}
\indent Indeed, let $\rho\rstr\bs{N}$ denote the restriction to a subalgebra $\bs{N}
\subseteq\bs{M}$ of a state $\rho$ on a larger
algebra $\bs{M}$.
Since
$\sum_{i_{n-1}}\mu_{i_0i_1\cdots i_{n-1}}=\mu_{i_0i_1\cdots  i_{n-2}}$,
when $n$ varies the states $\rho_{{\cal E}}^{[0,n-1]}$ form a compatible 
family, in the sense that%
\footnote{\[
\bs{A}^{[0,n-2]}_D\ni a \xrightarrow{\quad\rho_{{\cal
E}}^{[0,n-1]}\rstr\bs{A}^{[0,n-2]}_D\quad} 
\rho_{{\cal
E}}^{[0,n-1]}\left( a\otimes(\Id)_{n-1} 
\right) = \rho_{{\cal E}}^{[0,n-2]}\left( a
\right)\nonumber
\]}%
$\displaystyle
\rho_{{\cal E}}^{[0,n-1]}\rstr\bs{A}^{[0,n-2]}_D=\rho_{{\cal E}}^{[0,n-2]}$.
Then, the ``local states'' $\rho_{{\cal E}}^{[0,n-1]}$ define a
``global'' state $\rho_{{\cal E}}$ on
the infinite tensor product $\displaystyle \bs{A}_D^\infty$
so that $\rho_{{\cal E}}\rstr\bs{A}_D^{[0,n-1]}=\rho_{{\cal
E}}^{[0,n-1]}$.\\[-2ex] 
\begin{quote}
\begin{NNN}{}\ \\
$\displaystyle\bs{A}_D^\infty$ is a $D$-spin chain which is classical since 
the algebras at the integer sites consist of diagonal matrices.
The state $\rho_{{\cal E}}$ defines the statistical properties of such a 
chain, for instance the correlations among spins at different sites.
\end{NNN}
\end{quote}
\noindent
Furthermore, from~(\ref{KSE_1}) $T$-invariance of $\mu$ yields
\begin{align}
\begin{split}
\sum_{i_0}\mu_{i_0\,i_1\cdots i_{n-1}}&=\sum_{i_0}
\mu\left(E_{i_0}\cap T^{-1}(E_{i_1})\cap\cdots
\cap T^{-n+1}(E_{i_{n-1}})\right)\\
&=\mu\left({\cal X}\cap T^{-1}(E_{i_1})\cap\cdots
\cap T^{-n+1}(E_{i_{n-1}})\right)\nonumber\\
&=\mu\left(E_{i_1}\cap T^{-1}(E_{i_2})\cap\cdots
\cap T^{-n+2}(E_{i_{n-1}})
\right)\\
&=\mu_{i_1\,i_2\cdots i_{n-1}}
\end{split}
\label{SMaCSC_25}
\end{align}
It thus follows that the under the right-shift
$\displaystyle
\sigma:\bs{A}_D^\infty\mapsto\bs{A}_D^\infty$,
\begin{equation}
\sigma\pt{\bs{A}_D^{[p,q]}}=\bs{A}_D^{[p+1,q+1]}\ ,
\label{SMaCSC_3}
\end{equation}
the state $\rho_{{\cal E}}$ of the classical spin chain is translation
invariant:
\begin{equation}
\begin{split}
\rho_{{\cal E}}\circ\sigma\left(a_0\otimes\cdots\otimes a_{n-1}\right)
&=\rho_{{\cal E}}\left((\Id)_0\otimes(a_0)_1\otimes\cdots\otimes (a_{n-1})_n
\right)\\
&=\rho_{{\cal E}}\left(a_0\otimes\cdots\otimes a_{n-1}\right)\
\end{split}
\label{SMaCSC_4}
\end{equation}
Finally, denoting by $|j\rangle$ the basis vectors of the representation 
where the matrices $a\in\bs{A}_D$ are diagonal, that is
$\displaystyle a_m=\sum_{j_m=1}^D {(a_m)}_{j_m j_m}|j_m\rangle\langle
j_m|$, local states amount to
diagonal density matrices
\begin{equation}
\rho_{{\cal E}}^{[0,n-1]}=\sum_{\bs{i}\in\Omega_D^n}
\mu_{\bs{i}}\,
|i_0\rangle\langle i_0|\otimes|i_1\rangle\langle i_1|\otimes\cdots
\otimes|i_{n-1}\rangle\langle i_{n-1}|\ ,
\label{SMaCSC_5}
\end{equation}
and the Shannon entropy~\eqref{KSE_2} to the Von Neumann
entropy\footnote{%
See~\eqref{vonneumann} for the definition.}
\begin{equation}
S_\mu({\cal E}_{[0,n-1]})=-\Tr\left[\rho_{{\cal E}}^{[0,n-1]}\log
\rho_{{\cal E}}^{[0,n-1]}\right]\eqcol
H_\mu\left[{{\cal E}}_{[0,n-1]}\right]\ .
\label{SMaCSC_6}
\end{equation}
\section{Quantum Dynamical Entropies}\label{QDE}\vspace{6mm}

From an algebraic point of view, the difference between a triplet
$\tripQMoT$ (see Section~\ref{CDS}) describing a quantum dynamical
system and a triplet $\tripAAoT$ as in
Definitions~\ref{defalgST} is that $\omega$ and $\Theta$ are now a
$\Theta$--invariant state, respectively an automorphism over a
non--commutative (C* or Von Neumann) algebra of operators.\newpage
\begin{quote}\ \\[-5.5ex]
\begin{NNS}{}\ \\[-5.5ex]
\label{rem_41}
\begin{Ventry}{\mdseries a.}
\item[\mdseries a.]
In standard quantum mechanics the algebra ${\cal M}$ is the 
von Neumann algebra $B({\cal H})$ of all bounded linear operators 
on a suitable Hilbert space ${\cal H}$.
If ${\cal H}$ has finite dimension $D$, ${\cal M}$ is the algebra of 
$D\times D$ matrices.
\item[\mdseries b.]
The typical states $\omega$ are density matrices $\rho$, namely operators 
with positive eigenvalues $\rho_\ell$ such that
$\Tr(\rho)=\sum_\ell \; \rho_\ell=1$.
Given the state $\rho$, the mean value of any observable $X\in B({\cal H})$ 
is given by $\rho(X)\coleq\Tr(\rho X)$.
\item[\mdseries c.]
The $\rho_\ell$ in points (b.) are 
interpreted as the probabilities of finding the system in the
corresponding eigenstates. The uncertainty prior to the measurement is
measured by the Von~Neumann entropy of $\rho$:
\begin{equation}
H\pt{\rho}\coleq -\Tr
\pt{\rho\log\rho}= \sum_{\ell}\rho_\ell\log\rho_\ell\label{vonneumann}
\ \cdot 
\end{equation}
\item[\mdseries d.]
The usual dynamics on ${\cal M}$ is of the form $\Theta(X)=UXU^*$, where
$U$ is a unitary operator.
If one has a Hamiltonian operator that generates the continuous group 
$U_t=\exp{i\,t\,H/\hbar}$ then $U\coleq U_{t=1}$ and the time-evolution is
discretized by considering powers $U^j$.
\end{Ventry}
\end{NNS}
\end{quote}
\noindent 
The idea behind the notion of dynamical entropy is that information
can be obtained by repeatedly observing a system in the course of its
time evolution. Due to the uncertainty principle, or, in other words,
to non-commutativity, if observations are intended to gather
information about the intrinsic dynamical properties of quantum
systems, then  non-commutative extensions of the \co{KS}-entropy
ought first to decide whether quantum disturbances produced by
observations have to be taken into account or not.

Concretely, let us consider a quantum system described by a density
matrix $\rho$ acting on a Hilbert space $\c H$.  Via the wave packet
reduction postulate, generic measurement processes may reasonably
well be described by finite sets $\c Y = \{y_0, y_1,\ldots, y_{D-1}\}$
of bounded  operators $y_j\in \c B(\c H)$ such that $\sum_j y_j^* y_j
= \idty$. These sets are called \co{partitions of unity} ({\it p.u.},
for sake of shortness) and describe
the change in the state of the system caused by the corresponding
measurement process:
\begin{equation}
\label{18}
 \rho \mapsto \Gamma^*_{\c Y}(\rho) := \sum_j y_j\, \rho\, y^*_j.
\end{equation}
It looks rather natural to rely on partitions of unity to describe
the process of collecting information through repeated observations
of an evolving  quantum system~\cite{Ali94:1}. Yet, most of these
measurements interfere with the quantum evolution, possibly acting as
a source of unwanted extrinsic randomness. Nevertheless, the effect
is typically quantal and rarely avoidable.  Quite interestingly, as
we shall see later, pursuing these ideas leads to quantum
stochastic processes with a quantum dynamical entropy of their own, 
the \co{ALF}-entropy, that is also useful in a classical context.

An alternative approach~\cite{Con87:1} leads to the \co{CNT}-entropy.
This approach lacks the operational appeal of the
\co{ALF}-construction,  but is intimately  connected with the
intrinsic relaxation properties of quantum  systems~\cite{Con87:1,Nar92:1}
and possibly useful in the rapidly growing field of quantum
communication. The \co{CNT}-entropy is based on decomposing quantum
states rather than on reducing them as in~(\ref{18}).  Explicitly, if
the state $\rho$ is not a one dimensional projection, any partition
of unity $\c Y$ yields a decomposition
\begin{equation}
\label{19}
 \rho = \sum_j \tr \bigl(\rho\, y^*_jy_j\bigr)\, \frac{\sqrt\rho\,
 y^*_jy_j\sqrt\rho} {\tr\bigl(\rho\, y^*_jy_j\bigr)}.
\end{equation}
When $\Gamma^*_{\c Y}(\rho) = \rho$, reductions also provide decompositions, 
but not in general.
\subsection{Decompositions of states and
\co{CNT}--Entropy}\label{CNT}\vspace{3mm} 
The \co{CNT}-entropy is based on decomposing quantum states into
convex linear combinations of other states.  The information content
attached to the quantum dynamics is not based on modifications of the
quantum state or on perturbations of the time evolution. Let $(\c
M,\Theta,\omega)$ represent a quantum dynamical system in  the
algebraic setting and assume $\omega$ to be decomposable.  The
construction runs as follows.

\begin{itemize}
\item
 Classical partitions are replaced by finite dimensional
 C*-algebras     $\c N$ with identity embedded into $\c M$ by
 completely positive\footnote{%
 A completely positive map $\gamma$ is a map such that for every identity map
 $\Id_N:  \IC^N \mapsto \IC^N$, the tensor product
 $\gamma\otimes\Id_N$ is positive.},
 unity preserving (\co{cpu}) maps $\gamma: \c N
 \mapsto \c M$. Given $\gamma$, consider the cpu~maps  $\gamma_\ell
 := \Theta^\ell\circ\gamma$ that result from successive iterations of
 the dynamical automorphism $\Theta$, and associate to each of them an
 index set $I_\ell$. These index sets $I_\ell$ will be coupled to the
 cpu~maps $\gamma_\ell$ through the variational problem~\eqref{22}.
\item
 If $0\leq \ell < n$ then consider multi-indices $\bi i =
 (i_0,i_1,\ldots,i_{n-1}) \in I^{[0,n-1]} := 
 I_0\times \cdots\times I_{n-1}$ as labels of states $\omega_{\bi i}$ on
 $\c M$ and of weights $0<\mu_{\bi i}<1$  such that $\sum_{\bi i}
 \mu_{\bi i} = 1$ and $\omega = \sum_{\bi i} \mu_{\bi i}\,
 \omega_{\bi i}$. 
 These states are given by elements $0\leq x_{\bi
 i}^{\prime}\in {\c M}^{\prime}$, the commutant of $\c M$, such that
 $\sum_{\bi i} x_{\bi i}^{\prime} =  \idty_N$. Explicitly 
\begin{equation}
y\in {\c M} \longmapsto \omega_{\bi i} (y) := \frac{\omega (x_{\bi
 i}^{\prime} \, y)}{\omega (x_{\bi
 i}^{\prime})} \ , \quad \mu_{\bi i} := \omega (x_{\bi
 i}^{\prime}) \ \cdot \label{decomposition}
\end{equation}
The decomposition has be done with elements $x^{\prime}$ in the
commutant in order to ensure the positivity of the expectations $
\omega_{\bi i}$\footnote{%
Indeed ${\c M}\ni y\geq 0\Longrightarrow y=z^* z $ for some $z\in {\c
M}$; then it follows 
$\omega_{\bi i}(y)=\omega(x_{\bi i}^{\prime}z^* z )=\omega(z^* x_{\bi
i}^{\prime} z )\geq 0$, for $0\leq x_{\bi
 i}^{\prime}\in {\c M}^{\prime}$.}.
\item
 From $\omega = \sum_{\bi i} \mu_{\bi i}\, \omega_{\bi i}$, one
 obtains  subdecompositions $\omega = \sum_{i_\ell\in I_\ell}
 \mu_{i_\ell}^\ell\, \omega^\ell_{i_\ell}$,  where
 \begin{equation}
  \omega^\ell_{i_\ell} := \sum_{\substack{\bi i \\ i_\ell\text{
  fixed}}} \frac{\mu_{\bi i}}{\mu^\ell_{i_\ell}}\, \omega_{\bi i}
  \qquad\text{and}\qquad
  \mu_{i_\ell}^\ell := \sum_{\substack{\bi i \\ i_\ell\text{ fixed}}}
  \mu_{\bi i}.
 \label{20}
 \end{equation}
\item 
 Since $\c N$ is finite dimensional, the states $\omega \circ
 \Theta^\ell \circ  \gamma = \omega \circ \gamma$ and 
 $\omega^\ell_{i_\ell} \circ \Theta^\ell \circ \gamma$, have finite 
 von~Neumann entropies $S(\omega\circ \gamma)$ and  
 $S(\omega^\ell_{i_\ell} \circ \Theta^\ell \circ \gamma)$. With
 $\eta(x) := -x\ \log x$ if $0<x\le1$ and $\eta(0)=0$, one defines
 the $n$~subalgebra functional
 \begin{align}
  &H_\omega(\gamma_0, \gamma_1, \ldots, \gamma_{n-1}) 
  := \sup_{\omega = \sum_{\bi i} \mu_{\bi i}\, \omega_{\bi i}}\left\{
  \sum_{\bi i} \eta(\mu_{\bi i}) - \sum_{\ell=0}^{n-1} 
  \sum_{i_\ell\in I_\ell} \eta(\mu^\ell_{i_\ell}) \right.
 \nonumber \\
 &\quad+ \left.\sum_{\ell=0}^{n-1} \Bigl(S(\omega\circ \gamma_\ell)
   - \sum_{i_\ell\in I_\ell} \mu^\ell_{i_\ell}\,
    S(\omega^\ell_{i_\ell} \circ \gamma_\ell)\Bigr)\right\}.
 \label{22}
 \end{align}
\end{itemize}

We list a number of properties of $n$-subalgebra functionals,
see~\cite{Con87:1}, that will be used in the sequel:
\begin{itemize}
\item
 \co{positivity:} $0\le H_\omega(\gamma_0, \gamma_1, \ldots,
 \gamma_{n-1})$
\item
 \co{subadditivity:}\\[-2.8\baselineskip]
 \begin{eqnarray*}
 H_\omega(\gamma_0, \gamma_1, \ldots,
 \gamma_{n-1}) &\le& H_\omega(\gamma_0, \gamma_1, \ldots,
 \gamma_{\ell-1})\\
 &+& H_\omega(\gamma_\ell, \gamma_{\ell+1}, \ldots, 
 \gamma_{n-1})\ 
 \end{eqnarray*}
\item
 \co{time invariance:} $H_\omega(\gamma_0, \gamma_1, \ldots,
 \gamma_{n-1}) = H_\omega(\gamma_\ell, \gamma_{\ell+1}, \ldots,
 \gamma_{\ell+ n-1})$
\item
 \co{boundedness:} $H_\omega(\gamma_0, \gamma_1, \ldots,
 \gamma_{n-1}) \le n H_\omega(\gamma) \le n S(\omega \circ \gamma)$ 
\item
 The $n$-subalgebra functionals are invariant under 
 interchange and repetitions of arguments:
 \begin{equation}
  H_\omega(\gamma_0, \gamma_1, \ldots, \gamma_{n-1}) =
  H_\omega(\gamma_{n-1}, \ldots, \gamma_0, \gamma_0).
 \label{25}
 \end{equation}
\item
 \co{monotonicity:} If $i_\ell: \c N_\ell \mapsto \c N$, 
 $0\le\ell\le n-1$, are cpu~maps from finite dimensional
 algebras $\c N_l$ into $\c N$, then the  maps $\tilde\gamma_\ell :=
 \gamma \circ i_\ell$ are~cpu and
 \begin{equation}
  H_\omega(\tilde\gamma_0, \Theta\circ\tilde\gamma_1, \ldots, \Theta^{n-1}
  \circ \tilde\gamma_{n-1}) \le H_\omega(\gamma_0, \gamma_1, \ldots, 
  \gamma_{n-1}).
 \label{26}
 \end{equation}
\item 
 \co{continuity:} Let us consider for $\ell=0,1,\ldots,n-1$ a set of
 cpu~maps $\tilde\gamma_\ell: \c N\mapsto\c M$ such that 
 $\|\gamma_\ell - \tilde\gamma_\ell\|_\omega \le \epsilon$ for all
 $\ell$, where 
 \begin{equation}
  \|\gamma_\ell - \tilde{\gamma}_\ell\|_\omega := 
  \sup_{x\in\c N,\ \|x\|\le1}
  \sqrt{\omega\Bigl((\gamma_\ell(x) - \tilde\gamma_\ell(x))^*
  (\gamma_\ell(x) - \tilde\gamma_\ell(x))\Bigr)}\ .
 \label{27}
 \end{equation}   
 Then~\cite{Con87:1}, there exists $\delta(\epsilon)>0$ depending on the 
 dimension of the finite dimensional algebra $\c N$ and vanishing when 
 $\epsilon\to0$, such that  
 \begin{equation}
  \Bigl| H_\omega(\gamma_0, \gamma_1\ldots, \gamma_{n-1}) -
  H_\omega(\tilde{\gamma}_0, \tilde{\gamma}_1\ldots,
  \tilde{\gamma}_{n-1}) \Bigr| \le n\, \delta(\epsilon).
 \label{28}
 \end{equation}   
\end{itemize}

On the basis of these properties, one proves
the existence of the limit 
\begin{equation}
\label{new}
h^{\co{CNT}}_\omega(\theta,\gamma) :=
 \lim_n \frac{1}{n} H_\omega(\gamma_0, \gamma_1,\ldots, \gamma_{n-1})
\end{equation}
and defines~\cite{Con87:1}:\\[-3ex]
\begin{quote}
\begin{DDD}{}\ \label{defCNT}\\[-2ex]
The \co{CNT}-entropy of a quantum dynamical system 
$(\c M,\Theta,\omega)$ is
\begin{equation*}
 h_\omega^{\co{CNT}}(\Theta) := \sup_\gamma
 h^{\co{CNT}}_\omega(\Theta,\gamma)\ .
\end{equation*}
\end{DDD}
\end{quote}
\noindent
\subsection{Partitions of unit and \co{ALF}--Entropy}\label{AFE}\vspace{3mm}
The quantum dynamical entropy proposed in~\cite{Ali94:1} by Alicki and 
Fannes, \co{ALF}--entropy\footnote{L -- stands for Lindblad} for short, is
based on the idea that, in analogy 
with what one does for the metric entropy, one can model symbolically
the evolution of quantum systems by means of the
right shift along a spin chain. In the quantum case the
finite--dimensional matrix algebras at the various sites are not
diagonal, but, typically, full matrix algebras, that is the spin at
each site is a quantum spin.

This is done by means of p.u. $\c Y = \{y_0, y_1,\ldots,
y_{D-1}\}\subset{\cal M}_0\subset{\cal M}$, 
already defined in Section~\ref{QDE}; here ${\cal M}_0$ denotes a
$\Theta$--invariant subalgebra.
With ${\cal Y}$ and the state $\omega$ one constructs the $D\times D$ matrix
with entries $\omega(y_j^*y_i^{\phantom{*}})$;
such a matrix is a density matrix $\rho[{\cal Y}]$:
\begin{equation}
\rho{[{\cal Y}]}_{i,j}\coleq\omega(y_j^* y_i^{\phantom{*}})\ \cdot
\label{AFE_2}
\end{equation}
It is thus possible to define the entropy of a partition of unit as (compare~\eqref{SMaCSC_6}):
\begin{equation}
H_\omega[{\cal Y}]\coleq-\Tr\Big(\rho[{\cal Y}]\log\rho[{\cal Y}]\Big)\ \cdot
\label{AFE_3}
\end{equation}
Further, given two partitions of unit 
${\cal Y}=\Bigl(y_0,y_1,\ldots,y_D\Bigr)$,
${\cal Z}=\Bigl(z_0,z_1,\ldots,z_B\Bigr)$,
of size $D$, respectively $B$, one gets a finer partition  
of unit of size $BD$ as the set
\begin{equation}
\ \!\!\!\!\!{\cal Y}\circ {\cal Z}
\coleq\Big(
y_0 z_0,\cdots,y_0 z_B;
y_1 z_0,\cdots,y_1 z_B;\cdots;
y_D z_0,\cdots,y_D z_B \Big)\cdot
\label{AFE_4}
\end{equation}
After $j$ time--steps, ${\cal Y}$ evolves into  
$\Theta^j({\cal Y})\coleq\Big\{\Theta^j(y_1),\Theta^j(y_2),\cdots,
\Theta^j(y_D)\Big\}$. Since $\Theta$ is an automorphism,
$\Theta^j\pt{{\cal Y}}$ is a partition of unit;
then, one refines $\Theta^j({\cal Y})$, $0\leq j\leq n-1$,
into a larger partition of unit
\begin{alignat}{5}
{\cal Y}^{[0,n-1]}
& \coleq\Theta^{n-1}({\cal Y}) \ \; \circ
&\,
& \Theta^{n-2}({\cal Y}) \ \; \circ
&\,
& \cdots \ \; \circ \ \;
&\,
& \Theta({\cal Y})  \ \; \circ \ \; 
&\,
& {\cal Y}\cdot
\label{AFE_5}
\intertext{We shall denote the typical element of $\displaystyle
{\pq{{\cal Y}^{[0,n-1]}}}$ by}
{\pq{{\cal Y}^{[0,n-1]}}}_{\bs{i}} 
& = \Theta^{n-1}\pt{y_{i_{n-1}}^{\phantom{*}}}
&& \Theta^{n-2}\pt{y_{i_{n-2}}^{\phantom{*}}} 
&& \cdots 
&& \Theta(y_{i_1}^{\phantom{*}}) 
&& y_{i_0}^{\phantom{*}}\cdot
\label{AFE_55}
\end{alignat}
Each refinement is in turn associated with a density matrix
$\rho_{\cal Y}^{[0,n-1]}\coleq\rho\left[{\cal Y}^{[0,n-1]}\right]$
which is a state on the algebra 
$\displaystyle
{\bf M}_D^{[0,n-1]}\coleq\otimes_{\ell=0}^{n-1}{({\bf M}_D)}_\ell$,
with entries 
\begin{equation}
{\bigg[\rho\Big[{\cal Y}^{[0,n-1]}\Big]\bigg]}_{\bs{i},\bs{j}}\coleq
\omega\Big(
y_{j_0}^*\Theta\left(y_{j_1}^*\right)\cdots\Theta^{n-1}
\left(y_{j_{n-1}}^* y_{i_{n-1}}^{\phantom{*}}\right)
\cdots\Theta\left(y_{i_1}^{\phantom{*}}\right)y_{i_0}^{\phantom{*}}
\Big)\ \cdot
\label{AFE_6}
\end{equation}
Moreover each refinement has an entropy 
\begin{equation}
H_{\omega}\Big[{\cal Y}^{[0,n-1]}\Big] = -\Tr\Big(
\rho\Big[{\cal Y}^{[0,n-1]}\Big]\log
\rho\Big[{\cal Y}^{[0,n-1]}\Big]\Big)\ \cdot
\label{fabio3}
\end{equation}
The states $\rho_{\cal Y}^{[0,n-1]}$ are compatible%
: $\rho^{[0,n-1]}_{\cal Y}\rstr\;{\bf M}_D^{[0,n-2]}=
\rho^{[0,n-2]}_{\cal Y}$,
and define a global state
$\rho_{\cal Y}$ on the quantum spin chain 
$\displaystyle
{\bf M}_D^\infty\coleq\otimes_{\ell=0}^\infty({\bf M}_D)_\ell$.

Then, as in the previous Section, it is possible 
to associate with the quantum dynamical system $({\cal
M},\omega,\Theta)$ a symbolic dynamics which amounts to the
right--shift,
 $\displaystyle \sigma:{({\bf M}_D)}_\ell\mapsto{({\bf
M}_D)}_{\ell+1}$, along the quantum spin half--chain
(compare~\eqref{SMaCSC_3}).\\
Non-commutativity becomes evident when we check whether
$\rho_{\cal Y}$ is shift-invariant.
This requires%
\footnote{%
Moreover it is required that
$\displaystyle {\bigg[\rho\Big[{\cal
Y}^{[0,n-1]}\Big]\bigg]}_{\bs{i},\bs{j}} =
{\bigg[\rho\Big[\Theta\Big({\cal 
Y}^{[0,n-1]}\Big)\Big]\bigg]}_{\bs{i},\bs{j}}$, but this condition is
clearly satisfied from he fact that $\omega$ is invariant for the *morphism
$\Theta$. Then we derive:
\begin{align*}
{\bigg[\rho\Big[{\cal
Y}^{[0,n-1]}\Big]\bigg]}_{\bs{i},\bs{j}} = & \omega\Big(
y_{j_0}^*\Theta\left(y_{j_2}^*\right)\cdots\Theta^{n-1}
\left(y_{j_{n-1}}^* y_{i_{n-1}}^{\phantom{*}}\right)
\cdots\Theta\left(y_{i_1}^{\phantom{*}}\right)y_{i_0}^{\phantom{*}}
\Big)\\
( & \omega\circ\Theta)\Big(
y_{j_0}^*\Theta\left(y_{j_2}^*\right)\cdots\Theta^{n-1}
\left(y_{j_{n-1}}^* y_{i_{n-1}}^{\phantom{*}}\right)
\cdots\Theta\left(y_{i_1}^{\phantom{*}}\right)y_{i_0}^{\phantom{*}}
\Big)\\
& \omega\pt{\Theta\pt{
y_{j_0}^*\Theta\left(y_{j_2}^*\right)\cdots\Theta^{n-1}
\left(y_{j_{n-1}}^* y_{i_{n-1}}^{\phantom{*}}\right)
\cdots\Theta\left(y_{i_1}^{\phantom{*}}\right)y_{i_0}^{\phantom{*}}
}}\\
& \omega\pt{
\Theta\left(y_{j_0}^*\right)\Theta^2\left(y_{j_2}^*\right)\cdots\Theta^{n}
\left(y_{j_{n-1}}^* y_{i_{n-1}}^{\phantom{*}}\right)
\cdots\Theta^2\left(y_{i_1}^{\phantom{*}}\right)\Theta\pt{y_{i_0}^{\phantom{*}}}
}\\
= & {\bigg[\rho\Big[\Theta\Big({\cal 
Y}^{[0,n-1]}\Big)\Big]\bigg]}_{\bs{i},\bs{j}}\ \cdot
\end{align*}
}
$\omega\Big(\sum_\ell y_\ell^* x\:
y_\ell^{\phantom{*}}\Big)=\omega(x)$ 
for all $x\in{\cal M}$. Note that this is the case in which
$\rho \mapsto \Gamma^*_{\c Y}(\rho) :=\rho$ (see. equation~\eqref{18}).
For a comparison between classical and quantum spin chain properties, 
see table~\ref{ttt_1}.\\[2ex]
\begin{table}[H]
\begin{center}
\begin{tabular}{|| *{2}{p{3mm}@{\,}p{10mm}@{\,}p{52mm}|}|}
\hline
\multicolumn{3}{|| c |}{\rule[-2ex]{0pt}{5.5ex} \bf Classical System $({\cal X},\mu,T)$}
& \multicolumn{3}{ c ||}{\bf Quantum System $({\cal M},\omega, \Theta)$}\\
\hline
\hline
\multicolumn{3}{|| c |}{\rule[-1.5ex]{0pt}{6.5ex} $\displaystyle
\sum_{i_{n-1}}\mu_{i_0\,i_1\cdots i_{n-1}} = \mu_{i_0\,i_1\cdots i_{n-2}}$} &
\multicolumn{3}{ c ||}{$\displaystyle \sum_{\ell=1}^D
y_\ell^*y_\ell^{\phantom{*}}=\Id$ and $\Theta\pt{\Id} = \Id$}\\ 
\rule[-1ex]{0pt}{4.5ex} & $\displaystyle \Longrightarrow$ & the
local states $\rho_{\cal E}^{[0,n-1]}$ &
                          & $\displaystyle \Longrightarrow$ & the
local states $\rho_{\cal Y}^{[0,n-1]}$ \\
& \rule[-2ex]{0pt}{1ex} & form a compatible family&
&                       & form a compatible family\\
\hline
\multicolumn{6}{|| c ||}{\rule[-2.5ex]{0pt}{6.5ex} the local states
$\rho_{{\cal E}\pt{\cal Y}}^{[0,n-1]}$
 define a
global state $\rho_{{\cal E\pt{\cal Y}}}$ on the infinite tensor product}\\
\hline
\multicolumn{3}{|| c |}{\rule[-1.5ex]{0pt}{6.5ex} $\displaystyle
\sum_{i_{0}}\mu_{i_0\,i_1\cdots i_{n-1}} = \mu_{i_1\,i_2\cdots i_{n-1}}$} &
\multicolumn{3}{ c ||}{\underline{Non abelian} structure. of the algebra ${\cal M}$}\\
\rule[-5ex]{0pt}{4.5ex} & $\displaystyle \Longrightarrow$ &
\raggedright the
global state $\rho_{\cal E}$ \mbox{is translation invariant} &
                          & $\displaystyle \Longrightarrow$ &
absence of translation invariance for the global state $\rho_{\cal Y}$\\
\hline
\end{tabular}
\end{center}
\caption{Comparison between Classical and Quantum System}
\label{ttt_1}
\end{table}
In this case, the existence of a limit as in~\eqref{KSE_3} is not
guaranteed and one has to define
the \co{ALF}--entropy of $({\cal M},\omega,\Theta)$ as\\[-3ex]
\begin{quote}
\newpage
\begin{DDD}{}\ \label{defALF}\\[-5.5ex]%
\begin{subequations}
\label{AFE_7}
\begin{align}
h^{\co{ALF}}_{\omega,{\cal M}_0}(\Theta) &
\coleq\sup_{{\cal Y}\subset{\cal M}_0}
h^{\co{ALF}}_{\omega,{\cal M}_0}(\Theta,{\cal Y})\ ,
\label{AFE_7a} \\ 
\text{where}\qquad \qquad 
h^{\co{ALF}}_{\omega,{\cal M}_0}(\Theta,{\cal Y}) &  
\coleq\limsup_n \frac{1}{n} H_{\omega}\Big[{\cal Y}^{[0,n-1]}\Big]\ \cdot
\label{AFE_7b}
\end{align}  
\end{subequations}
\end{DDD}
\end{quote}
Like the metric entropy of a partition ${\cal E}$, also the
\co{ALF}--entropy of a partition of unit ${\cal Y}$ can be physically
interpreted as an asymptotic \enfasi{entropy production} relative to a
specific coarse--graining.
\section{Comparison of dynamical entropies}\label{}\vspace{6mm}
In this section we outline some of the main features of both quantum
dynamical entropies. 
The complete proofs of the above facts can be found
in~\cite{Con87:1} for the \co{CNT} and~\cite{Ali94:1,Ali96:1} for the
\co{ALF}-entropy. Here, we just sketch them, emphasizing those parts
that are important to the study of their classical limit.
\subsection{Entropy production in classical
systems}\label{}\vspace{3mm} 
Given a
dynamical system $\tripQMoT$, we will prove now that the \co{CNT}-
and the \co{ALF}-entropy  coincide with the Kolmogorov metric entropy when 
$\c M ={\cal A}_{\cal X}$ is the  Abelian von~Neumann algebra
$\Lspace{\infty}{\cal X}$ and $\Theta$ is a *automorphism of the same
kind of 
the ones defined in~(\ref{AoDCT_3}--\ref{AoDCT_3mod2}), that is
$\Theta^j\pt{f}\pt{\bs{x}}=f(T^j \pt{\bs{x}})$. \\[-3ex]
\begin{quote}
\begin{PPP}{}\ \\[-2ex]       \label{prop10}
 Let $\tripAAoT$ represent a classical dynamical
 system. Then, with the notations of the previous sections
 \begin{equation*}   
  h^{\co{CNT}}_{\omega_\mu}(\Theta) = h_\mu(T) =
  h^{\co{ALF}}_{\omega_\mu, {\cal A}_{\cal X}}(\Theta).
 \end{equation*}
\end{PPP}
\end{quote}
\noindent
\textbf{Proof of Proposition~\ref{prop10}:}\\[2ex]
\underline{\co{CNT-Entropy.}}\quad 
In this case, $h^{\co{CNT}}_{\omega_\mu}(\Theta)$ is computable by using
natural embedding  of  finite dimensional subalgebras of ${\cal A}_{\cal X}$
rather than generic cpu~maps  $\gamma$. Partitions $\c C
= \{ C_0, C_1,\ldots, C_{D-1} \}$ of $\c X$ can be
identified with the finite dimensional subalgebras $\c N_{\c C}\in\c
M_\mu$ generated by the  characteristic functions $\chi_{C_j}$ of the
atoms of the partition, with $\omega_\mu(\chi_C) = \mu(C)$. 
Also, the refinements $\c C^{[0,n-1]}$ of the evolving partitions $T^{-j}(\c
C)$  correspond to the subalgebras $\c N^{[0,n-1]}_{\c C}$ generated by 
$\chi_{C_{\bi i}} = \prod_{j=0}^{n-1} \chi_{T^{-j}(C_{i_j})}$.

Thus, if $\imath_{\c N_{\c C}}$ embeds $\c N_{\c C}$ into ${\cal A}_{\cal X}$,
then $\omega_\mu \circ \imath_{\c N_{\c C}}$ corresponds to
the state $\omega_\mu \restriction \c N_{\c C}$, which is obtained 
by restriction of $\omega_\mu$ to $\c N_{\c C}$ and
is completely determined by the expectation values  
$\omega_\mu(\chi_{C_j})$, $1\le j\le n-1$. 

Further, identifying the cpu~maps
$\gamma_\ell = \Theta^\ell \circ \imath_{\c N_{\c C}}$
with the corresponding 
subalgebras $\Theta^\ell(\c N_{\c C})$,
$h^{\co{CNT}}_{\omega_\mu}(\Theta) = h_\mu(T)$ 
follows from 
\begin{equation}
 H_\omega(\c N_{\c C}, \Theta(\c N_{\c C}),\ldots, \Theta^{n-1}(\c
 N_{\c C})) =  S_\mu(\c C^{[0,n-1]}), \quad\forall\, \c C,
\label{34}
\end{equation}
see~\eqref{KSE_3}.
In order to prove~\eqref{34}, we decompose the reference state as 
\begin{equation*}
 \omega_\mu = \sum_{\bi i} \mu_{\bi i}\, \omega_{\bi i}
 \qquad\text{with}\qquad
 \omega_{\bi i}(f) := \frac{1}{\mu_{\bi i}} \int_{\c X} \mu(dx)\, 
 \chi_{C_{\bi i}}(x)\, f(x)
\end{equation*}  
where $\mu_{\bi i} = \mu(C_{\bi i})$, see~(\ref{20}). Then, 
$\sum_{\bi i} \eta(\mu_{\bi i}) = S_\mu(\c C^{[0,n-1]})$.

On the other hand, 
\begin{equation*}
 \omega^\ell_{i_\ell}(f) =
 \frac{1}{\mu^\ell_{i_\ell}} \int_{\c X} \mu(dx)\,
 \chi_{T^{-\ell}(C_{i_\ell})}(x)\, f(x)
 \qquad\text{and}\qquad
 \mu^\ell_{i_\ell} = \mu(C_{i_\ell}).
\end{equation*}
It follows that  $\omega_\mu \circ \imath_{\c
N_{\c C}} = \omega_\mu \restriction \c N_{\c C}$ is the discrete  measure
$\{ \mu^\ell_0, \mu^\ell_1,\ldots, \mu^\ell_{n-1} \}$ for all 
$\ell=0,1,\ldots,n-1$ and, finally, that  $S(\omega^\ell_{i_\ell}
\circ \gamma_\ell) = 0$ as $\omega^\ell_{i_\ell} \circ j_\ell =
\omega^\ell_{i_\ell} \restriction  \Theta^\ell(\c N_{\c C})$ is a discrete
measure with values $0$ and $1$.
\smallskip

\underline{\co{ALF-Entropy.}} \quad The
characteristic functions of measurable 
subsets of ${\cal X}$ constitute a *subalgebra ${\cal N}_0\subseteq
{\cal A}_{\cal X}$; moreover, given a partition $\cal C$ of $\cal X$,
the characteristic functions $\chi_{C_\ell}$ of its atoms $C_\ell$, 
$\displaystyle {\cal N}_{\cal C}=\{\chi_{C_1},\chi_{C_2},\ldots,\chi_{C_D}\}$
is a partition of unit in ${\cal N}_0$.
From the definition of $\Theta$ it follows that \mbox{$\displaystyle
\Theta^j(\chi_{C_{\ell}})=\chi_{T^{-j}(C_\ell)}$} and
from~\eqref{omegamu} that
${\Big[\rho\big[{\cal N}_{{\cal C}}^{[0,n-1]}\big]\Big]}_{\bs{i},\bs{j}}
=\delta_{\bs{i},\bs{j}}\,\mu_{\bs{i}}$
(see~\eqref{KSE_1}),
whence $\displaystyle
H_\omega\big[{\cal N}_{{\cal C}}^{[0,n-1]}\big]
=S_\mu\big({\cal C}^{[0,n-1]}\big)$
(see~\eqref{KSE_2} and~\eqref{AFE_3}). In such a case, the $\limsup$
in~\eqref{AFE_7b} is actually a true limit and
yields~\eqref{KSE_3}.\hfill$\qed$\\[-2.5ex] 
\newpage
\begin{quote}
\begin{NNN}{}\ \\[-2ex]
In the particular case of the hyperbolic automorphisms of the torus,
we may restrict our attention to {\it p.u.}
whose elements
belong to the $\ast$-algebra ${\cal W}_{\text{exp}}$ of complex functions $f$ 
on $\IT$ such that the support of $\hat f$ is bounded~\cite{Ali96:1}:
\begin{equation*}
 h_\mu(T) = h^{\co{ALF}}_{(\omega_\mu, {\cal N}_0)}(\Theta) =
 h^{\co{ALF}}_{(\omega_\mu,{\cal W}_{\text{exp}})}(\Theta).
\end{equation*} 
Remarkably, the computation of the classical Kolmogorov entropy via the
quantum mechanical \co{ALF}-entropy yields a proof of~\eqref{Pesin1} that is much simpler than the 
standard ones~\cite{Arn68:1,Wal82:1}.
\end{NNN}
\end{quote}
\subsection{Entropy production in finite dimensional
systems}\label{}\vspace{3mm}  
The next case we are dealing with is characterized by
finite-dimensional algebra ${\cal M}$, as for the quantized hyperbolic 
automorphisms of the torus considered in Proposition~\ref{prop3cat};
in this case both 
the \co{CNT}- and the 
\co{ALF}-entropy are zero, see~\cite{Con87:1,Ali94:1}. Consequently, if we
decide to take the strict positivity of  quantum  dynamical entropies
as a signature of quantum chaos, quantized hyperbolic 
automorphisms of the torus cannot be called chaotic.\\[-2.5ex]
\begin{quote}
\begin{NNN}{}\ \\[-2ex]
Of course the latter observation depends on the quantum dynamical
entropy we are dealing with. There exist many alternative definitions
(different from \co{ALF} and \co{CNT}), and some of them need no to
be equal to zero for all quantum systems defined on a finite
dimensional Hilbert space: an interesting example is represented by
the \enfasi{Coherent States Entropy} introduced in~\cite{Slo94:1}.
\end{NNN}
\begin{PPP}{}\ \\[-2ex]
\label{prop4.3}
 Let $(\c M,\Theta,\omega)$ be a quantum dynamical system with
 $\c M$, a finite dimensional C*-algebra, then, 
 \begin{equation*}
  h^{\co{CNT}}_\omega(\Theta) = 0
  \qquad\text{and}\qquad
  h^{\co{ALF}}_{\omega,\c M}(\Theta) = 0.
 \end{equation*} 
\end{PPP}
\end{quote}
\noindent
\textbf{Proof of Proposition~\ref{prop4.3}:}\\[2ex]
\underline{\co{CNT-Entropy:}} as in the commutative case,
$h^{\co{CNT}}_\omega(\Theta)$ is computable by means of cpu~maps
$\gamma$ that are the natural embedding $\imath_{\c N}$ of
subalgebras $\c N\subseteq\c M$ into $\c M$. Since each
$\Theta^\ell(\c N)$ is obviously contained in the algebra  $\c
N^{[0,n-1]}\subseteq\c M$  generated by the subalgebras $\Theta^j(\c N)$,
$j=0,1,\ldots,n-1$, from the properties of the $n$-subalgebra
functionals $H$ and identifying again the natural embedding $\gamma_\ell
:= \Theta^\ell\circ \imath_{\c N}$ with the  subalgebras
$\Theta^\ell(\c N)\subseteq\c M$, we derive
\begin{alignat}{2}
  H_\omega(\gamma_0, \gamma_1,\ldots,\gamma_{n-1})
&=H_\omega(\c N, \Theta(\c N),\ldots, \Theta^{n-1}(\c N))\nonumber&&\\
  &\le H_\omega(\c N^{[0,n-1]}, \c N^{[0,n-1]},\ldots, \c N^{[0,n-1]})
&\quad&\text{by \co{monotonicity}\protect\footnotemark}\nonumber\\ 
  &\le H_\omega(\c N^{[0,n-1]}) &&\text{by ~\protect\eqref{25}}\nonumber\\
& \le S\bigl(\omega \restriction \c
  N^{[0,n-1]} \bigr) &&\text{by
\co{boundedness}}\nonumber\\ 
&\le \log N,&&\nonumber
\end{alignat}%
\protect\footnotetext{In order to match the notation
of~\protect\eqref{26}, all
 cpu maps $i_\ell$ comparing there can now be thought as the natural
embedding of $\c N$ into $\c N^{[0,n-1]}$.}%
\noindent{}where $\c M \subseteq \c M_N$. In fact, $\omega \restriction \c N$
amounts to a density matrix  with eigenvalues $\lambda_\ell$ and
von~Neumann entropy  $S(\omega \restriction \c N) = -\sum_{\ell=1}^d
\lambda_\ell \log\lambda_\ell \le \log d$. Therefore, for all $\c
N\subseteq\c M$, $h^{\co{CNT}}_\omega(\Theta, \c N) = 0$.
\smallskip

\underline{\co{ALF-Entropy:}} Let the state $\omega$ on $\c M_N$ be given by
$\omega(x) = \tr(\rho\,x)$, where $\rho$ is a density matrix in $\c
M_N$. Given a partition of unity $\c Y= {\pg{y_i}}_{i=1,2,\ldots,D}$,
the following cpu~map $\Phi_{\c Y}$
 \begin{equation}
\c M_D\otimes\c M_N
\ni M\otimes x \xrightarrow{\qquad\Phi_{\c Y}\qquad}
\Phi_{\c Y}(M\otimes x) := \sum_{i,j} y_i^*x\,y_j\, M_{ij}\in\c M_N
 \label{32}
 \end{equation}
can be used to define a state  
$\Phi_{\c Y}^*(\rho)$ on $\c M_D\otimes\c M_N$ which is dual to $\omega$:
\begin{equation*} 
 \Phi^*_{\c Y}(\rho)(M\otimes x) = \tr{\Bigl( \rho\, \Phi_{\c
 Y}(M\otimes x) \Bigr)},\quad M\in\c M_D,\ x\in\c M_N.
\end{equation*}
Since $\sum_{j=0}^D y^*_jy_j = \idty$, it follows\footnote{%
In the $\c M_D\otimes\c M_N$ space, $\Phi^*_{\c Y}(\rho)$ is
represented by the enlarged density matrix $\bv{\cal Y}\rho{\bv{\cal
Y}}^{\text{H}}$, where $\bv{\cal Y}$ denotes the $D$-dimensional vector
of $N\times N$ matrices $\pt{y_1,y_2,\ldots,y_D}$ and the superscript
``H'' stays for ``Hermitian conjugate''.}
that  
$\Phi^*_{\c Y}(\rho^k) = \bigl( \Phi^*_{\c Y}(\rho) \bigr)^k$.
Therefore, $\rho$ and $\Phi_{\c Y}^*(\rho)$ have the same spectrum,
apart possibly from the eigenvalue zero, and thus  the same
von~Neumann entropy. Moreover, $\Phi^*_{\c Y}(\rho) \restriction \c
M_D = \rho[\c Y]$ and  $\Phi^*_{\c Y}(\rho) \restriction \c M_N =
\Gamma^*_{\c Y}(\rho)$ as in~(\ref{18}). Applying the triangle
inequality~\cite{Ara70:1,Ohy93:1}
\begin{alignat}{2}
 S\bigl( \Phi^*_{\c Y}(\rho) \bigr) &\ge  \Bigl| S\bigl( \Phi^*_{\c
 Y}(\rho) \restriction \c M_D \bigr) - &&S\bigl( \Phi^*_{\c Y}(\rho)
 \restriction \c M_N \bigr) \Bigr|\quad,\text{ that is}\nonumber\\
 S\bigl(\rho\bigr) &\ge  \Bigl| S\bigl( \rho[\c Y] \bigr) \qquad- &&S\bigl(
 \Gamma^*_{\c Y}(\rho) \bigr) \Bigr|\ ,\nonumber
\end{alignat} 
that leads to $S(\rho[\c Y]) \le 2\log d$. Finally, as evolving
{\it p.u.} $\Theta^j(\c Y)$ and their ordered
refinements~(\ref{AFE_5}--\ref{AFE_55}) remain in  $\c M_N$, one gets 
\begin{equation}
 \limsup_k \frac{1}{k} H_\omega[\c Y^{[0,n-1]}] = 0,\quad \c
 Y\subset\c M_N\ \cdot\hfill\tag*{\qed}
\end{equation} 
From the considerations of above, it is clear that the main field of
application  of the \co{CNT}- and \co{ALF}-entropies are infinite
quantum systems, where the  differences between  the two come to the
fore~\cite{Ali95:1}. The former has been proved to be useful to connect
randomness with clustering  properties and  asymptotic commutativity.
A rather strong form of clustering and asymptotic Abelianness is
necessary to have a non-vanishing
\co{CNT}-entropy~\cite{Nar92:1,Nar95:1,Nes00:1}. 
In particular, the infinite dimensional quantization of the
automorphisms of the torus has vanishing \co{CNT}-entropy for most of
irrational  values of the deformation parameter $\phi$, whereas,
independently of the value of $\phi$, the \co{ALF}-entropy is always
equal to the positive Lyapounov exponent. These results reflect the
different perspectives upon which the two constructions are based. 
\section{An explicit construction: \co{ALF}--Entropy of Sawtooth
Maps}\label{}\vspace{6mm} 
In the following we develop a technique suited to compute and to
simplify the Von Neumann entropy $H_{\omega}\Big[{\cal Y}^{[0,n-1]}\Big]$
of~\eqref{fabio3} for the class of discrete classical systems
$\tripQSa$, whose continuous limit in $\tripASa$ has been shown
in Section~\ref{ClfSM}. 

For the class of discrete systems we are dealing with, one can not
define a metric entropy, being the measure a discrete one, instead we
can profitably use quantum dynamical entropies, although we are in a 
commutative case. Indeed, the only necessary ingredient to construct such
kind of entropies, is the algebraic description $\tripQSa$
and, in the \co{ALF} entropy computation, the use of a partition of unit.

The reason to choose the \co{ALF} entropy
instead of the \co{CNT} is the numerical compatibility of the former;
indeed the variational problem in
~\eqref{22} is apparently very hard to be attached numerically.

By remember Proposition~\ref{prop4.3}, we know that we cannot go to
compute neither $h^{\co{ALF}}_{\omega,{\cal M}_0}(\Theta)$ nor
$h^{\co{ALF}}_{\omega,{\cal M}_0}(\Theta,{\cal Y})$ of
Definition~\ref{defALF}, because these quantity are expected to be
zero. The analysis of entropy production will be performed in the next
Chapter, now we only set up the framework to compute it.

A useful partition of unit in 
$\tripASa$ is constructed by collecting a finite number $D$ of Weyl
operators 
$\widetilde{W}(\bs{r}_j)$ defined in~\eqref{RoeiW_4}, indexed by their
labels $\bs{r}_j$, as in the following\newpage
\begin{quote}\ \\[-7.5ex]
\begin{DDD}{}\label{partizione}\ \\
\noindent Given a subset $\Lambda$ of the lattice consisting of points
\begin{align}
\qquad {\bigg\{\bs{r}_j\bigg\}}_{j = 1}^D & \eqcol
\Lambda \subset 
{\pt{\IZ/N\IZ}}^2 \cdot \label{CoAFE_217}\ ,\\
\intertext{we shall denote by ${\cal
\widetilde{Y}}$ the partition of unit in
$\tripASa$ given by:}
{\cal \widetilde{Y}} = {\bigg\{\widetilde{y}_j\bigg\}}_{j = 1}^D & \coleq
{\pg{\frac{1}{\sqrt{D}}\;\widetilde{W}(\bs{r}_j)}}_{j = 1}^D\ \cdot
\label{CoAFE_3} 
\end{align}
\end{DDD}
\end{quote}
\noindent
From the above definition, the elements of the
refined partitions in~\eqref{AFE_55} take the form:
\begin{equation}
{\pq{{\cal \widetilde{Y}}^{[0,n-1]}}}_{\bs{i}} =
\frac{1}{N}\frac{1}{D^{\frac{n}{2}}}
\sum_{\bs{\ell} \in {(\ZNZ{N})^2}} 
e^{\frac{\:2\pi i}{N}\pq{
\bs{r}_{i_{n-1}}\cdot V_{\alpha}^{n-1} \pt{\bs{\ell}} + \,\cdots 
\,+\bs{r}_{i_1}\cdot V_{\alpha} \pt{\bs{\ell}} 
+ \bs{r}_{i_0}\cdot \bs{\ell}}}
\ket{\bs{\ell}}\bra{\bs{\ell}}\ \cdot
\label{CoAFE_31}
\end{equation}	
Then, the multitime correlation matrix $\rho_{\widetilde{\cal
Y}}^{\pq{0,n-1}}$ in~\eqref{AFE_6} has entries:
\begin{align}
{\bigg[\rho\Big[{\cal \widetilde Y}^{[0,n-1]}\Big]\bigg]}_{\bs{i},\bs{j}}
& = \frac{1}{N^2}\frac{1}{D^{n}}
\sum_{\bs{\ell} \in {(\ZNZ{N})^2}} 
e^{\frac{\:2\pi i }{N}
\overset{n-1}{\underset{p=0}{\sum}}
\pt{\bs{r}_{i_p}- \bs{r}_{j_p}} 
\cdot V_{\alpha}^{p} 
\pt{\bs{\ell}}}\ \ , \ \ V_{\alpha}^{0}\pt{\bs{\ell}} = {\Id}
\label{CoAFE_5}\\
& = \sum_{\bs{\ell} \in {(\ZNZ{N})^2}} 
\bkk{\bs{i}}{g_{\bs{\ell}}\pt{n}}\bkk{g_{\bs{\ell}}\pt{n}}{\bs{j}}\ , 
\label{CoAFE_6}\\
\text{with}\quad \bkk{\bs{i}}{g_{\bs{\ell}}\pt{n}}&\coleq
\frac{1}{N}\frac{1}{D^{\frac{n}{2}}}
e^{\frac{\:2\pi i }{N}
\overset{n-1}{\underset{p=0}{\sum}}
\bs{r}_{i_p} \cdot V_{\alpha}^{p} \pt{\bs{\ell}}}\in {\IC}^{D^n}\ .
\label{CoAFE_7}
\end{align}
The density matrix $\rho_{\widetilde{\cal
Y}}^{\pq{0,n-1}}$ can now be used
to numerically compute the Von Neumann entropy $H_{\tn}\Big[{\cal
Y}^{[0,n-1]}\Big]$ 
of~\eqref{fabio3};
however, the large dimension ($D^n\times D^n$) makes the computational
problem
very hard, a part for small numbers of
iterations. Our goal is to prove that another matrix (of fixed dimension $N^2
\times N^2$) can be used instead of $\rho_{\widetilde{\cal
Y}}^{\pq{0,n-1}}$. In particular, the next proposition can be seen as
an extension of the strategy that led us to prove
Proposition~\ref{prop4.3}\\[-2ex] 
\begin{quote}
\begin{PPP}{}\ \\[-5.5ex]
\begin{Ventry}{}\label{prop_41}
	\item[] Let ${\cal G}\pt{n}$ be the $N^2\times N^2$ matrix
	with entries
\begin{equation}
{\cal G}_{\bs{\ell}_1,\bs{\ell}_2}\pt{n} \coleq
\bkk{g_{\bs{\ell}_2}\pt{n}}{g_{\bs{\ell}_1}\pt{n}}
\label{CoAFE_1803}
\end{equation}
given by the scalar products of the vectors
$\ket{g_{\bs{\ell}}\pt{n}}\in {\cal H}_{D^n}=
{\IC}^{D^n}$ in~\eqref{CoAFE_7}.
Then, the entropy of the partition of unit ${\cal \widetilde
Y}^{[0,n-1]}$ with elements~\eqref{CoAFE_3} is given by:
\begin{equation}
H_{\tn}\pq{{\cal \widetilde
Y}^{[0,n-1]}}
= -{\Tr}_{{\cal H}_{N}^D}\Big({\cal G}\pt{n}\log{\cal G}\pt{n}\Big)
\label{CoAFE_10}
\end{equation}
\end{Ventry}
\end{PPP}
\end{quote}
\noindent 
\textbf{Proof of Proposition~\ref{prop_41}:}\\[2ex]
${\cal G}\pt{n}$ is hermitian and
from~\eqref{CoAFE_7} it follows that ${\Tr}_{{\cal H}_{N^2}}{\cal
G}\pt{n} =1$.\\
Let ${\cal H}\coleq {\cal H}_{D^n}\otimes{\cal H}_{N}^D$ and
consider the projection $\rho_{\psi} =\ket{\psi}\bra{\psi}$ onto
\begin{equation}
{\cal H}\ni\ket{\psi} \coleq \sum_{\bs{\ell} \in {(\ZNZ{N})^2}}
\ket{g_{\bs{\ell}}\pt{n}}\otimes \ket{\bs{\ell}}.
\label{CoAFE_80}
\end{equation}
We denote by $\Sigma_1$ the restriction of $\rho_{\psi}$ to the full
matrix algebra $M_1\coleq M_{D^n}\pt{\IC}$ and by $\Sigma_2$ the
restriction to $M_2\coleq M_{N^2}\pt{\IC}$. It follows that:
\begin{equation*}
\Tr_{{\cal H}_{D^n}}\pt{\Sigma_1 \cdot m_1} = 
\bra{\psi} m_1\otimes{\Id}_2\ket{\psi}
= \sum_{\bs{\ell} \in {(\ZNZ{N})^2}}
\bra{g_{\bs{\ell}}} m_1
\ket{g_{\bs{\ell}}}\ , \ \forall m_1\in M_1 \ \cdot
\end{equation*}
Thus, from~\eqref{CoAFE_6},
\begin{equation}
\Sigma_1 = \rho_{\cal \widetilde{Y}}^{\pq{0,n-1}} = 
\sum_{\bs{\ell} \in {(\ZNZ{N})^2}} 
\ket{g_{\bs{\ell}}\pt{n}}\bra{g_{\bs{\ell}}\pt{n}}\
\cdot\label{CoAFE_801}
\end{equation}
On the other hand, from
\begin{align*}
\Tr_{{\cal H}_{N}^D}\pt{\Sigma_2 \cdot m_2} & = 
\bra{\psi} {\Id}_1\otimes m_2\ket{\psi}\\
& = 
\sum_{\bs{\ell}_1,\bs{\ell}_2 \in {(\ZNZ{N})^2}}
\bkk{g_{\bs{\ell}_2}\pt{n}}{g_{\bs{\ell}_1}\pt{n}}
\langle\bs{\ell}_2
| m_2 |
\bs{\ell}_1\rangle
\ , \ \forall m_2\in M_2\ ,
\end{align*}
it turns out that $\Sigma_2 = {\cal G}\pt{n}$, whence the result
follows from Araki--Lieb's \mbox{inequality~\cite{Ara70:1}\hfill$\qed$}
\subsection{A simpler form for the ${\pg{T_\alpha}}$ subfamily of the
UMG}\label{}\vspace{3mm}   
\noindent We now return to the explicit computation of the density
matrix ${\cal G}\pt{n}$ in Proposition~\ref{prop_41}.
By using the transposed matrix $T_\alpha^{\text{tr}}$, the
vectors~\eqref{CoAFE_7} now read
\begin{align}
\bkk{\bs{i}}{g_{\bs{\ell}}\pt{n}} & =
\frac{1}{N\,D^{\frac{n}{2}}}
e^{\frac{\:2\pi i}{N}\;\bs{\ell}\cdot
\bs{f}_{\Lambda,\alpha}^{(n),N}\pt{\bs{i}}}
\label{CoAFE_710}\\
\bs{f}_{\Lambda,\alpha}^{(n),N}\pt{\bs{i}} & \coleq
\sum_{p=0}^{n-1}\;{\pt{T_\alpha^{\text{tr}}}}^{p} \,\bs{r}_{i_p}
\pmod{N} \label{CoAFE_720}
\end{align}
where we made explicit the various dependencies of~\eqref{CoAFE_720}
on $n$ the time--step, $N$ the inverse lattice--spacing, the chosen set
$\Lambda$ of $\bs{r}_j$'s and the $\alpha$ parameter of the dynamics
in ${\text{SL}}_2 {(\ZNZ{N})^2}$.

\noindent In the following we shall use the equivalence classes 
\begin{equation}
\pq{\bs{r}}  \coleq\pg{\bs{i}\in\Omega_{D}^{(n)} \;\Big|\; 
\bs{f}_{\Lambda,\alpha}^{(n),N}
\pt{\bs{i}}\equiv\bs{r}\in{(\ZNZ{N})^2}\pmod{N}}\ ,
\label{CoAFE_730}
\end{equation}
their
cardinalities $\# \pq{\bs{r}}$ and, in particular, the frequency
function $\nu_{\Lambda,\alpha}^{(n),N}$
\begin{equation}
\ZNZD\ni\bs{r} \longmapsto\nu_{\Lambda,\alpha}^{(n),N}\pt{\bs{r}}
\coleq 
\frac{\# \pq{\bs{r}}}{D^n}\ \cdot
\label{CoAFE_740}
\end{equation}
\\[-7.5ex]
\begin{quote}
\begin{PPP}{}\ \\[-7.5ex]
\begin{Ventry}{}\label{prop_42}
	\item[] The Von Neumann entropy of the refined (exponential)
	partition of unit up to time $n-1$ is given by:
\begin{equation}
H_{\tn}\pq{{\cal \widetilde Y}^{[0,n-1]}} =
- \sum_{\bs{r} \in {(\ZNZ{N})^2}} 
\nu_{\Lambda,\alpha}^{(n),N}\pt{\bs{r}}
\log\, \nu_{\Lambda,\alpha}^{(n),N}\pt{\bs{r}}\label{CoAFE_77}
\end{equation}
\end{Ventry}
\end{PPP}
\end{quote}
\noindent
\textbf{Proof of Proposition~\ref{prop_42}:}\\[2ex]
Using ~\eqref{CoAFE_710}, the matrix ${\cal
G}\pt{n}$ in Proposition~\ref{prop_41} can be written as:
\begin{align}
{\cal G}\pt{n} & = 
\frac{1}{D^n}\sum_{\bs{i}\in\Omega_D^n} 
\ket{f_{\bs{i}}\pt{n}}\bra{f_{\bs{i}}\pt{n}}\ ,
\label{CoAFE_421}\\
\bkk{\,\bs{\ell}\,}{f_{\bs{i}}\pt{n}} & =
\frac{1}{N}
e^{\frac{\:2\pi i}{N}\;
\bs{f}_{\Lambda,\alpha}^{(n),N}\pt{\bs{i}}\cdot\bs{\ell}}
\label{CoAFE_422}
\end{align}
The vectors $\ket{f_{\bs{i}}\pt{n}}\in {\cal H}_{N}^D = {\IC}^{N^2}$
are such that
$\displaystyle
\bkk{f_{\bs{i}}\pt{n}}{f_{\bs{j}}\pt{n}} = 
\delta_{
\bs{f}_{\Lambda,\alpha}^{(n),N}\pt{\bs{i}}
\;,\;
\bs{f}_{\Lambda,\alpha}^{(n),N}\pt{\bs{j}}
}^{\pt{N}}$,
where with $\delta^{\pt{N}}$ is the $N$--periodic Kronecker delta.
For sake of simplicity, we say that $\ket{f_{\bs{i}}\pt{n}}$ belongs
to the equivalence class $\pq{\bs{r}}$ in~\eqref{CoAFE_730} if
$\bs{i}\in\pq{\bs{r}}$; vectors in different equivalence classes are
thus orthogonal, whereas those in a same equivalence class
$\pq{\bs{r}}$ are such that
\begin{align}
\bra{\bs{\ell}_1}
\pt{\sum_{\bs{i}\in \pq{\bs{r}}} 
\ket{f_{\bs{i}}\pt{n}}\bra{f_{\bs{i}}\pt{n}}}
\ket{\bs{\ell}_2} & = \frac{1}{N^2}
\sum_{\bs{i}\in \pq{\bs{r}}}
e^{\frac{\:2\pi i}{N}\;
\bs{f}_{\Lambda,\alpha}^{(n),N}\pt{\bs{i}}\cdot
\pt{\bs{\ell}_1-\bs{\ell}_2}}\notag\\
& = D^n \, \nu_{\Lambda,\alpha}^{(n),N}\pt{\bs{r}} 
\;\bkk{\bs{\ell}_1}{\bs{e}\pt{\bs{r}}}
\bkk{\bs{e}\pt{\bs{r}}}{\bs{\ell}_2}\notag\\
\bkk{\bs{\ell}}{\bs{e}\pt{\bs{r}}} & = \frac{e^{
\frac{\:2\pi i}{N}\;\bs{r}\cdot\bs{\ell}}}{N}
\in {\cal H}_{N}^D\notag
\end{align}
Therefore, the result follows from the spectral decomposition\\[2ex]  
$\phantom{a} $\hspace{36mm}$\displaystyle
{\cal G}\pt{n} = 
\sum_{\bs{r} \in {(\ZNZ{N})^2}} 
\nu_{\Lambda,\alpha}^{(n),N}\pt{\bs{r}}
\ket{\bs{e}\pt{\bs{r}}}\bra{\bs{e}\pt{\bs{r}}}\cdot\hfill\qed$
\pagestyle{fancyplain}
\chapter{Classical/Continuous Limit of Quantum Dynamical
Entropies}\label{s7}\vspace{9mm} 
Proposition~\ref{prop4.3} confirms the intuition that finite
dimensional, discrete time, quantum dynamical systems, however
complicated the distribution  of their quasi-energies might be,
cannot produce enough information  over large times to generate a
non-vanishing entropy per unit time. This is due to the fact that,
despite the presence of almost random features over finite intervals,
the time evolution cannot bear random signatures if watched long
enough, because almost periodicity would always prevail
asymptotically.

However, this does not mean that
the dynamics may not be able to show a significant entropy rate over
finite interval of times, these being typical of the underlying
dynamics; all this Chapter is devoted to explore this phenomenon.

As already observed in the Introduction, in quantum chaos one deals
with quantized classically chaotic systems; there, one finds that
classical and quantum mechanics are both correct descriptions
over times scaling
with $\log\hbar^{-1}$. Therefore, the classical--quantum
correspondence occurs over times much smaller than the 
Heisenberg recursion time that typically scales as $\hbar^{-\alpha},\
\alpha>0$.  
In other words, for quantized classically chaotic systems, the
classical description has to be replaced by the quantum one much
sooner than for integrable systems.
\section{\co{CNT} and \co{ALF} Entropies on
$\tripQT$}\label{CNTALFoT}\vspace{6mm} 
In this section we take the
the \co{CNT} and the \co{ALF}-entropy as
good indicators of the degree of randomness of a quantum dynamical
system. Then, we show that underlying classical chaos plus
Hilbert space finiteness make a characteristic logarithmic time scale emerge 
over which these systems can be called
chaotic.
\subsection{\co{CNT}-entropy}\label{CNTE}\  \\[-8ex]
\begin{quote}
\begin{TT}{:} \label{TEOCNT}
 Let $\tripCT$ be a classical dynamical system 
 which is the classical limit of a sequence of finite dimensional
 quantum dynamical systems $\tripQT$. 
 We also assume that the dynamical localization
 condition~\ref{dynloc} holds. If
 \begin{enumerate}
 \item 
  $\c C = \{ C_0, C_1,\ldots, C_{D-1} \}$ is a finite measurable partition
  of $\c X$,
 \item 
  $\c N_{\c C} \subset \Lspace{\infty}{\cal X}$ is the finite
 dimensional subalgebra 
  generated by the characteristic functions $\chi_{C_j}$ of the atoms
  of $\c C$,
 \item 
  $\imath_{\c N_{\c C}}$ is the natural embedding of $\c N_{\c C}$
  into $\Lspace{\infty}{\cal X}$, ${\cal J}_{N\infty }$ the anti-Wick
  quantization map and 
  \begin{equation*}
   \gamma^\ell_{\c C} \coleq \Theta^\ell_N \circ {\cal J}_{N\infty}
   \circ \imath_{\c N_{\c C}},\quad \ell=0,1,\ldots,k-1,
  \end{equation*}
 \end{enumerate}
 then there exists an $\alpha$ such that
 \begin{equation*}
  \lim_{\substack{k,\ N\to \infty \\ k\le \alpha \log N}} \frac{1}{k} 
  \left| H_{\tn}(\gamma^0_{\c C}, \gamma^1_{\c C},\ldots, 
  \gamma^{k-1}_{\c C}) - S_\mu\bigl( \c C^{[0,k-1]} \bigr) \right| = 0.
 \end{equation*}
\end{TT}
\end{quote}
\noindent
\textbf{Proof of Theorem~\ref{TEOCNT}:}\\[2ex]
We split the proof in two parts:
\begin{enumerate}
\item 
 We relate the quantal evolution $\gamma^\ell_{\c C} = \Theta_N^\ell
 \circ {\cal J}_{N\infty} \circ \imath_{\c N_{\c C}}$ to the
 classical evolution $\tilde\gamma_{\c C}^\ell := {\cal J}_{N\infty}
 \circ \Theta^\ell \circ \imath_{\c N_{\c C}}$ using the continuity
 property of the entropy functional.
\item 
 We find an upper and a lower bound to the entropy functional that 
 converge to  the \co{KS}-entropy in the long time limit.
\end{enumerate}

We define for convenience the algebra $\c N_{\c C}^\ell := 
\Theta^\ell(\c N_{\c C})$ and the algebra $\c N^{[0,k-1]}_{\c C}$
corresponding to the refinements  $\c C^{[0,k-1]} =
\bigvee^{k-1}_{\ell=0} T^{-\ell}(\c C)$ which consist of
atoms $\c C_{\bi i} := \bigcap^{k-1}_{\ell=0} T^{-\ell}(C_{i_\ell})$
labeled by the multi-indices $\bi i = (i_0, i_1,\ldots, i_{k-1})$. 
Thus the algebra $\c N^{[0,k-1]}_{\c C}$ is generated by the
characteristic functions $\chi_{C_{\bi i}}$. 
\smallskip

\textbf{Step 1}
\smallskip

The maps $\gamma^\ell_{\c C} $ and $\tilde{\gamma}^\ell_{\c C}$
connect the quantum and classical time evolution. Indeed, using 
Proposition~\ref{prop3} 
\begin{equation*}
 k \le \alpha \log N \Rightarrow \|\Theta^k_N \circ  {\cal J}_{N\infty}
 \circ \imath_{\c N_{\c C}}(f) -  {\cal J}_{N\infty} \circ \Theta^k
 \circ \imath_{\c N_{\c C}}(f) \|_2 \le \varepsilon, 
\end{equation*}
or
\begin{equation*}
 k \le \alpha \log N \Rightarrow \| \gamma^k_{\c C} - 
 \tilde\gamma^k_{\c C} \|_2 \le \varepsilon 
\end{equation*}

This in turn implies, due to strong continuity,
\begin{equation*}
 \left| H_{\tn}(\gamma^0_{\c C}, \gamma^1_{\c C},\ldots, \gamma^{k-1}_{\c
 C}) - H_{\tn}(\tilde\gamma^0_{\c C}, \tilde\gamma^1_{\c C},\ldots,
 \tilde\gamma^{k-1}_{\c C}) \right| \le k \delta(\varepsilon)
\end{equation*}
with $\delta(\varepsilon)>0$ depending on the dimension of the space
$\c N_{\c C}$ and vanishing when $\varepsilon \to 0$. From now on we
can concentrate on the classical evolution and benefit from its
properties.\\[0.5ex]

\textbf{Step 2, upper bound}
\smallskip

We now show that 
\begin{equation*}
 H_{\tn}(\tilde\gamma^0_{\c C}, \tilde\gamma^1_{\c
 C},\ldots, \tilde\gamma^{k-1}_{\c C}) \le S_\mu(\c C^{[0,k-1]}).
\end{equation*}
Notice that we can embed $\c N^\ell_{\c C}$ into $\Lspace{\infty}{\cal X}$ by first
embedding it into $\c N^{[0,k-1]}_{\c C}$ with  $\imath_{
\c N^{[0,k-1]}_{\c C}\c N^\ell_{\c C}}$ and then embedding $\c
N^{[0,k-1]}_{\c C}$ into $\Lspace{\infty}{\cal X}$
with $\imath_{\c N^{[0,k-1]}_{\c C}}$:
\begin{equation*}
 \imath_{\c N^\ell_{\c C}} = \imath_{\c N^{[0,k-1]}_{\c C}} \circ
 \imath_{ \c N^{[0,k-1]}_{\c C}\c N^\ell_{\c C}}. 
\end{equation*}
We now estimate:
\begin{align}
 &H_{\tn}(\tilde\gamma^0_{\c C}, \tilde\gamma^1_{\c C},\cdots, 
 \tilde\gamma^{k-1}_{\c C})=
\nonumber \\ 
 &\quad = H_{\tn}({\cal J}_{N\infty} \circ \Theta^0 \circ
\imath_{\c N_{\c C}},\cdots, 
 {\cal J}_{N\infty} \circ \Theta^{k-1} \circ
 \imath_{\c N_{\c C}}) =
\nonumber \\
 &\quad = H_{\tn}({\cal J}_{N\infty} \circ 
\imath_{\c N^0_{\c C}},\cdots, 
 {\cal J}_{N\infty} \circ 
 \imath_{\c N^{k-1}_{\c C}}) =
\nonumber \\
 &\quad = H_{\tn}({\cal J}_{N\infty} \circ \imath_{\c N^{[0,k-1]}_{\c C}} 
 \circ \imath_{\c N^{[0,k-1]}_{\c C}\c N^0_{\c C}},\cdots, 
 {\cal J}_{N\infty} \circ \imath_{\c N^{[0,k-1]}_{\c C}}  \circ
 \imath_{\c N^{[0,k-1]}_{\c C}\c N^{k-1}_{\c C}}) 
\nonumber \le\\
 &\quad\le H_{\tn}({\cal J}_{N\infty} \circ \imath_{\c N^{[0,k-1]}_{\c
 C}},\cdots, {\cal J}_{N\infty} \circ  \imath_{\c N^{[0,k-1]}_{\c C}})\le
 \displaybreak  
\nonumber \\
&\quad\le H_{\tn}({\cal J}_{N\infty} \circ \imath_{\c N^{[0,k-1]}_{\c
 C}})\le
\nonumber \\
 &\quad\le S\left( \tau_N \circ {\cal J}_{N\infty} \circ \imath_{\c
 N^{[0,k-1]}_{\c C}} \right)\ \cdot
\end{align}
The first inequality follows from monotonicity of the entropy functional, the
second from invariance under repetitions~(see~\eqref{25}) and the third
from boundedness in terms 
of von~Neumann entropies. The state $\tau_N \circ {\cal J}_{N\infty} 
\circ \imath_{\c N^{[0,k-1]}_{\c C}}$ takes the values
\begin{align*}
 &\tau_N \bigl( {\cal J}_{N\infty}(\chi_{C_{\bi i}}) \bigr) = 
 \tau_N \Bigl( N \int_{\c X} \mu(\ud \bs{x})\,\chi_{C_{\bi i}}(\bs{x})\,
 |C_N^1(\bs{x})\> \< C_N^1(\bs{x})| \Bigr) \\
 &\quad= \int_{\c X} \mu(\ud \bs{x})\, \chi_{C_{\bi i}}(\bs{x})\, \< C_N^1(\bs{x}),
 C_N^1(\bs{x})\> = \omega_{\mu}(\chi_{C_{\bi i}}) = \mu(C_{\bi i}). 
\end{align*}
This gives, together with $S(\mu(C_{\bi i})) = S_\mu(\c C^{[0,k-1]})$,
the desired upper bound.\\[0.5ex]

\textbf{Step 2, lower bound}\\[0.5ex]

We show that $\forall\: \varepsilon>0$ there exists an $N^{\prime}$
such that 
\begin{equation*}
 H_{\tn}(\tilde\gamma^0_{\c C}, \tilde\gamma^1_{\c C},\ldots, 
 \tilde\gamma^{k-1}_{\c C}) \ge S_\mu(\c C^{[0,k-1]}) - k \varepsilon
\end{equation*}
will holds eventually for $N>N^{\prime}$.\\
As $H_{\tn}(\tilde\gamma^0_{\c C}, \tilde\gamma^1_{\c
C},\ldots,  \tilde\gamma^{k-1}_{\c C})$ is defined as a supremum over
decompositions of the state $\tau_N$, we can construct a lower
bound by picking a good decomposition. Consider the decomposition
$\tau_N = \sum_{\bi i} \mu_{\bi i}\, \omega_{\bi i}$ with
\begin{align*}
& \left\{
\begin{alignedat}{2}
 \omega_{\bi i}&:\phantom{=} \c M_N \ni x \mapsto \omega_{\bi i}(x) &&:=
 \frac{\tau_N\pq{ ({\cal J}_{N\infty}(\chi_{C_{\bi i}}))(x)}}
 {\tau_N \pq{{\cal J}_{N\infty}(\chi_{C_{\bi i}})}} \\
 \mu_{\bi i} &:= \tau_N \pq{{\cal J}_{N\infty}(\chi_{C_{\bi i}})}
  &&\phantom{:}= \mu\pt{C_{\bi i}}
\end{alignedat}
\right.
\intertext{and the subdecompositions $\tau_N =
 \sum_{j_\ell}\mu^\ell_{j_\ell}\, \omega^\ell_{j_\ell}$, \ \
 $\ell=0,1,\ldots,k-1$, with\protect\footnotemark}
& \left\{
\begin{alignedat}{2}
 \omega^\ell_{j_\ell}&:\phantom{=} \c M_N \ni x \mapsto
 \omega^\ell_{j_\ell}(x) &&:= 
 \frac{\tau_N \pq{({\cal J}_{N\infty}(\chi_{T^{-\ell}(C_{j_\ell})}))(x)}}
 {\tau_N \pq{{\cal J}_{N\infty}(\chi_{C_{j_l}})}} \\
 \mu^\ell_{j_\ell} &:= \tau_N \pq{{\cal J}_{N\infty}(\chi_{C_{j_\ell}})}
  &&\phantom{:}= \mu\pt{C_{j_\ell}}
\end{alignedat} 
\right.
\end{align*}
\footnotetext{In the derivation of following formulae we extensively
 use the relation 
\begin{equation*}
\sum_{\substack{\bi i \\ i_\ell\text{ fixed}}}
\chi_{C_{\bi
i}}\pt{\bs{y}}=\chi_{T^{-\ell}\pt{C_{i_\ell}}}\pt{\bs{y}}\ \cdot
\end{equation*}}
In comparison with~\eqref{decomposition}, it is not necessary to go to
the commutant for one can use the ciclicity property of the
trace\footnote{%
${\c M}_N\ni x\geq 0\Longrightarrow x= z z^* $ for some $z\in {\c
M}_N$; then it follows 
$\omega_{\bi i}(x)=\tau_N(y_{\bi i} z z^*)=\tau_N(z^* y_{\bi i}
z )\geq0$, 
for ${\c
M}_N \ni y_{\bi i}\coleq{\cal J}_{N\infty}(\chi_{C_{\bi i}})$ that is
obviously greater than zero; indeed for all $\ket{\psi}\in{\cal H}_N$
we have that $\bkkk{\psi}{{\cal J}_{N\infty}(\chi_{C_{\bi i}})}{\psi}=
N \int_{\c X} \mu(\ud \bs{x})\,\chi_{C_{\bi i}}(\bs{x})\,
 {\abs{\bkk{\psi}{C_N^1(\bs{x})}}}^2\geq 0$.
}.  
We then have:
\begin{equation*}
 H_{\tn}(\tilde\gamma^0_{\c C}, \tilde\gamma^1_{\c C},\ldots, 
 \tilde\gamma^{k-1}_{\c C}) \ge S_\mu(\c C^{[0,k-1]}) -\sum^{k-1}_{\ell=0}
 \sum_{i_\ell \in I_\ell} \mu^\ell_{i_\ell}\,  S(\omega^\ell_{i_\ell}
 \circ \tilde\gamma^\ell_{\c C}).
\end{equation*}
The inequality stems from the fact that $H_{\tn}(\tilde\gamma^0_{\c C}, 
\tilde\gamma^1_{\c C},\ldots, 
\tilde\gamma^{k-1}_{\c C} )$ is a supremum, whereas the middle terms
in the original definition of the entropy functional in~\eqref{22}
drop out because 
they are equal in magnitude but opposite in sign\footnote{%
Indeed \rule{0pt}{5ex}
$\displaystyle
  S(\tn\circ \tilde\gamma^\ell_{\c C})=\sum_{i_\ell\in I_\ell}
\eta(\mu^\ell_{i_\ell})\ \cdot
$}. For
$s=0,1,\ldots,D-1$, $\omega^\ell_{i_\ell} \circ \tilde\gamma^\ell_{\c
C}$ takes on the values 
\begin{equation*}
 \omega^\ell_{i_\ell}\pq{{\cal J}_{N\infty}(\chi_{T^{-\ell}(C_s)})} = 
 {\pt{\mu_{i_\ell}^{\ell}}}^{-1}\tau_N
 \pq{{\cal J}_{N\infty}(\chi_{T^{-\ell}(C_{i_\ell})})\,
 {\cal J}_{N\infty}(\chi_{T^{-\ell}(C_s)})}. 
\end{equation*}
Due to Proposition~\ref{prop2}, 
these 
converge to ${\pt{\mu\pt{C_{i_\ell}}}}^{-1}\omega_\mu(
\chi_{T^{-\ell}(C_{i_\ell})}\, \chi_{T^{-\ell}(C_s)}) = \delta_{s,i_\ell}$. 
This means that in
the limit the von~Neumann entropy will be zero. Or stated more carefully:
\[
\begin{CD}
\shadowbox{\begin{Beqnarray*}  
N^\prime\coleq\max_{s\in\pg{0,1,\cdots,D-1}}\pg{N_s}
\end{Beqnarray*}}\\
@AA{\text{We can determine } N^\prime \text{ by:}}A\\
\shadowbox{\begin{Beqnarray*}
\forall\:s\in\pg{0,1,\cdots,D-1}\ 
\exists\: N_s\in\IN \text { s.t. }N > N_s \Longrightarrow
\abs{\mu_{i_\ell}^{\ell}
\pt{\omega^\ell_{i_\ell}
\circ \tilde\gamma^\ell_{\c
C}}\pt{s}-\mu_{i_\ell}^{\ell}\delta_{s,i_\ell}}<
\delta_{\varepsilon^\prime}\mu_{i_\ell}^{\ell}\!\!\!\!\!\!\!\!\!\!\!\!
\end{Beqnarray*}}\\
@AA{\text{In correspondence to that }\delta_{\varepsilon^\prime}}A\\
\shadowbox{\begin{Beqnarray*}
\quad\text{\bf Uniform continuity of $\bs{\eta\pt{x}}$ function on
 $\bs{[0,1]}$ guarantees:}\\
 \forall\:\varepsilon^\prime>0,\
\exists\: \delta_{\varepsilon^\prime}>0\text{\ \ s.t. }\abs{
\pt{\omega^\ell_{i_\ell}
\circ \tilde\gamma^\ell_{\c
C}}\pt{s}-\delta_{s,i_\ell}}<\delta_{\varepsilon^\prime}\Longrightarrow
\qquad\\
\Longrightarrow\abs{
\eta\pq{\pt{\omega^\ell_{i_\ell}
\circ \tilde\gamma^\ell_{\c
C}}\pt{s}}-\eta\pq{\rule{0pt}{2.5ex}\delta_{s,i_\ell}}}<
\frac{\varepsilon^\prime}{D}\ ,    
\ \forall\:s\in\pg{0,1,\cdots,D-1}
\end{Beqnarray*}}\\
@VV{\text{summing over }s\in\pg{0,1,\cdots,D-1}}V\\
\shadowbox{\begin{Beqnarray*}  
S(\omega^\ell_{i_\ell}
 \circ \tilde\gamma^\ell_{\c C})\leq\varepsilon^\prime
\end{Beqnarray*}}\\
@VV{\text{that is}}V\\
\shadowbox{\begin{Beqnarray*}  
\sum^{k-1}_{\ell=0}
 \sum_{i_\ell \in I_\ell} \mu^\ell_{i_\ell}\,  S(\omega^\ell_{i_\ell}
 \circ \tilde\gamma^\ell_{\c C})\leq k  \varepsilon^\prime
\end{Beqnarray*}}
\end{CD}
\]
We thus obtain a lower bound.
\smallskip

Combining our results and choosing $\tilde N := \max(N,N')$, we
conclude
\begin{equation*}
 S_\mu(\c C^{[0,k-1]}) - k \varepsilon' - k \delta(\varepsilon) \le
 H_{\tn}(\gamma^0_{\c C}, \gamma^1_{\c C},\ldots, 
 \gamma^{k-1}_{\c C}) \le S_\mu(\c C^{[0,k-1]}) + k \delta(\varepsilon)\
 \cdot\hfill\tag*{\qed} 
\end{equation*}\newpage
\subsection{\co{ALF}-entropy}\label{ALFE}\vspace{3mm}\ \\[-8ex]
\begin{quote}
\begin{TT}{:} \label{TEOALF}
 Let $\tripCT$ be a classical dynamical system
 which is the classical limit of a sequence of finite dimensional
 quantum dynamical systems $\tripQT$. 
 We also assume that the dynamical localization
 condition~\ref{dynloc} holds. If\\[-5ex]
 \begin{enumerate}
 \item 
  $\c C = \{ C_0, C_1,\ldots, C_{D-1} \}$ is a finite measurable
  partition of $\c X$,
 \item 
  $\c Y_N = \{ y_0, y_1,\ldots, y_D \}$ is a bistochastic partition of
  unity, which is the quantization  of the previous partition, namely
  $y_i = {\cal J}_{N\infty}(\chi_{C_i})$  for $i\in\pg{0,1,\cdots,D-1}$  and
  $y_D := \sqrt{\idty-\sum_{i=0}^{D-1} y_i^*y_i}$,\\[-5ex]
 \end{enumerate} 
 then there exists an $\alpha$ such that
 \begin{equation*}
  \lim_{\substack{k,N\to\infty \\ k\le\alpha\log N}} \frac{1}{k}
  \left| H_{\tn}[\c Y^{[0,k-1]}] - S_\mu(\c C^{[0,k-1]}) \right| = 0.
 \end{equation*}
\end{TT}
\end{quote}
\noindent
\textbf{Proof of Theorem~\ref{TEOALF}:}\\[2ex]
First notice that $\c Y_N = \{ y_0, y_1,\ldots, y_D \}$ is indeed a 
bistochastic partition. We have
\begin{align*}
 &y_i^* = {\cal J}_{N\infty}(\chi_{C_i})^* =
 {\cal J}_{N\infty}(\overline{\chi}_{C_i}) =
 {\cal J}_{N\infty}(\chi_{C_i}) = y_i \\ 
 &0 \le {\cal J}_{N\infty}(\chi_{C_i})^2 = y_i^2 \le
 \gamma_{N \infty}(\chi_{C_i}^2) = \gamma_{N \infty}(\chi_{C_i})
\end{align*}
Summing the last line over $i$ from $0$ to $D-1$, we see that
$\sum_{i=0}^{D-1} y_i^2 \le \idty$, This means that $\{ y_0,
y_1,\ldots, y_{D-1} \}$ is not a partition of unity,  but we can use
this property to define an extra  element $y_D$ which completes it to
a bistochastic partition of unity,  $\c Y_N = \{ y_0, y_1,\ldots,
y_D \}$:
\begin{equation*}
 y_D := \sqrt{\idty-\sum_{i=0}^{D-1} y_i^*y_i}
\end{equation*}
The bistochasticity is a useful property because it implies
translation invariance of the state on the quantum spin chain, state which
arises during the construction of the \co{ALF}-entropy.

The density matrix $\rho[\c Y^{[0,k-1]}]$ of the refined partition
reads~(see \eqref{AFE_6})
\begin{align}
 \rho\left[ \c Y^{[0,k-1]} \right] 
 &= \sum_{\bi{i,j}} \rho\left[ \c Y^{[0,k-1]} \right]_{\bi{i,j}} |e_{\bi
 i}\> \< e_{\bi j}| \nonumber\\
 &= \sum_{\bi{i,j}} \tau_N \Bigl( y_{j_1}^* \Theta_N (y_{j_1}^*)\cdots
 \Theta_N^{k-1}(y_{j_k}^*) \Theta_N^{k-1}(y_{i_k}) \cdots
 \Theta_N (y_{i_1}) y_{i_1} \Bigr) 
 |e_{\bi i}\> \< e_{\bi j}|\label{explrhoAF}
\end{align}

Now we will expand this formula using the operators $y_i$ defined above, 
the quantities $K_{\ell}(\bs{x},\bs{y})$ defined in~(\ref{dyn-loc})
and controlling  the element $y_D$ as follows:
\begin{align}
 \|y_D\|_2^2 
 &= \Bigl\| \sqrt{\idty - \sum_{i=0}^{D-1} y_i^*y_i} \Bigr\|_2^2 =
 \tau_N \Bigl( \idty - \sum_{i=0}^{D-1} y_i^*y_i \Bigr) 
\nonumber\\
 & =\tau_N \Bigl( \sum_{i=0}^{D-1} y_i^*\pt{\idty - y_i} \Bigr) 
\nonumber\\
 &= \int \mu\pt{\ud\bs{y}} \mu\pt{\ud\bs{z}}\, \sum_{\substack{i,j=0 \\ {i\ne j}}}^{D-1}
\chi_{i}(\bs{y})\, \chi_{j}(\bs{z})\, N  |K_0(\bs{y},\bs{z})|^2.
\label{100}  
\end{align}
Thus, in the limit of large $N$, $N |K_0(\bs{y},\bs{z})|^2$ is just $\delta(\bs{y}-\bs{z})$ 
(see~(\ref{dyn-loc})) so 
that~(\ref{100}) tends to $\int \mu\pt{\ud\bs{z}}\, \sum_{i\ne j} \chi_{i}(\bs{z})\,
\chi_{j}(\bs{z}) = 0$ and we can  consistently
neglect those entries of $\rho[{\c Y}^{[0,k-1]}]$ containing
$y_D$.

By means of the properties of coherent states, 
we write out explicitly\footnote{%
Every elements of the p.u. is written in terms of C.S. as
$y_{j_\ell}=\int \mu\pt{\ud\bs{x}} \chi_{C_{j_\ell}}(\bs{x})
\ket{C_N^1\pt{\bs{x}}}\bra{C_N^1\pt{\bs{x}}}$: we make use of $\bs{y}_\ell$
and $\bs{z}_\ell$ as variables in the integral representation of
$y_{j_\ell}^*$, respectively $y_{i_\ell}$.}
the elements of the density matrix
in~\eqref{explrhoAF} 
\begin{align}
\nonumber
 \rho\left[\c Y^{[0,k-1]}\right]_{\bi{i,j}} 
& = \tau_N \Bigl( y_{j_1}^* \Theta_N (y_{j_1}^*)\cdots
 \Theta_N^{k-1}(y_{j_k}^*) \Theta_N^{k-1}(y_{i_k}) \cdots
 \Theta_N (y_{i_1}) y_{i_1} \Bigr) \nonumber\\
& = \tau_N \Bigl( \:y_{j_1}^* \:U_T \:y_{j_1}^* \:U_T \:\cdots \:U_T 
 \:y_{j_k}^* \:y_{i_k} \:U_T^* \:\cdots
 \:U_T^* \:y_{i_1} \:U_T^*\: y_{i_1}\: \Bigr) \nonumber\\
& = N^{2k-1}\text{\Large $\int$} \pt{\prod_{\ell=1}^k
\mu\pt{\ud\bs{y}_\ell} 
 \mu\pt{\ud\bs{z}_\ell}\,  \chi_{C_{j_\ell}}(\bs{y}_\ell)
 \,\chi_{C_{i_\ell}}(\bs{z}_\ell)}\ \times\nonumber\\
 \quad
 \times \ K_0(\bs{z}_1,\bs{y}_1 
& )\,
 \left(\prod_{p=1}^{k-1} K_1(\bs{y}_p,\bs{y}_{p+1})\right)\, 
 K_0(\bs{y}_{k},\bs{z}_{k})\,
 \left(\prod_{q=1}^{k-1} 
K_{-1}(\bs{z}_{k-q+1},\bs{z}_{k-q})
\right)\ \cdot
 \label{matr-el}
\end{align}
We now use that for $N$ large enough,  
\begin{equation}
\label{tel} 
 \Bigl| N\int \mu\pt{\ud\bs{y}}\,
 \chi_{C}(\bs{y})\, K_m(\bs{x},\bs{y})\, K_n(\bs{y},\bs{z}) -
 \chi_{T^{-m}C}(\bs{x})\, K_{m+n}(\bs{x},\bs{z}) \Bigr| \le \varepsilon_m(N)
 \, ,
\end{equation}
where $\varepsilon_m(N)\to 0$ with $N\to\infty$ uniformly in
$\bs{x},\bs{z}\in\c X$.
This is a consequence of the dynamical localization condition~\ref{dynloc}
and can be rigorously proven in the same way  as Proposition~\ref{prop1}.
However, the rough idea is the following: from the property 3.1.3 of coherent
states, one derives
\begin{align}
N\int \mu\pt{\ud\bs{y}}&\, \chi_{C}(\bs{y})\, K_m(\bs{x},\bs{y})\,
K_n(\bs{y},\bs{z})=\notag\\ 
&=
 N\int \mu\pt{\ud\bs{y}}\, 
\pq{1+\Bigl( \chi_{C}(\bs{y})-1\Bigr)}\, K_m(\bs{x},\bs{y})\,
K_n(\bs{y},\bs{z}) 
=\notag\\ 
&=K_{m+n}(\bs{x},\bs{z})
 +N\int \mu\pt{\ud\bs{y}}\, \Bigl(
 \chi_{C}(\bs{y})-1\Bigr)\, K_m(\bs{x},\bs{y})\, K_n(\bs{y},\bs{z})\
\cdot\label{evolKK} 
\end{align}
For large $N$ we look two cases:
\begin{itemize}
\item{$\bs{x}\not\in T^{-m}(C)$ -- } then the
condition~\ref{dynloc} makes the integral in~\eqref{tel} negligible
small, whereas the second term in the l.h.s of the same equation is
exactly zero;
\item{$\bs{x}\in T^{-m}(C)$ -- }  in this case it
is the second integral in formula~\eqref{evolKK} which can be
neglected, and using~\eqref{evolKK} in~\eqref{tel} we find negligibility. 
\end{itemize}
By applying ~(\ref{tel}) to the couples of products in~(\ref{matr-el})
one after the other, noting that every single integral
in~\eqref{matr-el} is less or equal to one, and using triangle
inequality for $\abs{\cdot}$, we finally arrive at the upper bound
\begin{equation*} 
 \abs{ \rho\left[\c Y^{[0,k-1]}\right]_{\bi{i,j}} - \delta_{\bi{i,j}}\, 
 \mu(C_{\bi i}) } \le \Bigl( 2 \sum_{m=1}^{k-1} \varepsilon_m(N) + 
 \varepsilon_0(N) \Bigr) =: \epsilon(N),  
\end{equation*}
where $C_{\bi i} := \bigcap_{\ell=1}^k T^{-\ell+1}C_{i_\ell}$ is an element 
of the partition $\c C^{[0,k-1]}$. 

We now set $\sigma\left[\c C^{[0,k-1]}\right] := \sum_{\bi i} \mu(C_{\bi
i}) |e_{\bi i}\>  \< e_{\bi i}|$ and use the following estimate: let
$A$ be an arbitrary matrix of dimension $d$ and let
$\{e_1, e_2,\ldots, e_d\}$ and $\{f_1, f_2,\ldots, f_d\}$ be two
orthonormal bases of $\Cx^d$, then
$\|A\|_1:=\tr |A| \le \sum_{i,j} |\< e_i,A\,f_j\>|$.
This yields 
\begin{equation*} 
\Delta(k):= \norm{ \rho\left[\c Y^{[0,k-1]}\right] - \sigma\left[\c C^{[0,k-1]}\right] }{1} 
 = \tr \Bigl| \rho\left[\c Y^{[0,k-1]}\right] - \sigma\left[\c C^{[0,k-1]}\right]
 \Bigr|  \le  D^{2k} \epsilon(N).
\end{equation*}
Finally, by the continuity of the von~Neumann entropy~\cite{Fan73:1}, we get 
\begin{align*}
 \left|S\pt{\rho\left[\c Y^{[0,k-1]}\right]} - S\pt{\sigma\left[\c C^{[0,k-1]}\right]}\right|
 \leq \Delta(k)\log D^k\,+\,\eta(\Delta(k))\ .
\end{align*}
Since, from $k \le \alpha \log N$, $D^{2k}\leq N^{2\alpha\log D}$, if
we want the bound $D^{2k}\epsilon(N)$  to converge to zero with
$N\to\infty$, the parameter $\alpha$ has to be chosen accordingly.
Then, the result follows because the von~Neumann entropy of $\sigma$
reduces to the Shannon entropy of the refinements of the classical
partition.\hfill$\qed$
\section{Numerical analysis of \co{ALF} Entropies in Discrete Classical
Chaos}\label{ALFEDCC}\vspace{6mm} 

Here in the following we are considering not the quantization of
classical systems, but their discretization; nevertheless, we have
seen that, under certain respects, quantization and discretization are
like procedures with the inverse of the number of states $N$ playing
the role of $\hbar$ in the latter case.

We are then interested to study how the classical continuous behaviour
emerges from the discretized one when $N\to\infty$; in particular,
we want to investigate the presence of characteristic time scales and
of ``breaking--times'' $\tau_B$, namely those times beyond which the
discretized systems cease to produce entropy because their granularity
takes over and the dynamics reveals in full its regularity. 

Propositions~\ref{prop_41} and~\ref{prop_42} afford useful means to attack
such a problem numerically. In the following we shall be concerned
with the time behavior of the 
entropy of partition of units as in Definition~\ref{partizione}, the
presence of breaking--times $\tau_B\pt{\Lambda,N,\alpha}$, and their
dependence 
on the set $\Lambda$, on the number of states $N$ and on the dynamical
parameter $\alpha$.

As we shall see, in many cases $\tau_B$ depends quite 
heavily on the chosen partition of unit; we shall then try to cook up
a strategy to find a $\tau_B$ as stable as possible upon variation of
partitions, being led by the idea that the ``true'' $\tau_B$ has to be
strongly related to the Lyapounov exponent of the underlying
continuous dynamical system.

Equations~\eqref{CoAFE_10} and~\eqref{CoAFE_77} allow us to compute
the Von Neumann entropy of the state $\rho_{\cal
\widetilde{Y}}^{\pq{0,n-1}}$; 
if we were to compute
the ALF--entropy according to the definitions~\eqref{AFE_7}, the
result would be
zero, in agreement with fact that
the Lyapounov exponent for a
system with a finite number of states vanishes. 
Indeed, it is sufficient to notice that
the entropy $H_{\tn}\pq{{\cal \widetilde
Y}^{[0,n-1]}}$ is bounded from above by the entropy of the tracial
state $\displaystyle \frac{1}{N^2}{\Id}_{N^2}$, that is by $2\log N$;
therefore the expression 
\begin{equation}
h_{\tn,{\cal W}_\infty}(\alpha,\Lambda,n)  
\coleq\frac{1}{n} H_{\tn}\Big[\widetilde{\cal Y}^{[0,n-1]}\Big],
\label{AEP_1}
\end{equation}
goes to zero with $n\longrightarrow 0$. It is for this reason that, in
the following, we will focus upon the temporal evolution of the
function $h_{\tn,{\cal W}_\infty}(\alpha,\Lambda,n)$
instead of taking its $\limsup$ over the number of iterations $n$.

In the same spirit, we will not take the supremum of~\eqref{AEP_1}
over all possible partitions $\widetilde{\cal Y}$ (originated by
different $\Lambda$); instead, we will study 
the dependence of 
$h_{\tn,{\cal W}_\infty}(\alpha,\Lambda,n)$
on different choices of partitions. 
In fact, if we vary over all possible choices of partitions of unit, we could
choose $\Lambda=\ZNZD$ in ~(\ref{CoAFE_217}), that is $D=N^2$;
then summation over all possible  
$\bs{r}\in\ZNZD$ would make the matrix elements 
$\displaystyle {\cal G}_{\bs{\ell}_1,\bs{\ell}_2}\pt{n}$ 
in~(\ref{CoAFE_1803}) equal to 
$\displaystyle \frac{\delta_{\bs{\ell_1},\bs{\ell_2}}}{N^2}$, whence 
$H_{\tn}\Big[\widetilde{\cal Y}^{[0,n-1]}\Big]=2\log N$.
\subsection{The case of the ${\pg{T_\alpha}}$ subfamily of the
UMG} \label{}\vspace{3mm} 
The maximum of $H_{\tn}$ is reached when the
frequencies~\eqref{CoAFE_740}
\begin{equation*}
\nu_{\Lambda,\alpha}^{(n),N}:{(\ZNZ{N})^2}\mapsto\pq{0,1}
\end{equation*}
become equal to $1/N^2$ over the torus: we will see that
this is indeed what happens to the frequencies
$\nu_{\Lambda,\alpha}^{(n),N}$ with $n\longrightarrow\infty$.
The latter behaviour
can be reached in various ways depending on:
\begin{itemize}
\item hyperbolic or elliptic regimes, namely on the dynamical
parameter $\alpha$;
\item number of elements ($D$) in the partition $\Lambda$;
\item mutual location of the $D$ elements $\bs{r}_i$ in $\Lambda$.
\end{itemize}
For later use we introduce the set of grid points with non--zero
frequencies 
\begin{equation}
\Gamma_{\Lambda,\alpha}^{(n),N} \coleq\pg{\frac{\bs{\ell}}{N}\ \bigg|\
\bs{\ell}\in{\pt{\IZ/N\IZ}}^2\ ,\ 
\nu_{\Lambda,\alpha}^{(n),N}\pt{\bs{\ell}}\neq 0}\ \cdot\label{gamma}
\end{equation}
\subsubsection{Hyperbolic regime with  $D$ randomly
chosen points $\bs{r}_i$ in $\Lambda$}\label{sez_511}\vspace{3mm} 
In the hyperbolic regime corresponding to
$\alpha\in{\IZ}\setminus\pg{-4,-3,-2,-1,0}$,
$\Gamma_{\Lambda,\alpha}^{(n),N}$ tends to 
increase its cardinality with the number of time--steps $n$.
Roughly speaking, there appear to be two distinct temporal patterns:
a first one, during which
\mbox{$\#\pt{\Gamma_{\Lambda,\alpha}^{(n),N}} \simeq D^n\leq N^2$} and
almost every 
$\nu_{\Lambda,\alpha}^{(n),N}\simeq D^{-n}$, followed by a second one
characterized by frequencies frozen to
$\nu_{\Lambda,\alpha}^{(n),N}\pt{\bs{\ell}} = \frac{1}{N^2}, 
\ \forall \bs{\ell}\in{(\ZNZ{N})^2}$.
The second temporal pattern is reached when, during the first one,
$\Gamma_{\Lambda,\alpha}^{(n),N}$ has covered the whole lattice and
$D^n\simeq N^2$.

From the point of view of the entropies, the first temporal regime is
characterized by
\begin{alignat}{2}
H_{\tn}(\alpha,\Lambda,n)  & \sim n\cdot \log D\qquad & ,\qquad
h_{\tn,{\cal W}_\infty}(\alpha,\Lambda,n)  & \sim \log D\ ,\nonumber
\intertext{while the second one by}
H_{\tn}(\alpha,\Lambda,n)  & \sim 2 \log N\nonumber\qquad & ,\qquad
h_{\tn,{\cal W}_\infty}(\alpha,\Lambda,n)  & \sim \frac{2 \log
N}{n}\ \cdot \nonumber
\end{alignat}
The transition between these two regimes occurs at $\bar{n} =
{\log}_D N^2$. However this time cannot be considered a realistic
breaking--time, as it too strongly depends on the chosen
partition.

Figure~\ref{lontani} (columns a and c) shows the mechanism clearly
in a temperature--like plot: hot points correspond to
points of 
$\Gamma_{\Lambda,\alpha}^{(n),N}$ and their number increases for small
numbers of iterations until the plot assume a \mbox{uniform green color for
large $n$.} 

The linear and stationary behaviors of
$H_{\tn}(\alpha,\Lambda,n)$ are apparent 
in fig.~\ref{uno}, where four different plateaus
($2\log N $) are reached for four different $N$, and in fig.~\ref{due},
in which four different slopes are showed for four different number of
elements in the partition. With the same parameters as in
fig.~\ref{due}, fig.~\ref{tre} shows the corresponding
entropy production $h_{\tn,{\cal W}_\infty}(\alpha,\Lambda,n)$.
\subsubsection{Hyperbolic regime with  $D$ nearest neighbors
$\bs{r}_i$ in $\Lambda$}\label{sub_512}\vspace{3mm} 
In the following, we will consider a set of points $\Lambda =
{\pg{\bs{r}_i}}_{i=1\ldots D}$ very close to each other, instances of
which are as below:

\begin{figure}[h]
\begin{center}
\begin{picture}(26,34)(-1,-9) 
\thicklines
\matrixput(0,0)(6,0){5}(0,6){5}{\circle{2}}
\matrixput(1,0)(6,0){4}(0,6){5}{\line(1,0){4}}
\matrixput(0,1)(6,0){5}(0,6){4}{\line(0,1){4}}
\put(12,12){\circle*{2}}
\put(6,12){\circle*{2}}
\put(12,6){\circle*{2}}
\put(12,18){\circle*{2}}
\put(18,12){\circle*{2}}
\put(0,-9){\makebox(24,8)[cc]{$D=5$}}
\end{picture}
\hspace{10mm}
\begin{picture}(26,34)(-1,-9) 
\thicklines
\matrixput(0,0)(6,0){5}(0,6){5}{\circle{2}}
\matrixput(1,0)(6,0){4}(0,6){5}{\line(1,0){4}}
\matrixput(0,1)(6,0){5}(0,6){4}{\line(0,1){4}}
\put(12,12){\circle*{2}}
\put(6,12){\circle*{2}}
\put(12,18){\circle*{2}}
\put(6,18){\circle*{2}}
\put(0,-9){\makebox(24,8)[cc]{$D=4$}}
\end{picture}
\hspace{10mm}
\begin{picture}(26,34)(-1,-9) 
\thicklines
\matrixput(0,0)(6,0){5}(0,6){5}{\circle{2}}
\matrixput(1,0)(6,0){4}(0,6){5}{\line(1,0){4}}
\matrixput(0,1)(6,0){5}(0,6){4}{\line(0,1){4}}
\put(12,12){\circle*{2}}
\put(6,12){\circle*{2}}
\put(12,18){\circle*{2}}
\put(0,-9){\makebox(24,8)[cc]{$D=3$}}
\end{picture}
\hspace{10mm}
\begin{picture}(26,34)(-1,-9) 
\thicklines
\matrixput(0,0)(6,0){5}(0,6){5}{\circle{2}}
\matrixput(1,0)(6,0){4}(0,6){5}{\line(1,0){4}}
\matrixput(0,1)(6,0){5}(0,6){4}{\line(0,1){4}}
\put(12,12){\circle*{2}}
\put(6,12){\circle*{2}}
\put(0,-9){\makebox(24,8)[cc]{$D=2$}}
\end{picture}
\caption{Several combinations of $D$ nearest neighbors in $\Lambda$
for different values $D$.}
\label{lattice}
\end{center}
\end{figure}
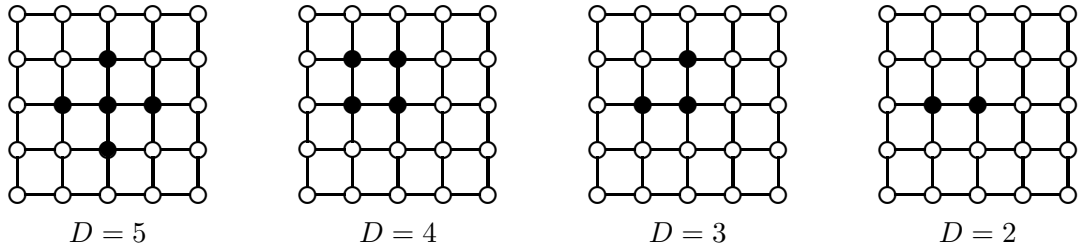
From eqs.~(\ref{CoAFE_730}--\ref{CoAFE_740}), the frequencies
$\nu_{\Lambda,\alpha}^{(n),N}\pt{\bs{\ell}}$ 
result proportional to 
how many strings have equal images $\bs{\ell}$, through the function
$\bs{f}_{\Lambda,\alpha}^{(n),N}$ in~\eqref{CoAFE_720}.
Due to the fact that
${\pq{T_\alpha}}_{11}={\pq{T_\alpha}}_{21}=1$, non--injectivity of
$\bs{f}_{\Lambda,\alpha}^{(n),N}$ occurs very frequently when
${\pg{\bs{r}_i}}$ are very close to each other. This is a dynamical
effect that, in continuous systems~\cite{Ali96:1}, leads to an entropy
production  
approaching the Lyapounov exponent. Even in the discrete case, during
a finite time interval though, 
$h_{\tn,{\cal W}_\infty}(\alpha,\Lambda,n)$ exhibits the same
behavior until $H_{\tn}$ reaches the upper bound $2\log N$.
From then on, the system behaves as described in
subsection~\ref{sez_511}, and the entropy production goes to zero as:
\begin{equation*}
h_{\tn,{\cal W}_\infty}(\alpha,\Lambda,n) \sim
\frac{1}{n}\quad\text{(see fig.~\ref{cinque}).}
\end{equation*}
Concerning figure~\ref{vicini} (column d), 
whose corresponding graph
for $h_{\tn,{\cal W}_\infty}(1,\Lambda,n)$ is 
labeled by $\triangleright$ in 
fig.~\ref{cinque}, we make the following consideration:
\begin{itemize}
\item for $n=1$ the red spot corresponds to five $\bs{r}_i$ grouped
as in fig.~\ref{lattice}.\\
In this case $h_{\tn,{\cal
W}_\infty}(1,\Lambda,1)=\log D=\log 5$;
\item for $n\in \pq{2,5}$ the red spot begins to stretch along the
stretching direction of $T_1$. In this case,
the frequencies $\nu_{\Lambda,\alpha}^{(n),N}$ are not constant on the
warm region: this leads to a decrease of
$h_{\tn,{\cal W}_\infty}(1,\Lambda,n)$;
\item for $n\in \pq{6,10}$ the warm region
becomes so elongated that it starts feeling the folding condition so that,
with increasing time--steps, it eventually fully covers the originally 
pale--blue space. In this case, the behavior of
$h_{\tn,{\cal W}_\infty}(1,\Lambda,n)$ remains the same
as before up to $n=10$;
\item for $n=11$, $\Gamma_{\Lambda,\alpha}^{(n),N}$
coincides with the whole lattice;
\item for larger times, the frequencies  $\nu_{\Lambda,1}^{(n),N}$ tend
to the constant value $\frac{1}{N^2}$ on almost every point of the
grid. In this case, the behaviour of the entropy production undergoes
a critical change (the crossover occurring at $n=11$) as
showed in fig.~\ref{cinque}.
\end{itemize}
Again, we cannot conclude that $n=11$ is a realistic breaking--time,
because once more we have strong dependence on the chosen partition
(namely from the number $D$ of its elements). For instance, in
fig.~\ref{cinque}, one can see that partitions with $3$ points 
reach their corresponding ``breaking--times''
faster than that with $D=5$; also they do it in an $N$--dependent way.

For a chosen set $\Lambda$ consisting of $D$ elements
very close to 
each other 
and $N$ very large,
$h_{\tn,{\cal W}_\infty}(\alpha,\Lambda,n)\simeq\log\lambda$
(which is the asymptote in the continuous case)
from a certain $\bar{n}$ up to a time $\tau_B$. Since this latter is
now partition independent, it can
properly be
considered as the
\underline{breaking--time} of the system; it is 
given by
\begin{equation}
\tau_B={\log}_{\lambda}N^2 \label{bbbttt}\cdot
\end{equation}
\begin{center}
\framebox[1.1\width][c]{
\begin{minipage}[c]{10cm}
\enfasi{\ \\[1ex]
It is evident from equation~\eqref{bbbttt} that if one knows
$\tau_B$ then also $\log\lambda$ is known. Usually, one is interested
in the latter which is a sign of the instability of the continuous
classical system. In the following we develop an algorithm which
allows us to extract $\log\lambda$ from studying the corresponding
discretized classical system and its ALF--entropy.\\[0.8ex]
\ }
\end{minipage}}
\end{center}
In working conditions, $N$ is not large enough to allow for $\bar{n}$
being smaller than $\tau_B$; what happens in such a case
is that $H_{\tn}\pt{\alpha,\Lambda,n}\simeq 2\log N$ before
the asymptote for $h_{\tn,{\cal
W}_\infty}(\alpha,\Lambda,n)$ is reached. Given $h_{\tn,{\cal
W}_\infty}(\alpha,\Lambda,n)$ for $n<\tau_B$, it is thus necessary to seek
means how to estimate the long time behaviour that one would have if
the system were continuous.
\begin{quote}
\begin{NNS}{}\ \label{rem_51}\\[-2ex]
When estimating Lyapounov exponents from
discretized hyperbolic classical systems, by using
partitions consisting of nearest neighbors, we have to take into
account some facts:
\begin{Ventry}{\mdseries a.}
\item[\mdseries a.] $h_{\tn,{\cal
W}_\infty}(\alpha,\Lambda,n)$ does not increase with $n$; therefore, if
$D<\lambda$, $h_{\tn,{\cal
W}_\infty}$ cannot reach the
Lyapounov exponent. 
Denoted by \framebox{$\log\lambda\pt{D}$} the asymptote that we
extrapolate from the data\footnote{$h_{\tn,{\cal
W}_\infty}(\alpha,\Lambda,n)$ may even equal $\log\lambda\pt{D}$ from
the start.}, in general we have \mbox{$\lambda\pt{D}\leqslant\log
D<\lambda$.} 
For instance, for
$\alpha=1$, $\lambda=2.618\ldots>2$ and partitions with \mbox{$D=2$} cannot
produce an entropy greater then $\log 2$; this is the case for the entropies
below the dotted line in figs.~\ref{due} e~\ref{tre}; 
\item[\mdseries b.] partitions with $D$ small but greater than
$\lambda$ allow $\log\lambda$ to be reached in a very short time and 
$\lambda\pt{D}$ is very close to $\lambda$ in this case;
\item[\mdseries c.] partitions with $D\gg\lambda$ require very
long time to converge to $\log\lambda$ (and so very large $N$) and,
moreover, it is not a trivial task to deal with them from a
computational point of view. On the contrary the entropy behaviour for
such partitions offers very good estimates of $\lambda$
(compare, in 
fig.~\ref{cinque}, $\triangleright$ with $\diamond$, {\scriptsize
$\bigtriangleup$}, $\circ$ and {\scriptsize $\Box$}) ;
\item[\mdseries d.] in order to compute $\lambda$ (and then $\tau_B$,
by~\eqref{bbbttt}), one can calculate $\lambda\pt{D}$ for increasing
$D$, until it converges to a stable value $\lambda$;
\item[\mdseries e.] due to number theoretical reasons, the UMG on
${\pt{\IZ/N\IZ}}^2$ present several anomalies. An instance of them 
is showed in fig.~\ref{vicini}
(col. f), where a partition with five nearest neighbors on a lattice
of $200\times200$ points confines the image of
$\bs{f}_{\Lambda,\alpha}^{(n),N}$ (under the 
action of a $T_{\alpha}$ map with $\alpha=17$) on a subgrid of the
torus. In this and analogous cases, there occurs an anomalous depletion of
the entropy production and no significant information is obtainable
from it.
To avoid this difficulties, in
Section~\ref{Sawtooth}
we will go beyond the UMG subclass considered so far and we will
include in our analysis the full family of Sawtooth Maps.
\end{Ventry} 
\end{NNS}
\end{quote}
\noindent
\subsubsection{Elliptic regime ($\alpha\in\pg{-1,-2,-3}$)}\vspace{3mm} 
One can show that 
all evolution matrices $T_{\alpha}$ are
characterized by the following property:
\begin{equation}
T_{\alpha}^2 = \bar{\alpha} \; T_{\alpha} - \Id \qquad,\qquad
\bar{\alpha}\coleq\pt{\alpha+2} \ \cdot\label{ell_1}
\end{equation}
In the elliptic regime $\alpha\in\pg{-1,-2,-3}$, therefore
$\bar{\alpha}\in\pg{-1, 0, 1}$ and 
relation~\eqref{ell_1} determines a periodic 
evolution with periods:
\begin{subequations}
\label{ell_2}
\begin{alignat}{2}
T_{-1}^{\phantom{-}3} & = -\Id & \qquad\qquad ( T_{-1}^{\phantom{-}6}
& = \Id )\label{ell_2a}\\ 
T_{-2}^{\phantom{-}2} & = -\Id & ( T_{-2}^{\phantom{-}4} & = \Id
)\label{ell_2b}\\  
T_{-3}^{\phantom{-}3} & = +\Id \label{ell_2c}\ \cdot&&
\end{alignat}
\end{subequations}
It has to be stressed that, in the elliptic regime, 
the relations~\eqref{ell_2} \underline{do not hold}
``modulo $N$'', instead they are completely independent from $N$.

Due to the high degree of symmetry in
relations (\ref{ell_1}--\ref{ell_2}), the frequencies
$\nu_{\Lambda,\alpha}^{(n),N}$ 
are different from zero only on a small subset of the whole lattice.

This behavior is apparent in fig.~\ref{lontani}~:~col. b, in which
we consider five randomly distributed $\bs{r}_i$ in $\Lambda$, and in
fig.~\ref{vicini}~:~col. e, in which the five $\bs{r}_i$ are grouped as
in fig.~\ref{lattice}. In both cases, the Von Neumann entropy
$\displaystyle H_{\tn}\pt{n}$ is not
linearly increasing with $n$, instead it assumes a
$\log$--shaped
profile (up to the breaking--time, see fig.~\ref{uno}).\\[-2ex]
\begin{quote}
\begin{NNN}{}\ \\[-2ex]
The last observation indicates how the entropy production analysis can be
used to recognize whether a dynamical systems is hyperbolic or not.
If we use randomly
distributed points as a partition, we observe that hyperbolic systems
show constant entropy production (up to the breaking--time),
whereas the others do not.

Moreover, unlike hyperbolic ones, elliptic systems do not change their
behaviour with $N$ (for 
reasonably large $N$) as clearly showed in fig.~\ref{uno}, in which
elliptic systems ($\alpha=-2$) with four different values of $N$ give
the same plot.
On the contrary, we have dependence on how rich is the chosen
partition, similarly to 
what we have for hyperbolic systems, as showed in
fig.~\ref{cinque}.
\end{NNN}
\end{quote}
\noindent
\subsubsection{Parabolic regime ($\alpha\in\pg{0, 4}$)}\vspace{3mm} 
This regime is characterized by  $\lambda = \lambda^{-1} = \pm 1$,
that is $\log\abs{\lambda} = 0$ (see Remark~\ref{Rem_21}, c.).
These systems behave as the hyperbolic ones
(see subsections~\ref{sez_511} and ~\ref{sub_512}) and this is true also
for the the general behavior of the entropy production, apart from the
fact that we never fall in the condition (a.) of Remark~\ref{rem_51}.
Then, for sufficiently large $N$, every partition consisting of $D$
grouped $\bs{r}_i$ will reach the asymptote $\log\abs{\lambda}=0$.
\subsection{The case of Sawtooth Maps} \label{Sawtooth}\vspace{3mm} 
From a computational point of view,
the study of the entropy production in the case of Sawtooth Maps
${S_{\alpha}}$ is more
complicated than for the
${T_{\alpha}}$'s.
The reason to study numerically these dynamical systems is twofold:
\begin{itemize}
	\item to avoid the difficulties described in Remark~\ref{rem_51}~(e.);
	\item to deal, in a way compatible with numerical computation
	limits, with the largest possible spectrum of accessible
	Lyapounov exponent. We know that 
	\begin{equation*}
	\alpha\in\IZ\bigcap\pg{\text{non elliptic domain}}\Longrightarrow
	\lambda^{\pm}\pt{T_{\alpha}} =
	\lambda^{\pm}\pt{S_{\alpha}} = 
	\frac{\alpha+2\pm\sqrt{(\alpha+2)^2-4}}{2}\ \cdot
	\end{equation*}
	In order to fit $\log\lambda_\alpha$ (
	$\log\lambda_\alpha$ being 
	the Lyapounov exponent corresponding to a
	given $\alpha$) via entropy production analysis, we need $D$
	elements in the partition (see points b. and c. of
	Remark~\ref{rem_51}) with $D\geq\lambda_\alpha$.
	Moreover, if we were to
	study the power of our method for \underline{different
	integer} values of $\alpha$ we would be forced 
	forced to use very large $D$, in which case we would need
	very long computing times in order to evaluate numerically the
	entropy production $h_{\tn,{\cal
	W}_\infty}(\alpha,\Lambda,n)$ in a reasonable interval of times
	$n$. 
	Instead, for Sawtooth Maps, we can fix the parameters
	$\pt{N,D,\Lambda}$ and study $\lambda_\alpha$ for $\alpha$
	confined in a small domain, but free to assume every real
	value in that domain.
\end{itemize}
In the following, we investigate the case of $\alpha$ in the hyperbolic
regime  
with $D$ nearest neighbors $\bs{r}_i$ in $\Lambda$, as done in
subsection~\ref{sub_512}. In particular, figures
(\ref{sei}--\ref{nove}) refer to the following fixed parameters:\\[-4.5ex]
\begin{center}
\begin{tabular}{%
@{}c%
@{}c%
@{}p{6mm}%
@{}c%
@{}c%
@{}c%
@{}p{6mm}%
@{}c%
@{}c%
@{}c%
@{}p{6mm}%
@{}c%
@{}c%
@{}c%
@{}p{6mm}%
@{}c%
@{}c%
@{}c%
@{}p{6mm}%
@{}c%
@{}c%
@{}c%
@{}}
\multicolumn{2}{c}{\rule[-1ex]{0pt}{6ex}} &
$N$ & $=$ & $38$ & 
\multicolumn{2}{c}{\quad} &
$;$ & \ &
\multicolumn{2}{c}{$n_{\text{max}}$} 
& $=$ & $5$ &
\multicolumn{2}{c}{\quad} &
$;$ & 
\multicolumn{2}{c}{\quad} &
$D$ & $=$ & $5$ &
$\;\;;$\\
$\Lambda$\rule[-1ex]{0pt}{7ex} & $\quad : \quad$ &
$\bs{r}_1$ & $=$ & $\displaystyle\begin{pmatrix}7\\8\end{pmatrix}$ & $\;\;,\;\;$ &
$\bs{r}_2$ & $=$ & $\displaystyle\begin{pmatrix}7\\9\end{pmatrix}$ & $\;\;,\;\;$ &
$\bs{r}_3$ & $=$ & $\displaystyle\begin{pmatrix}6\\8\end{pmatrix}$ & $\;\;,\;\;$ &
$\bs{r}_4$ & $=$ & $\displaystyle\begin{pmatrix}7\\7\end{pmatrix}$ & $\;\;,\;\;$ &
$\bs{r}_5$ & $=$ & $\displaystyle\begin{pmatrix}8\\8\end{pmatrix}$ & $\;\;;$\\
$\alpha$ \rule[-1ex]{0pt}{6ex}& $\quad : \quad$ &
\multicolumn{19}{c}{from $0.00$ to $1.00$ with an incremental step
	of $0.05$.} &
\end{tabular}\\[4ex]
\end{center}
First, we compute the Von Neumann entropy~\eqref{CoAFE_10}
using the (hermitian) matrix ${\cal
G}_{\bs{\ell}_1,\bs{\ell}_2}\pt{n}$ defined in 
~\eqref{CoAFE_1803}. This is actually a diagonalization problem:
once that the $N^2$ eigenvalues ${\pg{\eta_i}}_{i=1}^{N^2}$
are found, then 
\begin{equation}
H_{\tn}\pq{{\cal \widetilde Y}^{[0,n-1]}} = - \sum_{i=1}^{N^2}
\eta_{i}\log \eta_{i} \ \cdot\label{swtm_1} 
\end{equation}
Then, from~\eqref{AEP_1}, we can determine $h_{\tn,{\cal
W}_\infty}(\alpha,\Lambda,n)$. 
In the numerical example, the ($\Lambda$--dependent) breaking--time
occurs after $n=5$; for this reason we have chosen $n_{\text{max}}=5$.
In fact, we are interested in the region where the discrete
system behaves almost as a continuous one.

In figure~\ref{sei}, the entropy production is plotted for the chosen
set of $\alpha$'s: for very large $N$ (that is close to the
continuum limit, in which no breaking--time occurs) all 
curves (characterized by different $\alpha$'s) 
would tend to \mbox{$\log\lambda_\alpha$ with $n$.}

One way to determine the asymptote $\log\lambda_\alpha$ is to fit the
decreasing function $\displaystyle h_{\tn,{\cal
W}_\infty}(\alpha,\Lambda,n)$ over the range of data and
extrapolate the fit for $n\to\infty$. 
Of course we can not perform
the fit with polynomials, because every polynomial diverges in the 
$n\to\infty$ limit.

A better strategy is to compactify the time evolution by means of a
isomorphic positive function $s$ with bounded range, for instance:
\begin{equation}
\IN\ni n  \longmapsto
s_n \coleq \frac{2}{\pi}\arctan\pt{n-1}\in\pq{0,1}\ \cdot
\label{swtm_2} 
\end{equation}
Then, for fixed $\alpha$, in fig.~\ref{sette} we consider 
$n_{\text{max}}$ points
$\displaystyle {\pt{s_n\,,\,h_{\tn,{\cal
W}_\infty}(\alpha,\Lambda,n)}}$ and 
extract the 
asymptotic value of $\displaystyle h_{\tn,{\cal
W}_\infty}(\alpha,\Lambda,n)$ for $n\to\infty$, that is the value of $\displaystyle h_{\tn,{\cal
W}_\infty}\pt{\alpha,\Lambda,s^{-1}\pt{t}}$ for $t\to 1^-$, as
follows.

Given a graph consisting of $m\in\pg{2,3,\cdots,n_{\text{max}}}$
points, in our case
the first $m$ points of 
curves as in fig.~\ref{sette}, namely
\begin{equation*}
\pg{\pt{s_1\,,\,h_{\tn,{\cal
W}_\infty}(\alpha,\Lambda,1)},\pt{s_2\,,\,h_{\tn,{\cal
W}_\infty}(\alpha,\Lambda,2)}, \cdots,\pt{s_m\,,\,h_{\tn,{\cal
W}_\infty}(\alpha,\Lambda,m)}},
\nonumber
\end{equation*}
\vfill
\newpage
the data are fit by a Lagrange polynomial ${\cal P}^m \pt{t}$ (of
degree $m-1$) 
\begin{subequations}
\label{swtm_3}
\begin{align}
{\cal P}^m \pt{t} & = \sum_{i=1}^m \;P_i \pt{t}\label{swtm_3a}\\
\text{where}\qquad P_i \pt{t} & = \prod_{\substack{j = 1\\j\neq \,i}}^m
\;\frac{t - s_j}{s_i - s_j} \; h_{\tn,{\cal
W}_\infty}(\alpha,\Lambda,i)\ \cdot\label{swtm_3b}
\end{align}
\end{subequations}
The value assumed by this polynomial when $t\to 1^-$ (corresponding to
$n\to\infty$) will be the estimate (of degree $m$) of the
Lyapounov exponent, denoted by $l^m_\alpha$:
the higher the value of $m$, the more 
accurate the estimate. From~\eqref{swtm_3} we get:
\begin{equation}
l^m_\alpha \coleq {\cal P}^m \pt{t} {\Big|}_{t=1} = \sum_{i=1}^m \;
h_{\tn,{\cal 
W}_\infty}(\alpha,\Lambda,i)\;
\prod_{\substack{j = 1\\j\neq \,i}}^m 
\;\frac{1 - s_j}{s_i - s_j}\ \cdot
\label{swtm_4}
\end{equation}
The various $l^m_\alpha$ are plotted in figure~\ref{otto}
as functions of $m$ for
all considered $\alpha$. The convergence of
$l^m_\alpha$ with $m$ is showed in figure~\ref{nove},
together with the theoretical Lyapounov exponent $\log\lambda_\alpha$;
as expected, we find that the latter is the asymptote of
${\pg{l^m_\alpha}}_m$ with respect to the polynomial degree $m$. 

The dotted line in fig.~\ref{sette} extrapolates $21\ \alpha$--curves
in compactified time up to $t=1$ using five points
in the Lagrange polynomial approximation.
%
%
\begin{figure}[h]
\begin{center}
\begin{picture}(140,164)(0,0) 
\put(0,132){\includegraphics[width=\fotogram]{a__1.epsi}
\includegraphics[width=\fotogram]{b__1.epsi}
\includegraphics[width=\fotogram]{c__1.epsi}\hspace{\sepfoto}
\includegraphics[width=\fotogram]{a__8.epsi}
\includegraphics[width=\fotogram]{b__8.epsi}
\includegraphics[width=\fotogram]{c__8.epsi}}
\put(0,110){\includegraphics[width=\fotogram]{a__2.epsi}
\includegraphics[width=\fotogram]{b__2.epsi}
\includegraphics[width=\fotogram]{c__2.epsi}\hspace{\sepfoto}
\includegraphics[width=\fotogram]{a__9.epsi}
\includegraphics[width=\fotogram]{b__9.epsi}
\includegraphics[width=\fotogram]{c__9.epsi}}
\put(0,88){\includegraphics[width=\fotogram]{a__3.epsi}
\includegraphics[width=\fotogram]{b__3.epsi}
\includegraphics[width=\fotogram]{c__3.epsi}\hspace{\sepfoto}
\includegraphics[width=\fotogram]{a_10.epsi}
\includegraphics[width=\fotogram]{b_10.epsi}
\includegraphics[width=\fotogram]{c_10.epsi}}
\put(0,66){\includegraphics[width=\fotogram]{a__4.epsi}
\includegraphics[width=\fotogram]{b__4.epsi}
\includegraphics[width=\fotogram]{c__4.epsi}\hspace{\sepfoto}
\includegraphics[width=\fotogram]{a_11.epsi}
\includegraphics[width=\fotogram]{b_11.epsi}
\includegraphics[width=\fotogram]{c_11.epsi}}
\put(0,44){\includegraphics[width=\fotogram]{a__5.epsi}
\includegraphics[width=\fotogram]{b__5.epsi}
\includegraphics[width=\fotogram]{c__5.epsi}\hspace{\sepfoto}
\includegraphics[width=\fotogram]{a_12.epsi}
\includegraphics[width=\fotogram]{b_12.epsi}
\includegraphics[width=\fotogram]{c_12.epsi}}
\put(0,22){\includegraphics[width=\fotogram]{a__6.epsi}
\includegraphics[width=\fotogram]{b__6.epsi}
\includegraphics[width=\fotogram]{c__6.epsi}\hspace{\sepfoto}
\includegraphics[width=\fotogram]{a_13.epsi}
\includegraphics[width=\fotogram]{b_13.epsi}
\includegraphics[width=\fotogram]{c_13.epsi}}
\put(0,0){\includegraphics[width=\fotogram]{a__7.epsi}
\includegraphics[width=\fotogram]{b__7.epsi}
\includegraphics[width=\fotogram]{c__7.epsi}\hspace{\sepfoto}
\includegraphics[width=\fotogram]{a_14.epsi}
\includegraphics[width=\fotogram]{b_14.epsi}
\includegraphics[width=\fotogram]{c_14.epsi}}

\put(-6.42,164){\line(0,-6){164.7}}
\put(-0.82,164){\line(0,-6){164.7}}
\put(21.58,164){\line(0,-6){164.7}}
\put(43.98,164){\line(0,-6){164.7}}
\put(66.38,164){\line(0,-6){164.7}}
\put(67.78,164){\line(0,-6){164.7}}
\put(73.38,164){\line(0,-6){164.7}}
\put(95.78,164){\line(0,-6){164.7}}
\put(118.18,164){\line(0,-6){164.7}}
\put(140.58,164){\line(0,-6){164.7}}
\put(-6.48,164){\line(6,0){72.94}}
\put(-6.48,154){\line(6,0){72.94}}
\put(-6.48,-0.7){\line(6,0){72.94}}
\put(67.72,164){\line(6,0){72.94}}
\put(67.72,154){\line(6,0){72.94}}
\put(67.72,-0.7){\line(6,0){72.94}}
\put(-6.42,154){\makebox(5.6,10)[cc]{$\mathbf{n}$}}
\put(-0.82,154){\makebox(22.4,10)[cc]{$\boldsymbol{\alpha}\ \mathbf{= \phantom{-}1}$}}
\put(21.58,154){\makebox(22.4,10)[cc]{$\boldsymbol{\alpha}\ \mathbf{= -2}$}}
\put(43.98,154){\makebox(22.4,10)[cc]{$\boldsymbol{\alpha}\ \mathbf{= 17}$}}
\put(-0.82,164){\makebox(22.4,10)[cc]{(a)}}
\put(21.58,164){\makebox(22.4,10)[cc]{(b)}}
\put(43.98,164){\makebox(22.4,10)[cc]{(c)}}
\put(67.78,154){\makebox(5.6,10)[cc]{$\mathbf{n}$}}
\put(73.38,154){\makebox(22.4,10)[cc]{$\boldsymbol{\alpha}\ \mathbf{= \phantom{-}1}$}}
\put(95.78,154){\makebox(22.4,10)[cc]{$\boldsymbol{\alpha}\ \mathbf{= -2}$}}
\put(118.18,154){\makebox(22.4,10)[cc]{$\boldsymbol{\alpha}\ \mathbf{= 17}$}}
\put(73.38,164){\makebox(22.4,10)[cc]{(a)}}
\put(95.78,164){\makebox(22.4,10)[cc]{(b)}}
\put(118.18,164){\makebox(22.4,10)[cc]{(c)}}
\put(-6.42,131.5){\makebox(5.6,22)[cc]{$1$}}
\put(-6.42,109.5){\makebox(5.6,22)[cc]{$2$}}
\put(-6.42,87.5){\makebox(5.6,22)[cc]{$3$}}
\put(-6.42,65.5){\makebox(5.6,22)[cc]{$4$}}
\put(-6.42,43.5){\makebox(5.6,22)[cc]{$5$}}
\put(-6.42,21.5){\makebox(5.6,22)[cc]{$6$}}
\put(-6.42,-0.5){\makebox(5.6,22)[cc]{$7$}}
\put(67.78,131.5){\makebox(5.6,22)[cc]{$8$}}
\put(67.78,109.5){\makebox(5.6,22)[cc]{$9$}}
\put(67.78,87.5){\makebox(5.6,22)[cc]{$10$}}
\put(67.78,65.5){\makebox(5.6,22)[cc]{$11$}}
\put(67.78,43.5){\makebox(5.6,22)[cc]{$12$}}
\put(67.78,21.5){\makebox(5.6,22)[cc]{$13$}}
\put(67.78,-0.5){\makebox(5.6,22)[cc]{$14$}}
\end{picture}
\caption[Temperature--like plots showing the frequencies 
$\displaystyle \nu_{\Lambda,\alpha}^{(n),N}$
in two hyperbolic regimes and
an elliptic one,
for five randomly distributed $\bs{r}_i$ in
$\Lambda$ with $N=200$.]{Temperature--like plots showing the frequencies 
$\displaystyle \nu_{\Lambda,\alpha}^{(n),N}$
in two hyperbolic regimes (columns a and c) and
an elliptic one (col. b),
for five randomly distributed $\bs{r}_i$ in
$\Lambda$ with $N=200$. Pale--blue corresponds to
$\displaystyle \nu_{\Lambda,\alpha}^{(n),N} = 0$.
In the hyperbolic cases, $\displaystyle \nu_{\Lambda,\alpha}^{(n),N}$
tends to
equidistribute on ${\pt{\IZ/N\IZ}}^2$ with
increasing $n$ and becomes constant 
when the breaking--time is reached.}
\label{lontani}
\end{center}
\end{figure}
%
\begin{figure}[h]
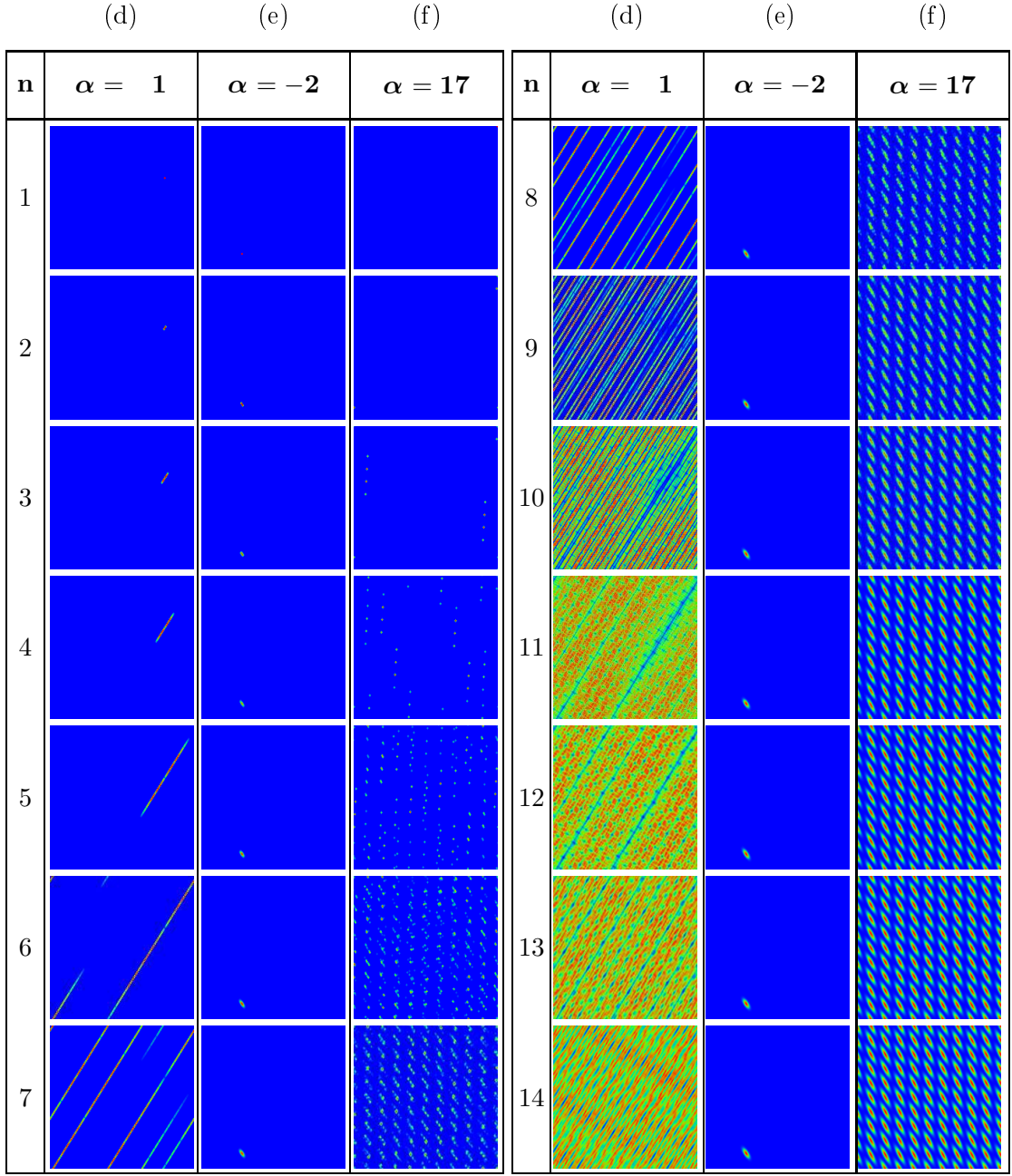

\begin{center}
\begin{picture}(140,164)(0,0) 
\put(0,132){\includegraphics[width=\fotogram]{d__1.epsi}
\includegraphics[width=\fotogram]{e__1.epsi}
\includegraphics[width=\fotogram]{f__1.epsi}\hspace{\sepfoto}
\includegraphics[width=\fotogram]{d__8.epsi}
\includegraphics[width=\fotogram]{e__8.epsi}
\includegraphics[width=\fotogram]{f__8.epsi}}
\put(0,110){\includegraphics[width=\fotogram]{d__2.epsi}
\includegraphics[width=\fotogram]{e__2.epsi}
\includegraphics[width=\fotogram]{f__2.epsi}\hspace{\sepfoto}
\includegraphics[width=\fotogram]{d__9.epsi}
\includegraphics[width=\fotogram]{e__9.epsi}
\includegraphics[width=\fotogram]{f__9.epsi}}
\put(0,88){\includegraphics[width=\fotogram]{d__3.epsi}
\includegraphics[width=\fotogram]{e__3.epsi}
\includegraphics[width=\fotogram]{f__3.epsi}\hspace{\sepfoto}
\includegraphics[width=\fotogram]{d_10.epsi}
\includegraphics[width=\fotogram]{e_10.epsi}
\includegraphics[width=\fotogram]{f_10.epsi}}
\put(0,66){\includegraphics[width=\fotogram]{d__4.epsi}
\includegraphics[width=\fotogram]{e__4.epsi}
\includegraphics[width=\fotogram]{f__4.epsi}\hspace{\sepfoto}
\includegraphics[width=\fotogram]{d_11.epsi}
\includegraphics[width=\fotogram]{e_11.epsi}
\includegraphics[width=\fotogram]{f_11.epsi}}
\put(0,44){\includegraphics[width=\fotogram]{d__5.epsi}
\includegraphics[width=\fotogram]{e__5.epsi}
\includegraphics[width=\fotogram]{f__5.epsi}\hspace{\sepfoto}
\includegraphics[width=\fotogram]{d_12.epsi}
\includegraphics[width=\fotogram]{e_12.epsi}
\includegraphics[width=\fotogram]{f_12.epsi}}
\put(0,22){\includegraphics[width=\fotogram]{d__6.epsi}
\includegraphics[width=\fotogram]{e__6.epsi}
\includegraphics[width=\fotogram]{f__6.epsi}\hspace{\sepfoto}
\includegraphics[width=\fotogram]{d_13.epsi}
\includegraphics[width=\fotogram]{e_13.epsi}
\includegraphics[width=\fotogram]{f_13.epsi}}
\put(0,0){\includegraphics[width=\fotogram]{d__7.epsi}
\includegraphics[width=\fotogram]{e__7.epsi}
\includegraphics[width=\fotogram]{f__7.epsi}\hspace{\sepfoto}
\includegraphics[width=\fotogram]{d_14.epsi}
\includegraphics[width=\fotogram]{e_14.epsi}
\includegraphics[width=\fotogram]{f_14.epsi}}

\put(-6.42,164){\line(0,-6){164.7}}
\put(-0.82,164){\line(0,-6){164.7}}
\put(21.58,164){\line(0,-6){164.7}}
\put(43.98,164){\line(0,-6){164.7}}
\put(66.38,164){\line(0,-6){164.7}}
\put(67.78,164){\line(0,-6){164.7}}
\put(73.38,164){\line(0,-6){164.7}}
\put(95.78,164){\line(0,-6){164.7}}
\put(118.18,164){\line(0,-6){164.7}}
\put(140.58,164){\line(0,-6){164.7}}
\put(-6.48,164){\line(6,0){72.94}}
\put(-6.48,154){\line(6,0){72.94}}
\put(-6.48,-0.7){\line(6,0){72.94}}
\put(67.72,164){\line(6,0){72.94}}
\put(67.72,154){\line(6,0){72.94}}
\put(67.72,-0.7){\line(6,0){72.94}}
\put(-6.42,154){\makebox(5.6,10)[cc]{$\mathbf{n}$}}
\put(-0.82,154){\makebox(22.4,10)[cc]{$\boldsymbol{\alpha}\ \mathbf{= \phantom{-}1}$}}
\put(21.58,154){\makebox(22.4,10)[cc]{$\boldsymbol{\alpha}\ \mathbf{= -2}$}}
\put(43.98,154){\makebox(22.4,10)[cc]{$\boldsymbol{\alpha}\ \mathbf{= 17}$}}
\put(-0.82,164){\makebox(22.4,10)[cc]{(d)}}
\put(21.58,164){\makebox(22.4,10)[cc]{(e)}}
\put(43.98,164){\makebox(22.4,10)[cc]{(f)}}
\put(67.78,154){\makebox(5.6,10)[cc]{$\mathbf{n}$}}
\put(73.38,154){\makebox(22.4,10)[cc]{$\boldsymbol{\alpha}\ \mathbf{= \phantom{-}1}$}}
\put(95.78,154){\makebox(22.4,10)[cc]{$\boldsymbol{\alpha}\ \mathbf{= -2}$}}
\put(118.18,154){\makebox(22.4,10)[cc]{$\boldsymbol{\alpha}\ \mathbf{= 17}$}}
\put(73.38,164){\makebox(22.4,10)[cc]{(d)}}
\put(95.78,164){\makebox(22.4,10)[cc]{(e)}}
\put(118.18,164){\makebox(22.4,10)[cc]{(f)}}
\put(-6.42,131.5){\makebox(5.6,22)[cc]{$1$}}
\put(-6.42,109.5){\makebox(5.6,22)[cc]{$2$}}
\put(-6.42,87.5){\makebox(5.6,22)[cc]{$3$}}
\put(-6.42,65.5){\makebox(5.6,22)[cc]{$4$}}
\put(-6.42,43.5){\makebox(5.6,22)[cc]{$5$}}
\put(-6.42,21.5){\makebox(5.6,22)[cc]{$6$}}
\put(-6.42,-0.5){\makebox(5.6,22)[cc]{$7$}}
\put(67.78,131.5){\makebox(5.6,22)[cc]{$8$}}
\put(67.78,109.5){\makebox(5.6,22)[cc]{$9$}}
\put(67.78,87.5){\makebox(5.6,22)[cc]{$10$}}
\put(67.78,65.5){\makebox(5.6,22)[cc]{$11$}}
\put(67.78,43.5){\makebox(5.6,22)[cc]{$12$}}
\put(67.78,21.5){\makebox(5.6,22)[cc]{$13$}}
\put(67.78,-0.5){\makebox(5.6,22)[cc]{$14$}}
\end{picture}
\caption[Temperature--like plots showing 
$\displaystyle \nu_{\Lambda,\alpha}^{(n),N}$
in two hyperbolic and
one elliptic regime, for five nearest neighboring $\bs{r}_i$
in $\Lambda$ ($N=200$).]{Temperature--like plots showing 
$\displaystyle \nu_{\Lambda,\alpha}^{(n),N}$
in two hyperbolic (columns d and f) and
one elliptic (col. e) regime, for five nearest neighboring $\bs{r}_i$
in $\Lambda$ ($N=200$). Pale--blue corresponds to 
$\displaystyle \nu_{\Lambda,\alpha}^{(n),N} = 0$.
When the system is chaotic, the frequencies tend to equidistribute on
${\pt{\IZ/N\IZ}}^2$ with
increasing $n$ and to approach, when the
breaking--time is reached,
the constant value $\frac{1}{N^2}$. Col. (f) shows how the dynamics
can be confined on a sublattice by a particular combination
$(\alpha,N,\Lambda)$ with a corresponding entropy decrease.}
\label{vicini}
\end{center}
\end{figure}
%
\newpage
%
\begin{figure}[h]
\begin{center}
\ \\[2ex]
\begin{picture}(130,70)(0,0) 
\put(10,80){\includegraphics[width=\altgraph,height=\larggraph,angle=-90]{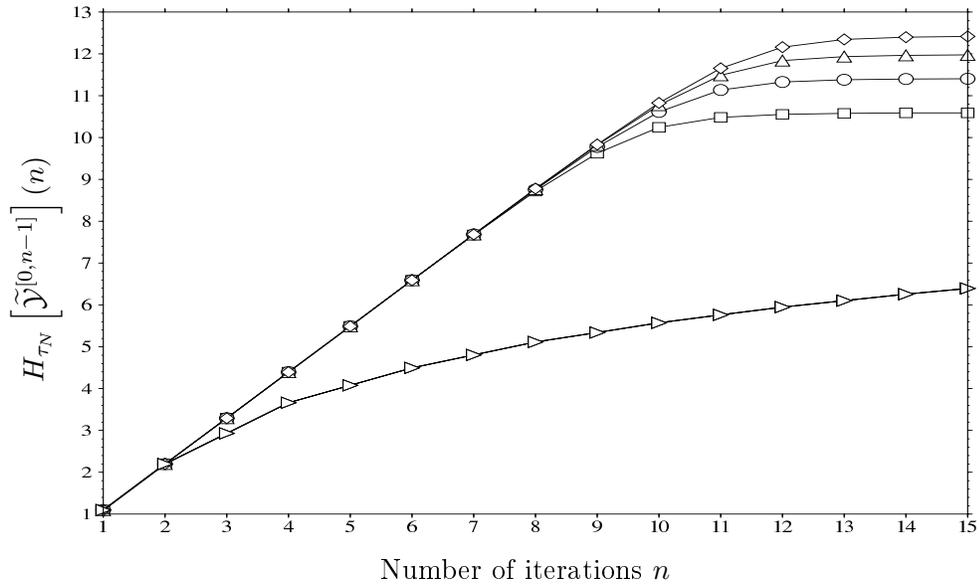}}
\begin{sideways}
\put(10,0){\makebox(70,10)[cc]{$\displaystyle H_{\tn}\pq{{\cal \widetilde
Y}^{[0,n-1]}}\pt{n}$}}
\end{sideways}
\put(0,0){\makebox(120,10)[cc]{Number of iterations $n$}}
\end{picture}
\caption[Von Neumann entropy $\displaystyle H_{\tn}\pt{n}$ 
in four hyperbolic and
four elliptic cases, for three
randomly distributed $\bs{r}_i$ in 
$\Lambda$.]{Von Neumann entropy $\displaystyle H_{\tn}\pt{n}$ 
in four hyperbolic ($\alpha = 1$ for $\diamond$, {\scriptsize
$\bigtriangleup$}, $\circ$, {\scriptsize $\Box$}) and
four elliptic ($\alpha = -2$ for $\triangleright$) cases, for three
randomly distributed $\bs{r}_i$ in 
$\Lambda$. Values for $N$ are: $\diamond = 500$, {\scriptsize
$\bigtriangleup$} $= 400$, $\circ = 300$ and {\scriptsize $\Box$} $=
200$, whereas the curve labeled by $\triangleright$ represents
four elliptic systems with $N\in\pg{200, 300, 400, 500}$.}
\ 
\label{uno}
\end{center}
\end{figure}
%
%
\begin{figure}[h]
\begin{center}
\ \\[0.7ex]
\begin{picture}(130,70)(0,0) 
\put(10,80){\includegraphics[width=\altgraph,height=\larggraph,angle=-90]{due.epsi}}
\begin{sideways}
\put(10,0){\makebox(70,10)[cc]{$\displaystyle H_{\tn}\pq{{\cal \widetilde
Y}^{[0,n-1]}}\pt{n}$}}
\end{sideways}
\put(0,0){\makebox(120,10)[cc]{Number of iterations $n$}}
\end{picture}
\caption[Von Neumann entropy $\displaystyle H_{\tn}\pt{n}$ 
in four hyperbolic cases, for $D$
randomly distributed $\bs{r}_i$ in 
$\Lambda$, with $N=200$.]{Von Neumann entropy $\displaystyle
H_{\tn}\pt{n}$  
in four hyperbolic ($\alpha = 1$) cases, for $D$
randomly distributed $\bs{r}_i$ in 
$\Lambda$, with $N=200$. Value for $D$ are: $\diamond = 5$,
\mbox{{\scriptsize 
$\bigtriangleup$} $= 4$}, $\circ = 3$ and {\scriptsize $\Box$} $=
2$. The dotted line represents $\displaystyle H_{\tn}\pt{n} =
\log\lambda \cdot n$ where $\log\lambda =
0.962\ldots$ is the Lyapounov exponent at$\alpha = 1$.}
\label{due}
\end{center}
\end{figure}
%
\newpage
%
\begin{figure}[h]
\begin{center}
\ \\[1.5ex]
\begin{picture}(130,70)(0,0) 
\put(10,80){\includegraphics[width=\altgraph,height=\larggraph,angle=-90]{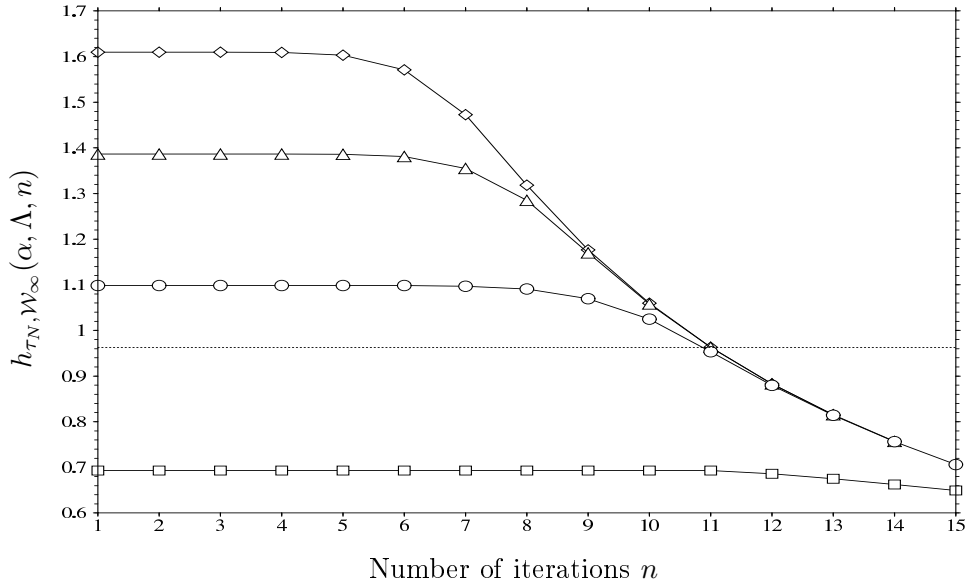}}
\begin{sideways}
\put(10,0){\makebox(70,10)[cc]{$\displaystyle h_{\tn,{\cal W}_\infty}(\alpha,\Lambda,n)$}}
\end{sideways}
\put(0,0){\makebox(120,10)[cc]{Number of iterations $n$}}
\end{picture}
\caption[Entropy production $\displaystyle h_{\tn,{\cal
W}_\infty}(\alpha,\Lambda,n)$ 
in four hyperbolic cases, for $D$
randomly distributed $\bs{r}_i$ in 
$\Lambda$, with $N=200$.]{Entropy production $\displaystyle
h_{\tn,{\cal 
W}_\infty}(\alpha,\Lambda,n)$ 
in four hyperbolic ($\alpha = 1$) cases, for $D$
randomly distributed $\bs{r}_i$ in 
$\Lambda$, with $N=200$. Values for $D$ are: $\diamond = 5$,
\mbox{{\scriptsize 
$\bigtriangleup$} $= 4$}, $\circ = 3$ and {\scriptsize $\Box$} $=
2$. The dotted line corresponds to the Lyapounov exponent $\log\lambda =
0.962\ldots$ at $\alpha = 1$.}
\ 
\label{tre}
\end{center}
\end{figure}
%
%
\begin{figure}[h]
\begin{center}
\ \\[0.2ex]
\begin{picture}(130,70)(0,0) 
\put(10,80){\includegraphics[width=\altgraph,height=\larggraph,angle=-90]{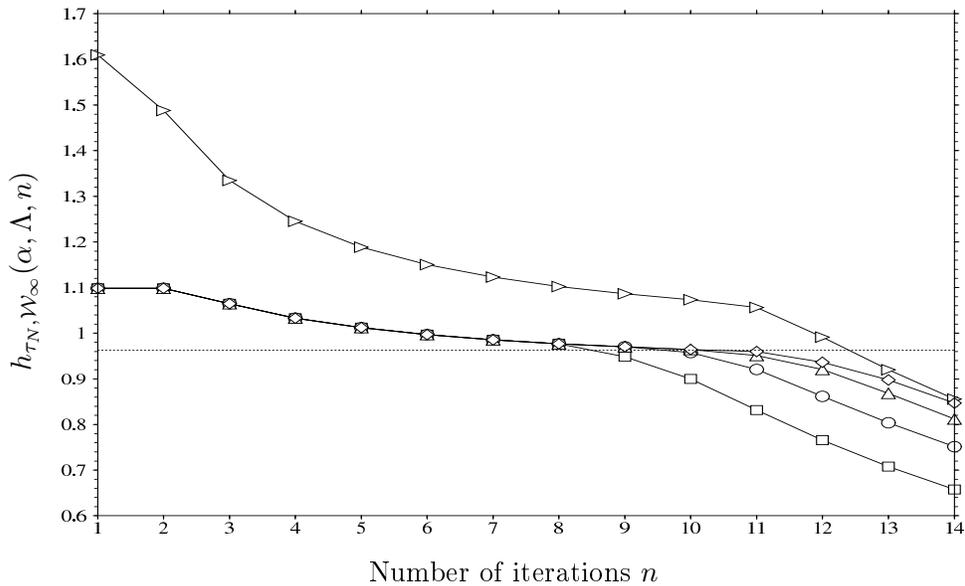}}
\begin{sideways}
\put(10,0){\makebox(70,10)[cc]{$\displaystyle h_{\tn,{\cal W}_\infty}(\alpha,\Lambda,n)$}}
\end{sideways}
\put(0,0){\makebox(120,10)[cc]{Number of iterations $n$}}
\end{picture}
\caption[Entropy production $\displaystyle h_{\tn,{\cal
W}_\infty}(\alpha,\Lambda,n)$ in five hyperbolic
cases, for $D$ nearest neighboring points $\bs{r}_i$
in $\Lambda$.]{Entropy production $\displaystyle h_{\tn,{\cal
W}_\infty}(\alpha,\Lambda,n)$ in five hyperbolic
($\alpha = 1$) cases, for $D$ nearest neighboring points $\bs{r}_i$
in $\Lambda$. Values for $\pt{N,D}$ are: $\triangleright= \pt{200,5}$,
$\diamond = \pt{500,3}$, \mbox{{\scriptsize 
$\bigtriangleup$} $= \pt{400,3}$}, $\circ = \pt{300,3}$ and
{\scriptsize $\Box$} $= \pt{200,3}$. The dotted line corresponds to
the Lyapounov exponent $\log\lambda = 0.962\ldots$ at
$\alpha = 1$ and represents the natural asymptote for all these curves
in absence of breaking--time.} 
\label{cinque}
\end{center}
\end{figure}
%
\newpage
%
\begin{figure}[h]
\begin{center}
\begin{picture}(130,70)(0,0) 
\put(10,80){\includegraphics[width=\altgraph,height=\larggraph,angle=-90]{quattro.epsi}}
\begin{sideways}
\put(10,0){\makebox(70,10)[cc]{$\displaystyle H_{\tn}\pq{{\cal \widetilde
Y}^{[0,n-1]}}\pt{n}$}}
\end{sideways}
\put(0,0){\makebox(120,10)[cc]{Number of iterations $n$}}
\end{picture}
\caption[Von Neumann entropy $\displaystyle H_{\tn}\pt{n}$ 
in four elliptic cases, for $D$
randomly distributed $\bs{r}_i$ in 
$\Lambda$, with $N=200$.]{Von Neumann entropy $\displaystyle
H_{\tn}\pt{n}$  
in four elliptic ($\alpha = -2$) cases, for $D$
randomly distributed $\bs{r}_i$ in 
$\Lambda$, with $N=200$. Value for $D$ are: $\diamond = 5$,
\mbox{{\scriptsize 
$\bigtriangleup$} $= 4$}, $\circ = 3$ and {\scriptsize $\Box$} $=
2$.}
\
\label{quattro}
\end{center}
\end{figure}
%
\begin{figure}[h]
\begin{center}
\begin{picture}(130,70)(0,0) 
\put(10,80){\includegraphics[width=\altgraph,height=\larggraph,angle=-90]{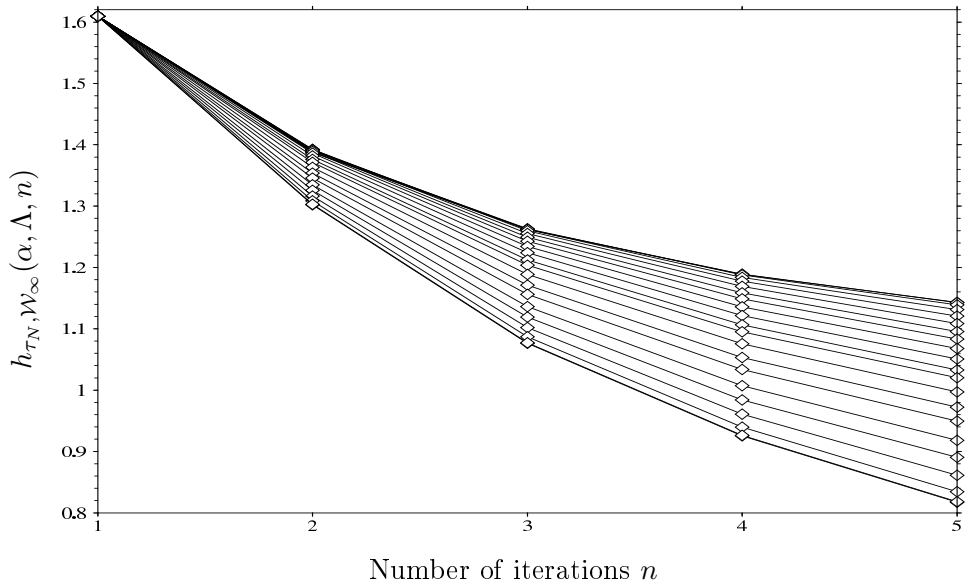}}
\begin{sideways}
\put(10,0){\makebox(70,10)[cc]{$\displaystyle h_{\tn,{\cal
W}_\infty}(\alpha,\Lambda,n)$}} 
\end{sideways}
\put(0,0){\makebox(120,10)[cc]{Number of iterations $n$}}
\end{picture}
\caption[Entropy production $\displaystyle h_{\tn,{\cal
W}_\infty}(\alpha,\Lambda,n)$  for $21$
hyperbolic Sawtooth maps, relative to a 
for a cluster of $5$ nearest neighborings points $\bs{r}_i$ in 
$\Lambda$.]{Entropy production $\displaystyle
h_{\tn,{\cal 
W}_\infty}(\alpha,\Lambda,n)$  for $21$
hyperbolic Sawtooth maps, relative to a 
for a cluster of $5$ nearest neighborings points $\bs{r}_i$ in 
$\Lambda$, with $N=38$.
The parameter $\alpha$ decreases from $\alpha=1.00$
(corresponding to 
the upper curve) to $\alpha=0.00$ (lower curve) through
$21$ equispaced steps.}
\label{sei}
\end{center}
\end{figure}
%
\newpage
%
\begin{figure}[h]
\begin{center}
\begin{picture}(130,70)(0,0) 
\put(10,80){\includegraphics[width=\altgraph,height=\larggraph,angle=-90]{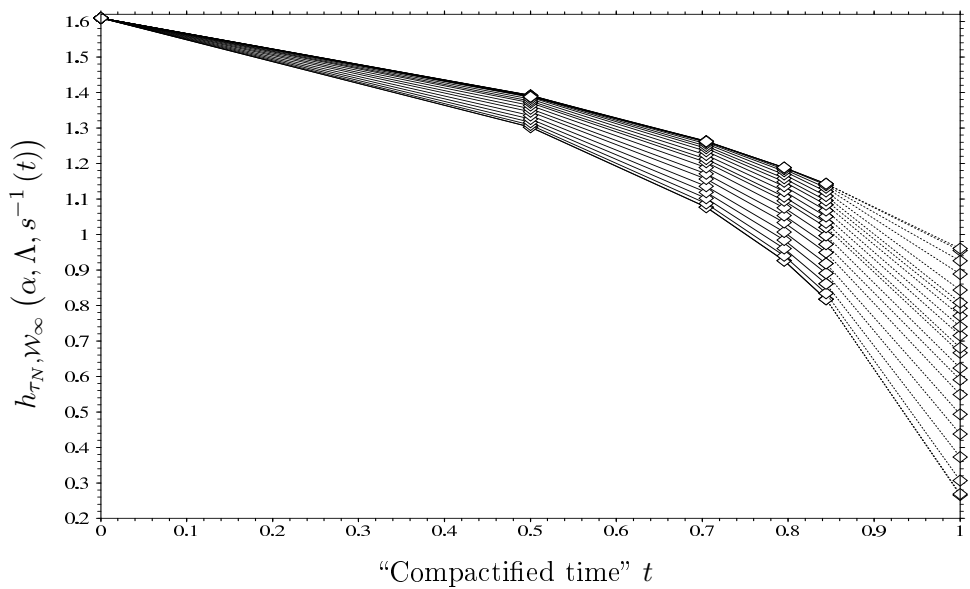}}
\begin{sideways}
\put(10,0){\makebox(70,10)[cc]{$\displaystyle h_{\tn,{\cal
W}_\infty}\pt{\alpha,\Lambda,s^{-1}\pt{t}}$}}
\end{sideways}
\put(0,0){\makebox(120,10)[cc]{``Compactified time'' $t$}}
\end{picture}
\caption[Entropy production $\displaystyle h_{\tn,{\cal
W}_\infty}(\alpha,\Lambda,n)$  for the $21$
hyperbolic Sawtooth maps of Figure~\ref{sei}, plotted
vs. the ``compactified time''.]{The solid lines correspond to $\displaystyle
{\pt{s_n\,,\,h_{\tn,{\cal 
W}_\infty}(\alpha,\Lambda,n)}}$, with $n\in\pg{1,2,3,4,5}$, for the
values of $\alpha$ considered in figure~\ref{sei}. Every
$\alpha$--curve is continued as a dotted line
up to $\pt{1,l^5_\alpha}$, where
$l^5_\alpha$ is the Lyapounov exponent extracted from the curve by
fitting all the five points via a Lagrange polynomial ${\cal P}^m\pt{t}$.}
\
\label{sette}
\end{center}
\end{figure}
%
\begin{figure}[h]
\begin{center}
\begin{picture}(130,70)(0,0) 
\put(10,80){\includegraphics[width=\altgraph,height=\larggraph,angle=-90]{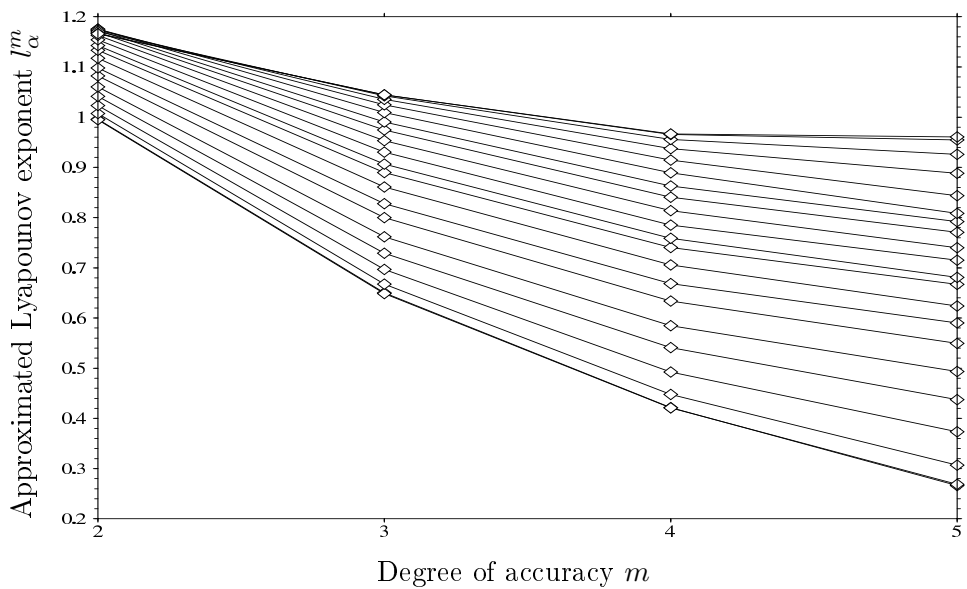}}
\begin{sideways}
\put(10,0){\makebox(70,10)[cc]{Approximated Lyapounov exponent $l^m_\alpha$}}
\end{sideways}
\put(0,0){\makebox(120,10)[cc]{Degree of
accuracy $m$}}
\end{picture}
\caption{Four estimated Lyapounov exponents $l^m_\alpha$ 
plotted vs. their degree of accuracy $m$ for the values of
$\alpha$ considered in figures~\ref{sei}~and~\ref{sette}.}
\label{otto}
\end{center}
\end{figure}
%
\newpage
%
\begin{figure}[h]
\begin{center}
\begin{picture}(130,70)(0,0) 
\put(10,80){\includegraphics[width=\altgraph,height=\larggraph,angle=-90]{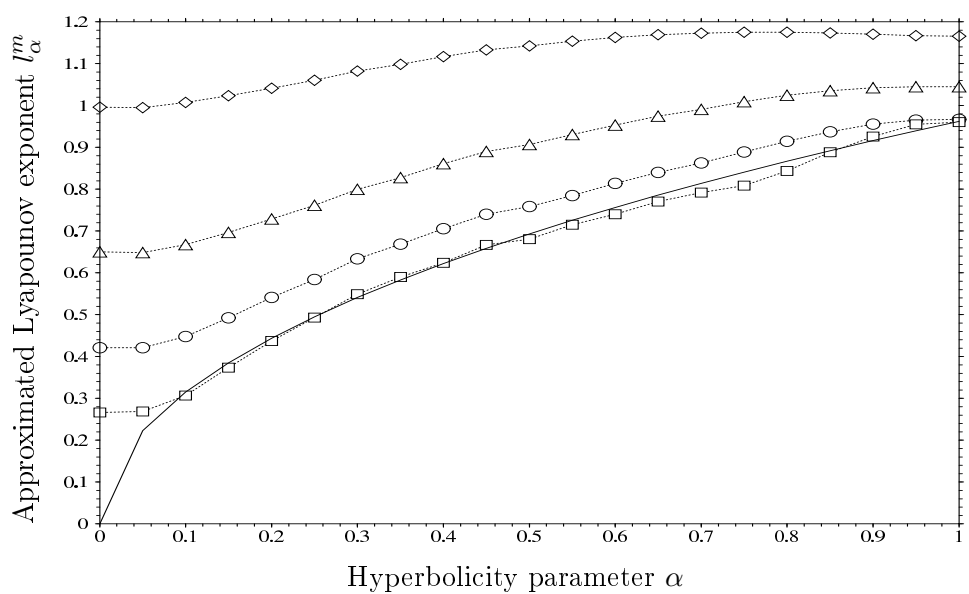}}
\begin{sideways}
\put(10,0){\makebox(70,10)[cc]{Approximated Lyapounov exponent
$l^m_\alpha$}}
\end{sideways}
\put(0,0){\makebox(120,10)[cc]{Hyperbolicity parameter $\alpha$}}
\end{picture}
\caption[Plots of the four estimated of Lyapounov exponents $l^m_\alpha$ of
figure~\ref{otto} 
vs. the considered values of
$\alpha$.]{Plots of the four estimated of Lyapounov exponents $l^m_\alpha$ of
figure~\ref{otto} 
vs. the considered values of
$\alpha$.
The polynomial degree $m$ is as follows: $\diamond = 2$,
\mbox{{\scriptsize 
$\bigtriangleup$} $= 3$}, $\circ = 4$ and {\scriptsize $\Box$} $=
5$. The solid line corresponds to the theoretical Lyapounov exponent 
$\displaystyle \log\lambda_\alpha=
\log\:(\alpha+2 + \sqrt{\alpha\:(\alpha+4)\,}\;) - \log 2
$.}
\label{nove}
\end{center}
\end{figure}
\appendix
\pagestyle{fancyplain}
\chapter{Non Overcompleteness of the set of
states of Section~\ref{CST2}} \label{app_A}\vspace{9mm}
The coherent state~\eqref{RoeiW_6} can be rewritten in the shorter form:
\begin{equation}
\vert \beta_N(\bs{x})\rangle= 
\sum_{\pt{\mu,\nu}\in{\pg{0,1}}^2}
\lambda_{\mu\nu}\pt{\bs{x}}
\ket{\floor{N x_1} + \mu,\floor{N x_2} + \nu}\ ,
\label{app1}
\end{equation}
whereas for the $\lambda$--coefficient we can write
\begin{equation}
\lambda_{\mu\nu}\pt{\bs{x}}= 
\cos\pq{\frac{\pi}{2}\pt{\mu-\bk{N x_1}}} 
\cos\pq{\frac{\pi}{2}\pt{\nu-\bk{N x_2}}}
\label{app2}
\end{equation}
Overcompleteness property of Definition~\ref{coh} can be expressed as
\begin{equation}
N^2 \int_{\c X}\mu(\ud\bs{x})\, 
\langle\bs{\ell}\ |\ \beta_N(\bs{x})\rangle
\langle \beta_N(\bs{x})\ |\ \bs{m}\rangle
= \delta^{(N)}_{\bs{\ell},\bs{m}},\qquad\forall\bs{\ell},\bs{m}\in\ZNZD
\label{app3}
\end{equation}
and this is exactly what we are going to check.
Let us define with $I_{\bs{\ell},\bs{m}}$ the l.h.s. of~\eqref{app3}
then, using~(\ref{app1}--\ref{app2}), we can write
\begin{multline}
I_{\bs{\ell},\bs{m}}\coleq
N^2 \!\!\!\!\!\!\!\!\!
\sum_{\pt{\mu,\nu,\rho,\sigma}\in{\pg{0,1}}^4}
\int_{0}^{1}\ud x_1\,
\int_{0}^{1}\ud x_2\;
\cos\pq{\frac{\pi}{2}\pt{\mu-\bk{N x_1}}} 
\cos\pq{\frac{\pi}{2}\pt{\nu-\bk{N x_2}}}\ \times\\
\times\ 
\cos\pq{\frac{\pi}{2}\pt{\rho-\bk{N x_1}}} 
\cos\pq{\frac{\pi}{2}\pt{\sigma-\bk{N x_2}}}\ \times\\
\times\ 
\Bigg\langle \ell_1,\ell_2\ \Bigg|\ 
\floor{N x_1} + \rho,\floor{N x_2} + \sigma
\Bigg\rangle 
\Bigg\langle 
\floor{N x_1} + \mu,\floor{N x_2} + \nu
\ \Bigg|\ m_1,m_2
\Bigg\rangle\ ,
\label{app4}
\end{multline}
\begin{multline}
\text{that is }I_{\bs{\ell},\bs{m}}\coleq
N^2 \!\!\!\!\!\!\!\!\!
\sum_{\pt{\mu,\nu,\rho,\sigma}\in{\pg{0,1}}^4}\ \times\\
\times\ 
\int_{0}^{1}\ud x_1\,
\cos\pq{\frac{\pi}{2}\pt{\mu-\bk{N x_1}}} 
\cos\pq{\frac{\pi}{2}\pt{\rho-\bk{N x_1}}} 
\delta^{(N)}_{\ell_1\:,\:\floor{N x_1} + \rho}\;
\delta^{(N)}_{\floor{N x_1} + \mu\:,\:m_1}\;\ \times\\
\times\ 
\int_{0}^{1}\ud x_2\;
\cos\pq{\frac{\pi}{2}\pt{\nu-\bk{N x_2}}}
\cos\pq{\frac{\pi}{2}\pt{\sigma-\bk{N x_2}}}
\delta^{(N)}_{\ell_2\:,\:\floor{N x_2} + \sigma}\;
\delta^{(N)}_{\floor{N x_2} + \nu\:,\:m_2}\ ,\notag
\end{multline}
or
$I_{\bs{\ell},\bs{m}}=\Gamma_{\ell_1,m_1}\times\Gamma_{\ell_2,m_2}$,
with $\Gamma_{p,q}$ defined by
\begin{align}
\Gamma_{p,q}
\coleq N \!\!&\!\!\!\sum_{\pt{\mu,\rho}\in{\pg{0,1}}^2}\ \times\notag\\
&\times\ \int_{0}^{1}\ud y\,
\cos\pq{\frac{\pi}{2}\pt{\mu-\bk{N y}}} 
\cos\pq{\frac{\pi}{2}\pt{\rho-\bk{N y}}} 
\delta^{(N)}_{p\:,\:\floor{N y} + \rho}\;
\delta^{(N)}_{q\:,\:\floor{N y} + \mu}\ =\notag\\
= N \!\!&\!\!\!\sum_{\pt{\mu,\rho}\in{\pg{0,1}}^2}\ \times\notag\\
&\times\ \int_{0}^{1}\ud y\,
\cos\pq{\frac{\pi}{2}\pt{\mu-\bk{N y}}} 
\cos\pq{\frac{\pi}{2}\pt{\rho-\bk{N y}}} 
\delta^{(N)}_{p-\rho\:,\:\floor{N y}}\;
\delta^{(N)}_{q-\mu\:,\:\floor{N y}}\ =\notag\\
= N \!\!&\!\!\!\sum_{\pt{\mu,\rho}\in{\pg{0,1}}^2}\ \times\notag\\
&\times\ \pg{\int_{0}^{1}\ud y\,
\cos\pq{\frac{\pi}{2}\pt{\mu-\bk{N y}}} 
\cos\pq{\frac{\pi}{2}\pt{\rho-\bk{N y}}} 
\delta^{(N)}_{p-\rho\:,\:\floor{N y}}}\;
\delta^{(N)}_{q-\mu\:,\:p-\rho}\cdot
\label{app8}
\end{align}
Defining the symbol $((s))\coleq\pg{t\in\ZNZn : t = s}$ (the element
in the residual class$\pmod{N}$ representing $s$), in order to have
the integrand of~\eqref{app8} different from zero we must have
$((p-\rho))\leq N y<((p-\rho))+1$ (note that in that range $\bk{N
y}=N y-((p-\rho))$), and~\eqref{app8} reads:
\begin{align}
\Gamma_{p,q}
= N \!\!&\!\!\!\sum_{\pt{\mu,\rho}\in{\pg{0,1}}^2}
\int_{\frac{((p-\rho))}{N}}^{\frac{((p-\rho))+1}{N}}\ud y
\notag\\ 
&\cos\pq{\frac{\pi}{2}\Big(\mu -N y+((p-\rho))\Big)} 
 \cos\pq{\frac{\pi}{2}\Big(\rho-N y+((p-\rho))\Big)}\;
\delta^{(N)}_{q-\mu\:,\:p-\rho}\cdot\notag\\
\intertext{Using now Werner trigonometric formula, we get}
\Gamma_{p,q}
= \qquad\frac{N}{2} \!\!&\!\!\!\sum_{\pt{\mu,\rho}\in{\pg{0,1}}^2}
\int_{\frac{((p-\rho))}{N}}^{\frac{((p-\rho))+1}{N}}\ud y\,
\cos\pq{\frac{\pi}{2}\pt{\mu + \rho}-\pi N y+\pi((p-\rho))}\;
\delta^{(N)}_{q-\mu\:,\:p-\rho}\ + \notag\\
 +\ \frac{N}{2} \!\!&\!\!\!\sum_{\pt{\mu,\rho}\in{\pg{0,1}}^2}
\int_{\frac{((p-\rho))}{N}}^{\frac{((p-\rho))+1}{N}}\ud y\,
\cos\pq{\frac{\pi}{2}\pt{\mu - \rho}}\;
\delta^{(N)}_{q-\mu\:,\:p-\rho}\ =\notag\\
\end{align}
\begin{align}
&=
\phantom{\sin}\phantom{\sin}\phantom{\frac{1}{2}}%
\!\!\!\!\!\!\!\!
\phantom{+}\frac{N}{2}\;
 \!\!\!\!\!\sum_{\pt{\mu,\rho}\in{\pg{0,1}}^2}
{\pt{-1}}^{((p-\rho))}
\cos\pq{\frac{\pi}{2}\pt{\mu + \rho}}\;
\int_{\frac{((p-\rho))}{N}}^{\frac{((p-\rho))+1}{N}}\ud y\,
\cos\pt{\pi N y}\;
\delta^{(N)}_{q-\mu\:,\:p-\rho}\ + \notag\\
&\phantom{=}
\phantom{\cos}\phantom{\cos}\phantom{\frac{1}{2}}%
\!\!\!\!\!\!\!\!
+\frac{N}{2}\;
 \!\!\!\!\!\sum_{\pt{\mu,\rho}\in{\pg{0,1}}^2}
{\pt{-1}}^{((p-\rho))}
\sin\pq{\frac{\pi}{2}\pt{\mu + \rho}}\;
\int_{\frac{((p-\rho))}{N}}^{\frac{((p-\rho))+1}{N}}\ud y\,
\sin\pt{\pi N y}\;
\delta^{(N)}_{q-\mu\:,\:p-\rho}\ + \notag\\
&\phantom{=}
\phantom{\sin}\phantom{\sin}\phantom{\frac{N}{2}}%
\!\!\!\!\!\!\!\!
+\frac{1}{2}\;
 \!\!\!\sum_{\pt{\mu,\rho}\in{\pg{0,1}}^2}
\cos\pq{\frac{\pi}{2}\pt{\mu - \rho}}\;
\delta^{(N)}_{q-\mu\:,\:p-\rho}\ =\notag\\
&=0
+\frac{1}{2}\;\phantom{\frac{1}{2\pi}}
 \!\!\!\!\!\!\!\!\!\!\sum_{\pt{\mu,\rho}\in{\pg{0,1}}^2}
\cos\pq{\frac{\pi}{2}\pt{\mu - \rho}}\;
\delta^{(N)}_{q-\mu\:,\:p-\rho}\ -
\frac{1}{2\pi}\;\phantom{\frac{1}{2}} 
 \!\!\!\!\! \!\!\!\!\!\sum_{\pt{\mu,\rho}\in{\pg{0,1}}^2}
{\pt{-1}}^{((p-\rho))}
\times\notag\\
&\phantom{=}
\phantom{\sin}\phantom{\sin}\phantom{\frac{N}{2}}%
\!\!\!\!\!\!\!\!
\times\sin\pq{\frac{\pi}{2}\pt{\mu + \rho}}\;
\bigg\{
\cos\Big[\pi ((p-\rho))+\pi\Big] - \cos\Big[\pi ((p-\rho))\Big]
\bigg\}
\delta^{(N)}_{q-\mu\:,\:p-\rho}\ = \notag\\
&=
\frac{1}{2}\;\phantom{\frac{1}{2}}
 \!\!\!\!\! \!\!\!\!\!\sum_{\pt{\mu,\rho}\in{\pg{0,1}}^2}
\delta^{(N)}_{q-\mu\:,\:p-\rho}
\Bigg\{
\frac{2}{\pi}\sin\pq{\frac{\pi}{2}\pt{\mu + \rho}}\;
+\cos\pq{\frac{\pi}{2}\pt{\mu - \rho}}
\Bigg\}\ \cdot\label{app15}
\intertext{$\displaystyle \text{For
}\pt{\mu,\rho}\in{\pg{0,1}}^2\text{, we have: }  
\begin{cases}
\cos\pq{\frac{\pi}{2}\pt{\mu - \rho}} & = \delta_{\mu,\rho}\\
\sin\pq{\frac{\pi}{2}\pt{\mu + \rho}} & = 1 - \delta_{\mu,\rho}
\end{cases}
$ thus from~\eqref{app15} we get:}
\Gamma_{p,q}
&=\frac{1}{2}\;\phantom{\frac{1}{2}}
 \!\!\!\!\! \!\!\!\!\!\sum_{\pt{\mu,\rho}\in{\pg{0,1}}^2}
\delta^{(N)}_{q-\mu\:,\:p-\rho}
\Bigg\{
\frac{2}{\pi}
\pt{1 - \delta_{\mu,\rho}}
\;+
\delta_{\mu,\rho}
\Bigg\}\ =\notag\\
&=\frac{1}{\pi}
\sum_{\pt{\mu,\rho}\in{\pg{0,1}}^2}
\delta^{(N)}_{q-\mu\:,\:p-\rho}
\ + \ \frac{1}{2}\;
\sum_{\rho\in{\pg{0,1}}}\pt{1-\frac{2}{\pi}}
\Bigg[\sum_{\mu\in{\pg{0,1}}}
\delta^{(N)}_{q-\mu\:,\:p-\rho}\;
\delta^{\phantom{(N)}}_{\mu\:,\:\rho}
\Bigg]\ =
\notag\\
&=\frac{1}{\pi}
\sum_{\pt{\mu,\rho}\in\pg{\pt{0,0},\pt{0,1},\pt{1,0},\pt{1,1}}}
\delta^{(N)}_{q-p\:,\:\mu-\rho}
\ +\ 
\frac{1}{2}\;
\sum_{\rho\in{\pg{0,1}}}\pt{1-\frac{2}{\pi}}
\delta^{(N)}_{q\:,\:p}\ =\notag\\
&=\frac{1}{\pi} 
\pt{\delta^{(N)}_{q-p\:,\:0}
+\delta^{(N)}_{q-p\:,\:-1}  
+\delta^{(N)}_{q-p\:,\:1}   
+\delta^{(N)}_{q-p\:,\:0}}  
\ +\delta^{(N)}_{q\:,\:p} - \frac{2}{\pi}\,\delta^{(N)}_{q\:,\:p}=
\notag\\
&=\delta^{(N)}_{q\:,\:p}+\frac{1}{\pi} 
\pt{\delta^{(N)}_{q\:,\:p+1}
+\delta^{(N)}_{q+1\:,\:p}}\label{app20}\ \cdot
\end{align}
Then
we can compute
$I_{\bs{\ell},\bs{m}}=\Gamma_{\ell_1,m_1}\times\Gamma_{\ell_2,m_2}$
that is different from $\delta^{(N)}_{\bs{\ell},\bs{m}}$, as expected
from equation~\eqref{app3}. Thus we conclude that the set $ \{\vert
\beta_N(\bs{x})\rangle 
\mid \bs{x}\in\IT\}$ does not satisfy the \co{overcompleteness}
property.
\newpage
\vspace{1mm}
\newpage
\pagestyle{fancyplain}
\chapter{Proofs of Lemmas~\ref{Lemma2} and~\ref{Lemma1}}
\label{app_B}\vspace{9mm} 
{\bf Proof of lemma~\ref{Lemma2}:}\\[2.5ex]
1) ~\eqref{lemma1_3} follow from
\begin{equation}
\norm{S_\alpha\cdot\bs{v}}{{\IR}^2}^2=
\bkkk{\bs{v}}{S_\alpha^\dagger S_\alpha^{\phantom{\dagger}}}{\bs{v}}
\label{lemma1_2} 
\end{equation}
Indeed the matrix $S_\alpha^\dagger S_\alpha^{\phantom{\dagger}}$ is
real, symmetric, positive, with determinant equal to one;
thus it has two orthogonal eigenvectors,
corresponding to two different positive eigenvalues, $\eta^2$ and
$\eta^{-2}$, depending only on $\alpha$, \mbox{with $\eta^2>1\ \forall\:
\alpha\in\IR$.}

\noindent The same argument can be used for the matrix
$S_\alpha^{-1}$: the matrix
${\pt{S_\alpha^{\phantom{\dagger}} S_\alpha^\dagger}}^{-1}$ has
the same eigenvalues $\eta^2$ and
$\eta^{-2}$.\\[2.5ex]
2) In order to prove~\eqref{t2r2tot}, it is
convenient to unfold $\IT$ and the discontinuity of $S_\alpha$ on the
plane $\IR^2$. This is most easily done as follows.
Points $A\in\IT={\IR}^2/{\IZ}^2$ are
equivalence
classes $\pq{\bs{a}}$ of points in $\IR^2$ such that
\begin{equation}
\pq{\bs{a}}\coleq\pg{\bs{a}+\bs{n}\ ,\ \bs{n}\in{\IZ}^2}\ ,\
\bs{a}\in{[0,1)}^2 \ \cdot\label{eqclass}
\end{equation}
\noindent Given $A,B\in\IT$, let $A^{b}\in\pq{\bs{a}}$ be 
the closest vector to $\bs{b}$ in the Euclidean norm
$\norm{\cdot}{\IR^2}$, namely that vector such that
\begin{equation}
d_{\IT}\pt{\pq{\bs{a}},\pq{\bs{b}}} = \norm{A^b-\bs{b}}{{\IR}^2}\ \cdot
\label{ldt2_1}
\end{equation}
Notice that
\begin{equation}d_{\IT}\pt{\pq{\bs{a}},\pq{\bs{b}}}=\norm{\bs{a}-\bs{b}}{{\IR}^2} \quad
\text{iff} \quad \norm{\bs{a}-\bs{b}}{{\IR}^2}\leq\frac{1}{2}
\label{normITIR}
\end{equation}
$2$a) $\pt{A,B}$ not crossing
$\gamma_{-1}$ means that the segment 
$\pt{A^b_{\phantom{2}},\bs{b}}$ does not intersect $\gamma_{-1}$.
Periodically covering the plane--${\IR}^2$ by squares ${[0,1)}^2$, 
the $\gamma_{-1}$-lines form a set of (parallel) straight lines $x_1 - x_2 =
n\in\IZ$; it follows that 
$\pt{A^b_{\phantom{2}},\bs{b}}$ does not cross
$\gamma_{-1}$ iff
\begin{equation}
\floor{A^b_1 - A^b_2} = \floor{b_1 - b_2}\ ,
\label{ldt2_2}
\end{equation}
where the integral part on the r.h.s. takes values $0,-1$,  depending
on which side of the diagonal $\gamma_{-1}$ the point $\bs{b}$ lies
within. Indeed, one can check that if any two points
$\bs{x},\bs{y}\in\IT$ lie on opposite sides with respect to
$\gamma_{-1}$ then they must violate the above
condition~\eqref{ldt2_2} on the integer part of the differences of
their components.\\
As $S_\alpha^{\pm}$ are not sensitive to the integer part of their
arguments, their actions are the same on all elements of the
equivalence classes~\eqref{eqclass}. Therefore,
\begin{align}
d_{\IT} & \pt{S_\alpha^{-1}\pt{A},S_\alpha^{-1}\pt{B}}  = 
d_{\IT} \pt{S_\alpha^{-1}\pt{[\bs{a}]},S_\alpha^{-1}\pt{[\bs{b}]}}  = 
d_{\IT}\pt{S_\alpha^{-1}\pt{A^b_{\phantom{2}}},S_\alpha^{-1}\pt{\bs{b}}}=
\notag
\intertext{(by~(\ref{AoDC_1d}--\ref{AoDC_1e}) and~\eqref{Gnbar_m2})}
& =\min_{\bs{m}\in{\IZ}^2}\norm{
\begin{pmatrix}
\bk{A^b_1 - A^b_2}\\
\bk{-\alpha\bk{A^b_1 - A^b_2} + A^b_2}
\end{pmatrix}
-
\begin{pmatrix}
\bk{b_1 - b_2}\\
\bk{-\alpha\bk{b_1 - b_2} + b_2}
\end{pmatrix}
+\bs{m}
}{{\IR}^2}
\notag
\intertext{(by using $\bk{x} = x - \floor{x}$)}
\begin{split}
& = \min_{\bs{m}\in{\IZ}^2}\bigg\|
\begin{pmatrix}
\pt{A^b_1 - A^b_2} - \pt{b_1 - b_2}\\
-\alpha\bk{A^b_1 - A^b_2} + A^b_2 + \alpha\bk{b_1 - b_2} - b_2
\end{pmatrix}
+\\
& \mspace{76.0mu}+ \begin{pmatrix}
\floor{b_1 - b_2} - \floor{A^b_1 - A^b_2} + m_1\\
\floor{-\alpha\bk{b_1 - b_2} + b_2} - \floor{-\alpha\bk{A^b_1 - A^b_2}
+ A^b_2} + m_2  
\end{pmatrix}
{\bigg\|}_{{\IR}^2}
\end{split}
\notag\\
\begin{split}
& =\min_{\bs{m^{\prime}}\in{\IZ}^2}\bigg\|
\begin{pmatrix}
\pt{A^b_1 - b_1} - \pt{A^b_2 - b_2}\\
-\alpha\pt{A^b_1 - A^b_2}  + \alpha\pt{b_1 - b_2} + \pt{A^b_2 - b_2}
\end{pmatrix}
+\\
& \mspace{229.5mu}+ \alpha
\begin{pmatrix}
0 \\
\floor{A^b_1 - A^b_2} - \floor{b_1 - b_2}
\end{pmatrix}
+\bs{m^{\prime}}
{\bigg\|}_{{\IR}^2}
\end{split}
\notag
\intertext{(because of~\eqref{ldt2_2})}
& =\min_{\bs{m^{\prime}}\in{\IZ}^2}\norm{
\begin{pmatrix}
1 & -1\\
-\alpha & 1+ \alpha
\end{pmatrix}
\begin{pmatrix}
A^b_1 - b_1\\
A^b_2 - b_2
\end{pmatrix}
+\bs{m^{\prime}}
}{{\IR}^2}
= d_{\IT}\pt{S_\alpha^{-1}\cdot\pt{A^b_{\phantom{2}} - \bs{b}},0}
\label{ldt2_8}
\end{align}

Applying point ($1$) of this lemma and using the hypothesis, we can
estimate  
\begin{align}
\norm{S_\alpha^{-1}\cdot\pt{A^b_{\phantom{2}} - \bs{b}}}{{\IR}^2}
& \leq \eta \norm{A^b_{\phantom{2}} - \bs{b}}{{\IR}^2}
= \eta\;d_{\IT}\pt{A,B} < \frac{1}{2}\ \cdot\label{ldt2_10}
\intertext{In particular, using~\eqref{normITIR},
the previous inequalities imply} 
d_{\IT}\pt{S_\alpha^{-1}\cdot\pt{A^b_{\phantom{2}} - \bs{b}},0}
& = \norm{S_\alpha^{-1}\cdot\pt{A^b_{\phantom{2}} - \bs{b}}}{{\IR}^2}\ 
\leq \eta\;d_{\IT}\pt{A,B} \ 
\cdot
\label{ldt2_11}
\end{align}
\noindent $2$b) Analogously, the union of
$\gamma_{0}$-lines constitute a set of 
straight lines $x_1= n\in\IZ$; 
Therefore the segment $\pt{A^b_{\phantom{2}},\bs{b}}$ does not cross
$\gamma_{0}$ iff
\begin{equation}
\floor{A^b_1} = \floor{b_1}\ \cdot
\label{ldt2_2p}
\end{equation}
Explicitly
\begin{align}
d_{\IT} & \pt{S_\alpha\pt{A},S_\alpha\pt{B}}  = 
d_{\IT} \pt{S_\alpha\pt{[\bs{a}]},S_\alpha\pt{[\bs{b}]}}  =
d_{\IT} \pt{S_\alpha\pt{A^b_{\phantom{2}}},S_\alpha\pt{\bs{b}}}  
\notag
\intertext{(by~\eqref{AoDC_1b} and~\eqref{Gnbar_m2})}
& =\min_{\bs{m}\in{\IZ}^2}\norm{\bk{S_\alpha\cdot
\begin{pmatrix}
\bk{A^b_1}\\
A^b_2
\end{pmatrix}}
-\bk{S_\alpha\cdot
\begin{pmatrix}
\bk{b_1}\\
b_2
\end{pmatrix}}
+\bs{m}
}{{\IR}^2}
\notag
\intertext{(since $\bk{x} = x - \floor{x}$)}
& =\min_{\bs{m}\in{\IZ}^2}\norm{S_\alpha\cdot
\begin{pmatrix}
\bk{A^b_1}-\bk{b_1}\\
A^b_2-b_2
\end{pmatrix}
-\floor{S_\alpha\cdot
\begin{pmatrix}
\bk{A^b_1}\\
A^b_2
\end{pmatrix}}
+\floor{S_\alpha\cdot
\begin{pmatrix}
\bk{b_1}\\
b_2
\end{pmatrix}}
+\bs{m}
}{{\IR}^2}
\notag\\
& =\min_{\bs{m^{\prime}}\in{\IZ}^2}\norm{S_\alpha\cdot
\pt{A^b-\bs{b}}
-S_\alpha\cdot
\begin{pmatrix}
\floor{A^b_1} - \floor{b_1}\\
0
\end{pmatrix}
+\bs{m^{\prime}}
}{{\IR}^2}
\notag
\intertext{Condition~\eqref{ldt2_2p} makes the second vector vanish
and we obtain} 
d_{\IT} & \pt{S_\alpha\pt{A},S_\alpha\pt{B}} 
= d_{\IT}\pt{S_\alpha\cdot\pt{A^b_{\phantom{2}} - \bs{b}},0}
\label{ldt2_8p}
\end{align}
The proof thus is exactly completed as before.\\[2.5ex]
3) We denote by $\displaystyle d_{\IT}\pt{\bs{x},\gamma} = 
\inf_{\bs{y}\in\gamma}
d_{\IT}\pt{\bs{x},\bs{y}}$ the distance of the point $\bs{x}\in
\IT$ from a curve $\gamma\in \IT$. Then, from
Definition~\ref{Gnbar_0} we have: 
\begin{equation} 
\bs{x}\in\overline{\gamma}_{p-1}\pt{\varepsilon}
\Longrightarrow \varepsilon \geq
d_{\IT}\pt{\bs{x},\gamma_{p-1}} 
= d_{\IT}\pt{\bs{x},\bs{y}^{\star}}\ ,\label{distg_1}
\end{equation} 
where $\bs{y}^{\star}$ 
is the nearest point to $\bs{x}$ belonging to $\gamma_{p-1}$.\\ 
We distinguish two cases:
\begin{quote}
\begin{Ventry}{$3^{\prime\prime}$)}
	\item[$3^{\prime}$)] 
\underline{The segment
	$\pt{\bs{x},\bs{y}^{\star}}$ does not cross
$\gamma_{-1}$}\\[2ex]
	(even if $\bs{y}^{\star}\in\gamma_{-1}$ or
	\mbox{$\bs{x}\in\gamma_{-1}$}, we are in a non--crossing
	condition).\\
	From~\eqref{distg_1} and point (2a), since
        $S_\alpha^{-1}\pt{\bs{y}^{\star}}\in\gamma_p$
        (see~\eqref{gammapiu}), we get
\begin{align} 
d_{\IT}\pt{S_\alpha^{-1}\pt{\bs{x}},\gamma_p}
& \leq
d_{\IT}\pt{S_\alpha^{-1}\pt{\bs{x}},
S_\alpha^{-1}\pt{\bs{y}^{\star}}}\notag\\
&\leq \eta\:d_{\IT}\pt{\bs{x},\bs{y}^{\star}}\leq
\eta\:\varepsilon\ \cdot
\label{distg_2} 
\end{align}
Therefore 
$\displaystyle
S_\alpha^{-1}\pt{\bs{x}}\in\overline{\gamma}_p\pt{\eta\,\varepsilon}$ 
	\item[$3^{\prime\prime}$)] \underline{The segment
	$\pt{\bs{x},\bs{y}^{\star}}$ crosses $\gamma_{-1}$}.\\[2ex]
In this case, there exists $\bs{z}\in\gamma_{-1}\cap\pt{\bs{x},\bs{y}^{\star}}$ such that 
\begin{gather} 
d_{\IT}\pt{\bs{x},\bs{y}^{\star}} =
d_{\IT}\pt{\bs{x},\bs{z}} + d_{\IT}\pt{\bs{z},\bs{y}^{\star}}\
\cdot \label{distg_4}
\intertext{Then, from~\eqref{distg_1} and~\eqref{distg_4},}
\varepsilon \geq d_{\IT}\pt{\bs{x},\bs{y}^{\star}} \geq
d_{\IT}\pt{\bs{x},\bs{z}}\ \cdot
\label{distg_5} 
\end{gather}
Since, according to~\eqref{gammapiu},
$S_\alpha^{-1}\pt{\bs{z}}\in\gamma_0$, from point (2a), we get 
\begin{equation} 
d_{\IT}\pt{S_\alpha^{-1}\pt{\bs{x}},\gamma_0}\leq
d_{\IT}\pt{S_\alpha^{-1}\pt{\bs{x}},S_\alpha^{-1}\pt{\bs{z}}}
\leq \eta\:\varepsilon\ ,
\label{distg_6} 
\end{equation}
that is $\displaystyle
S_\alpha^{-1}\pt{\bs{x}}\in\overline{\gamma}_0\pt{\eta\,\varepsilon}$.
\end{Ventry}
\end{quote}
4) From point ($3$), it follows that, when
$0\leq\varepsilon\leq\frac{1}{2}$, for $p\in{\IN}^+$,
\begin{equation}
\bs{x}\not\in\pt{
\overline{\gamma}_p\pt{\varepsilon}\cup
\overline{\gamma}_0\pt{\varepsilon}}
\Longrightarrow
S_\alpha\pt{\bs{x}}\not\in\overline{\gamma}_{p-1}\pt{\eta^{-1}\varepsilon}\
\cdot \label{ddg0_1}
\end{equation}
We prove by induction that, when
$0\leq\varepsilon\leq\frac{1}{2}$, for $m\in{\IN}^+$,
\begin{equation}
\bs{x}\not\in\bigcup_{p=0}^m\;
\overline{\gamma}_p\pt{\varepsilon}
\Longrightarrow
S_\alpha\pt{\bs{x}}\not\in
\bigcup_{p=0}^{m-1}\;
\overline{\gamma}_p
\pt{\eta^{-1}\varepsilon}\ \cdot.
\label{ddg0_2}
\end{equation}
For $m=1$,~\eqref{ddg0_2} follows from~\eqref{ddg0_1}; 
suppose~\eqref{ddg0_2} holds for $m=r$, then take
\begin{equation}
\bs{x}\not\in\bigcup_{p=0}^{r+1}\;
\overline{\gamma}_p\pt{\varepsilon}
\text{\ . This means that}\quad
\bs{x}\not\in\bigcup_{p=0}^{r}\;
\overline{\gamma}_p\pt{\varepsilon}\quad\text{and}\quad
\bs{x}\not\in\pt{\overline{\gamma}_{r+1}\pt{\varepsilon}\cup
\overline{\gamma}_0\pt{\varepsilon}}\ \cdot\nonumber
\end{equation}
Now, using the hypothesis and~\eqref{ddg0_1}, we get
\begin{equation}
\bs{x}\not\in\bigcup_{p=0}^{r+1}\;
\overline{\gamma}_p\pt{\varepsilon}
\Longrightarrow
S_\alpha\pt{\bs{x}}\not\in
\bigcup_{p=0}^{r-1}\;
\overline{\gamma}_p\pt{\eta^{-1}\varepsilon}
\quad\text{and}\quad
S_\alpha\pt{\bs{x}}\not\in
\overline{\gamma}_r\pt{\eta^{-1}\varepsilon}\ \cdot
\label{ddg0_4}
\end{equation}
Then~\eqref{ddg0_2} is proved for all $m\in{\IN}^+$. Now observe the
following: applying~\eqref{ddg0_2} to $S_\alpha\pt{\bs{x}}$ instead of
$\bs{x}$ one gets
\begin{align}
S_\alpha\pt{\bs{x}}\not\in\bigcup_{p=0}^{m-1}\;
\overline{\gamma}_p\pt{\eta^{-1}\varepsilon}
& \Longrightarrow
S_\alpha^2\pt{\bs{x}}\not\in
\bigcup_{p=0}^{m-2}\;
\overline{\gamma}_p
\pt{\eta^{-2}\varepsilon}\ \cdot
\intertext{By iterating~\eqref{ddg0_2}, with $m=n-1$, we deduce}
\bs{x}\not\in\bigcup_{p=0}^{n-1}\;
\overline{\gamma}_p\pt{\varepsilon}
& \Longrightarrow
S_\alpha^q\pt{\bs{x}}\not\in
\bigcup_{p=0}^{n-1-q}\;
\overline{\gamma}_p
\pt{\eta^{-q}\varepsilon}\ ,\ \forall \ 0\leq q < n\ \cdot\label{ddg0_5}
\intertext{In particular $\displaystyle S_\alpha^q\pt{\bs{x}}
\not\in\overline{\gamma}_0\pt{\eta^{-q}\varepsilon}$, which leads to
the lower bound} 
& d_{\IT}\pt{S_\alpha^q\pt{\bs{x}},\gamma_0}
>\eta^{-q}\varepsilon\ ,\ \forall \ 0\leq q < n \ ,\label{ddg0_6}
\end{align}
whence the result follows in view of Definitions~\eqref{Ualpha}
and~\eqref{Gnbar_1}.\\[2.5ex]
5) The prove is by induction;
we fix $n$ and choose $N>\widetilde{N}+3 =2\sqrt{2}n\eta^{2n} +
3$.\\[2ex] 
$\bs{p=0})$ \quad from
Definitions~\eqref{Ualpha} and~\eqref{Valpha}, it follows
\begin{equation}
d_{\IT}\pt{
\frac{U_\alpha^0\pt{N\bs{x}}}{N}\,,\,
\frac{V_\alpha^0\pt{\hat{\bs{x}}_N}}{N}}
 =d_{\IT}\pt{\bs{x}\,,\,\frac{\hat{\bs{x}}_N}{N}}<
\frac{1}{\sqrt{2}N}<
\frac{\sqrt{2}}{N}\label{prova5_20}\ ,
\end{equation}
Where the first inequality follows from~\eqref{nuovopt_2}
in~\eqref{Lemma1}, thus relation~\eqref{lemma1_1}
holds for $p=0$.\\[2ex]
$\bs{p=1})$ 
\begin{align}
\norm{\frac{U_\alpha\pt{N\bs{x}}}{N}-
\frac{V_\alpha\pt{\hat{\bs{x}}_N}}{N}}{{\IR}^2}
& \leq\norm{\frac{U_\alpha\pt{N\bs{x}}}{N}-
\frac{U_\alpha\pt{\hat{\bs{x}}_N}}{N}}{{\IR}^2} +
\norm{\frac{U_\alpha\pt{\hat{\bs{x}}_N}}{N}- 
\frac{V_\alpha\pt{\hat{\bs{x}}_N}}{N}}{{\IR}^2} 
\label{prova5_21}\ \cdot
\end{align}
Now, by definitions~\eqref{Ualpha} and~\eqref{Valpha}, we have for the second
term of~\eqref{prova5_21}
\begin{align}
\norm{\frac{U_\alpha\pt{\hat{\bs{x}}_N}}{N}- 
\frac{V_\alpha\pt{\hat{\bs{x}}_N}}{N}}{{\IR}^2} & =
\norm{\frac{1}{N}\bk{N\;S_\alpha\pt{
\frac{\hat{\bs{x}}_N}{N}
}}}{{\IR}^2} <
\frac{\sqrt{2}}{N}\label{prova5_2}\ ,
\end{align}
whereas for the first term in~\eqref{prova5_21}, \eqref{prova5_20}
together with the non--crossing condition with respect to $\gamma_0$,
 which is apparently fulfilled by
$\pt{\bs{x}\,,\,\frac{\hat{\bs{x}}_N}{N}}$, allow us to use
point (2b) of this Lemma and to get
\begin{align}
\norm{\frac{U_\alpha\pt{N\bs{x}}}{N}-
\frac{U_\alpha\pt{\hat{\bs{x}}_N}}{N}}{{\IR}^2} =
\norm{S_\alpha\pt{\bs{x}}-S_\alpha\pt{\frac{\hat{\bs{x}}_N}{N}}}{{\IR}^2}
\leq \eta\frac{\sqrt{2}}{N}
\label{prova5_22}\ \cdot
\end{align}
Thus, inserting~\eqref{prova5_2} and~\eqref{prova5_22}
in~\eqref{prova5_21}, we obtain
\begin{align}
\norm{\frac{U_\alpha\pt{N\bs{x}}}{N}-
\frac{V_\alpha\pt{\hat{\bs{x}}_N}}{N}}{{\IR}^2}
& \leq \pt{\eta+1}\frac{\sqrt{2}}{N}\label{prova5_24}\\
(\text{\footnotesize for $N>\tilde{N}+3$})\qquad& < \frac{\eta +
1}{2\:n\:\eta^{2n}} < 
\frac{1}{2}\ \cdot\label{prova5_3} 
\end{align}
But then the Euclidean norm equals the distance $\displaystyle
d_{\IT}\pt{\frac{U_\alpha\pt{N\bs{x}}}{N}\,,\,
\frac{V_\alpha\pt{\hat{\bs{x}}_N}}{N}}$; thus relation~\eqref{lemma1_1}
holds for $p=1$.\\[2ex]
$\bs{p=q-1\,,\ q\leq n})$ \quad Since
\begin{multline}
d_{\IT}\pt{\frac{U_\alpha^q\pt{N\bs{x}}}{N},
\frac{V_\alpha^q\pt{\hat{\bs{x}}_N}}{N}}\leq
d_{\IT}\pt{\frac{U_\alpha\pt{U_\alpha^{q-1}\pt{N\bs{x}}}}{N},
\frac{U_\alpha\pt{V_\alpha^{q-1}\pt{\hat{\bs{x}}_N}}}{N}}+\\
+d_{\IT}\pt{\frac{U_\alpha\pt{V_\alpha^{q-1}\pt{\hat{\bs{x}}_N}}}{N},
\frac{V_\alpha\pt{V_\alpha^{q-1}\pt{\hat{\bs{x}}_N}}}{N}}\ ,
\label{prova5_5}
\end{multline}
using~\eqref{Ualpha} in the first term and noting that, from
definitions~\eqref{Ualpha} and~\eqref{Valpha}, the second term is
less or equal to $\frac{\sqrt{2}}{N}$, we get
\begin{equation}
d_{\IT}\pt{\frac{U_\alpha^q\pt{N\bs{x}}}{N},
\frac{V_\alpha^q\pt{\hat{\bs{x}}_N}}{N}}\leq
d_{\IT}\pt{
S_\alpha\pt{\frac{U_\alpha^{q-1}\pt{N\bs{x}}}{N}},
S_\alpha\pt{\frac{V_\alpha^{q-1}\pt{\hat{\bs{x}}_N}}{N}}
}+
\frac{\sqrt{2}}{N}\ \cdot
\label{prova5_6}
\end{equation}
By the induction hypothesis we have:
\begin{align}
d_{\IT}\pt{
\frac{U_\alpha^{q-1}\pt{N\bs{x}}}{N},
\frac{V_\alpha^{q-1}\pt{\hat{\bs{x}}_N}}{N}}
& \leq\frac{\sqrt{2}}{N}\pt{\frac{\eta^{q}-1}{\eta-1}}
\label{prova5_7}\\
& \leq\frac{\sqrt{2}}{N}\;q\;\eta^{q-1}
\label{prova5_8}\\
(\eta>1,\ q\leq n\quad\Longrightarrow)\quad\quad
&<\eta^{q-2n}\;\frac{q}{n}\;\pt{\frac{1}{2}
\eta^{-1}} <\frac{1}{2}\eta^{-1}\ \cdot
\label{prova5_9}
\intertext{Moreover, from point ($4$) above with
$\varepsilon=\frac{\widetilde{N}}{2N}$, $0\leq q\leq n$ and $\eta>1$
we derive}  
d_{\IT}\pt{\frac{U_\alpha^{q-1}\pt{N\bs{x}}}{N},\gamma_0} & >
\frac{n\sqrt{2}\;\eta^{2n-q+1}}{N}  >
\frac{\sqrt{2}}{N}\;q\;\eta^{q-1} \ \cdot\label{prova5_10} 
\intertext{
Therefore, comparing~\eqref{prova5_10} with~\eqref{prova5_8}}
d_{\IT}\pt{
\frac{U_\alpha^{q-1}\pt{N\bs{x}}}{N},
\frac{V_\alpha^{q-1}\pt{\hat{\bs{x}}_N}}{N}}
& < d_{\IT}\pt{\frac{U_\alpha^{q-1}\pt{N\bs{x}}}{N},\gamma_0} \quad,\quad\forall q\leq
n\ \cdot
\label{prova5_11} 
\end{align}
Therefore, we deduce that the segment $\pt{
\frac{U_\alpha^{q-1}\pt{N\bs{x}}}{N},
\frac{V_\alpha^{q-1}\pt{\hat{\bs{x}}_N}}{N}}$ cannot cross the line
$\gamma_0$.
This condition, together with~\eqref{prova5_9}, allows
us to use point (2b) in~\eqref{prova5_6}
to finally estimate
\begin{equation}
d_{\IT}\pt{\frac{U_\alpha^q\pt{N\bs{x}}}{N},
\frac{V_\alpha^q\pt{\hat{\bs{x}}_N}}{N}}\leq
\eta\frac{\sqrt{2}}{N}\pt{\frac{\eta^{q}-1}{\eta-1}}
+
\frac{\sqrt{2}}{N} = \frac{\sqrt{2}}{N}\pt{\frac{\eta^{q+1}-1}{\eta-1}}
\label{prova5_12}
\end{equation}
and this concludes the proof.\hfill$\qed$\\[2.5ex]
{\bf Proof of lemma~\ref{Lemma1}:}\\[2.5ex]
a) In~\eqref{gammapiu}, we have defined $\gamma_p =
S_\alpha^{-p}\pt{\gamma_0}$ where  
$S_\alpha^{-1}\pt{\bs{x}}$ (and then also $S_\alpha^{-p}\pt{\bs{x}}$)
is a piecewise continuous mapping onto $\IT$ with jump--discontinuities 
across the $\gamma_p$ lines
due to the presence of
the function $\bk{\cdot}$ in~\eqref{AoDC_1d}. Away from the 
discontinuities, $S_\alpha^{-p}\pt{\bs{x}}$ behaves as the matrix action 
$S_\alpha^{-p}\cdot\bs{x}$, that is nothing but the action of the
Sawtooth Map in the tangent space. By integrating the evolution given
by $S_\alpha^{-p}\cdot\bs{x}$ on the tangent space along $\gamma_0$,
it follows that $l\pt{\gamma_p}$ is 
the length  of the
${\IR}^2$--segment
$\pg{\bs{x}\in{\IR}^2\ \Big|\ \bs{x}=S_\alpha^{-p}\cdot\pt{
\begin{smallmatrix}
0\\y
\end{smallmatrix}}
\ ,\ y\in [0,1)}$, which, in its turn, is 
the image of the (length one) segment
$\gamma_0$ under the matrix action given by
$S_\alpha^{-p}\cdot\bs{x}$.\\
Finally, using~\eqref{lemma1_3a} we get the result.\\[2.5ex]
b) Let $\overline{L}\pt{\varepsilon}$ denote the set of points having
distance from 
a segment of length $L$ smaller
or equal than $\varepsilon$: it has a volume (given by the Lebesgue
measure $\mu$) given by
\begin{equation*}
\mu\pt{\overline{L}\pt{\varepsilon}} = 2\,L\,\varepsilon +
\pi\varepsilon^2 \ , 
\end{equation*}
where the last term on the r.h.s. takes into account a small set of
points 
close to the extremity of the segment. Then~\eqref{lemma_1_2} 
follows from
~\eqref{lemma_1_1}.\\[2.5ex]
c) This follows from Definition~\eqref{Gnbar_1}:
\begin{alignat}{2}
\mu\pt{\overline{\Gamma}_n\pt{\varepsilon}} & = 
\mu\pt{\bigcup_{p=0}^{n-1}\;
\overline{\gamma}_p\pt{\varepsilon}}
&&\leq \sum_{p=0}^{n-1}\;\mu\pt{
\overline{\gamma}_p\pt{\varepsilon}}\ \cdot\nonumber
\intertext{Using~\eqref{lemma_1_2}, we can write:}
\mu\pt{\overline{\Gamma}_n\pt{\varepsilon}} 
&\leq 2\,\varepsilon\sum_{p=0}^{n-1}\eta^p + 
\sum_{p=0}^{n-1}\pi\,\varepsilon^2 
 &&= 2\,\varepsilon \,\frac{\eta^n-1}{\eta^{\phantom{p}}-1}
+ n\,\pi\,\varepsilon^2\ \cdot\nonumber
\intertext{Finally the estimate
$\displaystyle \frac{x^p-1}{x^{\phantom{p}}-1}\leq p\,x^p$, valid for
$x>1$, yields}
\mu\pt{\overline{\Gamma}_n\pt{\varepsilon}} 
&\leq 2\,\varepsilon \, n \, \eta^n +
n\,\pi\,\varepsilon^2
&&= \varepsilon\,n\pt{2\,\eta^n + \pi\,\varepsilon}\ \cdot\nonumber
\end{alignat}\\[2.5ex]
d) For every real number $t$, $0\leq\bk{N t
+\frac{1}{2}}=N t
+\frac{1}{2}-\floor{N t
+\frac{1}{2}}<1$; this leads to 
\begin{alignat}{3}
\abs{t-\frac{\floor{Nt+\frac{1}{2}}}{N}}&\leq\frac{1}{2N}
\label{nuovopt_1} \ &\ \quad&,\quad\ &\ 
\forall\;t&\in\IR  \cdot
\intertext{Using~\eqref{loc_c32}, Definition~\ref{Gnbar}, we write}
 d_{\IT}\pt{\bs{x}\;,\;\frac{\hat{\bs{x}}_N}{N}}&\leq\frac{1}{\sqrt{2}N}
\label{nuovopt_2}\ &\ \quad&,\quad\ &\ 
\forall\;\bs{x}&\in\IT \cdot 
\end{alignat}
If in the triangular inequality
\begin{align}
d_{\IT}\pt{\bs{x}\;,\;\bs{y}}&\leq
d_{\IT}\pt{\bs{x}\;,\;\frac{\hat{\bs{x}}_N}{N}} +
d_{\IT}\pt{\frac{\hat{\bs{x}}_N}{N}\;,\;\bs{y}}\qquad\forall\;\bs{y}\in\IT \qquad \label{nuovopt_3}\ ,
\intertext{we take the $\inf$ over $\bs{y}\in\Gamma_n$
of~\eqref{Gamman}, we get}
d_{\IT}\pt{\frac{\hat{\bs{x}}_N}{N}\;,\;\Gamma_n}
&\geq  
d_{\IT}\pt{\bs{x}\;,\;\Gamma_n} -
d_{\IT}\pt{\bs{x}\;,\;\frac{\hat{\bs{x}}_N}{N}} 
\label{nuovopt_4}\\
&\geq  
d_{\IT}\pt{\bs{x}\;,\;\Gamma_n}
\frac{1}{\sqrt{2}N}
\label{nuovopt_5}\ ,
\intertext{that is}
\bs{x}\in\pq{\overline{\Gamma}_n\pt{\varepsilon}}^{\circ}\quad&\Longrightarrow\quad
\frac{\hat{\bs{x}}_N}{N}
\in\pq{\overline{\Gamma}_n\pt{\varepsilon-\frac{1}{\sqrt{2}N}}}^{\circ} \label{nuovopt_6}
\end{align}
Therefore, from~\eqref{Gnbar_2}, if $\frac{\hat{\bs{x}}_N}{N}$
does not belong to
$\overline{\Gamma}_n\pt{\varepsilon-\frac{1}{\sqrt{2}N}}$. then the
corresponding $\bs{x}$ in~\eqref{nuovopt_6} belongs to ${G_n^N}\pt{\varepsilon-\frac{1}{\sqrt{2}N}}$.
Changing $\varepsilon-\frac{1}{\sqrt{2}N}\longmapsto\varepsilon$ we
obtain~\eqref{lemma_1_5}.\\[2.5ex]
e) Writing~\eqref{lemma_1_5} in terms of complementary sets, and
substituting $\varepsilon=\frac{\widetilde{N}}{2N}$, we get:
\begin{align}
\pq{{G_n^N}\pt{\frac{\widetilde{N}}{2N}}}^{\circ}&\subseteq 
\overline{\Gamma}_n\pt{\frac{\widetilde{N}}{2N}+ 
\frac{1}{\sqrt{2}N}}\ \text{and so}\label{nuovopt_7}\\
\mu\pt{\pq{{G_n^N}\pt{\frac{\widetilde{N}}{2N}}}^{\circ}}&\leq
\mu\pt{\overline{\Gamma}_n\pt{\frac{\widetilde{N}}{2N}+ 
\frac{1}{\sqrt{2}N}}}\label{nuovopt_8}\ \cdot
\intertext{Now we substitute $\frac{\widetilde{N}+\sqrt{2}}{2N}=\frac{\widetilde{N}}{2N}+ 
\frac{1}{\sqrt{2}N}$ in place of $\varepsilon$ in~\eqref{lemma_1_3} and
we get:}
\mu\pt{\pq{{G_n^N}\pt{\frac{\widetilde{N}}{2N}}}^{\circ}}&\leq
\frac{\widetilde{N}+\sqrt{2}}{2N}
\,n\pt{2\,\eta^n + \pi\,
\frac{\widetilde{N}+\sqrt{2}}{2N}
}\ \cdot\label{nuovopt_9}
\intertext{Finally we use:}
\text{inside brackets of r.h.s of ~\eqref{nuovopt_9}}\quad&:\quad\pi\,
\frac{\widetilde{N}+\sqrt{2}}{2N}<4<4\eta^n\quad,
\quad\forall\;N>\widetilde{N}\label{nuovopt_10}\,\\ 
\text{outside brackets of r.h.s of ~\eqref{nuovopt_9}}\quad&:\quad
\widetilde{N}+\sqrt{2}<2\widetilde{N}
\quad,
\quad\forall\;N>\widetilde{N}\label{nuovopt_10}
\end{align}
and this ends the proof.\hfill$\qed$
\chapter*{Outlook \& Perspective}
\lhead[\fancyplain{}{\bfseries\thepage}]%
      {\fancyplain{}{}}
\rhead[\fancyplain{}{\bfseries Outlook \& Perspective}]%
      {\fancyplain{}{\bfseries\thepage}}
\addcontentsline{toc}{chapter}{Outlook \& Perspective}

This project has been performed with the aim to inquire the footprint
of chaos present in classical dynamical systems even when some quantization
procedure maps these systems into quantum (or discrete) ones, with a 
finite number of states. The framework in which we moved is the
semi--classical analysis; we developed techniques of
quantization and discretization by using the well known Weyl or
Anti--Wick schemes of quantization, in particular we made use of
family of suitably defined Coherent States.

We used the entropy production as a
parameter of chaotic behaviour: in particular two notions of quantum dynamical
entropy have been used, namely the \co{CNT} and \co{ALF}
entropies, both reproducing the Kolmogorov entropy if applied to
classical systems.
\subsection*{Quantum Dynamical Systems}\vspace{3mm}
We have shown that both the \co{CNT} and \co{ALF} entropies
reproduce the Kolmogorov metric entropy in quantum systems too,
provided that we observe a strongly
chaotic system on a very short logarithmic time scale. However, due to the
discreteness of the spectrum of the quantizations, we know that
saturation phenomena will appear. It would be interesting to study
the scaling behaviour of the quantum dynamical entropies in the
intermediate region between the logarithmic breaking time and the
Heisenberg time. This will, however, require quite different
techniques than the coherent states approach.  
\subsection*{Discretized Dynamical Systems}\vspace{3mm}
We have considered discretized hyperbolic classical
systems on the torus by forcing them on a squared lattice with spacing
$\frac{1}{N}$. We showed how the discretization procedure is similar to
quantization; in particular, following the analogous case of the
classical limit $\hbar\longmapsto 0$, we have set up the theoretical
framework 
to discuss the continuous limit $N\longmapsto\infty$. Furthermore,
using the similarities between discretized and quantized classical
systems, we have applied the \co{ALF} entropy
to study the 
footprints of classical (continuous) chaos as it
is expected to reveal itself, namely through the presence of
characteristic time scales and corresponding breaking--times. Indeed,
exactly as in quantum chaos, a discretized hyperbolic system can
mimic its continuous partner only up to times which scale as $\log
N$, where $N^2$ is the number of allowed classical phase--point.\newpage
\singlespacing   		
\lhead[\fancyplain{}{\bfseries\boldmath\thepage}]%
      {\fancyplain{}{\bfseries\boldmath\rightmark}}
\rhead[\fancyplain{}{\bfseries\boldmath\leftmark}]%
      {\fancyplain{}{\bfseries\boldmath\thepage}}

\newpage \
\thispagestyle{empty} \
\newpage  
\thispagestyle{empty}
\begin{flushleft}
\rule{0pt}{1ex}\\
\vspace{2cm}
\rule{0pt}{1ex}\\
\small
Non senza fatica si giunge al fine.\\[0.7ex]
\textsc{GIROLAMO FRESCOBALDI},
Toccata IX,
\textit{Secondo Libro di Toccate}
(1627)
\end{flushleft}
\end{document}